\documentclass[11pt,a4paper]{article}
\usepackage{heppub}
\pdfoutput=1
\usepackage{braket,slashed,changepage,placeins,multirow,colortbl,makecell,subcaption,mathtools,array,xspace,enumitem, float}
\usepackage[svgnames]{xcolor}
\usepackage[T1]{fontenc}
\usepackage[utf8]{inputenc}
\usepackage{marginnote}
\usepackage[normalem]{ulem}

\newcommand{\hm}{\kappa}
\newcommand{\BoostInv}{\varsigma}
\newcommand{\FiDpre}{\frac{Q^2\BoostInv^2}{x}}
\newcommand{\img}{\mathrm{i}}
\newcommand{\chibar}{\bar{\chi}}

\newcommand{\cB}{\ensuremath{\mathcal{B}}}
\newcommand{\cC}{\ensuremath{\mathcal{C}}}

\newcommand{\cL}{\ensuremath{\mathcal{L}}}
\newcommand{\LQCD}{\ensuremath{\Lambda_{\rm QCD}}}
\newcommand{\cM}{\ensuremath{\mathcal{M}}}
\newcommand{\nbar}{\bar{n}} 

\newcommand{\cO}{\ensuremath{\mathcal{O}}} 

\newcommand{\psibar}{\bar{\psi}}
\newcommand{\bP}{\ensuremath{\mathbb{P}}}
\newcommand{\cP}{\ensuremath{\mathcal{P}}}

\newcommand{\cQ}{\ensuremath{\mathcal{Q }}}
\newcommand{\qbar}{\bar{q}}
\newcommand{\bR}{\ensuremath{\mathbb{R}}}

\newcommand{\cS}{\ensuremath{\mathcal{S}}}

\newcommand{\Tr}{\text{Tr}\,}

\newcommand{\Ubar}{\bar{U}}

\newcommand{\cW}{\ensuremath{\mathcal{W}}}
\newcommand{\xbar}{\bar{x}}
\newcommand{\Xbar}{\bar{X}}
\newcommand{\xibar}{\bar{\xi}}

\newcommand{\wdl}{\ensuremath{(W\!\!\cdot\!\! L)}}

\newcommand{\etaxybar}{{\bar\eta_{0}}}
\newcommand{\etaxy}{\eta_{0}}
\newcommand{\etaxylab}{\eta_{0}^{\rm lab}}
\newcommand{\etalabcut}{\eta_{\rm cut}^{\rm lab}}
\newcommand{\h}{h}

\newcommand{\gY}{{U}}
\newcommand{\gYa}{{\cal U}}
\newcommand{\ddslash}{{d\!\!{}^-}}
\newcommand{\IInt}{\int\!\!\!\!\!\int}
\newcommand{\deltaslash}{{\delta\!\!\!{}^-}}
\newcommand{\collapsel}{\Big\{\!\!\!\Big\{}
\newcommand{\collapser}{\Big\}\!\!\!\Big\}}
\newcommand{\scollapsel}{\{\!\!\{}
\newcommand{\scollapser}{\}\!\!\}}
\newcommand{\bcollapsel}{\bigg\{\!\!\!\bigg\{}
\newcommand{\bcollapser}{\bigg\}\!\!\!\bigg\}}

\newcommand{\SCETa}{\ensuremath{{\rm SCET}_{\rm I}}\xspace}

\def\nn{{\nonumber}}

\DeclareMathOperator*{\sumint}{%
\mathchoice%
{\ooalign{$\displaystyle\sum$\cr\hidewidth$\displaystyle\int$\hidewidth\cr}}
{\ooalign{\raisebox{.14\height}{\scalebox{.7}{$\textstyle\sum$}}\cr\hidewidth$\textstyle\int$\hidewidth\cr}}
{\ooalign{\raisebox{.2\height}{\scalebox{.6}{$\scriptstyle\sum$}}\cr$\scriptstyle\int$\cr}}
{\ooalign{\raisebox{.2\height}{\scalebox{.6}{$\scriptstyle\sum$}}\cr$\scriptstyle\int$\cr}}
}

\usepackage{tocloft}
\preprint{MIT-CTP 5746, LA-UR-25-21807}

\title{Effective Field Theory Factorization for Diffraction}

\author[a]{Kyle Lee,}
\author[a,b]{Stella T. Schindler,}
\author[a,c]{Iain W. Stewart}
\affiliation[a]{Center for Theoretical Physics -- a Leinweber Institute, Massachusetts Institute of Technology, Cambridge, MA 02139, USA}
\affiliation[b]{Theoretical Division, Los Alamos National Laboratory, Los Alamos, NM 87545, USA}
\affiliation[c]{University of Vienna, Boltzmanngasse 9, A-1090 Wien, Austria}

\emailAdd{kylel@mit.edu}
\emailAdd{schindler@lanl.gov}
\emailAdd{iains@mit.edu}

\abstract{
We derive a factorization formula for coherent and incoherent $ep$ diffraction using the soft collinear effective theory, utilizing multiple power expansion parameters to handle different kinematic regions.  This goes beyond the known hard-collinear diffractive factorization to address the small-$x$ Regge dynamics and Pomeron exchange from first principles. 
The effective field theory analysis also uncovers and factorizes an important irreducible incoherent background generated by color-nonsinglet exchange, 
dubbed
``quasi-diffraction'', for which we
calculate the associated Sudakov suppression.
For unpolarized scattering we show that there are four diffractive structure functions at leading power, and point out the importance of studying $F_{3,4}^D$ through asymmetries, in addition to $F_{2,L}^D$. 
For the quasi-diffractive background, we make model independent predictions for ratios of the corresponding structure functions in a perturbative kinematic region.
Our analysis also makes predictions for
six leading-power spin-dependent structure functions. 
Finally, we provide connections to diffractive parton distributions, and assess the Ingelman-Schlein model.
Our work lays a path for further QCD-based studies of diffraction.

}

\begin{document}
\maketitle
\allowdisplaybreaks

\section{Introduction}\label{sec:intro}
 
An important but not fully understood phenomenon at colliders is diffractive scattering, defined as a forward scattering process involving one or more hadron beams that produces final states exhibiting a large rapidity gap, an angular wedge devoid of particles.  Diffractive processes comprise a substantial portion of events at colliders \cite{UA4:1986igl, UA5:1986xrd, E710:1992agk, CDF:1993slg}, including approximately 10\% of the $ep$ cross section at HERA~\cite{ZEUS:1993vio, H1:1994ahk, H1:1997vke, ZEUS:1997ffu,ZEUS:2001tvy, H1:2006zyl, ZEUS:2008xhs, Aaron:2010aa}, up to 40\% of the $pp$ cross section at the LHC \cite{ATLAS:2012djz,ATLAS:2019asg}, and may account for up to 20\% of the $ep$ cross section at the planned Electron-Ion Collider (EIC) \cite{AbdulKhalek:2021gbh}. Historically, diffraction was among the earliest proposed phenomena~\cite{Feinberg1956, Chew:1961ev} for probing the high-energy/Regge/forward-scattering limit. In particular, diffraction enables the study of exchanged objects  with vacuum quantum numbers, such as the Pomeron, as well as soft (low energy) QCD processes. 
Diffraction also offers opportunities to investigate small-$x$ nuclear structure, including gluon saturation~\cite{Hentschinski:2022xnd, Gribov:1983ivg, Golec-Biernat:1998zce} and nuclear shadowing~\cite{Glauber:1955qq, Gribov:1968jf, Glauber:1970jm, Bauer:1977iq, Frankfurt:1988nt, Frankfurt:1998ym}, key elements for understanding initial heavy-ion states and dense gluonic dynamics.
Furthermore, diffraction provides a unique environment for new physics searches \cite{Khoze:2001xm, Albrow:2010yb, Baldenegro:2022kaa, Feng:2022inv, CMS:2023roj}, 
studying exotic Standard Model resonances~\cite{Ochs:2013gi, Fichet:2014uka, LHCb:2024smc}, and carrying out top and Higgs physics analyses~\cite{Tasevsky:2014cpa, CMS:2023naq}.
Experimentally, a thorough understanding of diffraction is crucial for tasks such as tracking luminosity at colliders, deciphering pile-up, and building accurate Monte Carlo generators~\cite{ATLAS:2012djz,Mieskolainen:2019jpv,Bopp:1998rc,Bierlich:2022pfr}. The impact of diffraction extends even beyond colliders, including to hadronic interactions in cosmic ray cascades~\cite{Engel:2011zzb,Grieder2010}.
Although existing models describe a range of diffraction data~\cite{Frankfurt:2022jns}, they are widely recognized as insufficient to capture the full scope of diffractive processes~\cite{Frankfurt:2022jns,CMS:2020dns,CDF:2001ijd,Kaidalov:2003xf,ATLAS:2012djz, ATLAS:2019asg}.
 
In this paper we apply top-down effective field theory (EFT) techniques based on the Soft Collinear Effective Theory (SCET)~\cite{Bauer:2000ew,Bauer:2000yr,Bauer:2001ct,Bauer:2001yt,Bauer:2002nz,Rothstein:2016bsq} to $ep$ diffractive processes. This enables us to derive a new Regge factorization formula for diffractive scattering, going beyond the known factorization for the hard interaction~\cite{Berera:1995fj,Collins:1997sr}. This general framework provides 
a universal field-theoretic method to describe a wide variety of diffractive 
processes.\\[-12pt]
 
There is a large body of literature on diffractive scattering; see for example the reviews~\cite{Amaldi:1976gr, Kaidalov:1979jz, Barone:2002cv, Arneodo:2005kd,Newman:2013ada, Frankfurt:2022jns,ParticleDataGroup:2022pth}.
Diffractive processes were first described in the 1950s \cite{Feinberg1956, Chew:1961ev}, long before the QCD Lagrangian was first written down in the early 1970s. Early studies described diffraction as being mediated by the exchange of Pomeron and Reggeon particles, which were later hypothesized to consist of gluons \cite{Nussinov:1975mw,Nussinov:1975qb, Low:1975sv}. The 1980s saw the construction of the widely-used Ingelman-Schlein model for diffraction \cite{Ingelman:1984ns},  which expresses the $ep$ diffractive cross section~\cite{H1:1997bdi,ZEUS:1995sar,Golec-Biernat:1995ryr,Jung:1993gf} as a product of the parton distribution function (PDF) of a struck parton inside a Pomeron (Reggeon) and another PDF-like distribution (flux factor) of the Pomeron (Reggeon) inside a proton.
Tools have also been separately developed to study small-$x$ dynamics, including 
Balitsky-Fadin-Kuraev-Lipatov (BFKL) evolution~\cite{Balitsky:1978ic, Kuraev:1977fs},
the Gribov-Levin-Ryskin/Mueller-Qiu (GLR-MQ) equation~\cite{Mueller:1985wy, Gribov:1983ivg},
the Color-Glass Condensate (CGC) formalism~\cite{McLerran:1993ni, McLerran:1993ka, McLerran:1994vd}, 
and the Balitsky-Kovchegov (BK)~\cite{Balitsky:1995ub,Kovchegov:1999yj}
and Balitsky-Jalilian Marian-Iancu-McLerran-Weigert-Leonidiv-Kovner (BJIMWLK)~\cite{Jalilian-Marian:1997qno,Jalilian-Marian:1997jhx,Iancu:2001ad} equations. 
Many of these advances have been combined with diffractive models to make phenomenological predictions~\cite{Kowalski:2008sa,AbdulKhalek:2021gbh, Frankfurt:2022jns,Golec-Biernat:1995ryr,Golec-Biernat:1999qor,Ingelman:1984ns}.  
The 1990s saw the successful factorization of the hard-collinear deep inelastic scattering (DIS) dynamics 
in an appropriate kinematic region of diffraction; here, diffractive $ep$ cross sections can be expressed as the convolution of a hard coefficient and a diffractive parton distribution function (dPDF), which obeys DGLAP evolution \cite{Berera:1994xh, Berera:1995fj,Collins:1997sr}. 
However,
the absence of a rigorous field-theoretic QCD factorization for the Regge limit characterizing the dynamics of the hadronic forward-scattering, or ``Regge factorization'', has been prominently noted in the literature~\cite{Berera:1994xh,Berger:1986iu, Berera:1995fj,Collins:1997sr,Arneodo:2005kd}. The derivation of the Regge factorization of diffraction is the main result of this paper.\\[-12pt]

\begin{figure}[t!] 
\begin{center}
	\includegraphics[width =5.6in]{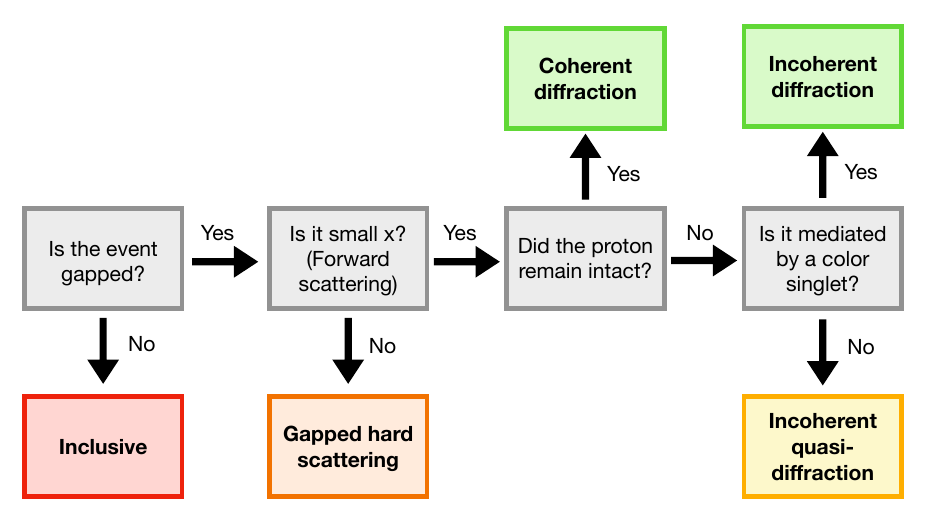}
\end{center}
\caption{Scheme for classifying rapidity-gapped processes in lepton-hadron colliders. 
}\label{fig:diffraction-chart}
\end{figure}

To begin, we give criteria for defining a diffractive event, 
as summarized in \fig{diffraction-chart}.
These criteria are chosen to distinguish the underlying diffractive process from backgrounds with similar or identical final-state signatures. The first criterion is straightforward~\cite{Frankfurt:2022jns}: a diffractive collision must involve at least one hadron or nucleus, such as in lepton-hadron or hadron-hadron scattering, and must exhibit a large rapidity gap between final-state hadronic particles. 
For instance, at the LHC, the typical lab-frame rapidity gap between nearby particles is approximately 0.15 \cite{ATLAS:2012djz}, so a diffractive event must display a significantly larger gap, typically $\gtrsim 3$.  
So far, these criteria are insufficient to separate diffraction from non-diffractive events with rapidity gaps. The key distinction is whether the initial hadron undergoes forward scattering. If it does, we refer to the event as diffractive. If it does not, we refer to the event as a gapped hard scattering event. For example, there exist non-forward hard DIS processes that produce beam and central jets with a rapidity gap \cite{Kang:2013nha,Dasgupta:2003iq, Kang:2012zr}; despite the gap, these are non-diffractive because they are caused by hard scattering rather than forward scattering.\footnote{Later, we will see that the type of hadronic scattering that occurs (hard or forward) fundamentally impacts the factorization structure of a process, which is why we prefer this terminology.}
We take small-$x$ and small-$t$ to enhance forward scattering in the $ep$ case, where $t$ is the invariant mass of the difference between the initial proton momentum and the final forward-scattered jet or proton.  In this region, the corresponding gapped hard scattering 
diagrams are small (power suppressed). 

Finally, diffraction is defined~\cite{Frankfurt:2022jns} as the process mediated by the exchange of a color-singlet state. Diffractive processes are further subcategorized by whether the initial-state hadron remains intact (coherent) or breaks apart (incoherent, also known as dissociation). Coherent scattering is always mediated by a color-singlet exchange between the incoming and outgoing proton.
However, incoherent forward scattering can be mediated by color-singlet states like Pomerons ($\cC$-even states) and Odderons ($\cC$-odd states), and color-nonsinglet states, like color-octet exchange.  
The fact that leading-power forward scattering amplitudes have contributions from various color channels has been well studied in the literature. (For a review of literature, see \refcite{Caron-Huot:2020grv,Gao:2024qsg} and references therein.\footnote{Note that the amplitude and diffractive communities sometimes mean something different by the word Reggeon. The amplitude community uses Reggeon to refer to the exchange of a color-octet state, whereas the diffraction community often uses Reggeon to denote a color singlet exchange that is not the Pomeron~\cite{Frankfurt:2022jns}.}) 
We coin the term {\it quasi-diffraction} to refer to incoherent color-nonsinglet contributions, which have an identical experimental signature to the desired color-singlet exchange diffractive events.  In principle, these quasi-diffractive contributions are suppressed by their propensity to produce additional final-state particles that can populate the rapidity gap. However, detectors cannot distinguish between events with very soft or no radiation in a gap, since there is always a detection threshold. The extent of this suppression thus depends on the precise experimental cuts used to identify diffractive events. Therefore, quasi-diffraction serves as a irreducible experimental background, whose size must be computed to determine the extent of this contamination. 

Often, the phrases ``diffraction'' and ``hard scattering'' are used in alternative ways to what we have described above. For example, Refs.~\cite{Berger:1986iu,Collins:1992cv} use the terminology ``diffractive hard scattering'' for events with forward hadronic interactions combined with a hard photon vertex, which leads to factorization involving diffractive structure functions.  This is a subset of what we refer to as diffraction.
Despite one of the vertices being hard, it is not a subset of the processes we refer to as gapped hard scattering, because the initial hadron primarily undergoes forward scattering interactions. 
A second example is the terminology ``diffractive hard exclusive scattering'' used in Refs.~\cite{Qiu:2022pla, Qiu:2024pmh}; this term is used to describe exclusive processes exhibiting a rapidity gap and at least one hard scattering vertex. However, 
no stringent kinematic requirement is enforced to   determine whether the initial hadron undergoes a hard or forward scattering interaction. Thus, in our terminology, these processes include contributions from both gapped hard scattering and diffraction. These processes bear resemblance to a well-studied class of gapped hard scattering processes involving jets, which are referred to as exclusive jet production or jet production with a jet veto~\cite{Berger:2010xi,Banfi:2012yh,Becher:2012qa,Tackmann:2012bt,Banfi:2012jm,Liu:2012sz,Jouttenus:2013hs,Stewart:2013faa,Shao:2013uba,Li:2014ria,Forshaw:2006fk,Becher:2023mtx}.  
The reasoning behind our terminology is that gapped hard scattering events have a different factorization structure than diffractive events; specifically, gapped hard scattering is insensitive to Regge- and Pomeron-type physics at leading power and low orders in perturbation theory.\\[-12pt]

Our focus here will be on inclusive diffraction at $ep$ colliders in the DIS region 
which has large $Q^2\gg \Lambda_{\rm QCD}^2$; namely, the process
\begin{align}
e + p \to e + X + Y\,, 
\end{align}
where $X$ represents the central hadronic radiation, and $Y$ denotes either a hadron $p$ or a jet-like multi-particle state. A large rapidity gap is required between $X$ and $Y$. The term ``inclusive diffraction'' reflects that the analysis is inclusive over the internal structure of the central system $X$: it may consist of a single jet or multi-pronged jets, provided they are sufficiently well separated from the system $Y$. 
In contrast, for inclusive DIS at small-$x$, one does not measure the momentum of the hadronic system $Y$, so the factorization is insensitive to the difference between diffractive, quasi-diffractive, and ungapped contributions.
Although we focus here on single-gapped inclusive diffraction~\cite{H1:2006zyl,Armesto:2019gxy,H1:1997bdi,ZEUS:1995sar,ZEUS:2009uxs,Zlebcik:2019tiu,H1:2006uea,Aaron:2010aa,ZEUS:2008xhs}, we remark that processes that identify multiple large rapidity gaps or additional tagged final-state particles within $X$ have been studied at HERA and are key targets at the EIC \cite{AbdulKhalek:2021gbh}. These include examples like diffractive dijet production~\cite{ZEUS:2007yji, H1:2010xdi, Altinoluk:2015dpi, Hatta:2016dxp, Guzey:2020gkk, Boer:2021upt,H1:2007oqt,H1:2011kou,H1:2014pjf,H1:2015okx}, diffractive multi-jet production~\cite{ZEUS:2001tvy}, diffractive open charm production \cite{Levin:1996vf, ZEUS:2003yug, H1:2006zxb, H1:2017bnb}, exclusive and semi-inclusive diffractive vector meson production \cite{Fraas:1969uwg, Brodsky:1994kf, Ivanov:2004ax, H1:2009cml, Mantysaari:2016jaz, Tu:2020ymk, Celiberto:2022dyf}, as well as diffractive semi-inclusive DIS (SIDDIS)~\cite{Hatta:2022lzj, Hatta:2024vzv, Bhattacharya:2024sck}. 
These processes all satisfy the criteria in \fig{diffraction-chart}, and are often explored in both the DIS and photoproduction regimes. 

In this paper, we develop a first-principles field-theoretic method to derive a full factorization formula for inclusive diffraction in $ep$ scattering for the first time.
Our analysis sets up a power counting and field theory framework for diffraction using SCET with Glauber interactions.  
The resulting forward-scattering factorization formula has terms associated to both diffraction and quasi-diffraction, forcing us to write the diffractive structure functions as a sum of the signal and an irreducible background,
\begin{align}\label{eq:signal-background}
  F_i^D =  F_i^{D\,{\rm diff}} + F_i^{D\,{\rm quasi}} \,.
\end{align}
Factorization decomposes these structure functions in terms of universal hadronic field theory matrix elements, of the form
$F_i^{D\,{\rm diff}} = B\otimes_{\perp} S_i $ and $F_i^{D\,{\rm quasi}}=B \otimes_{\perp} U\otimes_{\pm} S_i$.
Here $B$ describes the proton-to-forward proton/jet transition, $S_i$ contains the photon interaction and the central hadronic radiation, 
and these functions are tied together by multiple Glauber interactions which encode the Pomeron and other exchanges.
For quasi-diffraction, the additional function $U$ describes radiation in the rapidity gap that passes experimental cuts. The quasi-diffractive factorization formula enables us to quantify the size of this background; i.e., the associated Sudakov factor. 
For cases where $|t|\ll Q^2$, there is a further 
hard-collinear factorization of $S_i$ that separates out a DIS hard coefficient function, and yields Regge factorization for the diffractive parton distribution functions.

When accounting for all degrees of freedom, we find that the diffractive cross section can be expressed in terms of 18 independent structure functions: four that are unpolarized $F_{2,L,3,4}^D$, and 14 that are polarized (the same counting as in semi-inclusive DIS (SIDIS)~\cite{Bacchetta:2006tn}). 
At leading order in the forward-scattering power counting, the most common spin-independent diffractive structure functions $F_{2,L}^D$ are the same size as the often-ignored structure functions 
$F_{3,4}^D$. 
We show how to access $F_{3,4}^D$ using a Lorentz-invariant variable $\xbar$ and assess prospects for their measurement with a given experimental resolution in the lab frame. We prove that four corresponding polarized structure functions $F_{iP}^D$ and two additional antisymmetric structure functions $F_{4A}^D$ and $F_{4AP}^D$ survive at leading power.
We also argue that a larger region of phase space is interesting for probing diffraction, including  the region where $\mu^2\simeq -t, m_X^2\gg \LQCD^2$, where $m_X^2$ is the central jet invariant mass. 
In this region, we make model-independent predictions for quasi-diffractive structure functions; specifically, 
the ratios $(F_{3,4,L,4A}^D/F_{2}^D)^{\rm quasi}$ and $(F_{3P,4P,LP,4AP}^D/F_{2P}^D)^{\rm quasi}$ 
to leading order in $\alpha_s(\mu)$, including their dependence on two kinematic variables.
We also show that the diffractive signal requires at least two perturbative Glauber exchanges, whereas quasi-diffraction requires only one. This implies that the Sudakov suppression of the quasi-diffractive background is partly compensated by an $\alpha_s(\mu)$ suppression of the diffractive signal. 
In this region, more detailed predictions can be made for $F_i^{D\,{\rm diff}}$, though we leave some of the necessary calculations to future 
work.\\[-12pt]
 
The outline of this paper is as follows. In \sec{diffract-kinematics}, we describe the kinematics of $ep$ diffraction, including its Lorentz-invariant kinematic variables and bounds, structure function decomposition, relevant kinematic constraints, and prospects for experimental extraction of the structure functions $F_{3,4}^D$ that have not yet been measured.  We introduce EFT techniques in \sec{diffract-eft}, set up the appropriate degrees of freedom to study diffraction with SCET, and discuss five distinct power counting (expansion) parameters for diffraction. In \sec{diffract-factorization}, we derive a factorization formula for diffractive $ep$ scattering processes that separates the dynamics associated to the Regge and hard-collinear limits. Next, we study some key components in this factorization formula in greater depth. Specifically, in \sec{logarithms}, we discuss large logarithms appearing in diffraction, and we predict the size of the Sudakov double-logarithmic effects that suppress the quasi-diffractive background. Additionally, in \sec{soft-results}, we calculate the soft function describing the central rapidity region at leading order in $\alpha_s$. In \sec{predictions}, we discuss phenomenological applications of our results, and make model-independent testable predictions for quasi-diffractive structure functions that can be measured at HERA and the EIC. \Sec{comparisons} compares our results to existing factorization results and models in the literature. In \sec{diffract-outlook}, we provide a summary of our findings and concluding remarks, and we provide some guidance towards developing a deeper understanding of diffraction at the EIC and beyond.

\section{Diffractive $ep$ Scattering}
\label{sec:diffract-kinematics}

In this section, we provide an overview of the kinematics of diffraction, focusing on diffractive processes that involve an electron and a proton forward-scattering off of one another, producing a single central jet, as shown in \fig{diffraction}.  Notably, we bring attention to a number of important Lorentz invariants and structure functions that are less discussed in the literature.  
Our discussion of kinematics and structure functions applies equally well to gapped hard scattering and to (in)coherent diffraction. 

The outline of this section is as follows. In \sec{lorentz}, we provide a general discussion of the Lorentz invariants that appear in  
$ep\to eXY$ scattering, derive general bounds on 
these invariants that are not diffraction-specific, and provide a decomposition of momenta components in the Breit frame.
In \sec{structure-functions}, we give the Lorentz-invariant decomposition for the ten leading-power diffractive structure functions (leaving the description of the full basis of 18 structure functions to \app{structures}).
In \sec{diffractive-constraints}, we describe the additional kinematic constraints diffraction places on the Lorentz-invariant kinematic variables through requirements like a rapidity gap and forward scattering. Finally, we discuss prospects for measuring the largely-unexplored unpolarized structure functions $F_{3,4}^D$ at HERA and the EIC in \sec{experiments-xbar}.

\subsection{Kinematic variables and bounds}\label{sec:lorentz}

\begin{figure}
	\begin{center}
		\includegraphics[width = 2.4 in]{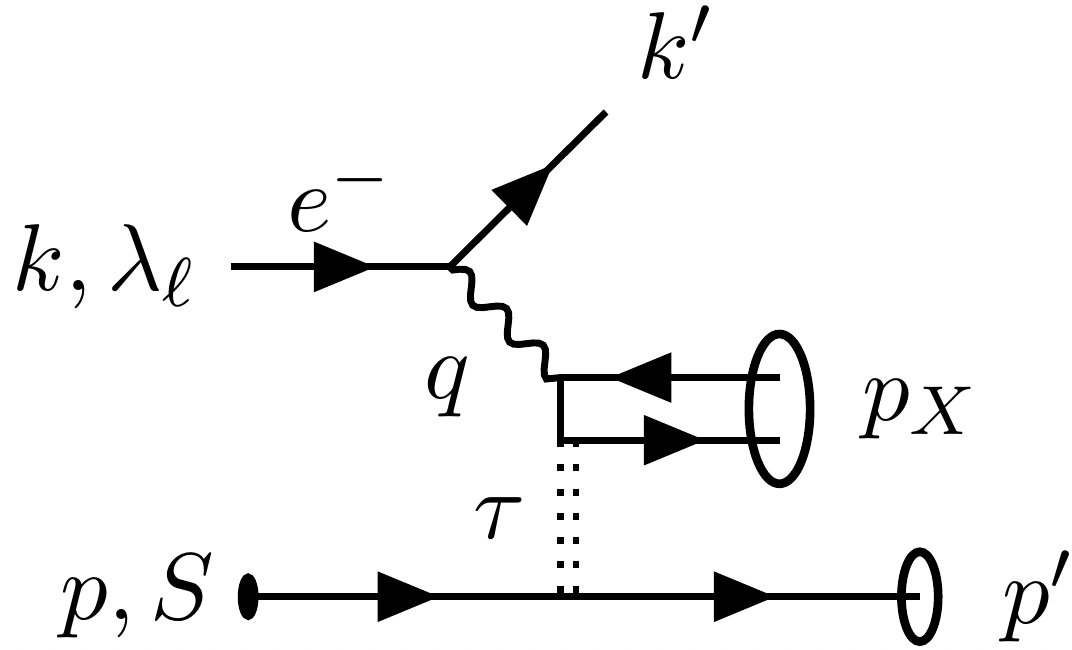}
		\qquad\quad
		\includegraphics[width = 2.4 in]{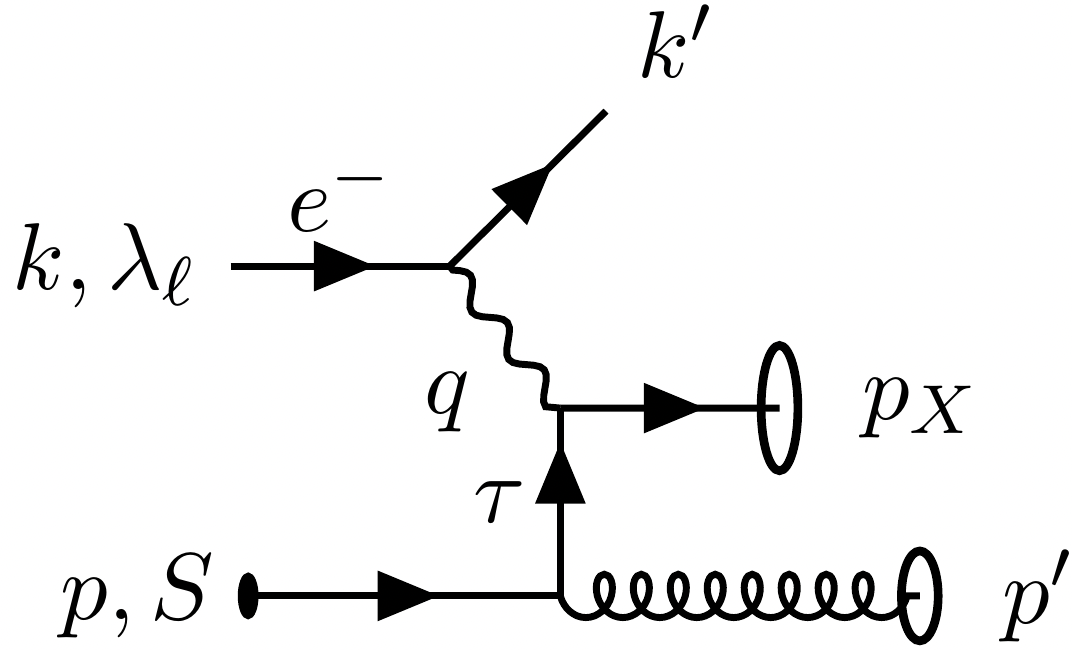}
		\\ \ \, (a) \hspace{2.6in} (b) \hfill \\
		\caption{
			$ep$ scattering producing a central jet $X$ with momentum $p_X$ and forward particles $Y$ with momentum $p'$, with a rapidity gap separating $p'$ and $p_X$.  
			(a) Partonic diagrams for the forward scattering case, where we have either incoherent diffraction with a multi-particle state $Y$ or coherent diffraction with a proton $Y$. (b) Leading diagram for the hard scattering case for moderate $x$. This is a subleading contribution to incoherent diﬀraction in the small-$x$ region.
			}
		\label{fig:diffraction}
	\end{center}
\end{figure}

Consider the diffractive (or hard scattering) process in \fig{diffraction}. We give the initial and final electrons momenta $k$ and $k'$ respectively, and let them transfer momentum $q = k-k'$ to the hadronic system via a photon. The initial proton carries momentum $p$, and the final hadronic states consist of central radiation $X$ with momentum $p_X$ (which could be a single particle, central jet, or uncollimated spray of particles confined to a well-defined region of phase space), and a forward final state $Y$ with momentum $p'$ (which is a proton in the case of coherent diffraction, and a multi-particle system that is jet-like in the case of incoherent diffraction). Thus, momentum conservation for the hadronic system is $q+p=p_X+p'$. 
The t-channel momentum transfer is then defined as $\tau = p-p' = p_X - q$. 
We also define the initial lepton helicity and proton spin vectors as $\lambda_\ell$ and $S$, respectively.

In diffractive $ep$ scattering, we measure four external momenta ($k$, $k'$, $p,$ and $p'$), three of which have vanishing invariant mass:
\begin{align}
	k^2 = k^{\prime 2} = m_e^2 \approx 0\,, \qquad \qquad p^2 = m_p^2 \approx 0\,.
\end{align}
Note that we include hadron mass corrections in \app{photon}. From the four external momenta, we can form exactly seven linearly independent Lorentz invariants. In this paper, we generally choose these seven invariants to be:
\begin{align}\label{eq:lorentz-invariants}
	Q^2 &= -q^2 \qquad &t&= \tau^2 \qquad &\xbar& = \frac{k\cdot \tau}{k\cdot p}
	\nonumber\\*
	x&= \frac{Q^2}{2p\cdot q} \qquad &m_Y^2 &= p^{\prime 2}
	\nonumber\\*
	y&= \frac{p\cdot q}{p\cdot k} \qquad &\beta& = \frac{Q^2}{2q\cdot \tau}
\end{align}
In the leftmost column, we see the familiar invariants $Q^2$, $x$, and $y$ from deep-inelastic scattering. The middle column contains diffractive invariants  that commonly appear in the literature that are based on the measured final-state hadron or jet momentum $p'$ \cite{Frankfurt:2022jns}. The rightmost column contains the important invariant $\xbar$, which becomes relevant when considering the full set of diffractive structure functions,
as we will see in \sec{structure-functions}. 

For concision in later mathematical expressions, it is useful to define several invariants that are not independent of those in \eq{lorentz-invariants}:
\begin{align}\label{eq:supplemental-lorentz-invariants}
	& W^2 = \, (p+q)^2 = \frac{1-x}{x}Q^2\,, 
    & & s = (p+k)^2 = \frac{Q^2}{xy}\,,
	& & m_X^2 = p_X^2 =  Q^2 \frac{1-\beta}{\beta} + t\,,
	\nonumber\\
	& \xi = x_{\mathbb{P}} =1-x_L=  \frac{q\cdot \tau}{p\cdot q} = \frac{x}{\beta}\,,
	&& z=  \, \frac{p\cdot p'}{p\cdot q} = \frac{x}{Q^2}(m_Y^2-t)\,,
\end{align}
For example, it is often convenient in incoherent diffraction to swap $m_Y^2$ in \eq{lorentz-invariants} out for $z$ in expressions where the combination $(m_Y^2-t)$ appears often. In addition, we note that coherent diffraction involves only 6 independent Lorentz invariants, as $m_Y^2 = m_p^2 \approx 0$. 

\subsubsection{Lightcone coordinate convention}\label{sec:diffract-lightcone}

In collider processes, the vast majority of momentum is concentrated near a lightcone, and thus it is convenient to decompose momenta onto a lightcone basis as
\begin{align}
  p^\mu = \nbar\cdot p\, \frac{n^\mu}{2} + n\cdot p\, \frac{\nbar^\mu}{2} + p_\perp^\mu \,,
\end{align}
where the basis vectors $n^\mu$ and $\nbar^\mu$ satisfy $n^2=\nbar^2=0$ with $n\cdot \nbar=2$.  For example, $n^\mu$ could have its 3-vector $\vec n$ correspond to the forward proton direction and $\nbar^\mu$ the opposite. The direction $\perp$ is then transverse to the beamline. 
We define notation for the projection onto lightcone momentum components as
\begin{align} \label{eq:pdecomp}
	p^\mu = (p^+,p^-,|\vec{p}_\perp|)= \left( {n\cdot p} ,\,  {\nbar\cdot p},\,  |\vec p_\perp| \right)\,.
\end{align}
Note that the final term in \eq{pdecomp} is quoted as the magnitude $|\vec{p}_\perp|$, and hence this choice of component notation does not capture the full perpendicular vector component of the momentum (but suffices for our focus on scaling). Minkowski and lightcone coordinates are related as
\begin{align}
	p^\pm = {p^t \mp p^z} \hspace{1 in} \vec{p}_\perp = (p^x,p^y) \,.
\end{align}
Dot products take the following form:
\begin{align}
	p \cdot q = \frac12 (p^+ q^- + p^- q^+) -  \vec q_\perp \cdot \vec p_\perp 
	\,,
\end{align}
under the lightcone convention used in this work. 

\subsubsection{Breit frame} \label{sec:Breit}

To study the dynamics of diffraction, it is useful to employ a number of frames in which the proton is energetic, such as the lab frame and the $e^-$--$p$ center-of-momentum (CM) frame (see \app{epCMframe}), the $\gamma^*$--$p$ CM frame (see \sec{gap-radiation}), and the Breit frame, which we discuss here. Defining light-like vectors $n^\mu = (0,2,0_\perp)$ and $\nbar^\mu = (2,0,0_\perp)$,  these frames have proton momentum $p^\mu \simeq p^- n^\mu/2$ with $p^- \gg p^+$.

The Breit frame provides a clean separation between leptonic and hadronic kinematic variables, and has
\begin{align}\label{eq:kp-vectors}
	q^\mu &= Q \, \Big(\frac{\nbar^\mu}{2} - \frac{n^\mu}{2}\Big) \,,
	&p^\mu&= p^- \frac{n^\mu}{2} + \frac{m_p^2}{p^-} \frac{\nbar^\mu}{2}  \,,
\end{align}
where $p^- = Q/x + x m_p^2/Q + \ldots\simeq Q/x$, dropping terms of ${\cal O}(m_p^2/Q^2)$. 
Using \eq{kp-vectors}, it is straightforward to express the components of the remaining momenta in \fig{diffraction} in terms of the Lorentz invariants in \eqs{lorentz-invariants}{supplemental-lorentz-invariants}:
\begin{align}\label{eq:four-vectors}
	q & = Q\Big(1,-1,\,0_\perp \,\Big) \,,
	\nonumber\\
	p_X & = Q \bigg(1-z,\, \frac{1-\beta}{\beta}-z,\, 
	\sqrt{{-t}/{Q^2} -z\big({1}/{\beta}-z\big)}
	\,\bigg) \,,
	\nonumber\\
	\tau & =Q\bigg(-\!z,\, \frac{1}{\beta}-z, \,
	\sqrt{{-t}/{Q^2} -z\big({1}/{\beta}-z\big)}
	\, \bigg)  \,,
	\nonumber\\
	p'
	& =   Q\bigg(z,\, \frac{1}{x}-\frac{1}{\beta}+z,\,
	\sqrt{-t/Q^2 -z\big({1}/{\beta}-z\big)}
	\,\bigg) \,,
	\nonumber\\
	p & = Q\Big(0,\frac{1}{x},\,0_\perp \,\Big) \,.
\end{align}
In this frame, we have $|\vec p_{X\perp}|=|\vec\tau_\perp| =|\vec p_\perp^{\,\prime}|$ because $q_\perp=0$.  Although we rarely need the decomposition of lepton momenta in this frame, we quote them here for completeness:
\begin{align} \label{eq:lepton-4-vectors}
	&k = Q\Big(\frac{1}{y},\frac{1-y}{y},\frac{\sqrt{1-y}}{y}\,\Big)
	\,,
	&&k' = Q\Big(\frac{1-y}{y},\frac{1}{y},\frac{\sqrt{1-y}}{y}\,\Big)\,.
\end{align}

\subsubsection{Bounds on variables} \label{sec:bounds}

From only positivity constraints, we can derive a number of frame-independent bounds on the linearly independent Lorentz invariants $(x,y,Q^2,t,m_Y^2,\beta,\xbar)$. We find it convenient to derive these bounds using the Breit frame defined in \sec{Breit}. We start out by noting that the classic DIS conditions also apply to diffraction in the DIS region, which is the focus of this paper:
\begin{align}\label{eq:xy-bounds}
	0 &\leq x \leq 1\,, 
	&0 &\leq  y  \leq 1\,,
	&\Lambda_{\rm QCD}^2& \ll Q^2 \leq s\,.
\end{align}
Next, we use that the jet momenta $p_X^+$ and $p^{\prime +}=-\tau^+$ are positive, which gives 
\begin{align}
& t  = \tau^+ \tau^- - \vec\tau_\perp^{\,2} \leq 0 \,,
& & 0\leq z\leq 1 \,.
\end{align}
An upper bound on $\beta$ arises from plugging $m_X^2 \geq 0$ into $\beta = Q^2/(Q^2+ m_X^2 - t)$, and a lower bound can be derived from  $\vec p_{X\perp}^{\,2} \geq 0$:
\begin{align} \label{eq:beta-bounds}
	& \frac{1}{z+(-t)/(zQ^2)} \le \beta \le \frac{1}{1+\frac{-t}{Q^2}} \le 1
\end{align}
This lower bound is stronger than $x\le \beta$ in the small-$x$ region of phase space that we focus on in this paper. 
Another possible lower bound comes from $p_X^2 + m_Y^2 \leq W^2$, yielding  $\beta \ge x/\big(1 + x (-t-m_Y^2)/Q^2\big)$, but this is a slightly weaker bound than \eq{beta-bounds} in the small-$x$ region of phase space.  For $\xbar$, we find that the most stringent bounds are
\begin{align} \label{eq:xbar-bound}
	-2 x \Delta  \le \bar x - \frac{x}{\beta} + (2-y) x z \le  2 x \Delta  \,,
\end{align}
where for convenience we define
\begin{align}\label{eq:Delta}
	\Delta^2 = (1-y)  (z^2-\frac{z}{\beta}-t/Q^2) \,.
\end{align}
The bound in \eq{xbar-bound} is derived below in \sec{photon-helicity-basis}.

Inverting \eq{beta-bounds} gives us bounds on $-t$: 
\begin{align}\label{eq:t-bounds}
	\frac{Q\sqrt{Q^2(\beta-x)^2+4 m_Y^2 x^2\beta^2}}{2 x^2\beta} -m_Y^2 -\frac{Q^2(\beta-x)}{2x^2\beta} 
	\leq -t \leq \frac{1-\beta}{\beta} Q^2  \,.
\end{align}
In the limit relevant for diffraction we can expand to leading order in $4x^2 m_Y^2 \ll Q^2$, giving
\begin{align}\label{eq:simp-t-bounds}
\frac{x }{(\beta-x)} m_Y^2 \lesssim -t \leq \frac{1-\beta}{\beta} Q^2
 = m_X^2 - t
\,.
\end{align}
Note that the hierarchies $m_Y^2 \leq -t$ and $m_Y^2 \geq -t$ are both allowed. 
The condition $\vec p_{X\perp}^{\,2}\geq 0$ gives an upper bound on the mass $m_Y^2$ for fixed $t$.
The lower bound comes from coherent diffraction, $m_Y^2 \simeq m_p^2 \sim \LQCD^2$, where the proton remains intact:
\begin{align}
	m_p^2 \le m_Y^2 \leq \frac{Q^2}{2 x \beta} + t 
	- \frac{Q^2}{x} \sqrt{\frac{1}{4\beta^2}+\frac{t}{Q^2}}
	\,.
\end{align}
In the limit most often studied in diffraction we can expand in $4\beta^2(-t)\ll Q^2$ to give
\begin{align}\label{eq:my-bounds}
	m_p^2 \le m_Y^2 \lesssim  \frac{(\beta-x)(-t)}{x} \,.
\end{align}
We reiterate that here, we have only used positivity and basic physical properties of $p'$ and $p_X$. A detailed discussion of conditions specific to diffraction (like imposing a rapidity gap and forward scattering) is taken up in \sec{diffractive-constraints}.

\subsection{Diffractive structure functions}\label{sec:structure-functions}


Now, we decompose the diffractive cross section in terms of structure functions. By doing so in a frame-independent manner, we facilitate connecting experimental measurements and theoretical computations by allowing convenient frame choices for each. 
The cross section for diffraction is given by 
\begin{align} \label{eq:sigmaLW0}
\mathrm{d} \sigma=\frac{\alpha^2}{\left(k \cdot p\right) Q^4} \frac{\mathrm{d}^3 k'}{2 E_{k'}} \frac{\mathrm{d}^3 p'}{2 E_{p'}} dm_Y^2\: L_{\mu \nu}\left(k, k'\right) W_D^{\mu \nu}\left(q, p,p'\right)\,,
\end{align}
where we have suppressed dependence on hadronic and leptonic spins for brevity. 
Making the cross section fully differential in our chosen six Lorentz-invariant variables, we have
\begin{align} \label{eq:sigmaLW}
\frac{d^6\sigma}{dx\, dQ^2\, d\beta\, dt\, d\bar x\, dm_Y^2} 
&= \frac{\pi \alpha^2 x y^2}{4 Q^6 \beta^2\, N_\sigma} \:
L_{\mu \nu}\left(k, k'\right) W_D^{\mu \nu}\left(q, p,p'\right)
\,,
\end{align}
where here we define 
\begin{align} \label{eq:Nsigma}
  N_\sigma = x \sqrt{(1-y)(z^2-z/\beta-t/Q^2) - (\bar{x}/x -1/\beta + (2-y) z)^2/4}\,.
\end{align}
The dependence on $N_\sigma$ arises from being differential in $\bar{x}$, and $\int d\bar{x}\, 1/N_\sigma = \pi$, as we discuss further below.

The diffractive hadronic and leptonic tensors appearing in \eqs{sigmaLW0}{sigmaLW} are
\begin{align} \label{eq:tensors}
W_D^{\mu\nu}(q, p, p') &
= \sumint_X \,\sumint_{Y}\: \delta^4(q + p - p' - p_X) \:
\big\langle p \big| J^{\dagger\,\mu}(0) \big| Y X\big\rangle
\big\langle Y X \big| J^\nu(0) \big| p \big\rangle
\delta(m_Y^2-p^{\prime\,2})
 ,\nn\\
L^{\mu\nu}(k,k') &
= \big\langle k \big| J_e^{\dagger \mu}(0) \big| k'\big\rangle
\big\langle k' \big| J_e^\nu(0) \big| k \big\rangle
\nn\\&
=  2 
\bigl[  k^\mu k^{\prime \nu} + k^\nu k^{\prime \mu}  - k\cdot k' \, g^{\mu\nu} + \img \lambda_e \epsilon^{\mu\nu \rho\sigma} k_\rho k_\sigma \bigr]
\,.
\end{align}
Note that our convention for $W_D^{\mu\nu}$ does not contain a $(2\pi)^4$.
Here, we have currents $J^\mu = \sum_q e_q \bar{\psi_q} \gamma^\mu \psi_q$ for quark flavors $q$ with electric charge $e_q$, and $J_\ell^\mu = - \bar{\psi_\ell} \gamma^\mu \psi_\ell$ for the lepton $\ell$. $\lambda_{\ell}=\pm 1$ is the longitudinal spin polarization of the incoming lepton. For the leptonic tensor, we have $\langle k' | J_e^\nu | k\rangle = \langle \ell'(k') | J_\ell^\nu | \ell(k)\rangle$.
The $\sumint_X$ includes a sum over states with different numbers of final-state particles $j\in X$ as well as an integration over their individual phase space, $\prod_j d^3p_j/(2E_j(2\pi)^3)$.
The $\sumint_{Y}$ includes the sum over particles $j\in Y$ as well as analogous integrals over their phase space, after extracting the overall $p'$ phase space integral in \eq{sigmaLW}. 
The meaning of the hadronic tensor depends on the process: for incoherent diffraction, $Y$ is a multi-particle jet-like state of momentum $p'$, so $\langle Y X | J^\nu | p\rangle = \langle Y(p') X | J^\nu| P(p)\rangle$.
The $\sumint_X$ is taken over all possible final states that $p_X$ could represent for particles outside $Y$ (as seen in \fig{diffraction} and its higher-order generalizations).  
We discuss definitions for $Y$ and $X$ in \sec{diffractive-constraints}. 
For coherent diffraction, $p'$ is a forward proton and $\langle Y X | J^\nu | p\rangle = \langle P(p') X | J^\nu| P(p)\rangle$; the $\sumint_X$ is taken over all other hadronic particles and $\sumint_{Y}\to 1$. We discuss further restrictions on $X$ due to the rapidity gap in \sec{diffractive-constraints}. 

The diffractive hadronic tensor depends on three measured momenta $\{p,q,p'\}$, just as in SIDIS~\cite{Bacchetta:2006tn}. (Here we prefer to use the equivalent choice of $\{p,q,\tau\}$.) Thus, these two processes share the same number of structure functions: 4 unpolarized and 14 that are $S$- and $\lambda_\ell$-dependent.
As far as we know, there have not been diffraction papers that consider all 18 structure functions relevant for the EIC, which has both leptonic and hadronic beam polarization.%
\footnote{
	Ref.~\cite{Arens:1996xw} identified all five $F_i^D$ relevant for HERA, involving a polarized $e^-$ beam and an unpolarized $p$ beam. 
	Ref.~\cite{Blumlein:2001xf} identified all four unpolarized $F_i^D$, and Ref.~\cite{Blumlein:2002fw} considered a subset of structure functions involving both lepton and hadron polarization. 
	Most papers only focus on $F_{2,L}^D$, which are analogous to the DIS structure functions $F_{2,L}$  \cite{AbdulKhalek:2021gbh, Frankfurt:2022jns}. 
	The other two unpolarized  functions $F_{3,4}^D$ become redundant under certain assumptions, including in the kinematic limit $p' = (1-\xi)p$~\cite{Blumlein:2001xf, Blumlein:2002fw} which makes $N_X X^\mu \propto U^\mu$, or vanish in models containing no photon helicity interference terms \cite{Arens:1996xw}.
The study of polarized structure functions is quite interesting, with HERA having had 40-60\% polarization of the lepton beam, and the EIC  expected to achieve 70\% polarization of both beams.
}
\Refcite{Arens:1996xw} formulated the diffractive structure functions relevant for HERA using a photon helicity basis, which we describe later in \sec{photon-helicity-basis}.  
Here, we favor an approach starting with Lorentz four-vectors. This approach has been used in~\refscite{Blumlein:2001xf,Blumlein:2002fw}, but with a non-orthogonal basis of tensor structures, whereas we find it advantageous to use an orthogonal basis in the style of the SIDIS analysis of \refcite{Donnelly:2021erx}.
Specifically, we use orthogonal vectors $\{q,U,X\}$ defined as
\begin{align}\label{eq:ux-vectors}
U^\mu = \frac{2x}{Q}\left(p^\mu - \frac{p\cdot q}{q^2}q^\mu \right)\,, \qquad\qquad 
X^\mu = \frac{1}{N_X} \left(V^\mu - \frac{U \cdot V}{U^2}U^\mu \right)\,,
\end{align}
where we define an auxiliary vector
$V^\mu = -\tau^{\mu} + q^\mu\, (\tau\cdot q) / q^2$
to simplify the expressions. 
These momenta are normalized as $X^2 = -1$ and $U^2 = +1$, so that
\begin{align}
N_X^2 = (U\cdot V)^2 - V^2 = \frac{Q^2\Delta^2}{1-y}\,,
\end{align} 
for $\Delta$ in \eq{Delta}.  Combining these formulas yields
\begin{align}
X^\mu &= \frac{1}{\sqrt{-t}} \frac{1}{\sqrt{1-x/\beta -tx^2/Q^2}}\left[ \frac{x}{\beta Q^2}(2\beta t x - Q^2)p^\mu + \frac{tx}{Q^2}q^\mu + \tau^\mu \right]
  \,.
\end{align}
Generally, we decompose the hadronic tensor into a sum of independent tensors $w_i^{\mu\nu}$ built out of these vectors multiplied by structure functions $F_i^D$:
\begin{align}\label{eq:hadronic-tensor-2}
W_D^{\mu\nu} = \sum_i w_i^{\mu\nu}(q,U,X,S)\, F_i^D (x,Q^2,\beta,t,m_Y^2)
\,.
\end{align}

\paragraph{Unpolarized beams.}
For the unpolarized case with a virtual photon, $w_i^{\mu\nu}$ are symmetric tensors satisfying $q_\mu w_i^{\mu\nu} =0=q_\nu w_i^{\mu\nu}$, which can be built from $(g^{\mu\nu}-q^\mu q^\nu/q^2)$, $U^\mu U^\nu$, $X^\mu X^\nu$ and $U^\mu X^\nu + X^\mu U^\nu$.  We choose $w_L^{\mu\nu}$ and $w_2^{\mu\nu}$ to be the standard tensors used in DIS, and choose the form of $w_{3,4}^{\mu\nu}$ to make clear their underlying differences from DIS,
\begin{align}\label{eq:tensor-structures}
w_L^{\mu\nu} &= \frac{1}{2x} \Big( g^{\mu\nu} - \frac{q^\mu q^\nu}{q^2}\Big) \,,  
&w_2^{\mu\nu}& = \frac{1}{2x} \Big( U^\mu U^\nu - g^{\mu\nu} +\frac{q^\mu q^\nu}{q^2}\Big) \,,
\nonumber\\
w_3^{\mu\nu} &= \frac{1}{2x} \Big( 2 X^\mu X^\nu - U^\mu U^\nu + g^{\mu\nu} 
   - \frac{q^\mu q^\nu}{q^2}\Big) \,,
&w_4^{\mu\nu}& = \frac{1}{2x}(U^\mu X^\nu + X^\mu U^\nu ) \,. 
\end{align}
Note that we work with the structure function $F_L^D= F_2^D - 2x F_1^D$ instead of the often-used $F_1^D$. The relation between $F_L^D$, $F_1^D$, and $F_2^D$ is the same as for inclusive DIS. 

Just like in DIS, it is useful here to define projectors $\cP_{i\, \mu\nu}$ that extract $F_i^D$ from $W_D^{\mu\nu}$:
\begin{align}\label{eq:projection}
\cP_{i\, \mu\nu} W_D^{\mu\nu} = F_i^D \,.
\end{align}
We form unpolarized projectors using a basis comprised of the symmetric tensors $g^{\mu\nu}$, $U^\mu U^\nu$, $X^\mu X^\nu$, and $(U^\mu X^\nu + U^\nu X^\mu)$. For $d=4$, we find
\begin{align}\label{eq:projectors}
\cP_L^{\mu\nu} &= 2 x\, U^\mu U^\nu \,, \hspace{0.5 in}
&\cP_2^{\mu\nu}& = -x\left(g^{\mu\nu} -3 U^\mu U^\nu\right) \,,
\nonumber\\
\cP_3^{\mu\nu} &= x\,(g^{\mu\nu} - U^\mu U^\nu+2 X^\mu X^\nu ) \,,
&\cP_4^{\mu\nu}& = -x\,( U^\mu X^\nu + X^\mu U^\nu)\,,
\end{align}
which satisfy $\cP_{i\mu\nu} w_j^{\mu\nu} =\delta_{ij}$.
For later convenience it is also useful to define tensors and projectors without the explicit factors of $x$,
\begin{align} \label{eq:primeprojectors}
  w_i^{\prime\mu\nu} \equiv  x\, w_i^{\mu\nu} \,,
  \qquad\qquad
 \cP_i^{\prime \mu\nu} \equiv \frac{1}{x}\, \cP_i^{\mu\nu} \,,
\end{align}
which by construction also satisfy $\cP^\prime_{i\mu\nu} w_j^{\prime\mu\nu} =\delta_{ij}$.

Next, we contract \eq{hadronic-tensor-2} with $L^{\mu\nu}$ in \eq{tensors}. For the unpolarized case, we have
\begin{align}\label{eq:unpolarized-coefficients}
L_{\mu\nu} W_D^{\mu\nu} 
&= \frac{2s}{y}\biggl[ -\frac{y^2}{2} F_L^D +  \left(1-y+\frac{y^2}{2}\right)F_2^D 
+ \bigg[ \frac{2(k\cdot X)^2 y^2}{Q^2 } - 1+y  \bigg] F_3^D 
\nn\\
&\qquad\quad + \frac{2y^2(k\cdot U)(k\cdot X)}{Q^2}F_4 ^D\biggr]\,.
\end{align}
We can write these coefficients in terms of Lorentz invariants:
\begin{align}\label{eq:ux-dot-product}
&k\cdot U = Q\frac{2-y}{2y} \,,
&&k\cdot X = Q^2\,\frac{x/\beta - \bar{x}- (2-y) x z }{2 N_X x y } \,.
\end{align}
Note that we employ the invariant $z$ here instead of $m_Y^2$ for simplicity.
Putting these ingredients together, we have 
\begin{align}
\label{eq:diffcross}
\frac{d^6\sigma}{dx\, dQ^2\, d\beta\, dt\, d\bar x\, dm_Y^2} 
&= \frac{\alpha^2 \pi }{2Q^4\beta^2\, N_\sigma} \Biggl\{ -\frac{y^2}{2} F_L^D +  \left(1-y+\frac{y^2}{2}\right)F_2^D 
\\
& \quad
+ \bigg[ \frac{2(k\cdot X)^2 y^2}{Q^2 } - 1+y  \bigg] F_3^D 
+ \frac{2y^2(k\cdot U)(k\cdot X)}{Q^2}F_4 ^D \Biggr\}
.\nn
\end{align}
Experimental measurements often integrate over small $m_Y<m_Y^{\rm cut}$, and this cumulative cross section is easily obtained from the differential one above.

If we integrate \eq{diffcross} over $\xbar$ with the limits in \eq{xbar-bound}, we find that
\begin{align}
\frac{d^5\sigma}{dx\, dQ^2\, d\beta\, dt\, dm_Y^2} 
&=\frac{\alpha^2 \pi^2 }{2Q^4\beta^2} \biggl[ -\frac{y^2}{2} F_L^D +  \left(1-y+\frac{y^2}{2}\right)F_2^D \biggr]\,,
\end{align}
and we lose access to $F_3^D$ and $F_4^D$.\footnote{Note that our $F_i^D$ differs from the usual form in the literature by a normalization factor of $\pi/4$~\cite{Diehl:1996st}. To match the $F_2^D$ and $F_L^D$ found in the literature, one needs to multiply our $F_i^D$ by $\pi/4$ and our cross section prefactor by $4/\pi$.} Note that the appropriate limits for the $\xbar$ integral
necessary to derive this result are given in \eq{xbar-bound}. 

\paragraph{Polarized beams.}
When we use polarized beams, we can construct fourteen additional independent structure functions $F_i^{D}$ involving $\lambda_\ell$ and $S$. Six of these fourteen structures are nonzero at leading power in diffraction, as we will see in \sec{diffract-factorization}. 
One of them corresponds to an antisymmetric tensor structure involving polarization of only the lepton beam,
\begin{align}\label{eq:antisymmetric-structure-lp}
	w_{4A}^{\mu\nu}=\frac{i}{2x}(U^\mu X^\nu - X^\mu U^\nu)\,.
\end{align}
Using \eqs{hadronic-tensor-2}{projection}, we can construct a corresponding antisymmetric projector 
\begin{align}\label{eq:P4A}
	&\mathcal{P}_{4A}^{\mu\nu}= ix(U^\mu X^\nu - X^\mu U^\mu)\,.
\end{align}
Contracting $w_{4AP}^{\mu\nu}$ with the leptonic tensor gives us 
\begin{align}
L_{\mu\nu} w_{4A}^{\mu\nu}
 = -\frac{2\lambda_e}{x}\epsilon^{\mu\nu\rho\sigma} U_\mu X_\nu k_{\rho} k'_{\sigma} 
 = - \text{sign}(\epsilon^{\mu\nu\rho\sigma} U_\mu X_\nu k_{\rho} k'_{\sigma} )
   \frac{2\lambda_e}{x}\frac{Q^2}{y}\sqrt{1-y -\frac{(k\cdot X)^2y^2}{Q^2}}\,,
\end{align}
where the overall sign depends on the sign of the contraction $\epsilon^{\mu\nu\rho\sigma} U_\mu X_\nu k_{\rho} k'_{\sigma}$. In the decomposition using photon helicity, described more in the next subsection and in~\app{photon}, this contraction can be written in terms of $\sin(\phi)$ as shown in~\eq{sine}.

The remaining five leading-power structure functions arise from simply multiplying the structures $i = \{1,2,3,4,4A\}$ in \eqs{tensor-structures}{antisymmetric-structure-lp} by a hadron spin term
\begin{align}\label{eq:polarized-structures-lp}
	w_{iP}^{\mu\nu} = \tilde{S}_T^{\,\prime} w_i^{\mu\nu}\,,
\end{align}
where we have defined 
\begin{align}\label{eq:tensor-structures-polarized}
	&\tilde{S}_T^{\,\prime} = \frac{\sqrt{-t}}{N_X} \,  \tilde S_T \,, 
	&\tilde{S}_T & = \frac{1}{(p\cdot q)\sqrt{-t}} \epsilon_{\alpha\beta\gamma\delta}\,p^\alpha q^\beta p^{\prime \gamma} S^\delta\,.
\end{align}
From \eq{polarized-structures-lp}, we see that it is straightforward to define polarized projectors $\cP_{iP}^{\mu\nu}$ and kinematic coefficients in the cross section from $L_{\mu\nu}W_D^{\mu\nu}$ in direct parallel to their unpolarized counterparts. 
Importantly, this generalizes \eq{diffcross} to
\begin{align}
\label{eq:diffcrosspol}
\frac{d^6\sigma}{dx\, dQ^2\, d\beta\, dt\, d\bar x\, dm_Y^2} 
&= \frac{\alpha^2 \pi }{2Q^4\beta^2\, N_\sigma} \Biggl\{ -\frac{y^2}{2} (F_L^D
+\tilde{S}_T^{\,\prime} F_{LP}^D) +  \left(1-y+\frac{y^2}{2}\right)(F_2^D + \tilde{S}_T^{\,\prime} F_{2P}^D) 
\nn\\
&\hspace{-2cm} + \bigg[ \frac{2(k\cdot X)^2 y^2}{Q^2 } - 1+y  \bigg] (F_3^D +\tilde{S}_T^{\,\prime} F_{3P}^D)
+ \frac{2y^2(k\cdot U)(k\cdot X)}{Q^2}(F_4 ^D + \tilde{S}_T^{\,\prime} F_{4P}^D) \nn\\
&\hspace{-2cm} -\text{sign}(\epsilon^{\mu\nu\rho\sigma} U_\mu X_\nu k_{\rho} k'_{\sigma} ) 2\lambda_e y \sqrt{1-y-\frac{(k\cdot X)^2 y^2}{Q^2}}  (F_{4A} ^D + \tilde{S}_T^{\,\prime} F_{4AP}^D) \Biggr\}
.
\end{align}

\subsubsection{Photon helicity basis}
\label{sec:photon-helicity-basis}

We began \sec{structure-functions} by presenting a frame-independent decomposition of $W^D_{\mu\nu}$. Here, we review a theoretical interpretation of the $F_i^D$'s based on their sensitivity to the virtual photon helicity, based on the Breit frame introduced in \sec{Breit}.  This methodology is well established in SIDIS~\cite{Gourdin:1972kro, Kotzinian:1994dv, Diehl:2005pc, Bacchetta:2006tn, Ebert:2021jhy} and was introduced to diffraction in \refcite{Arens:1996xw}, which considered only unpolarized protons, yielding five structure functions, one of which arises from the inclusion of a polarized lepton state. In the ensuing years, however, this approach has become less common in inclusive diffraction studies.

In the Breit frame, we can choose to align the exchanged photon $q$ with the $z$-axis, which helps us form a photon helicity basis. 
We choose the orthogonal vectors $U$ and $X$ in \eq{ux-vectors} to be aligned along $t$ and $x$, respectively.
This identification allows us to write the photon polarization vectors in terms of $U$ and $X$.
We label the structure functions associated with different photon polarizations as
\begin{align}
W_{\lambda \lambda'} \equiv \epsilon_{\lambda}^{* \mu}\epsilon_{\lambda}^{ \nu} W_{\mu\nu}\,,
\end{align}
where $\lambda \in \{+, -, 0\}$ represents the $\pm$ helicity formed by the two transverse polarization states or the longitudinal polarization state $0$. We can then express the structure functions in \eq{diffcross} as
\begin{align}
\label{eq:FinW}
F_L^D &=2x W^U_{00}\,,
& F_2^D &=2x\left(\frac{1}{2}(W^U_{++}+W^U_{--})+W^U_{00}\right)\,,\nn\\
F_3^D &=-x (W^U_{+-}+W^U_{-+})\,,
& F_4^D &=-\frac{x}{\sqrt{2}} \left(W^U_{0-}-W^U_{0+}+W^U_{-0}-W^U_{+0}\right)\,,
\end{align}
where the superscript $U$ indicates an unpolarized hadron beam. That is, $F_L^D$ involves the longitudinal polarization of the exchanged photon, whereas $F_2^D$ is a linear combination of transverse and longitudinal polarization terms. In contrast, $F_3^D$ involves interference between transverse polarizations and $F_4^D$ involves interference between longitudinal and transverse polarizations. Polarized hadron and lepton beams give access to five more photon helicity configurations $W_{\lambda\lambda'}$ and 14 more structure functions, which we review in detail in \app{photon}.

With the photon helicity decomposition, the coefficients of the structure functions $F_{3,4}^D$ have a characteristic dependence on the azimuthal angle $\phi$ of $p_\perp'$ with respect to the lepton plane. From \app{photon}, we have
\begin{align}  \label{eq:LWcosphi}
L_{\mu\nu} W^{\mu\nu}_D 
& = \frac{2s}{y}\biggl[-\frac{y^2}{2} F_L^D + \Bigl(1\!-\!y\!+\!\frac{y^2}{2}\Bigr)F_2^D 
 + (1\!-\!y) \cos(2\phi) F_3^D - (2\!-\!y)\sqrt{1\!-\! y} \cos\phi F_4^D\biggr]
 .
\end{align}  
The generalization of this result to include hadron mass corrections is provided in \app{photon}.
Comparing with \eqs{unpolarized-coefficients}{LWcosphi}, we can express $\cos\phi$ in terms of the Lorentz invariants defined in \sec{lorentz} as:
\begin{align}\label{eq:cosine}
\cos \phi
=  \frac{\bar{x} -x/\beta + (2-y) x z}{2 x \sqrt{(1-y)\left(z^2-z/\beta-t/Q^2 \right)}} 
= -\frac{y}{\sqrt{1-y}} \frac{k\cdot X}{Q}
\,,
\end{align}
Now, we can derive the bound on $\xbar$ provided in \eq{xbar-bound} from the bounds on $\cos\phi$:
\begin{align}\label{eq:xbar-limits}
-1 \leq \cos \phi \leq 1\,.
\end{align}
As explained in \app{photon}, $\sin\phi$ can also be expressed in a Lorentz-invariant form, which is relevant for the full spin-polarized cross section.  

The limits in \eq{xbar-limits} reveal why integrating over $\xbar$ causes $F_{3,4}^D$ to drop out of the cross section: integrating over $\xbar$ is equivalent to integrating over $\phi$, which removes the $\cos(2\phi)$ and $\cos(\phi)$ terms in \eq{LWcosphi}. Note that the variable change from $\phi$ to $\xbar$ is responsible for the Jacobian factor $N_\sigma$ in \eq{sigmaLW}, as $d/d\phi = 2 (1-\cos^2\phi)^{-1/2}\,d/d\cos\phi =  N_\sigma\, d/d\xbar$.

\subsection{Kinematic constraints on diffraction}
\label{sec:diffractive-constraints}

In \sec{bounds}, we placed bounds on our Lorentz invariants solely by considering positivity conditions. Next, we describe the kinematic constraints that we must impose for a process to be (quasi-)diffractive. 
One condition is that the central jet and hadronic system must have small invariant mass ($m_X^2$ and $m_Y^2$, respectively) when compared to the invariant mass of the entire hadronic system ($W^2$). This is required for defining $X$ and $Y$ as distinct and well-separated regions. Note that $X$ and $Y$ can each consist of a single hadron, collimated jets, or simply sprays of radiation confined to a particular angular region. We refer to this as the small invariant mass (or collimated jet) condition. 
Additionally, we must impose a rapidity gap between the particles in the regions $X$ and $Y$. Finally, we must add further constraints that enable us to distinguish between diffraction and gapped hard scattering. 

First, let us discuss the small invariant mass conditions.  
A minimal condition is that 
\begin{align}\label{eq:collimated-jets}
	&\LQCD^2 \lesssim m_X^2 \ll W^2\,,
	&&\LQCD^2 \lesssim m_Y^2 \ll W^2\,.
\end{align}
In coherent diffraction with $Y$ a proton and $m_Y^2\sim \LQCD^2$, we automatically saturate the lower bound of \eq{collimated-jets}, whereas in incoherent diffraction $Y$ is a multiparticle final state (such as a jet) and so $\LQCD^2 \ll m_Y^2$. Likewise, if $X$ is a single particle 
(such as a meson), we can saturate the lower bound $\LQCD^2 \sim m_X^2$, but if $X$ is a jet, then $\LQCD^2 \ll m_X^2$. 
A more stringent requirement that we will see shortly becomes applicable in the case of gapped hard scattering is to restrict the invariant masses to be smaller than that of the virtual photon  $Q^2$:
\begin{align}\label{eq:collimated-jets2}
	&\LQCD^2 \lesssim m_X^2 \ll Q^2\,,
	&&\LQCD^2 \lesssim m_Y^2 \ll Q^2\,.
\end{align}
Note that we in general do not impose this constraint on diffraction.
We will discuss the different situations that \eqs{collimated-jets}{collimated-jets2} describe shortly.

Second, we require that $p'$ and $p_X$ are separated by a rapidity gap in the lab frame, or equivalently, in any frame related to it by a boost along the collision axis, such as the $e^-$-$p$ CM frame. We discuss these frames in \app{epCMframe}. 
Rapidity is defined as $\mathcal{Y} = \frac{1}{2}\ln (p^-/p^+)$
for a particle of momentum $p$, and so we must have
\begin{align}\label{eq:rapidity-gap}
\gamma^2\frac{\xbar-xy}{y(1-z)}= \frac{p_X^{{\rm Lab}-}}{p_X^{{\rm Lab}+}}
\quad  \ll \quad 
\frac{p^{\prime {\rm Lab}-}}{p^{\prime {\rm Lab}+}}= \gamma^2 \frac{1-\xbar }{yz} 
\end{align}
for the hadronic systems $X$ and $Y$ to be well separated. Here the boost factor $\gamma = \sqrt{E_p/ E_e}$, with $E_p$ and $E_e$ the proton and electron energies; $\gamma = 1$ in the $e^-$-$p$ CM frame.

\paragraph{Distinguishing $X$ and $Y$.}
There are several distinct methods used in the experimental literature  on single-gapped inclusive diffraction  to impose a gap, and to specify which radiation belongs to the forward subsystem $Y$ and central subsystem $X$, as reviewed in \refscite{Arens:1996xw, Newman:2005wm,Lohr:2008zz, Monfared:2011xf}. 
These methods account for the fact that one can only experimentally detect particles in the gap above some minimum energy threshold, and that detector capabilities depend on the rapidity ${\cal Y}$ or equivalently the pseudo-rapidity $\eta$.
For example, the EIC will have a main detector spanning rapidities of approximately $-4 < \eta_{\rm Lab} < 4$, with more limited and targeted capabilities in the far forward and backward regions. Methods may also be of historical relevance when detector capabilities change with time, for example, HERA could not directly observe coherent diffraction until installing its forward proton spectrometer in 1995~\cite{H1:2012pbl}.

The primary methods are as follows~\cite{Newman:2005wm,Lohr:2008zz}.
First, there is the ``$m_X$ method,'' which exploits differences in the expected functional forms of diffractive and non-diffractive cross section distributions $d\sigma/dm_X^2$ to fit data and subtract off non-diffractive contributions. 
This uses the fact that
diffraction dominates the distribution at small $m_X^2$ values, 
since a small $m_X^2$ can force $Y$ into the forward region. 
For example, say we define $X$ to include all events in a rapidity range $|\eta_{\rm Lab}| < \eta_X^{\rm max}$ with a large $\eta_X^{\rm max}$ chosen by the upper limit of the central detector. Small $m_X$ ensures the radiation in this range is collimated, giving a gap between it and the remaining forward radiation in $Y$ which occurs for $\eta_{\rm Lab}>\eta_Y^{\rm min} = \eta_X^{\rm max}$. 
Second, in the ``rapidity gap method'', we require that any radiation in the rapidity gap region $\eta_{\rm min} < \eta_{\rm Lab} < \eta_{\rm max}$ has energy below the detection threshold, $E_{i,\rm Lab} < E_{\rm gap}$. At HERA, typically $E_{\rm gap}\sim 0.4\,{\rm GeV}$ \cite{ZEUS:2008xhs}. Here, the farthest-forward particle present in $X$ that is detected above this threshold has rapidity $< \eta_{\rm min}$. For the coherent case, $\eta_{\rm max}$ is the pseudo-rapidity of the forward proton, which is the only particle in $Y$. For the incoherent case, the particles in $Y$ have $\eta_{\rm Lab}>\eta_{\rm max}$, and the rapidity gap constraint restricts the particles in $X$.
Third, for the coherent case, we can carry out reconstruction of the scattered proton with far-forward detectors, 
which we refer to as the ``coherent forward detector method''.
In this case, diffractive events are selected by imposing an additional cut on $x$,  such as the bound $1-x_L=x/\beta \le 0.01$, which was used at HERA. 
In \sec{gap-radiation}, we discuss how the method used to ensure a rapidity gap influences the theoretical description.

In both the $m_X$ and rapidity gap methods, the cutoff $\etalabcut$ on rapidities of particles in Y
(equal to $\eta_X^{\rm max}$ or $\eta_{\rm max}$) places a constraint on the momentum $\tau$ exchanged between the $X$ and $Y$ systems.
One choice to ensure that the particles from the incoming proton are forward-scattered into $Y$, is to impose a hierarchy between the cutoff and rapidity of $p'$,
\begin{align}  \label{eq:Yandetacut}
\frac{p^{\prime {\rm Lab}-}}{p^{\prime {\rm Lab}+}}= \gamma^2 \frac{1-\xbar }{yz} = \frac{\gamma^2 s (1-\xbar)}{m_Y^2-t} \gg e^{2\etalabcut}\,.
\end{align}
For a given $\etalabcut$ this constrains the values of $m_Y^2-t$ that we can consider.

\paragraph{Gapped hard scattering versus diffraction.}
As illustrated in \fig{diffraction-chart}, we must be careful to distinguish our diffractive measurements from other gapped processes.
An important set of such events that are well studied in the collider physics community are exclusive jet cross sections, which arise from hard scattering and produce several jets with rapidity gaps of constrained radiation between them (see Refs.~\cite{Dasgupta:2003iq, Kang:2012zr, Kang:2013nha} for $e^-p$ examples).  Gapped hard scattering events satisfy  \eqs{collimated-jets2}{rapidity-gap}, and have a jet-veto constraint that restricts the amount of radiation between the jets. A factorization formula for gapped hard scattering was derived in \refcite{Kang:2013nha}, using a measurement of the 1-jettiness event shape $\tau_1$.\footnote{In DIS, $N$-jettiness refers to $N$ jets in addition to one beam jet. Therefore, 1-jettiness refers to a single jet $X$ in addition to the beam jet $Y$, creating two jets with a gap between them.} In the regime $\tau_1 \ll 1$, 1-jettiness roughly corresponds to a normalized sum of jet masses $\tau_1\simeq (m_Y^2 + m_X^2)/Q^2$. The 1-jettiness measurement constrains the radiation in the rapidity gap by measuring $m_X^2$ and $m_Y^2$, placing an upper bound on the energy of $m_X^2/Q \sim m_X^2/s \ll 1$. This approach to imposing a rapidity gap is similar in spirit to the $m_X$ method used in the diffraction literature.

The key distinction between gapped hard scattering and diffraction 
is the scaling of $x$ and $\beta$, as well as the manner in which 
the constraint $m_X^2\ll W^2$ is satisfied. Recall that $m_X^2=Q^2(1-\beta)/\beta+t$ and $W^2=Q^2(1-x)/x$.
In gapped hard scattering, the hadronic system and central jet are connected by a hard-scattering vertex, which takes $x\sim 1$, and 
$m_X^2\ll W^2$ is satisfied by having $(-t)\ll Q^2 \sim s$ 
and $\beta\to 1$, i.e. $(1-\beta) \ll 1$.   
In contrast, diffraction (and quasi-diffraction) are forward scattering processes, for which%
\footnote{To confirm that $x\ll 1$ and $(-t)\ll W^2$ correspond to forward scattering of the $p_X$ and $p'$ systems, we note that these conditions plus \eq{collimated-jets} imply $z= (1-x)\big({m_Y^2}/{W^2}-{t}/{W^2}\big)\ll 1$. Furthermore, $m_X^2/W^2 = \frac{x(1-\beta)}{\beta(1-x)} + t/W^2\ll 1$, which implies that $x\ll \beta$. We then examine \eq{four-vectors} and see that $\tau^+ \ll p_X^+$ since $z\ll 1$, that $\tau^-\ll p^{\prime\,-}$ since $x\ll 1$, and that $\vec\tau_\perp^{\,2} \gg |\tau^+\tau^-|$ since $z\lesssim -t/W^2$ and $x\ll \beta$. Thus, we have forward scattering kinematics. }
\begin{align}\label{eq:forward-scatter}
&x \ll 1\,,
&&-t \ll W^2\,.
\end{align}
Here we have $Q^2 \ll W^2$, and we satisfy \eq{collimated-jets} even when $m_X^2\sim m_Y^2 \sim Q^2$. That is, a diffractive process always obeys \eq{collimated-jets} but need not satisfy the stricter constraint \eq{collimated-jets2}; in constrast, gapped hard scattering always obeys \eq{collimated-jets2}.\footnote{It could be interesting to consider diffractive scattering with $x\ll 1$ together with  a threshold hard-scattering limit $(1-\beta)\ll 1$, which combines some aspects of the gapped hard scattering with the diffractive scattering. We do not do so here.}

Lastly, we note that diffractive scattering can also involve a hard scattering induced by the photon if we require an additional hierarchy $-t \ll Q^2 \ll W^2$, without imposing any additional hierarchy between $m_X, m_Y,$ and $Q$. 
This is not the same as gapped hard scattering, as it still involves forward scattering between $X$ and $Y$, satisfying \eq{forward-scatter}.
As we discuss later in~\sec{smalllambdat}, this is related to the hard-collinear factorization of diffraction in terms of diffractive parton distribution functions~\cite{Berera:1994xh, Berera:1995fj,Collins:1997sr}.

Altogether, the constraints in \eq{collimated-jets}, \eq{rapidity-gap}, \eq{forward-scatter},
and a potential further restriction on radiation in the rapidity gap specify the kinematics of diffraction (and quasi-diffraction).
In \sec{diffract-eft}, we discuss how these constraints enter when defining an effective field theory for diffraction.

\begin{figure}[t!]
	\begin{center}
		\includegraphics[width = 6 in]{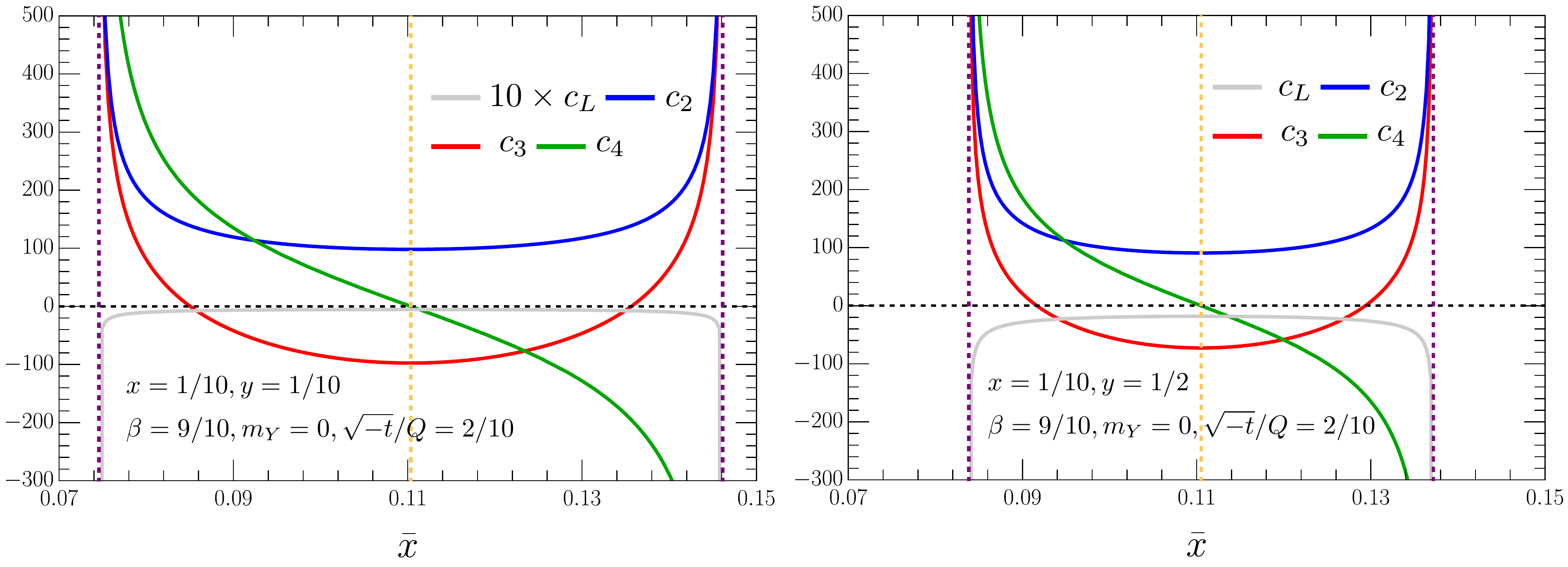}
		\caption{The $\xbar$ dependence of the coefficients $c_i$ of $F_i^D$ in \eq{diffcross} for coherent diffraction with $m_Y=0$.
We fix $(\sqrt{-t}/Q,\beta,x) = (0.2,0.9,0.1)$ in both plots, and take $y=\{0.1,0.5\}$ in the left and right panels, respectively. The coefficients $c_3$ and $c_4$ vanish when integrated over the full range of $\xbar$, whose size is proportional to $\Delta \approx \sqrt{-t}/Q$ and is centered at the point $x/\beta - (2 - y)xz$, which is plotted as a gold vertical dotted line. The purple dotted vertical lines indicate the lower and upper bounds of $\bar{x}$. The coefficients of $F_{3}^D$ (green) and $F_4^D$ (red) are large relative to that of $F_L^D$ (gray) at small $y$, which is shown at 10$\times$ scale in the left panel. }
		\label{fig:strcoeff}
	\end{center}
\end{figure}

\subsection{Experimental measurement of $F_3^D$ and $F_4^D$}
\label{sec:experiments-xbar}

The vast majority of the diffraction literature focuses only on $F_{2,L}^D$~\cite{Frankfurt:2022jns} and has paid very little attention to the structure functions $F_3^D$ and $F_4^D$. 
For example, most H1~\cite{H1:1995cha, H1:1997bdi, H1:2006uea, H1:2006zyl, H1:2007oqt, H1:2011jpo, H1:2012pbl, H1:2012xlc} and ZEUS \cite{ZEUS:1995sar, ZEUS:1996bqn, ZEUS:1997fox, ZEUS:1998suu, ZEUS:1998rvb, ZEUS:2002cih, ZEUS:2005vwg, ZEUS:2008xhs, ZEUS:2009uxs} studies
only consider $F_{2,L}^D$, provide measurements of $F_2^D$, and discuss the difficulty of extracting $F_L^D$.\footnote{Though we emphasize that measurements of $F_L^D$ do indeed exist; see \refcite{H1:2011jpo}.}
Notable exceptions include the theory work in \refscite{Arens:1996xw,Blumlein:2001xf,Blumlein:2002fw}, as well as a ZEUS measurement in \refcite{ZEUS:2004luu}, which found $F_{3,4}^D$ to be consistent with zero within experimental uncertainties for a certain subset of phase space. 
Here, we discuss prospects for measuring $F_{3,4}^D$ in archival HERA data and at the future EIC, highlighting the regions of phase space and binning needed to see these structures.
Unfortunately, more recent literature such as the EIC yellow book report or recent reviews on diffraction~\cite{AbdulKhalek:2021gbh,Frankfurt:2022jns} have also not discussed $F_3^D$ and $F_4^D$. 

We adopt a shorthand notation for the diffractive cross section in \eq{diffcross} by writing 
\begin{align} \label{eq:defineci}
\frac{d\sigma}{dx\,dQ^2\,d\beta \,dt\, d\xbar\, dm_Y^2} 
= \frac{\alpha^2}{Q^4} \sum_{i} c_i(y,\bar x,x,Q,\beta,t,m_Y) \: F_i^D\!(x,Q,\beta,t,m_Y)
\,.
\end{align}
In particular, we note that $\xbar$ and $y$ are purely leptonic variables, as is apparent from their $k$-dependence in \eq{lorentz-invariants}. Thus, they can be arguments of $c_i$ through contraction with the leptonic tensor, as shown in \eq{unpolarized-coefficients}, but do not appear in the hadronic structures $F_i^D$. Importantly, $c_{3,4}$ are the only coefficients with nontrivial dependence on $\xbar$  (i.e., beyond the $\xbar$ in the overall Jacobian factor $N_\sigma$). To estimate the relative contributions of each $F_i^D$ to the cross section, we can compare their coefficients, assuming the $F_i^D$'s themselves are roughly the same size.  In \fig{strcoeff}, we plot $c_i$ for typical values of a diffractive event, including for both a small-$y$ and a large-$y$ value. 
Importantly, $c_{3,4}$ are comparable in size to $c_2$, suggesting that $F_{3,4}^D$ can comprise a comparably large portion of the differential diffractive cross section as $F_2^D$, if $\bar x$ is measured. 
Note that picking a different choice of kinematics, such as $x$ and $-t/Q^2$, predominantly shifts the location of the vertical purple dashed lines and the scales on the  horizontal and vertical axes in the figure. For example, for $(x,\beta,\sqrt{-t}/Q^2)=(0.001,0.01,10)$, we have $0.086 <\bar x< 0.113$ and a very similar relative location of the curves.
We discuss the size of $c_i$ and $F_i^D$ in greater detail in \sec{predictions}.

To study $F_{3,4}^D$, we need experimental measurements that can resolve the value of $\xbar$ in multiple bins. As discussed, $\bar{x}$ is a Lorentz-invariant variable. It is also related to the Breit frame angle $\phi$, as shown in \eq{cosine}. The coefficients $c_3$ and $c_4$ correspond to the $\cos(2\phi)$ and $\cos(\phi)$ asymmetries, as discussed in Sec.~\ref{sec:photon-helicity-basis}, and thus in any frame $F_3^D$ and $F_4^D$ could (for example) be resolved by integrating the cross section weighted with $\cos(\phi)$ and $\cos(2\phi)$, respectively (with $\phi$ expressed in terms of $\xbar$).\footnote{The antisymmetric structure functions $F_{4A}^D$ and $F_{4AP}^D$ arising from a polarized lepton beam also come with a $\sin(\phi)$ asymmetry, as seen in \app{photon}. For convenience, we only consider $F_3^D$ and $F_4^D$ here. Note that the discussion here about $\bar{x}$ resolution here also applies to the antisymmetric structure functions.}

\begin{figure}
	\begin{center}
		\includegraphics[width = 2.75 in]{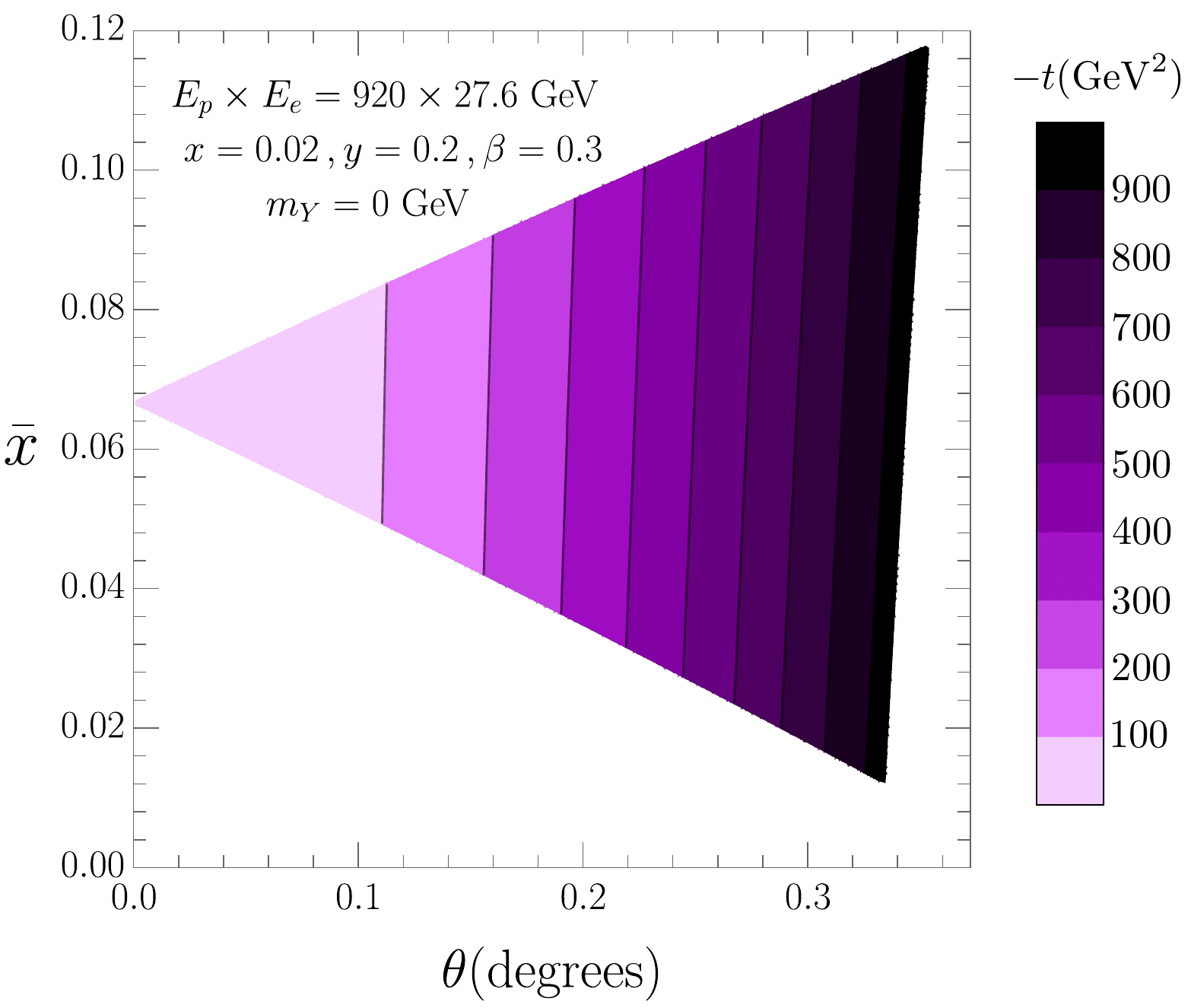}\qquad 
		\includegraphics[width = 2.8in]{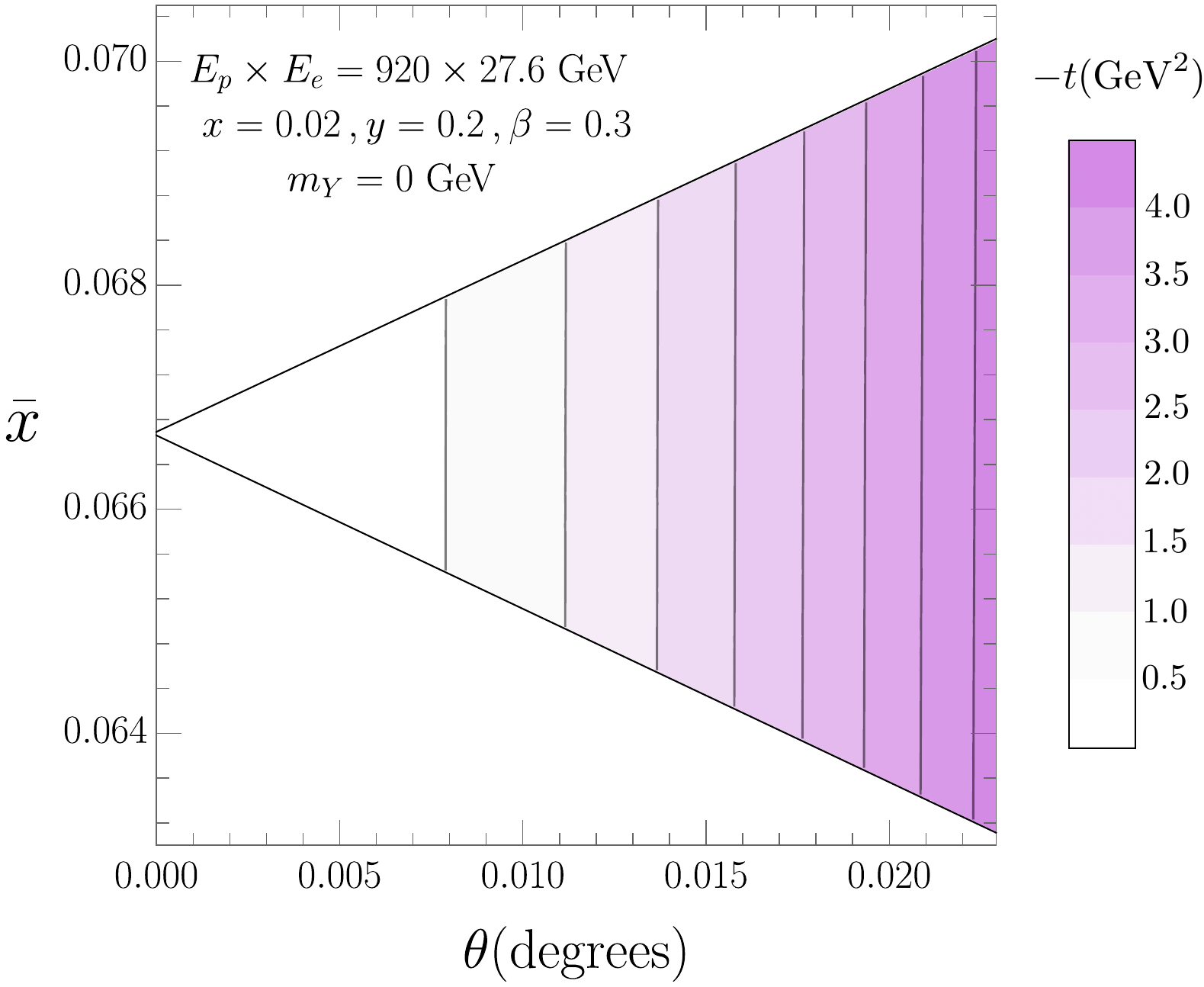}
	\end{center}
	\caption{Range of allowed values for $\xbar$  at HERA relative to the polar angle $\theta$ measured in the lab frame, with shading indicating the corresponding value of $|t|$. In both panels we fix  $(x,y,\beta,m_Y) = (0.02,0.2,0.3,0)$  and the beams carry HERA energies $(E_p,E_e) = (920,27.6)$ GeV. The left panel shows the full range of possible $t$ values. The right panel zooms in to the small-$t$ regime that was the focus at HERA. Note that a finer $\theta$-resolution is required to resolve $\bar{x}$ at smaller $-t$. 
	}\label{fig:xbartheta}
\end{figure}

\begin{figure}
	\begin{center}
		\includegraphics[width = 2.7 in]{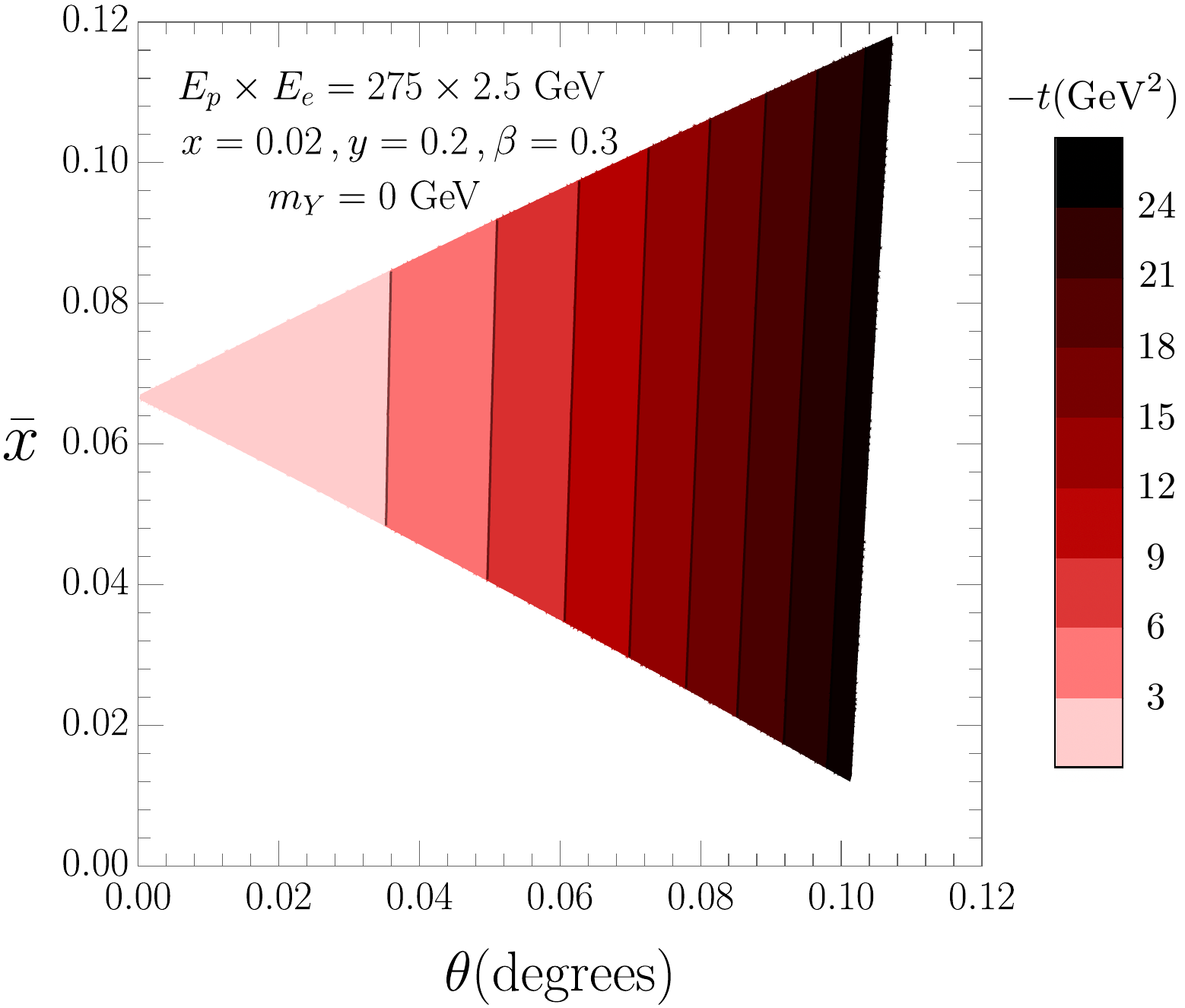}\qquad
		\includegraphics[width = 2.7 in]{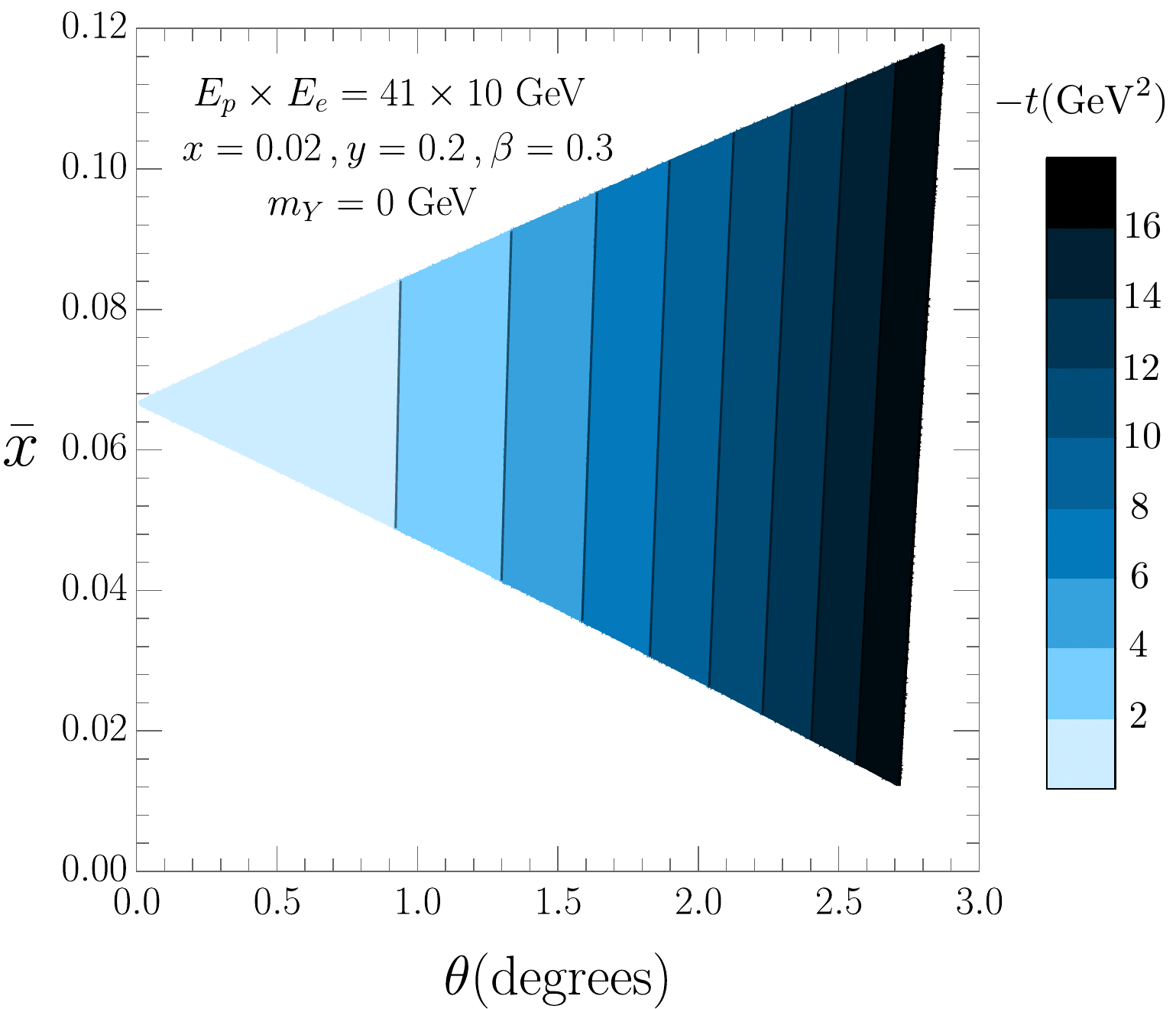}
	\end{center}
	\caption{Range of allowed values of $\xbar$ at the EIC relative to the polar angle $\theta$ measured in the lab frame, with shading indicating the corresponding value of $|t|$. In both panels we fix $(x,y,\beta,m_Y) = (0.02,0.2,0.3,0)$. In the left panel the beam energies are $(E_p,E_e) = (275,2.5)$ GeV, while in the right panel they are $(E_p,E_e) = (41,10)$ GeV. Importantly, the right panel provides a parameter choice for which the outgoing hadronic state $Y$ is at an angle $\theta > 2^{\circ}$ that is observable in the main EIC detector.  
	}\label{fig:xbar-theta2}
\end{figure}

We now discuss the experimental capabilities required to resolve $\bar{x}$, and provide a potential explanation for why the measurement in~\refcite{ZEUS:2004luu} may not have been able to resolve a nonzero $F_{3,4}^D$.
In the lab frame, for diffractive kinematics, we can express $\xbar$ as
\begin{align}\label{eq:xbar-theta}
	\bar{x} &= 1 - \frac{m_Y^2-t}{s\gamma^2 \tan^2 \frac{\theta}{2} +m_Y^2}\,,
\end{align}
where $\gamma = \sqrt{E_p/E_e}$, $s = 4E_pE_e$, and $\theta$ is the polar angle of $p'$ relative to the beamline.
In \app{epCMframe}, we provide the derivation of this formula and explain the approximations it depends on, which are valid for diffractive scattering.
We can use the bounds in \eq{xbar-bound} to determine the values $\xbar$ can take on at any fixed $(x,y,Q^2,t,\beta,m_Y)$; notably, these values are centered at $(x/\beta - (2-y)xz)$ and have a width of $4x\Delta$, with $\Delta$ given in \eq{Delta}. 

In \figs{xbartheta}{xbar-theta2}, we plot the range of $\xbar$ values accessible at a given polar angle $\theta$, when we vary over the range of all possible $t$ values allowed by the bounds in \eq{simp-t-bounds}. We fix the other Lorentz invariants to $(x,y,\beta,m_Y) = (0.1,0.2,0.3,0)$, and consider beam energies within experimental scenarios: $40 < E_p < 275$ GeV and $2.5 < E_e < 18$ GeV for the EIC, as well as $460 <E_p  <920$ GeV and $E_e = 27.6$ GeV for HERA.
Many parameter choices achieve $\xbar$ ranges in which $F_{3,4}^D$ are both resolvable and interesting. 

Generally, we need finer $\theta$-resolution to observe multiple $\xbar$ bins when we decrease the value of $(E_e, x,y,m_Y^2,-t)$ or increase $(E_p,\beta)$, due to a combination of \eq{xbar-theta} and the bounds on $t$ and $\beta$ in \sec{bounds}. In \fig{xbartheta}, we illustrate the increasing difficulty of measuring multiple $\xbar$ bins (and thus measuring $F_{3,4}^D$) as we decrease the value of $|t|$ at characteristic HERA beam energies. As we lower $|t| \lesssim $ a few GeV, a case often examined in the literature, we must have sub-milli-radian resolution of $\theta$. 
This is a potential explanation for why ZEUS did not observe $F_{3,4}^D$, as \refcite{ZEUS:2004luu} studied events with
$0.075\,{\rm GeV}^2 \leq -t \leq 0.35 \,{\rm GeV}^2$
and  $2$ GeV $ \leq Q \leq 10$ GeV for their high-$Q$ bin.  
In \fig{xbar-theta2}, we illustrate two different EIC beam energy choices, both of which give significantly greater access to $F_{3,4}^D$. Importantly, for some energy choices, $\theta$ is so large as to be measurable in the main detector, which reaches angles as low as $2^{\circ}$.
Given the strong emphasis in the EIC literature \cite{AbdulKhalek:2021gbh} on opportunities to improve existing measurements of $F_L^D$ \cite{H1:2011jpo}, the prospect of measuring the even more accessible and unexplored structure functions $F_{3,4}^D$ is quite exciting.

\section{Effective Field Theory for Diffraction and Quasi-Diffraction}\label{sec:diffract-eft}

A key theoretical tool in QCD is factorization: given a complicated process involving multiple different energy scales, it is often possible to cleanly isolate these scales from one another and calculate the dynamics at each scale separately. Early work successfully carried out factorization of many important processes to all orders in perturbation theory by working at the level of arbitrary individual Feynman diagrams in QCD~\cite{Collins:1989gx, Collins:2011zzd}. As the range and precision of collider experiments rapidly progressed in the ensuing years, so too grew the need for new theoretical tools to handle increasingly complex processes. Effective field theory (EFT) is a technique that organizes multiscale dynamics at the level of fields, rather than Feynman diagrams. To form an EFT, we power-expand the QCD Lagrangian, leading to a new EFT Lagrangian comprised of only the modes relevant for the problem at hand, which simplifies factorization, perturbative calculations, resummation of logarithms, and analysis of power corrections in complicated problem. Often, a single EFT is applicable to a wide swath of processes.  We emphasize that no information from QCD is lost in this technique: because an EFT is a power expansion, it is systematically improvable, to arbitrary precision. 

\subsection{Overview of Glauber SCET}

In a particle collision, much of the momentum is concentrated collinear to the beamline in the forward and backward directions, though collisions can also produce soft (low-energy) particles radiating away from the beamline in arbitrary directions. The soft-collinear effective theory SCET~\cite{Bauer:2000ew, Bauer:2000yr, Bauer:2001ct, Bauer:2001yt} with Glauber operators~\cite{Rothstein:2016bsq} is an EFT that helps theorists cleanly separate these soft and collinear dynamics, and has enabled precision calculations of many processes \cite{Gross:2022hyw}, such as SIDIS \cite{Ebert:2021jhy}, event shapes~\cite{Bauer:2008dt, Becher:2008cf,Abbate:2010xh,Stewart:2010tn}, jet production and substructure \cite{Kang:2020xyq,Lee:2024icn,Fleming:2007qr, Schwartz:2007ib, Bauer:2008dt, Becher:2008cf, Abbate:2010xh,  Kelley:2010fn, Ellis:2010rwa, Bauer:2011uc,Feige:2012vc,Lee:2022uwt}, heavy particle production and decays \cite{Beneke:2002ph,Bauer:2002aj,Beneke:2003pa,Bauer:2001cu,Mantry:2003uz,Bauer:2004tj}, as well as forward scattering and Regge phenomena \cite{Moult:2017xpp, Bhattacharya:2021zsg, Moult:2022lfy}. In \sec{power-counting}, we explicitly show why a version of the Glauber \SCETa is the EFT relevant for diffraction. 

The SCET Lagrangian describes the interactions of particles carrying hard, collinear, soft and/or ultrasoft, and Glauber-scaled momenta, whose precise definitions we provide in \tab{scet-modes}. To formulate a SCET Lagrangian, we integrate out all modes from QCD that do not carry one of the above scalings. The fields that appear in the SCET Lagrangian are fermions and gluons carrying collinear and soft momenta; we give their notation in \tab{scet-fields}, and express their momenta in terms of a so-called power counting parameter $\lambda \ll 1$, which is often a ratio of energy scales. Note that in standard QCD, a single particle corresponds to a single field. In SCET, a given type of particle is described by multiple different fields, each corresponding to a different momentum scaling. Note also that a given process may have one or multiple collinear directions $n_i$. Two collinear directions are considered to be distinct if $n_1\cdot n_2 \gg \lambda^2$. Particles carrying (offshell) Glauber momenta manifest in the SCET Lagrangian as transverse momentum potentials in operators $\cO$ connecting collinear and soft modes. 
Note that here we discuss the dynamic EFT Lagrangian for Glauber exchange between SCET fields~\cite{Rothstein:2016bsq}. 
Historically, approaches in SCET have also been used with Glaubers as a background field, to couple to a dense medium~\cite{Idilbi:2008vm, DEramo:2010wup,Ovanesyan:2011xy,Ovanesyan:2011kn,
Vaidya:2020cyi,Vaidya:2020lih,Vaidya:2021mly,Vaidya:2021vxu,Mehtar-Tani:2024smp}.

\begin{table}\def\arraystretch{1.1}
\begin{center}
\begin{tabular}{|c|c|}
	\hline
	\rowcolor{WhiteSmoke} {\bf SCET mode} & {\bf Momentum scaling for $(n\cdot p,\nbar\cdot p,|\vec{p}_\perp|)$} 
    \\ \hline
	Hard & $\sqrt s (1,\,0,\,0)$\\
	Collinear & $\sqrt s(1,\,\lambda^2,\,\lambda)\,$ or $\sqrt s(\lambda^2,1,\lambda)$\\
	Soft & $\sqrt s(\lambda,\,\lambda,\,\lambda)$\\
	Ultrasoft & $\sqrt s(\lambda^2,\,\lambda^2,\,\lambda^2)$\\
	Glauber & $\sqrt s(\lambda^a,\,\lambda^b,\,\lambda)\,$, for $a+b>2$\\
	\hline
\end{tabular}
\end{center}
\caption{Modes in SCET. Here $\sqrt{s}$ is the overall energy of the process, and $\lambda \ll 1$ is the power-counting parameter, and is often a ratio of two momentum  scales.}\label{tab:scet-modes}
\end{table}

At leading power in $\lambda$, the SCET Lagrangian factorizes cleanly into terms describing hard ($h$), collinear quark ($n_i\xi$), collinear gluon ($n_i g$), and soft ($s$) fields individually. After making the BPS field redefinition~\cite{Bauer:2001ct}, the coupling of ultrasoft ($us$) fields can also be separated in leading-power Lagrangians.  In addition, there is a so-called Glauber ($G$) Lagrangian term that mediates interactions between collinear and soft particles:
\begin{align}\label{eq:scet-lagrangian}
\cL^{(0)}_{\rm SCET}(\xi_{n_i},\,\psi_s,\, A_{n_i},\, A_s)  
= \cL_h^{(0)} + \sum_{n_i} {\cal L}_{n_i}^{(0)}
+ \cL_s^{(0)} + \cL_{us}^{(0)} + \cL_G^{(0)}\,.
\end{align}
Here, the sum is taken over all distinct collinear directions $n_i$ appearing in a given process. The hard Lagrangian is expressed in terms of hard operators $\cO^h_i$ and Wilson coefficients $C^h_i$ as
\begin{align}\label{eq:hard-lagrangian}
\cL_h = \sum_i C_i^h \otimes \cO_i^h \,,
\end{align}
and the relevant leading-order operators $\cO_i^h(\{\xi_{n_i},\,A_{n_i}\},\, \psi_s,\,A_s,\cdots)$ depend on the process being studied (they will not be needed for our description of diffraction).
The collinear Lagrangian splits into pieces for quarks and gluons, ${\cal L}_{n_i}^{(0)}=\cL_{n_i\,\xi}^{(0)} + \cL_{n_i\,g}^{(0)}$.
The Lagrangian for a quark collinear to the direction $n$ takes the form:
\begin{align}\label{eq:collinear-quark-lagrangian}
\cL_{n\xi}^{(0)} = e^{-ix\cdot\cP} \xibar_n \Big( in\cdot D_n + i\slashed{D}_{n\perp}\frac{1}{i\nbar\cdot D_n}i\slashed{D}_{n\perp}\Big)\frac{\slashed\nbar}{2}\xi_n
\,.
\end{align}
The $n$-collinear gluon Lagrangian can be expressed as:
\begin{align}\label{eq:collinear-gluon-lagrangian}
\cL_{ng}^{(0)}  = \frac{1}{2g^2}\Tr\Big([iD^\mu_{ns},iD^\nu_{ns}][iD^{\mu}_{ns},iD^\nu_{ns}] \Big) + \cL_{g,gf}^{(0)}
\,.
\end{align}
The gauge-fixing term $\cL_{g,gf}^{(0)}$  is given in \refcite{Bauer:2001yt}. 
The (ultra)soft Lagrangian has the same structure as the full QCD Lagrangian, albeit in terms of SCET (ultra)soft fields:
\begin{align}\label{eq:soft-lagrangian}
&\cL_s^{(0)} = \psibar_s i\slashed{D}_s \psi_s - \frac{1}{4}G_{s\,\mu\nu}G^{\mu\nu}_s
\,,
&&\cL_{us}^{(0)} = \psibar_{us} i\slashed{D}_{us} \psi_{us} - \frac{1}{4}G_{us\,\mu\nu}G^{\mu\nu}_{us}\,.
\end{align}
Here, $G^{\mu\nu}_{(u)s}$ denotes the standard gluon field strength tensor, defined with (ultra)soft gluon fields $A_{(u)s}$. 
The Glauber Lagrangian \cite{Rothstein:2016bsq} is
\begin{align}\label{eq:glauber}
\cL_G^{(0)} = e^{-ix\cdot\cP}\sum_{n,\nbar}\sum_{i,j=q,g} \cO^{ij}_{ns\nbar} + e^{-ix\cdot \cP}\sum_n \sum_{i,j=q,g}\cO_{ns}^{ij}
\,,
\end{align}
where the Glauber operators connecting 2 and 3 rapidity sectors $\cO_{ns}^{ij}$ and $\cO^{ij}_{ns\nbar} $ are discussed in \sec{diffract-glaubers}. Since each ${\cal O}_h^i$ appears only once, the only term in the SCET Lagrangian that causes factorization violation is ${\cal L}_G^{(0)}$. A number of covariant derivatives appear in the SCET Lagrangians above:
\begin{align}
iD_n^\mu &= i\partial_n^\mu + gA_n^\mu 
\,, \nonumber\\
\partial_n^\mu &= \cP^\mu + \frac{\nbar^\mu}{2}in\cdot\partial
=  \frac{n^\mu}{2} \bar\cP + \cP_\perp^\mu + \frac{\nbar^\mu}{2}in\cdot\partial 
\,, \nonumber\\
in\cdot D_n &= in\cdot \partial + gn\cdot A_n 
\,.
\end{align}
Here, $\cP^\mu$ is a label derivative operator and $i\partial^\mu$ is the standard derivative operator, which pick out the large $\cO(1,\lambda)$ and small $\cO(\lambda^2)$ parts of a momentum, respectively. 
While \eq{scet-lagrangian} has many ingredients, the universal understanding of how these various fields communicate simplifies the derivation of factorization formulas in QCD. 

\begin{table}\def\arraystretch{1.1}
	\begin{center}
		\begin{tabular}{|c|c|c|}
			\hline
			\rowcolor{WhiteSmoke} {\bf SCET field} & {\bf Particle Type} & {\bf Scaling for $(n\cdot p,\nbar\cdot p,|\vec{p}_\perp|)$} 
			\\ \hline
			$\xi_{n}$ & Fermion collinear to the $n^\mu$ direction & $(\lambda^2,1,\lambda)$ \\
			$A_{n}$ & Gluon collinear to the $n^\mu$ direction & $(\lambda^2,1,\lambda)$ \\
			$\psi_s$ & Fermion with soft scaling & $(\lambda,\lambda,\lambda)$ \\
			$A_s$ & Gluon with soft scaling & $(\lambda,\lambda,\lambda)$ \\
			$\psi_{us}$ & Fermion with ultrasoft scaling & $(\lambda^2,\lambda^2,\lambda^2)$\\
			$A_{us}$ & Gluon with ultrasoft scaling & $(\lambda^2,\lambda^2,\lambda^2)$ \\
			\hline
		\end{tabular}
	\end{center}
	\caption{SCET fields. Note that a single type of particle (fermion or gluon) can be described by multiple different fields in SCET, depending on the momentum it carries. This differs from the QCD Lagrangian, where a single particle is described by a single field.} \label{tab:scet-fields}
\end{table}

\subsubsection{Glauber operators}\label{sec:diffract-glaubers}

The quark and gluon fields that appear in the SCET Lagrangian can be broken down further into building block operators, which we list in \tab{building-blocks}. The relationship between the building blocks and the fields in \tab{scet-fields} are:
\begin{align}\label{eq:building-block-relations}
	\chi_n &= W_n^\dagger \xi_n 
	&\cB_{n\perp}^\mu &= \frac{1}{g}\Big[W_n^\dagger iD_{n\perp}^\mu W_n \Big]\nonumber\\
	\psi_s^n &= S_n^\dagger\psi_s 
	&\cB_{s\perp}^{n\mu} &= \frac{1}{g}\Big[S_n^\dagger iD^\mu_{s\perp}S_n \Big]
\end{align}
where $W_n$ and $S_n$ are Wilson lines in the fundamental representation:
\begin{align}
	W_n &= 
	W_n(\nbar\cdot A_n)  =  P \exp\Big[ig\int_{-\infty}^0 ds\, \nbar\cdot A_n(x+\nbar s)\Big]
	\,,  \nonumber\\
	S_n &= S_n(n\cdot A_s)  =  P \exp\Big[ig\int_{-\infty}^0 ds\, n\cdot A_s(x+n s)\Big] \,, 
\end{align}
and $P$ indicates path ordering of color matrices.
It is also possible to define adjoint versions of the gluon building blocks, which we distinguish by using color indices:
\begin{align}\label{eq:adjoint-building-blocks}
	\cB_{n\perp}^\mu &= \cB_{n\perp}^{\mu A} T^A\,,
	&	\cB_{s\perp}^{n\mu}&= \cB_{s\perp}^{n\mu A} T^A\,.
\end{align}
There are likewise corresponding adjoint Wilson lines $\cS$ and $\cW$, defined by 
\begin{align}
	W_n^\dagger T^A W_n &= \cW_n^{AB} T^B\,,
	&S_n^\dagger T^A S_n &= \cS_n^{AB} T^B\,.
\end{align}

\begin{table}\def\arraystretch{1.2}
	\begin{center}
		\begin{tabular}{|c|c|c|}
			\hline
			\rowcolor{WhiteSmoke}  & {\bf Building block} & {\bf Scaling}
			\\ \hline
			Transverse label operator & $\cP_{\perp}^\mu$ &  $\lambda$ 
			\\ 
			Collinear gluon with Wilson line ${\cal W}_n$ & $\cB_{n\perp}^{\mu A}$& $\lambda$
			\\ 
			Collinear fermion with Wilson line $W_n$ & $\chi_n$ &  $\lambda$
			\\ \hline
			Soft gluon with ${\cal S}_n$ Wilson line& $\cB_{s\perp}^{n\mu A}$ & $\lambda$
			\\ 
			Soft gluon field strength & $G_s^{\mu\nu}$ & $\lambda^2$
			\\
			Soft fermion with $S_n$ Wilson line& $\psi_s^n$ & $\lambda^{3/2}$
			\\ \hline
		\end{tabular}
	\end{center}
	\caption{Building block operators in SCET and their scaling in the power counting $\lambda \ll 1$. The fundamental representation operators relate to the SCET fermion and gluon fields in \tab{scet-fields} by \eq{building-block-relations}. We also define adjoint versions of the gluon operators in \eq{adjoint-building-blocks}.}\label{tab:building-blocks}
\end{table}

We can now write down the Glauber operators that appear in \eq{glauber}. The two-rapidity-sector Glauber operators for a collinear direction $n$ are: 
\begin{align}\label{eq:2Rglauber-operators}
\cO_{ns}^{ij} &= \cO_{n}^{iB}\frac{1}{\cP_\perp^2}\cO_s^{j_{n}B}
\,,
\end{align}
where $i,j$ are taken as either $q$ or $g$. The corresponding two-rapidity operators for the combination $\nbar s$ are the same as above, albeit with $n\to \nbar$. In addition, there are three-rapidity-sector operators given by 
\begin{align}\label{eq:glauber-operators3}
&\cO_{ns\nbar}^{ij} = \cO_n^{iB}\frac{1}{\cP_\perp^2}\cO_s^{BC}\frac{1}{\cP_\perp^2}\cO_{\nbar}^{jC}\,.
\end{align}

We will see later that only the two-rapidity-sector ${\cal O}_{ns}^{ij}$ operators are needed for the analysis in this paper, and hence we only provide the constituent operators needed for their construction:
\begin{align}
\label{eq:2secop}
\cO_n^{qB} &=\chibar_n T^B \frac{\slashed \nbar}{2} \chi_n
\,,
& \cO_n^{gB} &= if^{BCD}\cB_{n\perp\mu}^C \Big[\frac14 \nbar\!\cdot\!(\cP\!+\!\cP^\dagger)\Big]\cB_{n\perp}^{D\mu} 
\,, \\
\cO_s^{q_nB} &= 8\pi\alpha_s(\psibar_s^n T^B \frac{\slashed{n}}{2} \psi_s^n)  
\,,
& \cO_s^{g_nB} &=8\pi\alpha_s i f^{BCD}\cB_{s\perp\mu}^{nC}\Big[\frac14 n\!\cdot\!(\cP \!+\! \cP^\dagger) \Big] \cB_{s\perp}^{nD\mu}
\,. \nn   
\end{align}
Note that here $\alpha_s$ is grouped with the soft operators because soft loops are associated with their renormalization group evolution~\cite{Rothstein:2016bsq}.

The Glauber Lagrangian freely sums over the quark and gluon operators, so for later convenience we also define operators which are their sums,
\begin{align}
& \cO_n^{B} = \cO_n^{qB} + \cO_n^{gB} \,,
&& \cO_s^{B} = \cO_s^{q_nB} + \cO_s^{g_nB} \,.
\end{align}	
This allows us to write
\begin{align} \label{eq:2Rglaubops}
\cO_{ns} \equiv \sum_{i,j=q,g} \cO_{ns}^{ij} &= \cO_{n}^{B}\frac{1}{\cP_\perp^2}\cO_s^{B} 
\,.
\end{align}

\subsection{Power counting for diffraction}\label{sec:power-counting}

To construct the EFT that describes a given physical process, one must first enumerate what scales can come into play and determine their relative magnitudes. This reveals what power expansions one can carry out, and in turn how to treat the process. 
In (quasi-)diffraction, there are a large number of scales, and a correspondingly significant number of relevant power counting parameters:
\begin{align}\label{eq:power-counting-params}
&\lambda = \frac{Q}{\sqrt{s}}\,,
&\lambda_t& = \frac{\sqrt{-t}}{Q}\,,
&\rho & = \frac{m_Y}{\sqrt{-t}}\,,
&\lambda_g& = \gamma \, e^{-\etalabcut} \,,
&\lambda_\Lambda& = \frac{\LQCD}{\sqrt{-t}}  \,.
\end{align}
In this section, we will see that the diffractive constraints in  \eq{collimated-jets}, \eq{rapidity-gap}, and \eq{forward-scatter} always impose that $\lambda \ll 1$. The other parameters in \eq{power-counting-params} have far more freedom to vary, leaving us with numerous cases to examine. 
The parameter $\lambda_g$ arises from having asymmetric beam energies $\gamma^2=E_p^{\rm lab}/E_e^{\rm lab}\neq1$ and a lab frame rapidity cut $\etaxylab$ that distinguishes the regions $X$ and $Y$, which we discuss in detail in \sec{gap-radiation}.

Let us derive the scaling of the $\lambda_i$'s (and in turn the scaling of the Lorentz invariants in \sec{lorentz}) from the ground up. Our first step is to note that we do not need to assign a scaling to the invariants $y$ and $\xbar$, as they only appear in cross section coefficients $c_i$, not in the structure functions $F_i^D$ that contain the QCD dynamics we will factorize.
To prevent superfluous factors of $\xbar$ and $y$ from appearing in our analysis, 
we work in the Breit frame.\footnote{This can be contrasted with the analysis of small-$x$ inclusive DIS in \refcite{Neill:2023jcd}, which was carried out in the $ep$ CM frame. There, it was convenient to assign a scaling $y\sim x\sim\lambda$ because $y$ appeared in the components $(q^+,q^-,q_\perp)$ in this frame, even though the final results were independent of this choice.}

Next, we impose the forward scattering constraint $x\ll 1$ in \eq{forward-scatter}. Using $xy = Q^2/s=\lambda^2$, and the fact that we do not assign a scaling to $y$ (effectively $y\sim 1$), we have
\begin{align}\label{eq:scaling1}
	x \sim \lambda^2 \ll 1 \,.
\end{align}
Combining this with $Q = \sqrt{sxy} \sim \lambda\, \sqrt{s}$ and  using \eq{supplemental-lorentz-invariants} implies 
\begin{align}
	W^2\sim  s \,,
\end{align} 
indicating that both $s$ and $W^2$ correspond to the hardest scale in (quasi-)diffraction. 
The other forward scattering condition $-t\ll W^2$ in \eq{forward-scatter}  implies that 
\begin{align}\label{eq:forward-scattering-condition2}
	&\frac{(-t)}{W^2} \sim \frac{(-t)}{s} = \lambda^2 \lambda_t^2  \ll 1
	\,,
\end{align}
which is trivially satisfied when $\lambda_t \lesssim 1$, but is a nontrivial constraint in diffractive events with $\lambda_t\gg 1$  (see e.g. the HERA measurements \refscite{ZEUS:1999ptu, Cox:1999hw, ZEUS:2002vvv, H1:2003ksk, H1:2006ogl, H1:2008jdn, ZEUS:2009ixm}). 

Next, we examine the implications of the small invariant mass conditions in \eq{collimated-jets}, $m_X^2,m_Y^2 \ll W^2\sim s$. Comparing the Lorentz invariants from \eqs{lorentz-invariants}{supplemental-lorentz-invariants} with the definition of our power counting parameters, we find that 
\begin{align}  \label{eq:lamtconstraints}
& \frac{m_Y^2}{s} = \lambda^2 \lambda_t^2 \rho^2  \ll 1  \,,
&&z \sim \lambda^2 \lambda_t^2 (1+\rho^2) \ll 1\,,
\end{align}
where we have used \eq{forward-scattering-condition2}. Finally, we use that $m_X^2/W^2= \frac{x(1-\beta)}{\beta(1-x)} + t/W^2\ll 1$ implies that 
$x\ll \beta/(1-\beta)$, and thus
\begin{align}\label{eq:beta-bound-x}
	\lambda^2 \sim x \ll \beta \,,
\end{align}
which holds for both $\beta\sim \lambda$ and $\beta\sim 1$. (We do not consider more exotic
scalings for $\beta$, like $\beta\sim \lambda^{3/2}$.)  
The kinematic restriction $m_X^2\ge 0$ implies an upper bound on $\lambda_t$ given by
\begin{align}  \label{eq:lamtbetaconstraint}
\lambda_t^2 \le (1-\beta)/\beta \,.
\end{align}
For $\beta\sim 1$, this bounds $\lambda_t \lesssim 1$, implying that the forward scattering constraint in \eq{scaling1} implies the one in \eq{forward-scattering-condition2}.
We could have a large $\lambda_t\gg 1$ while still ensuring $\lambda_t \ll \lambda^{-1}$ if we choose a small $\beta\sim \lambda^2 \ll 1$ or $\beta\sim \lambda_t^{-2}\ll 1$.
For example, we could have $\lambda_t^2 \sim \lambda^{-1} \gg 1$ and still satisfy \eq{forward-scattering-condition2}, which in this case is an independent condition. 
The $\lambda_t\ll 1$ case is most commonly studied in the diffraction theory literature, likely due to its relationship to hard-collinear factorization; however, diffraction does not require $\lambda_t \ll 1$. 

We also consider how the lower bounds in \eq{collimated-jets}, $\LQCD^2 \lesssim m_X^2, m_Y^2$, constrain $\lambda_\Lambda$. When $X$ is a jet, the condition $\LQCD^2\ll m_X^2$ gives $\lambda_\Lambda^2 \ll (1-\beta)/\beta - \lambda_t^2$. In contrast, if $X$ is a single particle with $m_X^2\sim \LQCD^2$, then
$\lambda_\Lambda^2 \sim (1-\beta)/\beta - \lambda_t^2$.
Likewise, when $Y$ is a multi-particle state (incoherent diffraction), we have $\lambda_\Lambda^2 \ll \rho^2\lambda_t^2$. When $Y$ is a proton (coherent diffraction), we have 
$\lambda_\Lambda^2\sim \rho^2\lambda_t^2$.

Note that these power counting results tell us that in the (quasi-)diffractive region, the hadronic system and central jet are always well separated by a rapidity gap. In particular, since $z\ll 1$ and $\xbar \sim x\ll 1$ from \eq{xbar-bound}, then the rapidity gap condition in \eq{rapidity-gap} in the lab frame becomes
\begin{align}\label{eq:rapidity-gap1}
 x \Big(\frac{\xbar}{x}-y\Big)  
\quad  \ll \quad 
  \frac{1}{z} 
\,,
\end{align}
which is satisfied.   
Applying instead the Breit frame results in \eq{four-vectors} to the calculation of rapidities, we see that
\begin{align} \label{eq:rapgap_pc}
	\frac{p_X^-}{p_X^+} \sim \frac{1-\beta}{\beta} \ll \frac{1}{x} \ll 
	\frac{p^{\prime -}}{p^{\prime +}} \sim \frac{1}{x z}
	\,,
\end{align}
which is also satisfied.
That is, we have rapidity gaps in both the Breit and the lab frames. 
We discuss the exact relation between rapidity gaps in the lab and Breit frames further in \app{epCMframe}.

Finally, we revisit our discussion from \sec{diffractive-constraints} and examine how the power counting parameters in \eq{power-counting-params} behave in gapped hard scattering. In this case, we power expand in $\lambda_t \ll 1$ 
and choose $Q^2\sim W^2\sim s$, which implies that $\lambda\sim 1$ and $x\sim 1$.
Additionally, this gapped hard scattering has two collimated jets (or hadrons) $X$ and $Y$. Therefore we have $m_Y^2/Q^2 = \lambda_t^2\rho^2 \ll 1$ and thus $z\ll 1$. However, the other collimated jet condition gives $m_X^2/Q^2 = (1-\beta)/\beta + t/Q^2 \ll 1$ and so here $\beta$ must be close to one, $(1-\beta)\ll 1$. Nonetheless, the hard scattering Lorentz invariants still satisfy the rapidity gap condition between jets in \eq{rapidity-gap}. In summary, the key distinction between gapped hard scattering and (quasi-)diffraction is that the former has $\lambda \sim 1$ and the latter has $\lambda \ll 1$.

\subsubsection{Scaling of momenta in the Breit frame}\label{sec:Breit-scaling}

Let us examine how momenta scale in (quasi-)diffraction. 
Our focus will be on the scaling in $\lambda$ and $\lambda_t$, though we will initially also track dependence on $\rho$ and $\beta$. 
To simplify our analysis, we take $\lambda_\Lambda\sim 1$, and thus do not initially try to distinguish the scales $-t$ and $\LQCD^2$. We can do so without loss of generality, as later on we will see that it is straightforward to consider further expansions in $\lambda_\Lambda$.
Likewise, we postpone discussion of the $\lambda_g$ expansion until the next section.  
Writing \eqs{kp-vectors}{four-vectors} in terms of the power counting parameters in \eq{power-counting-params}, we find
\begin{align}\label{eq:four-vector-scales}
k ,\,&k' \sim \sqrt{s} (\lambda,\lambda,\lambda) \,,
&&p\sim \sqrt{s} (0,\lambda^{-1},0) \,,
\\
&q \sim\sqrt{s}(\lambda,\,\lambda,\, 0) \,,
&&p' \sim\sqrt{s}\Big( \lambda^3 \lambda_t^2 (1+\rho^2) , \lambda^{-1}, \lambda \lambda_t	\Big)
\,,\nonumber\\
&p_X \sim \sqrt{s}(\lambda,\, \lambda/\beta,\,  \lambda\lambda_t)
 \,,
&&\tau \sim\sqrt{s}\Big(\lambda^3 \lambda_t^2 (1+\rho^2), \lambda/\beta, \lambda \lambda_t	\Big)
\,. \nonumber
\end{align}
Although we have kept factors of $\beta$  in \eq{four-vector-scales} for completeness, we henceforth simply take $\beta \sim 1$. It is clear that the modes in \eq{four-vector-scales} bear close resemblance to the SCET modes in \tab{scet-modes}. In particular, we see that $q$ and $p_X$ have soft scaling, $p$ and $p'$ have $n$-collinear scaling with $p_i^+ \ll p_i^\perp \ll p_i^-$, 
and $\tau$ has Glauber momentum scaling with $\tau^+\tau^- \ll \vec\tau_\perp^{\, 2}$.  Thus, Glauber SCET is the appropriate EFT for analyzing diffraction, and we can describe each particle or jet in terms of its corresponding field in \sec{diffract-glaubers}. 
Due to our frame choice we have a non-standard scaling for collinear momenta. This leads to a modified scaling for the $n$-collinear fields and building blocks, as shown in \tab{building-blocks2}. The soft fields still follow the counting shown in \tab{scet-fields}.

\begin{table}\def\arraystretch{1.2}
	\begin{center}
      \begin{tabular}{|c|c|c|}
			\hline
			\rowcolor{PeachPuff} {\bf SCET field}  & {\bf $(n\cdot p,\nbar\cdot p,|\vec{p}_\perp|)$} 
			\\ \hline
			$\xi_{n}$ &  $(\lambda^3,\lambda^{-1},\lambda)$ \\
			$A_{n}$ &  $(\lambda^3,\lambda^{-1},\lambda)$ \\
			\hline
		\end{tabular} \qquad
		\begin{tabular}{|c|c|}
			\hline
			\rowcolor{MistyRose}  {\bf Building block} & {\bf Scaling}
			\\ \hline
			 $\cP_{\perp}^\mu$, $\cB_{n\perp}^{\mu A}$ &  $\lambda$ 
			\\ 
			$\chi_n$ &  $\lambda^{1/2}$ 
            \\
 \hline
		\end{tabular}
	\end{center}
	\caption{Field and building block $\lambda$-scalings for the $n$-collinear modes in \eq{four-vector-scales}. These are modified from \tabs{scet-fields}{building-blocks} due to our frame choice here.}\label{tab:building-blocks2}
\end{table}

Since we have Glauber exchange between soft and $n$-collinear fields, the relevant operators are those in \eq{2Rglaubops}.
The derivation of a factorized (quasi-)diffractive cross section forms the focus of \sec{diffract-factorization}. 
Additionally, we remark that as we alluded in \sec{structure-functions}, it is now apparent that in diffraction, the $p_i^+$ and $p_i^\perp$ lightcone components of $p$ and $p'$ are not multiples of one another, implying that the assumption $p' = (1-\xi)p$ does not hold, which will become relevant in the full leading-power factorized description of diffraction.  

It is straightforward to find the $(+,-,\perp)$ scaling in  $\lambda$ for $U^\mu$ and $X^\mu$ in \eq{ux-vectors},
\begin{align}\label{eq:ux-scaling}
U^\mu&\sim (1,1,0) 
\,,
& X^\mu&\sim (0,0,1)
\,.
\end{align}
If we examine the diffractive cross section in \eq{diffcross} and compare the structure function prefactors $c_i$ to one another, we see that they are all the same size with respect to the hadronic power counting, $c_i/c_2 \sim {\cal O}(\lambda^0)$, because $y\sim 1$ and $k\cdot U\sim k\cdot X\sim Q\sim \sqrt{s}\lambda$.  
We also observed this when plotting $c_i$ in \fig{strcoeff}. We note that the hadronic structure functions $F_i^D(x,Q^2,\beta,t,m_Y^2)$ are independent of $\bar{x}$ and $y$. Thus, when we carry out an expansion of the structure functions using Glauber SCET, we can apply our results to any value of $y$ (e.g. $y=10^{-3}$ or $y=1/2$) and any value of $\bar x$, within their physical bounds. 
In contrast, working out the scaling of the $F_i^D$'s themselves is a dynamical question, which we can only answer after carrying out their factorization. 

\subsection{Ultrasoft radiation in the gap}
\label{sec:gap-radiation}

\begin{figure}
	\begin{center}
		\includegraphics[width = 2.9 in]{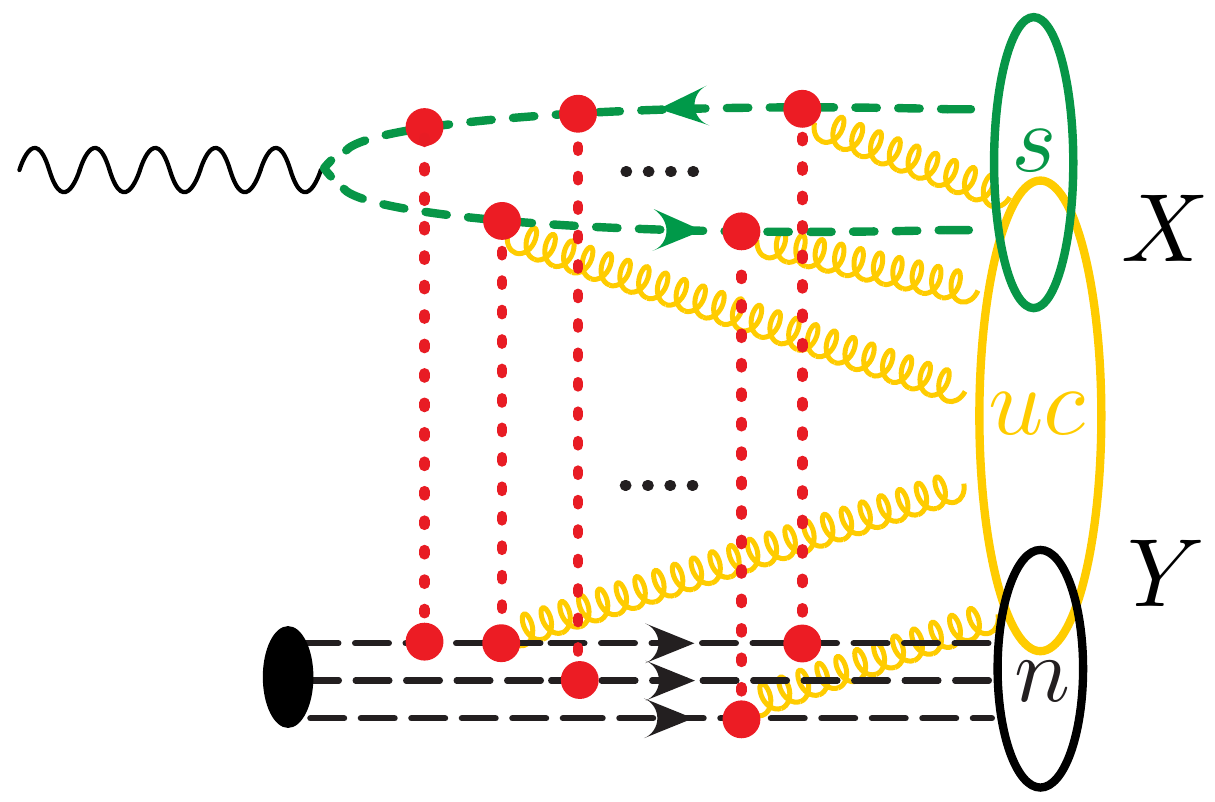}
	\end{center}
	\caption{Picture showing a typical quantum field theory amplitude with multiple Glauber exchange, with soft, $n$-collinear, and ultra-collinear (uc) radiation into hadronic states $X$, $Y$, and/or the gap.
	}\label{fig:gap-radiation}
\end{figure}

Next, we must add to our formalism a description of softer radiation that is below the detection threshold in the gapped region.
This radiation will affect theoretical predictions, particularly for quasi-diffraction.
Up until this point, we have focused on a hadronic system $Y$ with total momentum $p'$ and a central system $X$ with total momentum $p_X$. These systems contain forward and soft particles, respectively, but also can include low-energy particles, with momenta denoted by $p_{uc}$,
that contribute to both the gapped and non-gapped regions,
as shown in \fig{gap-radiation}. 
We use $p_n'$ to refer to the $n$-collinear radiation in $Y(p')$ that does not enter the gap, and $p_{X_s}$ to refer to the soft radiation in $X(p_X)$ that does not enter the gap. These scale in the Breit frame as:
\begin{align}\label{eq:four-vector-scales-new}
&p_{X_s} \sim \sqrt{s}(\lambda,\, \lambda/\beta,\,  \lambda\lambda_t)
\,,
&&p'_n \sim\sqrt{s}\Big( \lambda^3 \lambda_t^2 (1+\rho^2) , \lambda^{-1}, \lambda \lambda_t	\Big)
\,.
\end{align}
Note that the overall scalings for $p_X$ and $p'$ in \eq{four-vector-scales} remain correct.

The radiation with momentum $p_{uc}$, on the other hand, is most straightforward to analyze in the $\gamma_{\nbar}$-collinear frame, which is defined by the choice
\begin{align} \label{eq:gamma-nbar-frame}
&\bar q^\mu_{\gamma_{\nbar}\text{-frame}} 
  =   \sqrt{ys}\, \frac{\nbar^\mu}{2} - \frac{Q^2}{\sqrt{ys}}\, \frac{n^\mu}{2} \,,
&& \bar p^\mu_{\gamma_{\nbar}\text{-frame}} = \sqrt{ys}\, \frac{n^\mu}{2}\,.
\end{align}
This frame is related to the Breit frame in \eq{four-vectors} by a simple boost along the $z$-axis, with
\begin{align} \label{eq:Breit2gamcollin}
& \bar p_{i,\gamma_{\nbar}\text{-frame}}^+ =  \frac{p_{i,\text{Breit}}^+ }{ \sqrt{x}}
\,,
&& \bar p_{i,\gamma_{\nbar}\text{-frame}}^- =  \sqrt{x}\, p_{i,\text{Breit}}^-
\,.
\end{align}
Below we adopt a more compact notation: we drop the subscripts ``Breit'' and ``$\gamma_{\nbar}$-frame''  above,  
but put bars over momenta to indicate the $\gamma_{\nbar}$-collinear frame.
In \app{epCMframe}, we give further detail on the relationship between the Breit and lab frames; note that we explicitly include the subscript ``lab'' when referring to quantities in the lab frame. 

The key attribute of the $\gamma_{\nbar}$-collinear frame is that $X$ and $Y$
have large momentum components that are parametrically the same size, with $\bar p_{X_s}^+ \sim \bar p_n^{\,\prime -}\sim \sqrt{y s} \sim \lambda^0$.
\Eq{Breit2gamcollin} implies that in terms of the power counting parameters in \eq{power-counting-params}, modes in the $\gamma_{\nbar}$-collinear frame scale as follows (setting $\rho\sim 1$ for simplicity): 
$\bar q\sim \sqrt{ys}(1,\lambda^2,0)$ is $\nbar$-collinear, 
and so is $\bar p_{X_s} \sim \sqrt{ys}(1,\lambda^2/\beta,\lambda\lambda_t)$. 
We still have Glauber scaling for $\bar\tau\sim \sqrt{ys} (\lambda^2\lambda_t^2,\lambda^2/\beta,\lambda\lambda_t)$. 
The proton $p$ is $n$-collinear with $\bar p\sim \sqrt{ys}(0,1,0)$, as is 
$\bar p^{\,\prime}_n \sim \sqrt{ys}  (\lambda^2\lambda_t^2,1,\lambda\lambda_t)$. 
These mode identifications hold true for any $\lambda_t$, as $\lambda\lambda_t\ll 1$. In this frame, the relevant hard scale is $\sqrt{ys}$. 
The modes all scale proportionally to $\sqrt y$, and we again do not need to assign a power counting to $y$ itself.%
\footnote{Note that in the $\gamma_{\nbar}$-collinear frame, we could also factorize this process using  the $n$-$\nbar$ Glauber operators in \eq{glauber-operators3}; see also \refcite{Rothstein:2016bsq}. In this case the restriction on radiation in the gap (which allows for ultrasoft but not soft radiation) enables us to replace $\cO_s^{AB} \propto \cP_\perp^2 \delta^{AB}$, so that these operators reduce to an identical form as \eq{2Rglauber-operators}, just with a renaming of the fields. Thus, such an analysis does not differ from the above. Nevertheless, we find it more intuitive to use the Breit frame.}  

In the $\gamma_{\nbar}$-collinear  frame, $X$ and $Y$ are back-to-back, and the only relevant low-energy radiation that populates the gap is that which can modify the masses $m_X$ and $m_Y$. Taking this radiation to have momentum $\bar p_{uc}^{\mu}$ with $\bar p_{uc}^+ \ll \bar p_{X_s}^+$ and $\bar p_{uc}^-\ll \bar p_n^{\prime -}$, we have $m_X^2 \sim (\bar p_{X_s} + \bar p_{uc})^2 \sim \bar p_{X_s}^+ \bar p_{uc}^-\sim \lambda^2/\beta$ and $m_Y^2 \sim (\bar p_n^{\prime\,2}+\bar p_{uc})^2\sim \bar p_n^{\prime\,-} \bar p_{uc}^+\sim \lambda^2 \lambda_t^2\rho^2$, 
so $\bar p_{uc}$ is isotropic in its $\lambda$ scaling, with $\bar p_{uc}^-\sim \bar p_{uc}^+\sim\lambda^2$.
Thus, we take this radiation to be ultrasoft for its $\lambda$ scaling (and collinear for its $\lambda_t$ scaling)
\begin{align}
\label{eq:pucbar}
\bar p_{uc}^{\mu}&\sim \sqrt{ys} \bigl(\lambda^{2}\lambda_t^2, \lambda^{2},
    \lambda^{2}\lambda_t \bigr)\,.
\end{align}

\begin{figure}
	\begin{center}
 		\includegraphics[width = 2.9 in]{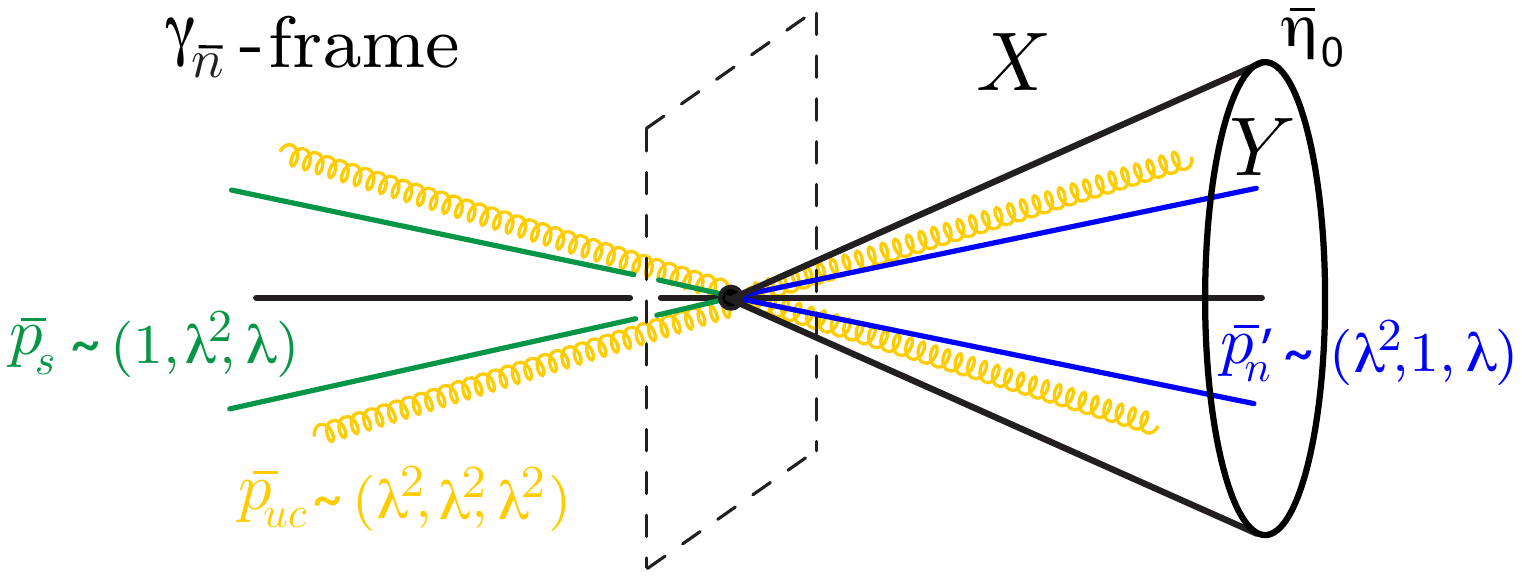}
\ 
 		\includegraphics[width = 2.9 in]{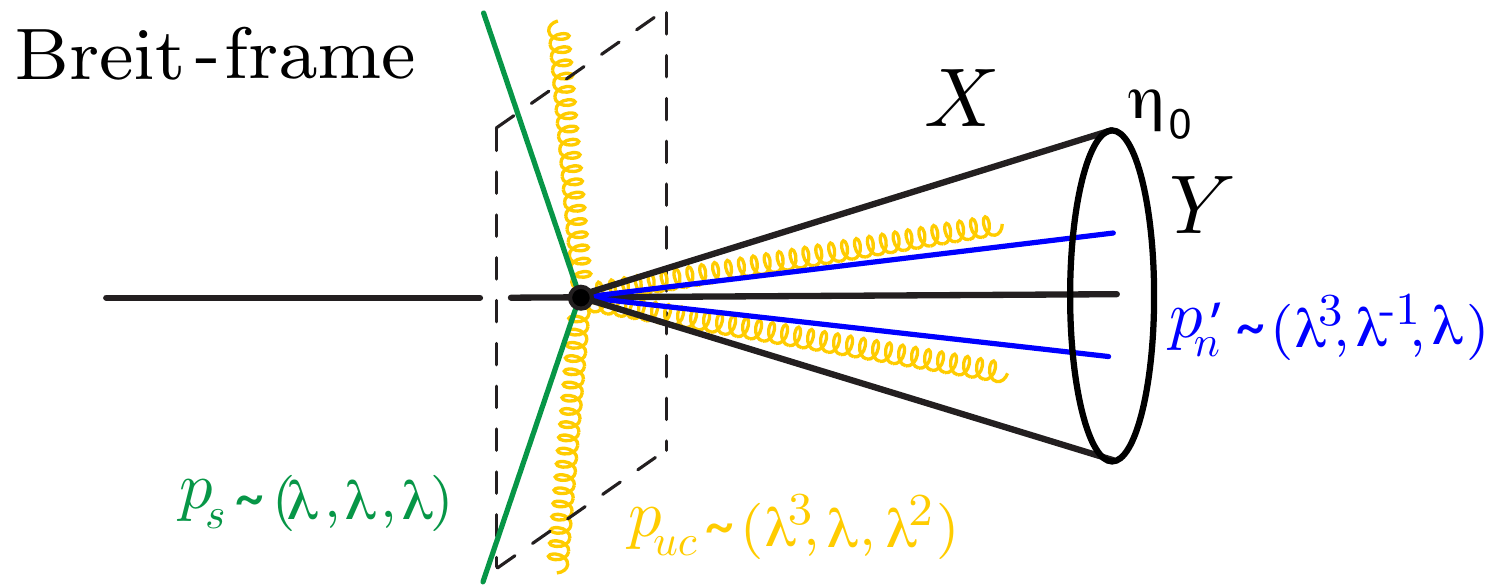}
	\end{center}
 \vspace{-0.6cm} (a)\hspace{7.1cm} (b) \\[-10pt]
\caption{
		Comparison of radiation in (a) the $\gamma_{\nbar}$-frame and (b) the Breit frame using the power counting in $\lambda$. The $n$- and $\nbar$-collinear directions points towards the positive and negative horizontal axes, respectively. The dashed boundary in the transverse plane separates two hemispheres. 
		In both frames, a rapidity cut (black cone) separates particles belonging to the hadronic system $Y$ (inside cone) and the central system $X$ (outside cone).  In the Breit frame, the system $Y$ contains forward hadronic particles $p_n'$ (blue), and the system $X$  contains central soft particles $p_s$ (green). Both $X$ and $Y$ can contain ultracollinear (uc) particles $p_{uc}$ (yellow) that can populate the gapped region below the experimental detection threshold. 
}
\label{fig:frame-radiation}
\end{figure}

We can use \eq{Breit2gamcollin} to transform our $\gamma_{\nbar}$-frame results back to the Breit frame discussed in \sec{Breit}.
In the Breit frame, radiation with momentum $p_{uc}^\mu$ carries an overall ultrasoft scaling $\lambda^2$, multiplied by a collinear boost $\sim \lambda$. 
We refer to this as ultrasoft-collinear (uc) scaling:
\begin{align}\label{eq:gap-radiation-Breit}
	p_{uc}^\mu \sim \sqrt{ys}\: \big( \lambda^3\lambda_t^2, \lambda, \lambda^2\lambda_t \big) \,,
\end{align}
where the collinear $\lambda$ scaling arises from $\sqrt{x}\sim \lambda$. We illustrate how these modes look in the Breit and $\gamma_{\nbar}$ frames in \fig{frame-radiation}.
The leading-power interactions among uc fermions and uc gluons are described by an ultrasoft Lagrangian expressed as a function of uc fields, i.e. simply replacing ${\cal L}_{us}^{(0)}[\psi_{us},A_{us}] \to {\cal L}_{uc}^{(0)}[\psi_{uc},A_{uc}]$ in \eq{soft-lagrangian}. 

\begin{figure}   
	\begin{center}
 		\includegraphics[width = 0.49\textwidth]{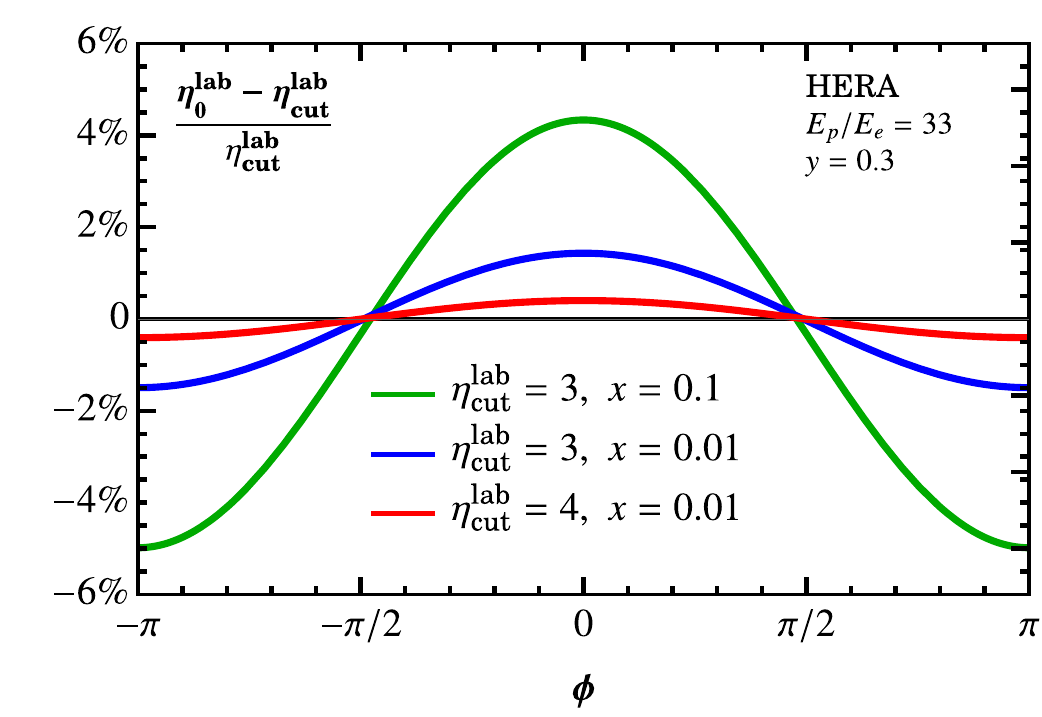}
 		\includegraphics[width = 0.49\textwidth]{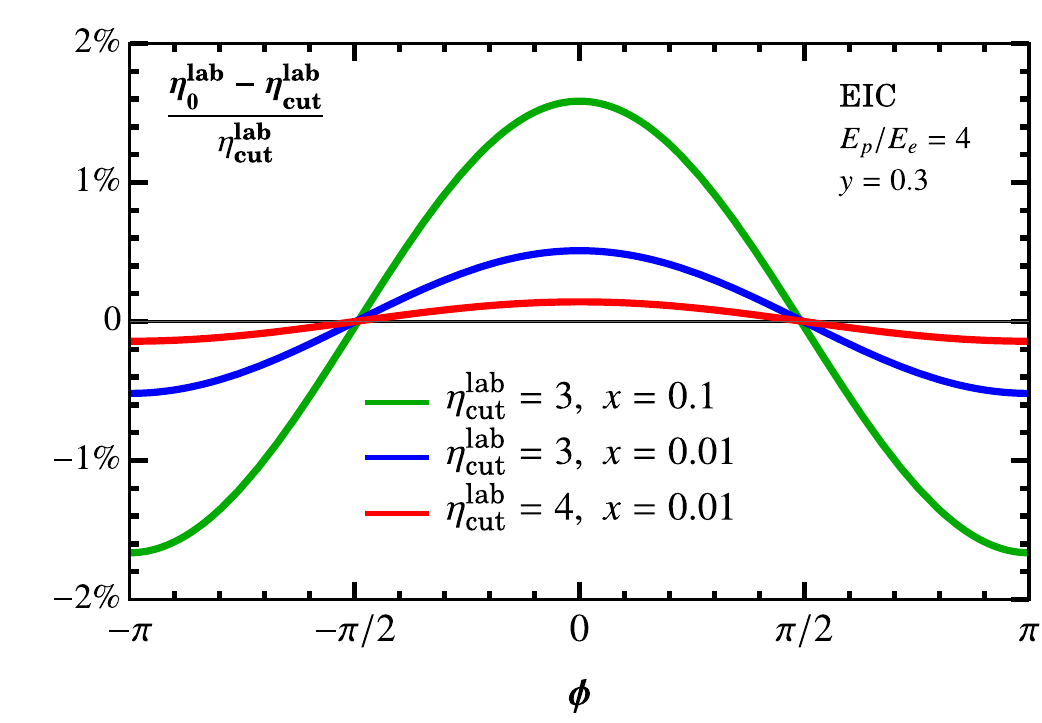}
	\end{center}
\caption{ 
Plot of $(\etaxylab-\etalabcut)/\etalabcut$ to compare the rapidity boundaries that define $X$ and $Y$ in the lab frame ($\etalabcut$) and the $\gamma_{\nbar}$-frame ($\etaxybar$ converted to a lab frame $\etaxylab$). For HERA kinematics we take $E_p=920\,{\rm GeV}$ and $E_e=27.6\,{\rm GeV}$ (left panel), while for EIC kinematics we choose $E_p=41\,{\rm GeV}$ and $E_e=10\,{\rm GeV}$ in order to minimize the ratio (right panel). 
	}\label{fig:etacut} 
\end{figure}

Next, we explore how our description of gap radiation relates to experimental methods for identifying diffractive processes.
\paragraph{Small $m_X$ method.}
Experimentally, it is natural to distinguish the regions $X$ and $Y$ by a rapidity cut $\etalabcut$ in the lab frame, taking $p_i^{\rm lab}\in X$ when $\eta_i^{\rm lab}<\etalabcut$, and $p_i^{\rm lab}\in Y$ when $\eta_i^{\rm lab}> \etalabcut$. This closely resembles the $m_X$-cut method for identifying diffractive processes discussed in \sec{diffractive-constraints}, as the gap is only imposed indirectly by the measurement of a small $m_X$. 
Theoretically, it is more natural to define $X$ and $Y$ by a rapidity cut $\etaxybar$ in the $\gamma_{\bar n}$-frame. 
Here, we demonstrate that these two definitions are identical to quite high accuracy.
First, we remark that
since we are considering low-mass hadronic particles here, their pseudo-rapidity and rapidity are approximately equal, $\eta_i = \frac{1}{2}\ln (p_i^-/p_i^+)$.
 With a rapidity cut in the $\gamma_{\bar n}$-frame we 
have
\begin{align}
\label{eq:bdrygammanbar}
 \bar p_i\in X \text{\ \ when\ \ } \bar\eta_i < \etaxybar\,,
 \qquad &\text{and\qquad} 
 \bar p_i \in Y \text{\ \ when\ \ } \bar\eta_i > \etaxybar \,.
\end{align}
From \eq{Breit2gamcollin}, we see that the boundary between $X$ and $Y$ in the Breit frame is given by
\begin{align}
  \etaxy = \etaxybar - \frac{1}{2}\ln x \,,
  \qquad \text{with\quad } X:\ \eta_i<\etaxy
  \quad \text{and\quad } Y:\ \eta_i >\etaxy
  \,,
\end{align} 
as shown in \fig{frame-radiation}. Likewise, we use the results of \app{epCMframe} to express the boundary between $X$ and $Y$ in the lab frame as
\begin{align} 
\label{eq:etabdrylab} 
 \frac{\gamma^2}{y}e^{2\etaxybar} =   e^{2\etaxylab}
  -2 \sqrt{\frac{x\gamma^2(1-y)}{y}} \cos \bigl(\phi^{\rm lab} \bigr)\:  e^{\etaxylab}   + \frac{x\gamma^2(1 -y)}{y} 
  \,,
\end{align}
where $\gamma^2 = E_p/E_e$, the ratio of proton and electron lab frame energies, and
\begin{align}
 X:\ \eta_{i}^{\rm lab} <\etaxylab \bigl(\phi_{i}^{\rm lab} \bigr)
  \qquad \text{and }\qquad
 Y:\ \eta_{i}^{\rm lab} >\etaxylab \bigl(\phi_{i}^{\rm lab} \bigr)
 \,.
\end{align}
Unlike the simple lab-frame rapidity cut $\etalabcut$ that is natural for experiments, we find that the theoretical cut $\etaxylab$ is a nontrivial function of the azimuthal angle $\phi^{\rm lab}$ and the quantity $x\gamma^2(1-y)/y = x E_p(1-y)/(yE_e)$. 
For diffraction at HERA and the EIC, we typically have that $E_p/E_e$ is large, $x$ is small,
and the rapidity cut is large and lies
in the forward region, e.g. $\etalabcut>5$ (or $\etaxybar>5$ and $\etaxylab > 5$). 
For sufficiently large $\etalabcut$,
the last two terms in \eq{etabdrylab} are small and subleading. 
That is, if we choose
\begin{align} \label{eq:choose-etaxybar} 
  \etaxybar = \etalabcut + \frac12 \ln\frac{y}{\gamma^2} 
  \qquad \longrightarrow\qquad 
  \etaxylab \simeq \etalabcut \,. 
\end{align}
That is, this choice of $\etaxybar$ causes the experimentally- and theoretically-motivated lab frame rapidity bounds to match to very good approximation.  
In \fig{etacut}, we compare $\etalabcut$ and the $\etaxylab$ from \eq{etabdrylab} with $\etalabcut=3$ or $4$, demonstrating that 
these definitions agree to better than 5\% or even to better than 1\%, depending on the choice of kinematics.
At smaller $E_p/E_e$ and $x$, or larger $\etalabcut$, the agreement improves further. 
Thus, the factorization we derive in this work, using a cut $\eta_0=\eta_{\rm cut}^{\rm lab}+ \frac12 \ln[y/(x\gamma^2)]$, is directly applicable for data 
that separates $X$ and $Y$ by measuring a small $m_X$ in a region defined up to some forward rapidity cut.

For future purposes, it is useful to define a shorthand notation for the combination of factors
\begin{align} \label{eq:hdefn}
  \h = e^{-\etaxybar} = \frac{\gamma}{\sqrt{y}}  \, e^{-\etalabcut} \,
  \sim\, \lambda_g
 \,.
\end{align}
The restriction on rapidities in \eq{Yandetacut} that constrains the $n$-collinear particles to be well within $Y$ becomes
\begin{align}  \label{eq:beamconstraint}
  z\ll h^2\,,\ \  \text{or alternatively}\ \ \lambda^2 \lambda_t^2 \ll \lambda_g^2 \,.  
\end{align}
This means that for a given $m_Y^2-t$, there is a restriction on the choice of $\etalabcut$. It is useful to write this constraint as
\begin{align} \label{eq:hconstraint}
   m_Y^2 -t  \ll \frac{h^2 Q^2}{x} = s \gamma^2 e^{-2\etalabcut} \,.
\end{align}
Intuitively, this is due to the fact that for larger $-t$, the forward-scattered $n$-collinear hadronic system has a larger transverse kick. This requires increasing the size of $Y$ and thus taking smaller values of $\etalabcut$.  
To satisfy this constraint while still leaving an appreciable range for $-t$, one can consider using smaller values of $\etalabcut$ than those used in past HERA analyses.  For example, with HERA kinematics the choices $\etalabcut = \{ 3, 4, 5, 6\}$ give the bounds $ \sqrt{m_Y^2 -t}  \ll \{ 92, 34, 12, 4.6 \}$\,GeV, respectively.  For a large $\etalabcut=6$ we are thus forced to work in a nonperturbative region where $m_Y^2-t \sim \Lambda_{\rm QCD}^2$, while for smaller $\etalabcut$ there can still be an appreciable perturbative region for $m_Y^2-t$.
The values $\etalabcut=3,4$ were used in \fig{etacut}, which increases the allowed values of $m_Y^2-t$. This comes at the expense of introducing $\phi$-dependence in the rapidity boundary from the translation between the lab and Breit frames; these are at the level of other power-suppressed contributions, and hence can be ignored.

We can also consider a constraint that ensures that the soft particles are fully in $X$. Since $p_{uc}^+ \ll p_{X_s}^+$, this constraint can be imposed by demanding $p_X^-/p_X^+ \ll e^{2\eta_0}$, which we can write as $xh^2 \ll \beta/(1-\beta)$.  With $\beta\sim 1$, we can express this more simply as 
\begin{align} \label{eq:softhconstraint}
 xh^2 \ll 1 \,, \ \  \text{or alternatively,}\ \ \lambda^2\lambda_g^2 \ll 1
  \,.
\end{align}

\paragraph{Rapidity boundary in the Breit frame.}
If experimentally the rapidity cut defining $X$ and $Y$ were instead imposed in the Breit frame,
through $h_0\equiv e^{-\eta_0}$, then the condition in \eq{hconstraint} becomes
\begin{align} \label{eq:hconstraint2}
  x z\ll h_0^2  \qquad \text{as} \qquad   m_Y^2 -t  \ll \frac{h_0^2 Q^2}{x^2}  \,.
\end{align}
Although this is more challenging experimentally and leads to a joint bound on $x^2(m_Y^2-t)$, it does not rely on the approximation between rapidity boundaries shown in \fig{etacut}.
  
\paragraph{Rapidity gap method.} A second way to define $X$ and $Y$ is the rapidity gap method from \sec{diffractive-constraints}, which requires a gapped region $S: \eta_{\rm min}^{\rm lab} <  \eta_i^{\rm lab} < \eta_{\rm max}^{\rm lab}$ in the lab frame, with a detection threshold $E_{i}^{\rm lab} < E_{\rm gap}^{\rm lab}$.  Here, $Y$ consists of particles with rapidity $\eta_i^{\rm lab} > \eta_{\rm max}^{\rm lab}$, so we have $S\subset X$.  Transforming this constraint to a Breit frame built from an incoming lepton with momentum in the $x$-$z$ plane, we find
\begin{align}\label{eq:rapgap-method}
  \sqrt{\frac{x\gamma^2}{y}}  p_i^-
   + \sqrt{\frac{y}{x\gamma^2}}\Big( 1+\frac{x\gamma^2(1-y)}{y}\Big) p_i^+ 
   - 2 \sqrt{\frac{x\gamma^2(1-y)}{y}} 
  |\vec p_{i\perp}| \cos\phi_i 
< 2 E_{\rm gap}^{\rm lab} 
  \,.
\end{align}
Due to the hierarchy in uc momentum components in \eq{gap-radiation-Breit}, if $x\gamma^2\sim 1$ then \eq{rapgap-method}  gives the constraint $\sqrt{x\gamma^2/y}\, p_{uc\,i}^- < 2 E_{\rm gap}^{\rm lab}$, whereas if $x\gamma^2\sim x$ then \eq{rapgap-method} gives  $\sqrt{x\gamma^2/y}\, p_{uc\,i}^- + \sqrt{y/(x\gamma^2)}\, p_{uc\,i}^+< 2 E_{\rm gap}^{\rm lab}$. 
Therefore, the uc radiation is sensitive to three regions: region $X/S$ and region $Y$, corresponding to uc radiation contributing to measurements of $m_X$ and $m_Y$, respectively, as well as the gapped region $S\subset X$, which is sensitive to $E_{\rm gap}^{\rm lab}$. 
Thus, the rapidity gap method introduces sensitivity to an additional scale $E_{\rm gap}^{\rm lab}$, which is typically taken to be $\lesssim 1\,{\rm GeV}$. The full treatment of this additional scale leads to a more complicated factorization that can involve additional expansions and modes.  In \sec{diffract-factorization}, we briefly comment on this case but do not carry out a full analysis.

\paragraph{Coherent forward detector method.}
Diffractive processes can also be identified by the presence of a single proton in the forward detector. In this case, the division between $X$ and $Y$ can be taken at the boundary of the forward detector,  $\eta_{\rm cut}^{\rm lab}\simeq 6$, similar to the small-$m_X$ method. Thus, the factorization we derive below is applicable to this method.

\paragraph{SCET operators for gap radiation.} To discuss how uc modes enter into Glauber SCET, let us denote the uc radiation state with momentum $p_{uc}$ as $Z_{uc}$, and let us analyze the case where $p_X^- \sim p_{uc}^-$ and $p^{\prime\,+}\sim p_{uc}^+$. 
(It is straightforward to make further expansions to obtain results for $p_X^- \gg p_{uc}^-$ and/or $p^{\prime\,+}\gg p_{uc}^+$.) 
By comparing which momentum components are the same order in the power counting,
the scaling of $p_{uc}^\mu$ implies that at leading order in the power expansion, only  $p_{uc}^+\sim p_{n}^+$ can influence the momentum of $Y$, and only $p_{uc}^-\sim p_{X_s}^-\sim \tau^-$ 
can influence the momentum of $X$.  In particular, we can decompose
\begin{align}\label{eq:gap-radiation-def}
  p^{\prime} = p_n^{\prime} + p_{uc}^Y \,,\qquad\quad
  p_X = p_{X_s} + p_{uc}^X \,.
\end{align}
To leading power, $p^{\prime -} = p_n^{\prime -}$, $p_\perp^{\prime} = p_{n\perp}^\prime$, $p_X^+= p_{X_s}^+$, and $p_{X}^{\perp} = p_{X_s}^\perp$, but we must be careful to track subleading components in certain parts of our analysis, as they are necessary to preserve the full momentum conservation.  

In the ${\cal L}_n$ SCET Lagrangian, uc gluons only couple to $n$-collinear fields through their $n\cdot A_{uc}$ component, and in the ${\cal L}_s$ SCET Lagrangian they only couple to the soft fields through their $\nbar \cdot A_{uc}$ component. These interactions are eikonal and can be decoupled from the $n$-collinear and soft Lagrangians by suitable BPS field redefinitions~\cite{Bauer:2001yt}, 
\begin{align}
	& \chi_n \to \gY_n \chi_n \,, 
	&& {\cal B}_{n\perp}^A \to \gYa_n^{AB} \, {\cal B}_{n\perp}^B \,, 
	\nn\\
	& \psi_s^n \to \gY_{\nbar} \psi_s^n \,,
	&& {\cal B}_{s\perp}^{nA} \to \gYa_{\nbar}^{AB} \, {\cal B}_{s\perp}^{nB} 
 	 \,,
\end{align}
where the Wilson lines are defined by
\begin{align}
	&  \gY_n(x) = \bar P \exp\Big( -ig \int_0^\infty\!\! ds\: n\cdot A^A_{uc}(ns + x) T^A \Big) \,,
	\nn\\
	&  \gYa_n(x) = \bar P \exp\Big( -ig \int_0^\infty\!\! ds\: n\cdot A^A_{uc}(ns + x) {\cal T}^A \Big) \,.
\end{align}
Here, $T^A$ and ${\cal T}^A$ are color generators in the fundamental and adjoint representations, respectively, and $\bar P$ denotes anti-path-ordering of the color matrices.
Simply switching $n\to \nbar$ gives the analogous definitions for $\gY_{\nbar}$ and $\gYa_{\nbar}$.  This field redefinition removes the uc interactions from  ${\cal L}_n$ and ${\cal L}_s$, but introduces uc interactions into the Glauber Lagrangian ${\cal L}_G^{(0)}$, including for the Glauber operators ${\cal O}_{ns}^{ij}$ in \eq{2Rglauber-operators} relevant for the gap radiation.  Using the color representation identities for Wilson lines
\begin{align}
	&  \gY_n^\dagger T^B \gY_n = \gYa_n^{BB'} T^{B'} \,, 
	&& (-i f^{BCD}) \gYa_n^{CC'}  \gYa_n^{DD'} = \gYa_n^{BB'} (-i f^{B'C'D'}) 
	\,,
\end{align}
we find that the Glauber operators $\cO_n^{qB}$ and $\cO_n^{gB}$ both have a single $\gYa_n^{DB}$, while the Glauber operators $\cO_s^{q_nC}$ and $\cO_s^{g_nC}$ both have a single $\gYa_{\nbar}^{DC}$. 
Thus, the final result from the uc field redefinitions is independent of whether we have quark or gluon operators, and is simply 
\begin{align}\label{eq:2Rglaubopsnew}
	\cO_{ns} &= 
	\gYa_n^{DB} \gYa_{\nbar}^{DC}
	\Big( \cO_{n}^{B} \frac{1}{\cP_\perp^2}\cO_s^{C} \Big)
	\,.
\end{align}
This gives the modification to the $ns$ Glauber operators in \eq{2Rglaubops} from the coupling of uc modes.
These uc modes contribute radiation to both the $X$ and $Y$ systems.

\section{Factorization of Diffraction and Quasi-Diffraction}\label{sec:diffract-factorization}

Now, we carry out a factorization of diffraction and quasi-diffraction. The similar underlying nature of these processes allows us to factorize both at the same time, using the exact same procedure. In \sec{power-counting} and \sec{gap-radiation}, we proved the necessary and sufficient constraints on our power counting parameters $\{\lambda,\,\lambda_t,\,\rho,\,\lambda_g,\,\lambda_\Lambda\}$ in \eq{power-counting-params} for a process to be (quasi-)diffractive.  We start in \sec{lambda-fact} by factorizing (quasi-)diffraction in the case with the minimal assumption needed for a (quasi-)diffractive process; namely, $\lambda \ll 1$ with no scaling assumptions on the other parameters in \eq{power-counting-params}; i.e. taking $\{\lambda_t,\, \rho,\,\lambda_g,\,\lambda_\Lambda\}\sim 1$. 
In \sec{smalllambdat}, we discuss the additional hard-collinear factorization that arises if we take $\lambda_t\ll 1$, 
and use this to present a Regge factorization formula for diffracitve PDFs.
We do not consider here the additional factorization that can arise from taking $\lambda_t\gg 1$ and $\lambda\lambda_t\ll 1$.   
We will consider some implications of the $\lambda_\Lambda \ll 1$ expansion in \sec{soft-results}.

\subsection{Factorization for $\lambda \ll 1$}\label{sec:lambda-fact}

Our primary objective is to factorize the (quasi-)diffractive structure functions for an unpolarized initial proton ($i=2,L,3,4$). 
Our factorization straightforwardly generalizes to the spin-polarized case, as we will see later in this section.
From \eqs{tensors}{projection}, the full, unexpanded definition of these structure functions in QCD is
\begin{align} \label{eq:hadtensor}
	F_i^D(x,Q^2,&\beta,t,m_Y^2) 
	= \cP_{i\,\mu\nu} \sumint_{X,Y} \, 
	\delta^4(q \!+\! p \!-\! p' \!-\! p_X) \,
	\big\langle p \big| J^{\mu}(0)\big| XY \big\rangle
	\big\langle XY  \big| J^\nu(0) \big| p \big\rangle
	\delta(m_Y^2-p^{\prime\,2})
	\nn\\
	&= x\, \cP^\prime_{i\,\mu\nu}   \sumint_{X,Y}  \,
	\int\!\! \ddslash^dz\, e^{iz\cdot q} \:
	\big\langle p \big| J^{\mu}(z) \big| XY \big\rangle
	\big\langle XY \big| J^\nu(0) \big| p \big\rangle
	\delta(m_Y^2-p^{\prime\,2})
	\,,
\end{align}
where $\ddslash^dz = d^dz/(2\pi)^d$ 
and we switch to $\cP^{\prime}_{i\,\mu\nu}$ from \eq{primeprojectors}.
We use dimensional regularization with $d=4-2\epsilon$ to handle any ultraviolet or infrared divergences. Note that we  study the bare cross section first, before switching to renormalized quantities. 

\paragraph{Momentum regions and operators for the $\lambda$ expansion.}
Let us consider the power counting parameters defined in \eq{power-counting-params}, setting $\lambda \ll 1$ and keeping  $\{\lambda_t, \rho,\beta \} \sim 1$.  Then, the Breit frame scaling results in \eqs{four-vector-scales}{four-vector-scales-new} simplify to 
\begin{align}\label{eq:four-vector-scales2}
	&q \sim\sqrt{s}(\lambda,\,\lambda,\, 0) \,,
	&&p_{X_s} \sim \sqrt{s}(\lambda,\, \lambda,\,  \lambda )\,,
	&&\tau \sim\sqrt{s}(\lambda^3 , \lambda, \lambda )
	\nonumber\\
	&p\sim \sqrt{s} (0,\lambda^{-1},0)\,,
	&&p'_n \sim\sqrt{s}( \lambda^3 , \lambda^{-1}, \lambda )\,.
\end{align}
As noted in \sec{Breit-scaling}, $q$ and $p_{X_s}$ carry soft scaling ($s$), $\tau$ is Glauber ($G$), and $p$ and $p'_n$ are $n$-collinear. 
We also must include radiation in $X$ and $Y$ that is soft enough to pass the rapidity gap restrictions, which from \eq{gap-radiation-Breit} carries ultrasoft-collinear (uc) 
momentum
\begin{align}
	p_{uc}^\mu \sim \sqrt{s} \lambda^2 \: ( \lambda, \lambda^{-1} , 1)
	\sim \sqrt{s} (\lambda^3, \lambda, \lambda^2) \,.
\end{align}

Let us analyze \eq{hadtensor} at leading power in SCET, using the scaling of particles in \eq{four-vector-scales2}. The hadronic current $J^\mu$ attaches to a soft quark as in \fig{diffraction}a, and hence in SCET can be replaced by a purely soft current,  
\begin{align}\label{eq:diffract-current}
  J^\mu(z) \to
  J_s^\mu(z) =  \sum_q e_q \bar\psi_{s}^{(q)}(z) \gamma^\mu \psi_{s}^{(q)}(z)
\,,
\end{align}
where $q$ indicates quark flavor. To study the hadronic states, we use an interaction picture where the factorizable components of the SCET Lagrangian, ${\cal L}_n^{(0)} + {\cal L}_s^{(0)} + {\cal L}_{uc}^{(0)}$, evolve the quantum states; meanwhile, we write out the non-factorizable component ${\cal L}_G^{(0)}$ as insertions with the hadronic currents. 
States in SCET can be organized such that they consist of particles from a single sector of the factorizable Lagrangian, or are direct product states from multiple sectors. 
In (quasi-)diffraction, $|p\rangle$ is entirely $n$-collinear at leading power. In contrast, $Y$ can contain both $n$-collinear and uc particles, and $X$ can contain both soft and uc particles. We factorize $|XY\rangle$ into product states as 
\begin{align}\label{eq:hadronic-state-fact}
| X Y \rangle =  | X_s \rangle \: | Y_{n} \rangle \:  | Z_{uc} \rangle 
  \,,
\end{align}
where $X_s\subset X$ contains soft particles, $Y_n\subset Y$ describes $n$-collinear particles, and $Z_{uc}$ involves uc particles coming from both $X$ and $Y$; see \fig{gap-radiation}.

To study the effects of Glauber exchange in diffraction, it is convenient to write the relevant terms of the Glauber action in \eq{2Rglaubopsnew} as
\begin{align} \label{eq:LGmulti}
\int d^d y \, {\cal L}_G^{(0)} &= \int\! [d\tilde y] 
\int\!\! \frac{\ddslash^{d'}\!\tau_\perp'}{\tau_\perp^{\prime\,2}} \:
{\cal O}_n^B(\tilde y,\tau_\perp')\: 
{\cal O}_s^C(\tilde y,-\tau_\perp') \:
\mathbb{U}_{n\nbar}^{BC}(\tilde y)
 \,.
\end{align}
Here, we abbreviate $\tilde y^\mu = n\cdot y \nbar^\mu/2 + \nbar\cdot y n^\mu/2$ and $[d\tilde y]= \frac12 dy^+ dy^-$, and we take the transverse integral over $d'=d-2$ dimensions. We Fourier-transform the Glauber operators into transverse momentum space as ${\cal O}_n^A(y^\pm,\tau_\perp')=\int d^{d'}\!y_\perp \exp(-iy_\perp\cdot\tau_\perp') {\cal O}_n^A(y)$ and similarly for ${\cal O}_s^A(y^\pm,-\tau_\perp')$. We also introduce a shorthand for the product of adjoint Wilson lines:
\begin{align} \label{eq:UU}
 \mathbb{U}_{n\nbar}^{BC}(\tilde y)
   \equiv \gYa_n^{DB}(\tilde y)\,\gYa_{\nbar}^{DC}(\tilde y)  \,.
\end{align}
The internal fields in ${\cal O}_n^B$ and ${\cal O}_s^C$ exactly conserve the $p_n^-\sim \lambda^0$ and $p_s^+\sim \lambda$ momenta, so the locality of these operators at $y^\pm$ in \eq{LGmulti} encodes momentum conservation for $\tau^-\sim \sqrt{s} \lambda$ and $\tau^+\sim \sqrt{s}\lambda^3$.  Note that $\tau^-$ is a leading momentum for the soft operator, but subleading to the $n$-collinear operator, as $\tau^- \ll p^{\prime\,-}$. Likewise, $\tau^+$ is a leading momentum for the $n$-collinear operator, but subleading for the soft operator, as $\tau^+ \ll p_X^+, q^+$. For now, we allow soft and $n$-collinear operators to fully conserve momentum in subleading momentum components, but we will soon carry out multipole expansions of these operators. 

\paragraph{Factorization.} 
A breadth of past literature on SCET informs our factorization procedure for diffraction. 
Our ability to carry out a systematic factorization is enabled by \refcite{Rothstein:2016bsq}, which points out in $\S8$ that order by order in insertions of the Glauber Lagrangian, we can factorize soft and collinear interactions to all orders in $\alpha_s$. 
A useful benchmark for our factorization comes from the Glauber SCET factorization of small-$x$ DIS in \refcite{Neill:2023jcd}, which studied squared matrix elements of Glauber operators involving a single exchange on each side of the cut, to sum logs of $x$ at NLL order. 
(Compared to this small-$x$ DIS analysis, our diffractive case involves the additional $\mathbb{U}_{n\nbar}$, and hence is in a \SCETa type theory. Additionally, for simplicity we treat the transverse coordinates in momentum space from the start.)  
To inform the all-orders renormalization of our factorization, we use elements from the forward $2\to 2$ scattering amplitudes analysis in \refcite{Gao:2024qsg}, which carried out an all-orders factorization and renormalization using Glauber Lagrangians.
We also note certain parallels in techniques between our work and \refcite{Stewart:2023lwz}, which carries out an all-orders factorization for DIS off a heavy nucleus in the limit where Glaubers interact with color-uncorrelated partons in different nucleons, in order to show how saturation arises in an top-down EFT framework. Our factorization analysis in this section treats the Glauber Lagrangian to all orders for diffractive scattering off a DIS proton target, with only a small-$x$ expansion.

To start, we combine the structure functions in \eq{hadtensor}, currents in \eq{diffract-current}, states in \eq{hadronic-state-fact}, and Glaubers in \eq{LGmulti} with techniques from \refcite{Rothstein:2016bsq} to get:
\begin{align}  \label{eq:factstep1}
	& F_i^D = x \!\!\!\! \sumint_{X_s,Y_{n},Z_{uc}} 
	 \sum_{N,N'=1}^\infty \frac{(-\img)^N}{N!} \frac{(+\img)^{N'}}{N'!}
	 \int \prod_{i=1}^N  \frac{\ddslash^{d'}\!\!\tau_{i\perp}}{\tau_{i\perp}^2} 
	  [d\tilde y_i] \:
	\prod_{j=1}^{N'} \frac{\ddslash^{d'}\!\!\tau'_{j\perp}}{\tau_{j\perp}^{\prime\,2}} 
	  [d\tilde y'_j]
	  \:\delta\big(m_Y^2 - (p_n'+p_{uc}^Y)^2\big)
	\nn \\
	&\times
	\big\langle p \big|  
	\cO_n^{A_1}(\tilde y_1,\tau_{1\perp})\cdots \cO_n^{A_N}(\tilde y_N,\tau_{N\!\perp}) 
	\big| Y_{n} \big\rangle 
	\big\langle Y_{n} \big| 
	\cO_n^{A'_1}(\tilde y_{1}',\tau_{1\perp}^\prime) \cdots 
	\cO_n^{A'_{N'}}(\tilde y'_{N'},\tau_{N'\!\perp}^\prime) 
	\big| p \big\rangle 
	\nn \\
	&\times 
	\big\langle 0 \big| 
	\mathbb{U}_{n\nbar}^{A_1 B_1}(\tilde y_1) 
	   \cdots \mathbb{U}_{n\nbar}^{A_N B_N}(\tilde y_N)\,
	\big| Z_{uc} \big\rangle
	\big\langle Z_{uc} \big| 
	\mathbb{U}_{n\nbar}^{A_1' B_1'}(\tilde y_1') 
	   \cdots \mathbb{U}_{n\nbar}^{A'_{N'} B'_{N'}}(\tilde y'_{N'}) 
	\big| 0\big\rangle 
	\nonumber\\
	& \times
	\cP^\prime_{i\,\mu\nu} \int\!\! \ddslash^dz\, e^{iz\cdot q}\,
	\big\langle 0 \big| \bar T\,
	J_s^\mu(z) \cO_s^{B_1}(\tilde y_1,-\tau_{1\perp}) \cdots
	\cO_s^{B_N}(\tilde y_N,-\tau_{N\!\perp})
	\big| X_s \big\rangle  
	\nn\\
	& \qquad\quad \times
	\big\langle X_s \big| T\,
	J_s^\nu(0) \cO_s^{B_1'}(\tilde y'_1,-\tau_{1\perp}^\prime) \cdots 
	\cO_s^{B'_{N'}}(\tilde y'_{N'},-\tau_{N'\!\perp}^\prime) 
	\big| 0 \big\rangle 
	\,. 
\end{align}
Here $T$ and $\bar T$ are time and anti-time ordering operators, respectively,
$p_n'$ is the momentum of the state $Y_n$, and $p_{uc}^Y$ is the momentum of the ultracollinear radiation that ends up in the hadronic object $Y$.
Recall that to formulate the differential diffractive cross section we extracted a phase space integral for the momentum $p'$ of $Y$, which in this factorized form should be viewed as extracted from the $Y_n$ phase space integrations.
In this form, the $n$-collinear matrix elements (2nd line), uc matrix elements (3rd line), and soft matrix elements (4th and 5th lines) have been factorized, albeit with infinite sums over the number of Glauber exchanges to the left and right of the cut ($N$ and $N'$ respectively).  
Note that we must have at least one Glauber exchange on each side of the cut to get a non-zero result, so $N,N'\ge 1$.  This is clear from the particles produced in the central region which have $p_X^\pm >0$, which is allowed due to $q^+>0$ and $\tau^- >0$. 

To derive the final factorization formula we carry out a number of steps to manipulate \eq{factstep1}, which we discuss in order:
\begin{enumerate}[leftmargin=*]
\item First, we use translation invariance to factor out transverse-momentum-conserving $\delta$-functions from the collinear matrix elements. This imposes $\sum_{i=1}^{N} \tau_{i\perp} = -\tau_\perp$ and 
$\sum_{j=1}^{N'} \tau'_{j\perp} = \tau_\perp$ (where $\tau_\perp \equiv p_\perp-p'_\perp$), and leaves a free integration over the transverse momentum of one operator in each matrix element. We capture this by introducing the notation
\begin{align} 
  \bar {\cal O}_n^{A}(\tilde y)\equiv \! \int\! \ddslash^{d'}\!\! \tau''_{\perp}\, {\cal O}_n^{A}(\tilde y, \tau''_{\perp}) \,.
\end{align}

\item Second we exploit the {\em collapse} of Glauber interactions, which causes a significant simplification of graphs involving multiple Glaubers.  Contracting soft and collinear fields that generate propagators between Glauber vertices and carrying out the rapidity-regulated $\pm$ integrals for Glauber loops causes instantaneity in the 
Glauber vertices' $\pm$ light-cone coordinates~\cite{Rothstein:2016bsq},
setting $y_i^\pm = y_1^\pm$ and $y_i^{\prime \pm} = y_1^{\prime \pm}$. 
This also sets the coordinate dependences in the uc matrix element to these values. 
While the combinatorics of the propagator contractions eliminate a $1/(N! N'!)$ factor, it is regenerated by the collapsed integrals. 
Although a Glauber rapidity regulator $\eta'$ is important at intermediate steps and for generating the correct $1/N!$ prefactors, the collapsed answer does not have leftover dependence on the regulator.  (A different rapidity regulator $\eta$ is still needed for the soft and collinear matrix elements; see Ref.~\cite{Moult:2022lfy}.)  
When neighboring Glaubers are attached to the same soft and collinear lines in a graph, the collapse eliminates the intermediate propagators between them. When Glaubers are attached to different lines in a graph there is still a removal of propagators, but the collapse modifies the leftover external propagators such that they now carry a sum of various Glauber light-cone and transverse energies. See Ref.~\cite{Rothstein:2016bsq} for examples.

\end{enumerate}
After these first two steps we have
\begin{align}  \label{eq:factstep2}
	& F_i^D = x\!\!\! \sumint_{X_s,Y_{n},Z_{uc}} 
	 \sum_{N,N'=1}^\infty \IInt^\perp_{(N,\,N')} 
	 \int 
	  [d\tilde y_1] \:  [d\tilde y'_1]
	  \:\delta\big(m_Y^2 - (p_n'+p_{uc}^Y)^2\big)
	 \\
	&\times
	\big\langle p \big|  
	 \collapsel\text{\footnotesize $\prod_{i=1}^{N-1}$} 
	   \cO_n^{A_i}(\tilde y_1,\tau_{i\perp}) \bar \cO_n^{A_N}(\tilde y_1) 
	   \!\collapser \big| Y_{n} \big\rangle 
	\big\langle Y_{n} \big| 
	 \collapsel \text{\footnotesize $\prod_{j=1}^{N'-1}$}
	  \cO_n^{A'_j}(\tilde y_{1}',\tau_{j\perp}^\prime) \bar \cO_n^{A'_{N'}}(\tilde y_{1}') 
	  \!\collapser
	\big| p \big\rangle 
	\nn \\
	&\times 
	\big\langle 0 \big| 
	\mathbb{U}_{n\nbar}^{A_1 B_1}(\tilde y_1) 
	   \cdots \mathbb{U}_{n\nbar}^{A_N B_N}(\tilde y_1)\,
	\big| Z_{uc} \big\rangle
	\big\langle Z_{uc} \big| 
	\mathbb{U}_{n\nbar}^{A_1' B_1'}(\tilde y_1') 
	   \cdots \mathbb{U}_{n\nbar}^{A'_{N'} B'_{N'}}(\tilde y'_{1}) 
	\big| 0\big\rangle 
	\nonumber\\
	& \times
	\cP^\prime_{i\,\mu\nu} \int\!\! \ddslash^dz\, e^{iz\cdot q}\,
	\big\langle 0 \big| \bar T\,
	J_s^\mu(z)  \collapsel\! \text{\footnotesize $\prod_{i=1}^{N}$} 
	   \cO_s^{B_i}\!(\tilde y_1,-\tau_{i\perp}) \!\!\collapser 
	\big| X_s \big\rangle  
	\big\langle X_s \big| T\,
	J_s^\nu(0) 
	  \collapsel\! \text{\footnotesize $\prod_{j=1}^{N'}$} 
	   \cO_s^{B_j'}\!(\tilde y_1^{\prime},-\tau_{j\perp}^\prime) \!\!\collapser
	\big| 0 \big\rangle 
	\,. \nn
\end{align}
Here, we follow \refcite{Gao:2024qsg} and use a compact notation for the transverse momentum integrals
\begin{align} \label{eq:iint_perp}
	  \IInt^\perp_{(N,\,N')} 
	  \!\! \equiv 
	  \frac{(-\img)^{N}(\img)^{N'}}{N!\, N'!}\!
	  \int\! \prod_{i=1}^{N}\! \frac{\ddslash\!^{d'}\!\tau_{i\perp}}{\vec\tau_{i\perp}^{\,2}} 
	   \, \deltaslash^{d'}\!\Bigl(\sum_i \tau_{i\perp} + \tau_\perp\Bigr) 
	  \int\! \prod_{j=1}^{N'} \frac{\ddslash\!^{d'}\!\tau'_{j\perp}}{{\vec\tau_{j\perp}^{\,\prime 2}}} \,  \deltaslash^{d'}\!\Bigl(\sum_j {\tau}^{\prime}_{j\perp}- \tau_\perp\Bigr) 
	 \,,
\end{align}
where $\deltaslash^{d'}(\tau_{\perp}) = (2\pi)^{d'} \delta^{d'}(\tau_{\perp})$.
Note that to the order we work in the $\lambda\ll 1$ expansion, $\tau_\perp^2 = -\vec\tau_\perp^2 = t$. 
The brackets $\{ \! \! \{ ... \} \! \! \}$ in \eq{factstep2} indicate that the operators inside have undergone collapse, and hence appear at the same $\pm$ lightcone coordinates.  Note that these operators can still contract independently with different fields, such as the two quark fields in the currents $J^\mu$ and $J^\nu$. 
We provide explicit calculations involving these collapsed operator products in \sec{soft-results}.
For example, if $N'=2$ and we consider a generic contraction in the proton matrix element, then there are four 
different contractions in the final soft matrix element. The collapse brackets are normalized such that for $N=1$,
\begin{align}
  \collapsel\! \bar \cO_n^{A}(\tilde y_1) \! \collapser 
    &= \bar \cO_n^{A}(\tilde y_1)
   \,,
  & \collapsel\! \cO_s^{B}\!(\tilde y_1,-\tau_{\perp}^\prime) \!\collapser 
    &= \cO_s^{B}\!(\tilde y_1,-\tau_{\perp}^\prime) 
  \,.
\end{align}

\begin{figure}
	\begin{center}
		\includegraphics[width = 4in]{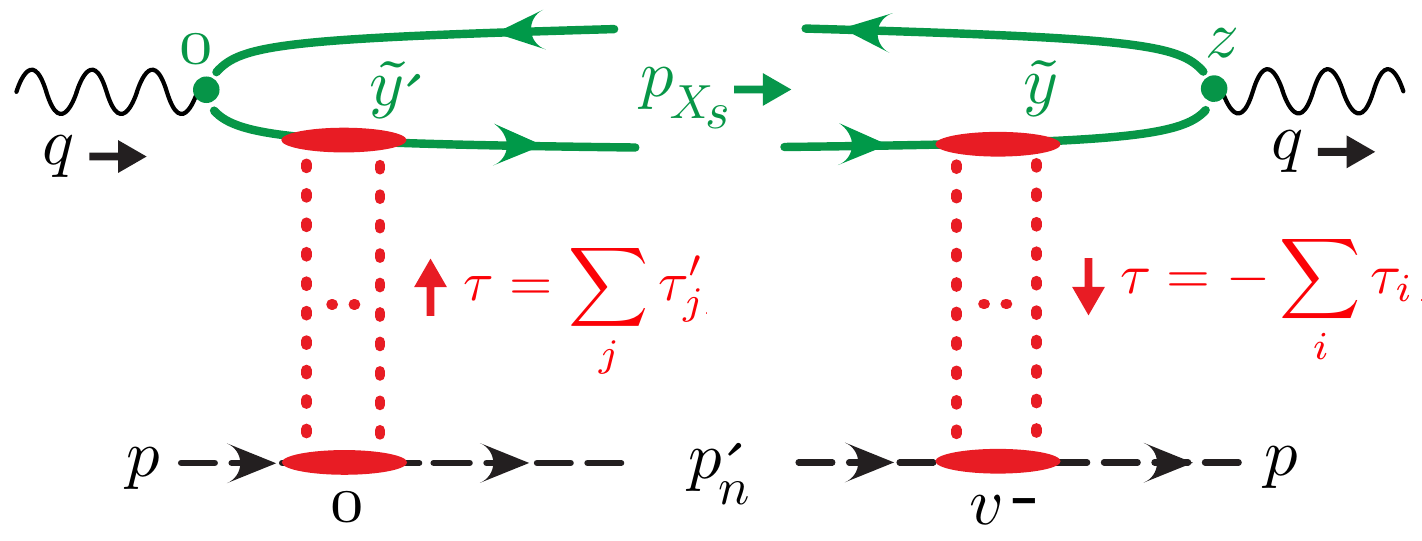}
	\end{center}
\vspace{-0.3cm}
\caption{\label{fig:CoordMom}
Spacetime coordinates and momenta in the factorization formula for diffractive structure functions.
For simplicity, we only show Glaubers attaching to a single line in $S$ and $B$ and do not show any radiation in $U$. Note that our convention is such that both $\tau_{j\perp}'$ and $\tau_{i\perp}$ are incoming momenta for the soft function (on both sides of the cut).  }
\end{figure}

The remaining steps involve carrying out appropriate multipole expansions for the soft and collinear matrix elements:
\begin{enumerate}[leftmargin=*]

\item[3.] As a third step, we remove the $y_1^-$ and $y_1^{\prime\,-}$ dependences of the $n$-collinear and uc operators by carrying out a multipole expansion of the soft function. To see this, we first factor these dependences out of the collinear and uc matrix elements using translation invariance. This leads to a phase $\exp[\frac{i}{2}(y_1^- - y_1^{\prime -}) (p^+-p_n^{\prime +}-p_{uc}^+)]$, which feeds momenta into the soft matrix element. However, we can drop this phase because its momentum is much smaller than the intrinsic soft particle momenta encoded by soft fields depending on $y_1^-$ and $y_1^{\prime\,-}$; namely, $(p^+-p_n^{\prime +}-p_{uc}^+)\sim \sqrt{s}\lambda^3 \ll p_{X_s}^+\sim \sqrt{s}\lambda$. 

\item[4.] Fourth, we factor out the $y_1^+$ and $y_1^{\prime\,+}$ dependences of the $n$-collinear and uc operators, which provides a nontrivial injection of momenta into the soft function. 
Recall that the external kinematics fix the forward scattering condition $p^{\prime -}_n\simeq p^- \sim \lambda^{-1}$, whereas for the residual momentum $p^- - p_n^{\prime -}  = \tau^- + p_{uc}^{Y-}\sim \lambda$. Using translation invariance, we factor out this small residual momentum from the collinear matrix elements and the momentum $p_{uc}^-$ from the uc matrix element, giving the phase $\exp[-\frac{i}{2}(y_1^{\prime +}-y_1^+)(\tau^-+ p_{uc}^{Y-}-p_{uc}^-)] =
\exp[-\frac{i}{2}(y_1^{\prime +}-y_1^+)(\tau^- - p_{uc}^{X-})]$.  To leading order in $\lambda$, we can set $\tau^-=Q/\beta$  in the Breit frame, but we leave it as $\tau^-$ to emphasize the boost invariance of our final results. Since $\tau^- - p_{uc}^{X-}\sim \sqrt{s}\lambda$ is the same size as $p_{X_s}^-\sim \sqrt{s}\lambda$, this phase injects this momentum into the soft matrix element. We rewrite this phase with two extra integration variables as
\begin{align}
 e^{-\frac{i}{2}(y_1^{\prime +}-y_1^+)(\tau^- - p_{uc}^{X-})} 
  = \int\!\! dk_s^- dp_g^- \: \delta(\tau^- - p_g^- - k_s^-) \: \delta(p_g^- - p_{uc}^{X-}) \:
   e^{-\frac{i}{2}(y_1^{\prime +}-y_1^+) k_s^-} \,,
\end{align}
in order to factor pieces of the expression together with the uc and soft matrix elements.

\item[5.] Fifth,
we manipulate the $\delta$-function on the first line of \eq{factstep2}. Using \eq{gap-radiation-def}, we can write its argument to leading power as 
\begin{align}
	m_Y^2 - (p-\tau)^2 \simeq m_Y^2-t + p^- \tau^+
	  =p^-( p^{\prime +} + p^+ - p_n^{\prime +} - p_{uc}^{Y+}) \,. 
\end{align}
This enables us to write the $\delta$-function as a suitable phase
\begin{align}
  \delta(m_Y^2 - (p_n'+p_{uc}^Y)^2) 
  &\simeq \frac{1}{2p^-} 
   \int dv^-\: e^{\frac{i}{2} v^- (p^{\prime +}+p^+ -p^{\prime +}_n-p_{uc}^{Y+})}   
   \\
  & =\! \int\!\! dk_n^+ dp_g^+\: \delta(p^{\prime +}-k_n^+-p_{g}^+)\: \delta(p_g^+-p_{uc}^{Y+})
   \! \int\!\! \frac{dv^-}{2p^-}\: e^{\frac{i}{2} v^- (k_n^+ +p^+-p_n^{\prime +})}   
  . \nn
\end{align}
We could set $p^{\prime +}=Qz$ in the Breit frame, but we choose to leave it as $p^{\prime +}$ to emphasize the boost invariance of our final results. In the second line, we have introduced two extra integration variables to factor pieces of the expression together with the uc and $n$-collinear matrix elements. 
We then use $\exp(i v^- (p^+-p_n^{\prime +})/2)$ to shift the (collapsed) collinear operators in a matrix element to the coordinate $v^-$.   

\end{enumerate}
Finally, we let $\tilde y_1=\tilde y$ and $\tilde y_1' =\tilde y'$. This gives
\begin{align}  \label{eq:factstep3}
  & F_i^D = x\!\!\! \sum_{N,N'=1}^\infty \! \IInt^\perp_{(N,\,N')} 
  \int\!\! dk_n^+ dk_s^- dp_g^+ dp_g^-\: 
  \delta( p^{\prime +} - k_n^+ - p_g^+) \: \delta(\tau^- - k_s^- - p_g^-)
  \nn \\*
   & \times \!\!
  \int\!\!  \frac{dv^-}{2p^-}  e^{\frac{i}{2} v^- k_n^+}  \sumint_{Y_{n}}   
  \big\langle p \big| 
  \collapsel\text{\footnotesize $\prod_{i=1}^{N-1}$} 
   \cO_n^{A_i}(v^-,\tau_{i\perp}) \bar \cO_n^{A_N}(v^-)    \!\collapser 
   \big| Y_{n} \big\rangle 
   \big\langle Y_{n} \big|
   \collapsel \text{\footnotesize $\prod_{j=1}^{N'-1}$}
  \cO_n^{A'_j}(0,\tau_{j\perp}^\prime) \bar \cO_n^{A'_{N'}}(0) 
  \!\collapser
   \big| p \big\rangle 
   \nn \\*
&\times 
\sumint_{Z_{uc}} \delta\bigl(p_g^+-p_{uc}^{Y+}\bigr)\, \delta\bigl(p_g^--p_{uc}^{X-}\bigr)\,
  \big\langle 0 \big| \bar T\,
\,\text{\footnotesize $\prod_{i=1}^{N}$}\, \mathbb{U}_{n\nbar}^{A_i B_i}(0)\, 
\big| Z_{uc} \big\rangle \big\langle Z_{uc} \big| T\,
 \,\text{\footnotesize $\prod_{j=1}^{N'}$}\, \mathbb{U}_{n\nbar}^{A'_j B'_j}(0)\, 
	\big| 0\big\rangle 
\nn\\*
   & \times  \cP^\prime_{i\,\mu\nu} 
  \int\!\! \ddslash^{d}\!z\:  [d\tilde y]\,[d\tilde y^{\prime}] \: e^{i z\cdot q}\,
  e^{\frac{i}{2} (y^+ - y^{\prime +}) k_s^-} \,
  \sumint_{X_s} 
\big\langle 0 \big| \bar T\,
J_s^\mu(z)  \collapsel\! \text{\footnotesize $\prod_{i=1}^{N}$} 
   \cO_s^{B_i}\!(\tilde y,-\tau_{i\perp}) \!\!\collapser 
\big| X_s \big\rangle 
 \nn\\*
 &\qquad \times   
\big\langle X_s \big| T\,
J_s^\nu(0) 
  \collapsel\! \text{\footnotesize $\prod_{j=1}^{N'}$} 
   \cO_s^{B_j'}\!(\tilde y^{\prime},-\tau_{j\perp}^\prime) \!\!\collapser
  \big| 0 \big\rangle 
  \: 
  \,,
\end{align}
where the second line involves purely $n$-collinear matrix elements, the third line involves purely uc matrix elements, and the last two lines involve purely soft matrix elements.  We depict the spacetime and momentum flows for this expression in \fig{CoordMom}.

The $p_g^+$ and $p_g^-$ momenta in the uc matrix element correspond to components in the Breit frame regions $Y$ and $X$ respectively, as shown in \fig{frame-radiation}b. Below, we find it convenient to relate these to hemisphere regions, separated by the dashed boundary shown in \fig{frame-radiation}b. The boost that takes the $X$-$Y$ boundary to the hemisphere partition is given by $p_H^+ = x^{-1/2} e^{\etaxybar} p^+$ and $p_H^-=x^{1/2} e^{-\etaxybar} p^-$. 
We can implement this by writing
\begin{align} \label{eq:tohemi}
 & x^{1/2} e^{-\etaxybar}  \delta\bigl(p_g^+-p_{uc}^{Y+}\bigr)\, 
     x^{-1/2} e^{\etaxybar} \delta\bigl(p_g^--p_{uc}^{X-}\bigr) 
\nn\\
  &\quad
 = \delta\bigl( x^{-1/2} e^{\etaxybar} p_g^+-  p_{uc}^{\,{\cal H}_Y +}\bigr)\,  
      \delta\bigl(x^{1/2} e^{-\etaxybar} p_g^--  p_{uc}^{\,{\cal H}_X -}\bigr) 
 \nn\\
 &\quad 
 = \delta\Bigl( \frac{1}{h\sqrt{x}}\,  p_g^+ 
     - p_{uc}^{\,{\cal H}_Y +}\Bigr)\,  
   \delta\Bigl( h\sqrt{x} \,  p_g^-
     - p_{uc}^{\,{\cal H}_X -}\Bigr) 
   \,,
\end{align}
where $p_{uc}^{{\cal H}_Y+}$ and $p_{uc}^{{\cal H}_X-}$ are momentum components in hemisphere regions,
\begin{align}
  p_{uc}^{{\cal H}_Y+} &= \sum_i p_{i\,H}^+ \,\theta\bigl(p_{i\,H}^- - p_{i\,H}^+\bigr) 
  \,,
 & p_{uc}^{{\cal H}_X-} &= \sum_i p_{i\,H}^- \,\theta\bigl(p_{i\,H}^+ - p_{i\,H}^-\bigr) 
  \,.
\end{align}
To agree more closely with how experimental measurements distinguish $X$ and $Y$, here we fix $\etaxybar$ using \eq{choose-etaxybar}, and write the answer in terms of  $e^{-\etaxybar}= h=\gamma e^{-\etalabcut}/\sqrt{y}$ from \eq{hdefn}, with $\gamma^2=E_p/E_e$ the ratio of proton and electron energies in the lab frame.

Next, we decompose the octet color indices by introducing sums over irreducible color representations (irreps), again following the notation of \refcite{Gao:2024qsg}.
Letting $\alpha = A_1 \cdots A_m$ be the indices for $m$ gluons, the decomposition into irreps involves a sum over projectors 
$P_{m\, R}^{\alpha \alpha'}$, which project an $m$-gluon color state onto a subspace $R$, and whose construction can be found in \cite{Keppeler:2012ih}. These projectors satisfy completeness and orthogonality relations
\begin{align}\label{eq:proj_compl}
  \delta_{m}^{\alpha\alpha'} &= \sum_R P_{m\, R}^{\alpha\alpha'}
  = \sum_R P_{m\, R}^{\alpha\alpha''} P_{m\, R}^{\alpha''\alpha'}
  \,,
  &&
  P_{m\, R}^{\alpha_1 \alpha_2}\, P_{m\, R'}^{\alpha_2\alpha_3} = \delta_{RR'} P_{m\, R}^{\alpha_1 \alpha_3}.
\end{align}
In \eq{factstep2}, we utilize two copies of $\delta_{N}^{\alpha\alpha'}$ and two copies of $\delta_{N'}^{\alpha\alpha'}$ to project the $\{A_i\}$, $\{B_i\}$, $\{A'_i\}$, and $\{B'_i\}$ color indices. For example, the matrix elements on the RHS of the cut give
\begin{align}
  & ({\cal O}_n \cdots {\cal O}_n)^{\alpha} 
  ({\cal U}\cdots {\cal U})^{\alpha\beta} 
  ({\cal O}_s \cdots {\cal O}_s)^{\beta} 
 \nn\\
  &\quad 
 = \sum_{R,R'}  
  ({\cal O}_n \cdots {\cal O}_n P_{NR})^{\alpha''} 
  (P_{NR}\,{\cal U}\cdots {\cal U}P_{NR'} )^{\alpha''\beta''} 
  (P_{NR'} {\cal O}_s \cdots {\cal O}_s)^{\beta''} 
  \,.
\end{align}
Note that the vacuum state $|0\rangle$, proton state $|p\rangle$, and currents $J^\mu$ are all color singlets, but the states $|Y_{n}\rangle$ and $|X_s\rangle$ inherit the color representation of the operator in their corresponding matrix elements. Since these same states appear once in each of the $n$-collinear and soft matrix elements, the color representations $R$ and $R'$ must have the same dimension on both sides of the cut. This implies $R_A=R_{A'}$ and $R_B = R_{B'}$. However, these representations can be formed from a different number of gluon indices, so we can still have $N\ne N'$.   We therefore denote the common color representations by $R_A^{N\!N'}$ and $R_B^{N\!N'}$.  

\begin{figure} 
	\begin{center}
		\includegraphics[width = 2.75 in]{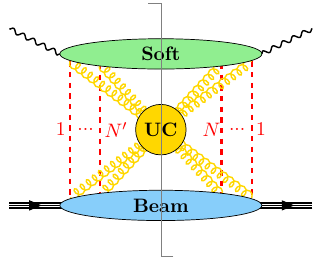}
	\end{center}
\caption{\label{fig:factthm}
Factorization of diffractive and quasi-diffractive $ep$ scattering for $\lambda\ll 1$. Here, we see the convolution of a soft function describing the central jet, a beam function describing the hadronic system, and an ultrasoft-collinear (UC) function describing radiation penetrating the gap. Red dashed lines indicate Glauber ($G$) modes.
}
\end{figure}

Using these results, we further tidy up \eq{factstep3} by defining bare beam, soft, and uc functions containing the associated factorized matrix elements
\begin{align} \label{eq:FDfact}
F_i^D
&= \FiDpre \!\!
  \sum_{N,N'=1}^\infty  \sum_{ \{R_{X}\} }\! \IInt^\perp_{(N,N')}  
  \int\!\! dk_n^+ dk_s^- dp_g^+ dp_g^-
   \: \delta\bigl( p^{\prime +} - k_n^+ - p_g^+\bigr) \, 
    \delta\bigl( \tau^- - k_s^- -p_g^- \bigr)
 \nn\\*
 & \quad\ \ \times
   B_{(N,N')}^{R_A^{N\!N'}} 
    \bigl(k_n^+ p^-, \{\tau_{i\perp},\tau_{j\perp}'\}, t \bigr) 
  \  U_{(N,N')}^{R_A^{N\!N'} R_B^{N\!N'}} 
\Bigl( \frac{1}{h\sqrt{x}}\,p_g^+, h\sqrt{x}\, p_g^-,\mu \Bigr)
  \nn \\*
& \quad \ \ \times
 S_{i(N,N')}^{R_B^{N\!N'}} 
    \bigl(k_s^-q^+, \{\tau_{i\perp},\tau_{j\perp}'\}, Q,t \bigr)  \,,
\end{align}
Here, $\BoostInv=x\,(\bar n\cdot p) (n\cdot q)/Q^2$ enables us to define $B$ and $S$ in a manner that is boost invariant with respect to the $n$-$\nbar$ axis; in the Breit frame, $\BoostInv=1$.
More succinctly, carrying out the $k_s^-$ and $k_n^+$ integrals,
\begin{align} \label{eq:FDfactsimp}
F_i^D &=  \FiDpre \!\!
  \sum_{N,N'=1}^\infty   \sum_{ \{R_{X}\} }\! \IInt^\perp_{(N,N')}  
  \int\!\!  dp_g^+ dp_g^-\ 
   B_{(N,N')}^{R_A^{N\!N'}} 
    \Bigl(p^-(p^{\prime +}-p_g^+), \{\tau_{i\perp},\tau_{j\perp}'\}, t \Bigr) 
 \nn\\*
 & \quad \times
    U_{(N,N')}^{R_A^{N\!N'} R_B^{N\!N'}} 
\Bigl( \frac{1}{h\sqrt{x}}\,p_g^+, h\sqrt{x}\, p_g^-,\mu \Bigr)
 \
 S_{i(N,N')}^{R_B^{N\!N'}} 
    \Bigl( q^+(\tau^- -p_g^- ), 
        \{\tau_{i\perp},\tau_{j\perp}'\}, Q,t  \Bigr) 
  . 
\end{align}
For later convenience, we define the shorthand notation
\begin{align}\label{eq:FDfact-shorthand}
F_i^D\,\equiv & \sum_{N,N'}\sum_{R} B_{(N,N')}^{R_A^{N\!N'}} \otimes_\perp S_{i(N,N')}^{R_B^{N\!N'}} \otimes_{\pm} U_{(N,N')}^{R_A^{N\!N'} R_B^{N\!N'}}\nn\\
  \equiv &  \sum_{N,N'}\sum_{R} (-i)^N (+i)^{N'}B_{(N,N')}^{R_A^{N\!N'}} \tilde{\otimes}_\perp S_{i(N,N')}^{R_B^{N\!N'}} \otimes_{\pm} U_{(N,N')}^{R_A^{N\!N'} R_B^{N\!N'}}\,.
\end{align}
In the second line we explicitly factored out $(-i)^N (+i)^{N'}$ from the $\otimes_{\perp}$ perpendicular-momentum convolution, which will be used in a discussion below.
The soft function $S$ is a matrix element describing the production of the central jet from the electromagnetic current and soft Glauber operators.  $S$ is the only term in the factorization that depends on the choice of structure function $i$. The beam function $B$ describes the hadronic system: it is either a matrix element representing the transition of a proton into a multi-particle state through Glauber exchanges (incoherent), or a proton that remains intact in the presence of the Glauber exchanges (coherent). 
The gap function $U$ describes ultrasoft-collinear radiation that enters the rapidity gap and passes any imposed experimental restrictions. 
These functions are tied together by transverse momenta from the Glauber exchanges and by the fact that while collinear radiation only contributes to $Y$ and soft radiation only contributes to $X$, the uc radiation can contribute to both.  

The collinear beam function in \eq{FDfact} projected onto the color representation $R_A$ is 
\begin{align}  \label{eq:Bdefn}
 & B_{(N,N')}^{R_A^{N\!N'}} 
  \bigl(p^- k_n^+, \{\tau_{i\perp},\tau_{j\perp}'\}, t\bigr) 
  =  \frac{1}{(\nbar\cdot p)^2} 
  \int\! \frac{dv^-}{2p^-}\:
  e^{\frac{i}{2} v^- k_n^+ }  \sumint_{Y_{n}}  
  \\*
 & \ \times
    \big\langle p \big| \!
   \collapsel\text{\footnotesize $\prod_{i=1}^{N-1}$} 
   \cO_n^{A_i}(v^-\text{\small$\frac{n}{2}$},\tau_{i\perp}) \bar \cO_n^{A_N}(v^-\text{\small$\frac{n}{2}$}) 
   \!\collapser P_{N R_A}
   \big| Y_{n} \big\rangle 
   \big\langle Y_{n} \big|  P_{N' R_{A}}
  \collapsel \text{\footnotesize $\prod_{j=1}^{N'-1}$}
  \cO_n^{A'_j}(0,\tau_{j\perp}^\prime) \bar \cO_n^{A'_{N'}}(0) 
  \!\collapser
   \big| p \big\rangle 
   .  \nn
\end{align}
Here the $1/(\nbar \cdot p)$ factors make $B$ boost invariant along the $n$-$\nbar$ axis.
We use this boost invariance and power counting to determine the form of the first argument of $B$. 
Due to the presence of the state $Y_n$, $B$ also depends implicitly on $h$, despite the fact that it is not shown as one of its arguments.
Note that while this beam function encodes the fundamental hadronic dynamics, \eq{Bdefn} is not identical to any known PDF or related object like transverse momentum distributions (TMDs) \cite{Boussarie:2023izj} or generalized parton distributions (GPDs) \cite{Muller:1994ses, Ji:1996nm, Diehl:2003ny}.%
\footnote{Since our $B$ is differential in $t$, it gives more direct information about the dynamics of the initial-state hadron and struck parton than the analogous beam function in small-$x$ inclusive DIS~\cite{Neill:2023jcd}.} In particular, even for the color singlet $R_A^{NN'}=1$, there are still an infinite number of functions indexed by $N,N'\ge 2$.
The soft function with projection onto color representation $R_B$ is 
\begin{align} \label{eq:Sdefn}
& S_{i(N,N')}^{R_B^{N\!N'}} 
   \bigl(q^+ k_s^-, \{\tau_{i\perp},\tau_{j\perp}'\}, Q,t \bigr) 
  =  \Big(\frac{Q}{n\cdot q}\Big)^{2} \,\cP^\prime_{i\,\mu\nu} \!\!
  \int\! [d \tilde y] \, [d\tilde y^{\prime}]  \, \ddslash^{d}\!z
  \ 
  e^{\frac{i}{2} (y^+ - y^{\prime +}) k_s^-}  e^{ i z\cdot q}\,
   \\*
 &\ \: \times
  \sumint_{X_s} \big\langle 0 \big| \bar T\,  J_s^\mu(z) 
  \collapsel\! \text{\footnotesize $\prod_{i=1}^{N}$} 
   \cO_s^{B_i}\!(\tilde y,-\tau_{i\perp}) \!\!\collapser 
  P_{N R_B} \big| X_s \big\rangle  
   \big\langle X_s \big| P_{N'R_{B'}} T\,
   J_s^\nu(0) 
  \collapsel\! \text{\footnotesize $\prod_{j=1}^{N'}$} 
   \cO_s^{B_j'}\!(\tilde y^{\prime},-\tau_{j\perp}^\prime) \!\!\collapser
   \big| 0 \big\rangle
   . \nn
\end{align}
Here the factors of $Q/n\cdot q$ make $S$ boost invariant along the $n$-$\nbar$ axis, and  $Q/n\cdot q=1$ in the Breit frame.
We use boost invariance and power counting to determine the first argument of $S$. 
The presence of the state $X_s$ implies that $S$ depends implicitly on $h$.
Finally, we have the ultrasoft-collinear function, which allows a transition between color representations $R_A$ and $R_B$, and uses hemispheres to define momenta $\bar p_{uc}^{X-}$ and $\bar p_{uc}^{Y+}$ in the regions $X$ and $Y$:
\begin{align}  \label{eq:Udefn}
 & U_{(N,N')}^{R_A^{N\!N'} R_B^{N\!N'}}\!\! 
  \bigl( p_H^{+}, p_H^{-}\bigr)
   = \sumint_{Z_{uc}} \delta\Bigl( p_H^+ -  p_{uc}^{{\cal H}_Y+}\Bigr)\,  
      \delta\Bigl( p_H^- -  p_{uc}^{{\cal H}_X-}\Bigr)
 \\*
 &\quad \times 
  \big\langle 0 \big| P_{N R_A} \bar T\,
 \text{\footnotesize $\prod_{i=1}^{N}$}\, \mathbb{U}_{n\nbar}^{A_i B_i}(0) \,
 P_{N R_B} \big| Z_{uc} \big\rangle
 \big\langle Z_{uc} \big| P_{N' R_{A'}} T\,
 \text{\footnotesize $\prod_{j=1}^{N'}$}\, \mathbb{U}_{n\nbar}^{A'_j B'_j}(0)
  P_{N' R_{B'}} \big| 0\big\rangle 
  .  \nn
\end{align}
We recall from \eq{UU} that $ \mathbb{U}_{n\nbar}^{A B}$ is the product of adjoint Wilson lines.
In the arguments of eqs.~\eqref{eq:Bdefn} through \eqref{eq:Udefn}, we suppress dependence on regulators, though all three functions depend on $\epsilon$, and $B$ and $S$ depend on an additional rapidity regulator $\eta$. 

In \sec{diffract-kinematics}, we used Lorentz invariance to show that $F_i^D$ is not a function of $y$, which seems to be contradicted by the the $y$-dependence inside $h$ in \eq{FDfactsimp}. This is not a contradiction because we spoil Lorentz invariance when defining $X$ and $Y$ with a rapidity cut in the lab frame, and this breaking is encoded by the appearance of $h$. 
If we instead imposed a fixed rapidity cut $\etaxybar$ in the $\gamma_{\nbar}$-frame, we would have $h=e^{-\etaxybar}$ and thus would retain $y$-independence of $F_i^D$.
This is also true if we impose a fixed rapidity cut in the Breit frame with a constant $h_0$, see \eq{hconstraint2}, where the factorization formula simplifies through the replacement $h\sqrt{x} \to h_0$.
For a lab frame rapidity cut $\etalabcut$ it is convenient to write 
\begin{align}
  h\sqrt{x} = \frac{x \gamma\sqrt{s}}{Q} e^{-\etalabcut} \,,
\end{align}
which makes the dependence on collider variables explicit. 

Setting $d'=2$ in \eq{FDfact}, we see that $F_i^D$ has mass dimension $-4$, which  equals the dimension of $\IInt^\perp_{(N,\,N')}$ in \eq{iint_perp}. 
The function $S$ is dimensionless; $B$ has dimension $-2$, which compensates for the dimension of the prefactor; whereas $U$ has mass dimension $-2$, which compensates for the $+2$ dimension from $dp_g^+\, dp_g^-$.

From~\eq{diffcrosspol}, we 
recall that the diffractive structure functions $F_i^D$
must be real, and we can reorganize  \eq{FDfact} to make this apparent in the factorized expression. 
Taking complex conjugates,  we find that $B_{(N,N')}^* = B_{(N',N)}$, $S_{i(N',N)}^* = S_{i(N,N')}$, and $U_{(N,N')}^* = U_{(N',N)}$. 
These equalities must be understood with the color indices and arguments associated with matrix elements to left and right of the cut appropriately swapped. For example, $\Bigl(B_{(N,N')}^{R_A^{N\!N'}} 
  \bigl(p^- k_n^+, \{\tau_{i\perp},\tau_{j\perp}'\}, t\bigr)\Bigr)^*  = B_{(N',N)}^{R_A^{N'\!N}}  \bigl(p^- k_n^+, \{\tau_{j\perp}',\tau_{i\perp}\}, t\bigr)$, where the transverse momenta on the right-side of the cut of $B_{(N,N')}$ are identified with those of the left-side of the cut of $B_{(N',N)}$. 
We can now group different terms in~\eq{FDfact} as
\begin{align}
F_i^D=& \sum_{N}\sum_{R} (-i)^N (+i)^{N}B_{(N,N)}^{R_A^{N\!N}} \tilde{\otimes}_\perp S_{i(N,N)}^{R_B^{N\!N}} \otimes_{\pm} U_{(N,N)}^{R_A^{N\!N} R_B^{N\!N}}\nn\\
&+ \sum_{N<N'}\sum_{R} \Bigg[(-i)^N (+i)^{N'}B_{(N,N')}^{R_A^{N\!N'}} \tilde{\otimes}_\perp S_{i(N,N')}^{R_B^{N\!N'}} \otimes_{\pm} U_{(N,N')}^{R_A^{N\!N'} R_B^{N\!N'}}\nn\\
&\hspace{1.9cm}+(-i)^{N'} (+i)^{N}B_{(N',N)}^{R_A^{N'\!N}} \tilde{\otimes}_\perp S_{i(N',N)}^{R_B^{N'\!N}} \otimes_{\pm} U_{(N',N)}^{R_A^{N'\!N} R_B^{N'\!N}}\Bigg]\,,
\end{align}
where each term is explicitly real by complex conjugation properties, manifestly demonstrating that $F_i^D$ is real, as expected.

\paragraph{Renormalization.}
Next, we consider the impact of renormalization on our bare factorization in \eq{FDfact}. Using \refcite{Gao:2024qsg} and knowledge of renormalization in the di-hemisphere mass distribution factorization~\cite{Fleming:2007xt}, we find that $B$, $S$, and $U$ are renormalized multiplicatively in their momentum function space (i.e., with convolutions over all momentum arguments that appear as integrals in \eq{FDfact}), and diagonal in the reduced color representations $R_A^{N\!N'}$ and $R_B^{N\!N'}$.   $B$ and $S$ exhibit standard ultraviolet renormalization in the $\overline{\rm MS}$ scheme that leaves dependence on the invariant mass renormalization scale $\mu$, as well as rapidity renormalization that leaves dependence on a (mass-dimension-1) rapidity scale $\nu$. To treat rapidity renormalization, we follow \refscite{Chiu:2011qc,Chiu:2012ir}. $U$ has no rapidity divergences, so the renormalized $U$ is only $\mu$ dependent. Renormalization group (RG) consistency in SCET implies that the rapidity anomalous dimensions of $B$ and $S$ are equal and opposite, while the $\mu$ anomalous dimensions of $B$, $S$, and $U$ sum to zero when combined with suitable integrations. These considerations lead to a renormalized factorization formula, 
\begin{align} \label{eq:FDfactRen}
F_i^D
&=\FiDpre \!\!
 \sum_{{\footnotesize N,N'=1}}^\infty 
 \! \sum_{ \{R_{X}\} }\! \IInt^\perp_{(N,N')}  
  \int\!\! dp_g^+ dp_g^-\ 
   B_{(N,N')}^{R_A^{N\!N'}} 
    \Bigl( m_Y^2-t -p^- p_g^+, \{\tau_{i\perp},\tau_{j\perp}'\}, t, 
     \mu,\frac{\nu}{Q/x} \Bigr) 
 \nn\\*
 &\!\! \times\!
  U_{(N,N')}^{R_A^{N\!N'} R_B^{N\!N'}} \!\!\!
 \Big( \frac{1}{h\sqrt{x}}\,p_g^+, h\sqrt{x}\, p_g^-,\mu \Big)
   S_{i(N,N')}^{R_B^{N\!N'}} \!
    \Bigl(\frac{Q^2}{\beta}\!-\!q^+ p_g^-, \{\tau_{i\perp},\tau_{j\perp}'\}, Q,t,\mu,
  \frac{\nu}{Q/\beta} \Bigr) 
   . 
\end{align}
Here, we substitute Breit frame variables with Lorentz invariants: $p^- p^{\prime+}=m_Y^2-t$ and $q^+ \tau^- = Q^2/\beta$. 
The beam function gains dependence on the proton momentum $p^-=Q/x$ due to boost invariance breaking by the rapidity regulator in the argument $\frac{\nu}{Q/x}$.

\paragraph{Diffraction~and~quasi-diffraction.} Experimental cross sections include contributions from both diffraction (color-singlet $R=1$)  and quasi-diffraction (all $R\neq 1$ channels), as $F_i^D =  F_i^{D\,{\rm diff}} + F_i^{D\,{\rm quasi}}$.
\Eq{FDfactRen} allows us to clarify the distinction between these two contributions.  

Let us begin by discussing diffraction, which proceeds via a color-singlet channel. This occurs for all $N,N' \ge 2$, and can occur with multiple copies for larger $N$ and $N'$. The factorization in \eq{FDfact} simplifies significantly for diffraction: the uc function becomes trivial for color-singlet exchange, irrespective of the value of $N$ and $N'$, to all orders in $\alpha_s$:
\begin{align}
	 U_{(N,N')}^{1\, 1}\bigl( \bar p_g^+, \bar p_g^-\bigr) 
	   =  \delta(\bar p_g^+) \delta(\bar p_g^-) \,.
\end{align}
Simply put, at leading power there is automatically no ultrasoft-collinear radiation into the gap for diffraction. This occurs because the color-singlet projection cancels out all Wilson lines ${\cal U}_n$ and separately all Wilson lines ${\cal U}_{\nbar}$.
For example, for $N=2$, we have
\begin{align}
	  P_{2\, 1} {\cal U}_n^{C_1 A_1 } {\cal U}_{\nbar}^{C_1B_1}
	             {\cal U}_n^{C_2 A_2 } {\cal U}_{\nbar}^{C_2B_2}
	  &\propto \delta^{A_1 A_2} {\cal U}_n^{C_1 A_1 }{\cal U}_n^{C_2 A_2}
	      {\cal U}_{\nbar}^{C_1B_1} {\cal U}_{\nbar}^{C_2B_2}
	= \delta^{B_1 B_2}
	   \,.
\end{align} 
Similar algebra eliminates the Wilson lines for any $(N,N')$ for singlet exchange.
Thus, for diffraction we have a much simpler renormalized factorization formula 
\begin{align} \label{eq:FDfactdiff}
	F_i^{D\,{\rm diff}}\!
	&= \FiDpre \!\!
	\sum_{{\footnotesize N,N'=1}}^\infty
	  \sum_{R^{N\!N'}=1} \IInt^\perp_{(N,N')}  
	  \!\! B_{(N,N')}^{R^{N\!N'}}
	\Bigl(m_Y^2-t , \{\tau_{i\perp},\tau_{j\perp}'\}, t,\frac{\nu}{Q/x}\Bigr) 
	 \nn\\*
	 & \qquad\qquad\qquad\qquad \quad  \times
	 S_{i(N,N')}^{R^{N\!N'}} 
	  \Bigl(\frac{Q^2}{\beta}, \{\tau_{i\perp},\tau_{j\perp}'\}, Q,t, \frac{\nu}{Q/\beta} \Bigr) 
	 . 
\end{align}
Here, $B$ and $S_i$ are separately $\mu$-independent because there is no uc function; i.e., there is no RG evolution in $\mu$ between them, as they exist at the same invariant mass scale in our analysis so far. 
The dependence on parameters like $h$ drops out and only enters through the definition of the $X$ and $Y$ states.

Quasi-diffraction, in contrast, proceeds via color-nonsinglet channels, which occur for all $N,N'\ge 1$. In this case, the uc function is nontrivial, so quasi-diffraction is described by the full renormalized factorization formula in \eq{FDfactRen} minus the singlet channel:
\begin{align} \label{eq:FDfactquasi}
	& F_i^{D\,{\rm quasi}}
	\!= \FiDpre \!\!\!\!
	\sum_{{\footnotesize N,N'=1}}^\infty 
	  \sum_{R_{X}\ne 1}\! \IInt^\perp_{(N,N')}  
	\int\!\! dp_g^+ dp_g^-\,
	B_{(N,N')}^{R_A^{N\!N'}} \!
	\Bigl(m_Y^2-t \!-\! p^-p_g^+, \{\tau_{i\perp},\tau_{j\perp}'\}, t,\mu,\frac{\nu}{Q/x} \Bigr) 
	\nn\\*
	& \ \ \ \ \  \times
	U_{(N,N')}^{R_A^{N\!N'} R_B^{N\!N'}} \!
	\Bigl( \frac{1}{h\sqrt{x}}\,p_g^+, h\sqrt{x}\, p_g^-,\mu \Bigr)
	\: S_{i(N,N')}^{R_B^{N\!N'}} \!
	\Bigl(\frac{Q^2}{\beta}\!-\! q^+p_g^-, \{\tau_{i\perp},\tau_{j\perp}'\}, Q,t,\mu, \frac{\nu}{Q/\beta} \Bigr) 
	. 
\end{align}
Thus, the quasi-diffractive process gives a non-trivial contribution to the total diffractive cross section, and even with cuts, is a nonzero irreducible background to the color singlet diffractive process.
Note that $h$ is the only explicit parameter that scales with the power counting $\lambda_g$ from \eq{power-counting-params}.
We discuss the magnitude of quasi-diffraction relative to diffraction in \sec{predictions}.

We emphasize that the results in \eqs{FDfactdiff}{FDfactquasi} apply for the experimental $m_X$-method and coherent forward detector method discussed in \secs{diffractive-constraints}{gap-radiation}. For the rapidity gap method, an additional scale $E_{\rm gap}^{\rm lab}$ enters due to the constraint $E_i^{\rm lab}<E_{\rm gap}^{\rm lab}$ for $\eta_{\rm min}^{\rm lab}<\eta_i^{\rm lab}< \eta_{\rm max}^{\rm lab}$, which modifies the allowed radiation in $X$. Therefore, $E_{\rm gap}^{\rm lab}$, $\eta_{\rm min}^{\rm lab}$, and $\eta_{\rm max}^{\rm lab}$ would appear as additional arguments for the functions $S$ and $U$. 
Although it is tractable to analyze this method in our setup, we do not do so here.

The diffractive factorization in \eq{FDfactdiff} includes contributions from all singlet Glauber gluon exchanges, including different singlet channels (Pomeron, odderon, etc.), as well as both perturbative and non-perturbative contributions. 
When $\sqrt{-t} \gg \LQCD$, we say that a perturbative (hard) Pomeron contribution occurs when $\tau_{i\perp} \sim \sqrt{-t} \gg \LQCD$ for all $i$, which we can calculate with an expansion in $\alpha_s(\sqrt{-t})$. A non-perturbative Pomeron contribution occurs when $\tau_{i\perp} \sim \LQCD$ for one or more $i$, in which case there is no coupling constant suppression from adding the corresponding Glauber exchange operator.  We integrate over the $\tau_{i\perp}$ variables, so both of these contribute for $\sqrt{-t} \gg \LQCD$. For the situation where $\sqrt{-t}\sim \LQCD$, all Glauber gluons are nonperturbative with $\tau_{i\perp} \sim \LQCD$, corresponding to the so-called soft Pomeron.  Quasi-diffraction likewise has perturbative and non-perturbative contributions for various color channels. We explore this separation of perturbative and nonperturbative contributions further in \sec{soft-results}.

So far we have only carried out the factorization that arises from the $\lambda\ll 1$ expansion. In \sec{smalllambdat}, we consider an additional expansion in $\lambda_t\ll 1$ that further factorizes $S_i$.  We leave exploration of the re-factorization of $B$ for $\lambda_{\Lambda}=\LQCD/\sqrt{-t}\ll 1$ to a future publication.

\paragraph{Spin-dependent structure functions.}

From our $\lambda\ll 1$ factorization, we can now understand why the six polarized structure functions listed in \eq{polarized-structures-lp} contribute to (quasi-)diffraction at leading power, but the remaining eight in \app{structures} do not. Due to factorization, the proton spin $S$ only enters the beam function. The beam function is a scalar in Lorentz space because the Glauber exchanges have a fixed structure and do not transmit spin information. Thus, we can decompose $B$ into two terms:
\begin{align} \label{eq:Bdecomp} 
  \widehat B_{(N,N')}^R = B_{(N,N')}^{R} + \tilde S_T \: B_{(N,N')}^{T R} \,,
\end{align}
In the Breit frame, $\tilde S_T = \epsilon_{\perp}^{\alpha\beta} \tau^\perp_\alpha S^T_\beta/\sqrt{-t}$ with $\epsilon_\perp^{\alpha\beta}\equiv \epsilon^{\alpha\beta\gamma\delta} \nbar_\gamma n_\delta / 2$, and hence is an allowed structure in $B$. 
Here $B_{(N,N')}^{R}$ appears for the unpolarized structure functions $F_{L,2,3,4}^D$ and the antisymmetric structure function $F_{4A}^D$, with the hadron beam unpolarized and the lepton beam polarized, and $B_{(N,N')}^{T R}$ appears for hadron-spin-polarized structure functions $\propto \tilde S_T$. 
Thus, only the structure functions $F_{iP}^D$ with simple multiplicative hadronic spin dependence can be nontrivial at this order, and we predict that their leading-power factorization formulas are given by
\begin{align} \label{eq:FiPD}
  F_{iP}^D &= \frac{N_X}{\sqrt{-t}}\: 
     F_i^D \ \text{with}\ B_{(N,N')}^{R} \to B_{(N,N')}^{T R} \,,
\end{align}
with $i=L,2,3,4,4A$ and $F_i^D$ given by eqs.~(\ref{eq:signal-background}), (\ref{eq:FDfactdiff}), and (\ref{eq:FDfactquasi}). 

In contrast, the structure functions in \app{structures} with tensor coefficients $w_i^{\mu\nu}$ for $i=5$-$8$ or $i=5A$-$8A$ are proportional to either $\bar U$ or $\bar X$. In the Breit frame, we have
\begin{align}
  \bar X^\mu = \epsilon_\perp^{\mu\alpha} S^T_\alpha \,, \qquad
  \bar U^\mu = \frac{n^\mu+\nbar^\mu}{2}\: \tilde S_T^{\prime} 
     - \epsilon_\perp^{\mu\alpha} \tau^\perp_\alpha \: \frac{S_L}{\gamma N_X} 
  \,,
\end{align}
where $S_L$ is the $z$-component of the spin vector $S^\mu$ as defined in \eq{s-components}.
These structure functions will include prefactors $\epsilon_\perp^{\mu\alpha} k^\perp_\mu S^T_\alpha$ or $\epsilon_\perp^{\mu\alpha} k^\perp_\mu p^\prime_\alpha$, and are impossible to generate at leading power due to the scalar nature of $B$, which does not allow for a contraction of transverse Lorentz indices with the soft function $S_i^{\mu\nu}$.

\subsection{Hard-collinear factorization of $S^{\mu\nu}$ for $\lambda \ll 1 $ and $\lambda_t \ll 1$}
\label{sec:smalllambdat}

Next, we consider the case where we keep $\lambda \ll 1$ and ${\rho,\beta}\sim 1$ as in \sec{lambda-fact}, but now impose $\lambda_t =\sqrt{-t}/Q\ll 1$. This $\lambda_t \ll 1$ expansion will induce a further factorization beyond \eqs{FDfactRen}{FDfactquasi}.
We can deduce from the dependence on scales how this additional factorization will play out for (quasi-)diffraction. Examining \eq{FDfactRen}, we see that only the soft function $S$ depends on both $t$ and $Q^2$, which means that a $\lambda_t \ll 1$ expansion can impact its structure. Since the beam function $B$ only depends on $t$, $m_Y^2$ and $\Lambda_{\rm QCD}^2$, it will remain unaffected by a $\lambda_t \ll 1$ expansion. 

To make this more precise, let us examine how the momenta in eqs.~\eqref{eq:four-vector-scales}, \eqref{eq:four-vector-scales-new}, 
and \eqref{eq:gap-radiation-Breit} scale in $\lambda, \lambda_t$:
\begin{align}\label{eq:four-vector-scales3} 
&q \sim\sqrt{s}\lambda(1,\,1,\, 0) 
 \,,
&&p_{X_s} \sim \sqrt{s}\lambda(1,\, 1,\,  \lambda_t )
 \,,
&&\tau \sim\sqrt{s}\lambda(\lambda^2\lambda_t^2 , 1, \lambda_t ) 
 \,,
\nonumber\\
&p\sim \sqrt{s} \lambda(0,\lambda^{-2},0)
 \,,
&&p'_n \sim\sqrt{s}\lambda( \lambda^2\lambda_t^2 , \lambda^{-2}, \lambda_t )
 \,,
&&p_{uc}^\mu \sim \sqrt{s}\lambda \big( \lambda^2\lambda_t^2, 1, \lambda\lambda_t \big) 
\,.
\end{align}
The combination $\sqrt{s} \lambda \sim Q$ appears as a prefactor in all the momenta and serves as an overall hard scale for the problem. The modes $\tau$ and $q$ have collinear and hard scalings in the parameter $\lambda_t$, respectively. This implies that the momentum $p_{X_s}$ in the soft-$\lambda$ region with $q+\tau = p_X = p_{X_s} + p_{uc}^X$ has contributions from both types of modes. A $\lambda_t$-refactorization of the soft function entails separating loop and outgoing momenta with these hard-$\lambda_t$ and collinear-$\lambda_t$ scalings.

Our factorization procedure here closely parallels the hard-collinear factorization used in inclusive DIS. Specifically, we can 
refactorize the soft function in \eq{Sdefn} into a hard function $\mathcal{H}$ and a soft-collinear (sc) function $S_{\rm c}$.  From the EFT perspective, $\mathcal{H}$ encodes the hard-$\lambda_t$ momentum region and serves as the Wilson coefficient for $S_{\rm c}$, which encodes the collinear-$\lambda_t$ momenta.  For this analysis it is convenient to write the first argument of the $S_{i(N,N')}^{R^{N\!N'}}$ soft function as $Q^2 u/\beta$, where $u=1$ for diffraction, and $u=1-p_g^-/\tau^-$ for quasi-diffraction.  The minus component of both modes scales as $\mathcal{O}(\lambda_t^0)$, so they are coupled by a convolution integral:
\begin{align}
\label{eq:refactS}
S_{i(N,N')}^{R^{N\!N'}} &\Bigl( \frac{Q^2 u}{\beta} , \{\tau_{k\perp},\tau_{\ell\perp}'\}, Q,t,\mu,\frac{\nu}{Q/\beta} \Bigr) \\
 &= \frac{\beta^2}{Q^2u^2}
 \sum_{\kappa} \int_{\beta}^u \frac{d\zeta}{\zeta}\: \mathcal{H}^{\kappa}_{i}\Bigl(\frac{\beta}{\zeta}, Q,\mu\Bigr)\:  
  S^{\kappa; R^{N\!N'}}_{\rm c(N,N')}\left( \frac{\zeta}{u}, \{\tau_{k\perp},\tau_{\ell\perp}'\},t,
  \frac{\nu}{Q/\beta},\mu \right)
  \times \left[1+\mathcal{O}(\lambda_t)\right]
 \nn\\
  &\equiv \frac{\beta^2}{Q^2u^2}\, \mathcal{H}_i^\kappa \otimes_{-\:} 
    S^{\kappa; R^{N\!N'}}_{\rm c(N,N')}
 \times \left[1+\mathcal{O}(\lambda_t)\right]\,,\nn
\end{align}
where $u>\beta$ since $p_{X_s}^->0$. 
Here ${\cal H}_i^\kappa$ captures the hard $\lambda_t$-regime and $S_{\rm c}$ describes the collinear-$\lambda_t$ regime, and encodes the color and functional dependence of $S^{\mu\nu}$ that talks to $B$ (and $U$).
Physically, it is easiest to understand the form of the convolution in \eq{refactS} in the Breit frame. The original $S_i$ depends on two minus-momenta, through $Q^2u/\beta= q^+ (\tau^- u)$ and $Q^2= q^+ (-q^-)$.  We can view \eq{refactS} as describing the $(N,N')$ Glaubers which enter the soft function with total minus-momentum $\tau^- u$, splitting into a quark or gluon $\kappa$ that carries momentum $p_\kappa^-$, which then participates in the hard scattering. The $S_{\rm c}$ function depends on the momentum fraction $p_\kappa^-/(u\tau^-)=\zeta/u $, while the hard function 
${\cal H}_i^\kappa$ depends on the momentum fraction $(-q^-)/p_\kappa^- = (-q^-/\tau^-)(\tau^-/p_\kappa^-)= \beta/\zeta$, and thus \eq{refactS} factorizes the dependence on the two original minus-momenta into individual functions.
Recall that $\beta= -q^-/\tau^-$ in the Breit frame.
The last line in \eq{refactS} introduces a shorthand $\otimes_{-}$ for the $\zeta$ momentum fraction convolution.

\paragraph{Results for dPDFs.} In the 1990s, \refscite{Berera:1995fj, Collins:1997sr} established a hard-collinear factorization framework for diffraction with $\lambda_t = \sqrt{-t}/Q \ll 1$, which resembles the factorization approach used in inclusive DIS and our hard-collinear factorization of $S$ in \eq{refactS}.
This factorization for coherent diffraction is written as~\cite{Frankfurt:2022jns}
\begin{align}\label{eq:collins-factorization}
 \frac{1}{x}
 F^D_{2/L} = \sum_\kappa \int_\beta^1 \frac{d\zeta}{\zeta} H^{\kappa}_{2/L}\Big(\frac{\beta}{\zeta}, Q,\mu\Big)  f_\kappa^{D\,{\rm coh}}\left(\zeta,\frac{x}{\beta},t,\mu\right)+ \cO\left( \frac{-t}{Q^2},\frac{ \LQCD^2}{Q^2} \right) \,.
\end{align}
Here,  $H^{\kappa}_{2/L}$ is the same hard coefficient as for inclusive DIS, and $f_\kappa^D$ is the diffractive PDF (dPDF) for a parton of type $\kappa$. The $\mu$-dependence of $H^{\kappa}_{i}$ and $f_\kappa^D$ is governed by DGLAP evolution, which relates diffractive structure functions at different $Q$. The dPDF factorization formula in~\eq{collins-factorization} forms the basis of many HERA global fits ~\cite{H1:2006zyl,ZEUS:2009uxs,Goharipour:2018yov,Khanpour:2019pzq,Monfared:2011xf,Salajegheh:2022vyv} and is the subject of many model calculations and generalizations \cite{Hatta:2024vzv,Maktoubian:2019ppi,Golec-Biernat:2007mao,Golec-Biernat:2001gyl,Hebecker:1997gp}. 

The literature frequently notes the lack of a first-principles description of how the hadronic sector forward-scatters; i.e., the Regge factorization of $f_i^D$~\cite{Collins:1997sr,Berera:1995fj,Collins:2001ga,Berger:1986iu,Arneodo:2005kd}, which corresponds to our $\lambda\ll 1$ expansion.
We can resolve this issue by combining our $\lambda\ll 1$ factorization from \sec{lambda-fact} with the $\lambda_t\ll 1$ factorization of the soft function in \eq{refactS}. Note that beyond just the $F_{2,L}^D$ often considered in diffraction, our results also apply to $F_{3,4}^D$ and to quasi-diffractive processes. 
Substituting \eq{refactS} into \eq{FDfactdiff} and~\eq{FDfactquasi}, we obtain
\begin{align} \label{eq:FactDPDF}
\frac{1}{x}\, F_i^{D\,{\rm diff}} 
  &=   \sum_{\kappa} \int_{\beta}^1 \frac{d\zeta}{\zeta}\: 
     \mathcal{H}^{\kappa}_{i}\Bigl(\frac{\beta}{\zeta}, Q,\mu\Bigr)
     f_\kappa^{D\,{\rm diff}}\Bigl(\zeta, \frac{x}{\beta}, m_Y^2-t,t,\mu \Bigr) 
\times \big[ 1 + {\cal O}(\lambda_t) \big]
  \,,\nn\\
 \frac{1}{x}\,  F_i^{D\,{\rm quasi}} 
  &=   \sum_{\kappa} \int_{\beta}^1 \frac{d\zeta}{\zeta}\: 
     \mathcal{H}^{\kappa}_{i}\Bigl(\frac{\beta}{\zeta}, Q,\mu\Bigr)
     f_\kappa^{D\,{\rm quasi}} \Bigl(\zeta, \frac{x}{\beta}, m_Y^2-t,t,\mu \Bigr) 
  \times \big[ 1 + {\cal O}(\lambda_t) \big]
  \,, 
\end{align}
where using a shorthand $m^2=m_Y^2-t$, the diffractive PDF is
\begin{align} \label{eq:FDfactdifflambdat}
 f_\kappa^{D\,{\rm diff}}\Bigl(\zeta, \frac{x}{\beta}, m^2,t,\mu \Bigr) 
 &= \BoostInv^{2}\!\!
 \sum_{{\footnotesize N,N'=1}}^\infty
  \sum_{R^{N\!N'}=1} \frac{\beta^2}{x^2} \IInt^\perp_{(N,N')} 
  \!\! 
S^{\kappa; R^{N\!N'}}_{\rm c(N,N')}\left(\zeta, \{\tau_{k\perp},\tau_{\ell\perp}'\},t,
   \frac{\nu}{Q/\beta},\mu\right)
 \nn \\
&\ \ \times 
  B_{(N,N')}^{R^{N\!N'}}
  \Bigl(m^2, \{\tau_{k\perp},\tau_{\ell\perp}'\}, t, \frac{\nu}{Q/x}\Bigr)
 \times \big[ 1 + {\cal O}(\lambda) \big]
 ,
\end{align}
and the quasi-diffractive PDF is
\begin{align} \label{eq:FDfactquasilambdat}
f_\kappa^{D\,{\rm quasi}} &\Bigl(\zeta, \frac{x}{\beta}, m^2,t,\mu \Bigr) 
 = 
   \BoostInv^{2}\!\!
  \sum_{{\footnotesize N,N'=1}}^\infty 
  \sum_{ R_{X}\ne 1 }\! 
  \IInt^\perp_{(N,N')}  
  \int\!\! dp_g^+ dp_g^-\:
  U_{(N,N')}^{R_A^{N\!N'} R_B^{N\!N'}} \!
  \Bigl( \frac{p_g^+}{h\sqrt{x}}, h\sqrt{x}\, p_g^-,\mu \Bigr)
 \nn \\*
 &\times
\frac{\beta^2}{x^2}\left(\frac{\tau^- }{\tau^--p_g^-}\right)^2\: 
S^{\kappa; R_B^{N\!N'}}_{\rm c(N,N')}\left(\zeta\frac{\tau^-}{\tau^--p_g^-}, \{\tau_{k\perp},\tau_{\ell\perp}'\},t,\frac{\nu}{Q/\beta},\mu\right) 
 \,\theta\big(\tau^-(1-\zeta)-p_g^-\big) 
 \nn\\*
 &\times
   B_{(N,N')}^{R_A^{N\!N'}} \!
    \Bigl(m^2 - p^-p_g^+, \{\tau_{k\perp},\tau_{\ell\perp}'\}, t,\mu,\frac{\nu}{Q/x} \Bigr) 
 \times \big[ 1 + {\cal O}(\lambda) \big]
 \,.
\end{align}
Note that the nontrivial $x/\beta$ dependence in the (quasi-)diffractive PDF is only induced by the sum of large rapidity logarithms between the soft-collinear and beam functions.  
The total dPDF is given by the sum of these two contributions%
\begin{align}
f_\kappa^{D}\Bigl(\zeta,\frac{x}{\beta},m^2,t,\mu\Bigr)
   &=  f_\kappa^{D\,{\rm diff}}\Bigl(\zeta,\frac{x}{\beta},m^2,t,\mu\Bigr)  + f_\kappa^{D\,{\rm quasi}} \Bigl(\zeta,\frac{x}{\beta},m^2,t,\mu\Bigr)
  \,.
\end{align}
These diffractive results can also be decomposed into coherent and incoherent contributions, see~\sec{coherent-and-beams} and~\sec{quasibkgnd}.
This result marks the first explicit Regge factorization of diffractive PDFs and of the new quasi-diffractive PDFs in \eq{FDfactquasilambdat}.  
Note that the analysis above is based on the consistency of results obtained separately in the $\lambda\ll 1$ and $\lambda_t\ll 1$ limits, rather than a more rigorous derivation based on an operator analysis, which we leave to future work.

\paragraph{Relationship to DIS.} Inclusive DIS factorizes as $F_i = x \int_x^1 d\xi/\xi\, H_i^{\kappa}(x/\xi,Q,\mu) f_\kappa(\xi,\mu)$. In constrast, the hard coefficient in diffraction depends on $\beta=Q^2/2q\cdot \tau$ instead of $x=Q^2/2q\cdot p$, since the momentum fraction $\zeta=p_\kappa^-/\tau^-$ is relative to the incoming Glauber momentum $\tau^-=Q/\beta$ rather than the incoming proton momentum $p^-=Q/x$. (Recall that we always expand in $\lambda\lambda_t \ll 1$, in which case  $\tau^-/p^- = x/\beta$.) A further distinction between DIS and diffraction is that when $\lambda_\Lambda =  \Lambda_{\rm QCD}/\sqrt{-t}\ll 1$, it is possible to perturbatively match $S_{\rm c}\otimes_\perp B $ onto simpler nonperturbative matrix elements (though we do not do so here). Of course, if $\lambda_\Lambda \sim 1$, $S_{\rm c}$ and $B$ become fully nonperturbative just like PDFs. 

Inclusive and diffractive DIS have identical tensor structures for $w_2^{\mu\nu}$ and $w_L^{\mu\nu}$; see \eq{tensor-structures}.  Thus, we expect that the hard coefficients $\mathcal{H}^{\kappa}_{2/L}$ and $H^\kappa_{2/L}$ of diffraction and DIS are identical functions,
\begin{align}
\label{eq:identityH}
\mathcal{H}^{j}_{2/L}(\zeta,Q,\mu)  = H_{2/L}^{j}(\zeta,Q,\mu) \,.
\end{align} 
For $F_2$ and $F_L$, these matching coefficients at NLO are~\cite{Collins:2011zzd,Ellis:1996mzs,Moch:1999eb}:
\begin{align}\label{eq:hardcoeff}
	H^{q_j[0]}_2(\zeta,Q,\mu) &=e_j^2 \delta(1-\zeta)\,,
	\nn  \\
	H^{q_j[1]}_L(\zeta,Q,\mu) &=\frac{\alpha_s C_F e_j^2}{\pi}  \zeta\,,
	\nn  \\
	H^{q_j[1]}_2(\zeta,Q,\mu) &=\frac{\alpha_s C_F e_j^2}{4\pi}  \left[4\left(\frac{\ln(1-\zeta)}{1-\zeta}\right)_+-3\left(\frac{1}{1-\zeta}\right)_+-2(1+\zeta)\ln(1-\zeta)\right.
	\nn\\
	&\qquad\left.-2\frac{1+\zeta^2}{1-z}\ln \zeta+6+4\zeta-\left(\frac{2\pi^2}{3}+9\right)\delta(1-\zeta)\right]\,,
	\nn  \\
	H^{g[1]}_2(\zeta,Q,\mu) &=\sum_j \frac{\alpha_s T_F e_j^2}{2\pi}  \left[ (1-2\zeta+2\zeta^2)\ln\frac{Q^2(1-\zeta)}{\mu^2 \zeta}-1+8\zeta-8\zeta^2\right]\,,
	\nn  \\
	H^{g[1]}_L(\zeta,Q,\mu) &= \sum_j \frac{\alpha_s T_F e_j^2}{2\pi} 4\zeta(1-\zeta)
  \,,
\end{align}
where the superscript $[n]$ indicates the result at ${\cal O}(\alpha_s^n)$ and $\sum_j$ sums over quark and anti-quark flavors. For inclusive DIS, $H^{j}_{2/L}$ is known at N$^3$LO for both neutral and charged currents~\cite{Vermaseren:2005qc,Moch:2007gx,Moch:2008fj}. In \sec{soft-lo-1g-lt}, we perturbatively compute $S_{\rm c}$ for $N=N'=1$ and verify \eq{identityH}. We leave the explicit operator definitions of $S_{\rm c}$  to a future paper.

We can do a consistency check between the hard-collinear factorization of diffractive DIS in~\eq{collins-factorization} and inclusive DIS with our prefactors and the same hard coefficients. The two hadronic tensors are defined as
\begin{align}
W_D^{\mu\nu}(q, p, p') 
&= \sumint_X  \delta^4(q + p - p' - p_X) \:
\big\langle p \big| J^{\dagger\,\mu}(0) \big| p' X\big\rangle
\big\langle p' X \big| J^\nu(0) \big| p \big\rangle\,,\nn\\
W_{\rm DIS}^{\mu\nu}(q, p) 
&= \frac{1}{2}(2\pi)^3 \sumint_X \: \delta^4(q + p - p_X) \:
\big\langle p \big| J^{\dagger\,\mu}(0) \big| X\big\rangle
\big\langle X \big| J^\nu(0) \big| p \big\rangle\,,
\end{align}
where we restrict to coherent diffraction for convenience. In other words,
\begin{align}
\frac{1}{2}\int \frac{d^3p'}{2E_p'} W_D^{\mu\nu}(q, p, p')  =  W_{\rm DIS}^{\mu\nu}(q, p) \,.
\end{align}
The phase space of $p'$ includes a non-diffractive region. However, the hard-collinear factorization does not depend on this condition, and so the comparison is valid. Parameterizing $W_D^{\mu\nu}$ and $W^{\mu\nu}$ according to~\sec{structure-functions}, we find
\begin{align}
\label{eq:intF2DtoF2}
\frac{1}{2} \int \frac{d^3p'}{2E_{p'}} F_{2/L}^D = \frac{\pi}{4}\int_x^1 d\beta \int dt\, \frac{x}{\beta^2} F_{2/L}^D = F_{2/L}\,.
\end{align}
Hard-collinear factorization gives
\begin{align}
\frac{1}{x}F_{2/L}^D &= \sum_\kappa \int_\beta^1 \frac{d\zeta}{\zeta} H^{\kappa}_{2/L}\Big(\frac{\beta}{\zeta}, Q,\mu\Big)  f_\kappa^{D\,{\rm coh}}\left(\zeta,\frac{x}{\beta},t,\mu\right)+ \cO\left( \frac{-t}{Q^2},\frac{ \LQCD^2}{Q^2} \right) \,,\nn\\
\frac{1}{x}F_{2/L}  &= \sum_\kappa \int_x^1 \frac{d\zeta}{\zeta} H^{\kappa}_{2/L}\Big(\frac{x}{\zeta}, Q,\mu\Big)  f_\kappa\left(\zeta,\mu\right)+ \cO\left( \frac{ \LQCD^2}{Q^2} \right) \,.
\end{align}
Therefore,~\eq{intF2DtoF2} implies that
\begin{align} \label{eq:hardrel}
\frac{\pi}{4}\int_x^1\!\!\! d\beta\, dt\, \frac{x}{\beta^2}  \int_\beta^1 \frac{d\zeta}{\zeta} 
 & H^{\kappa}_{2/L}\Big(\frac{\beta}{\zeta}, Q,\mu\Big)  
   f_\kappa^{D\,{\rm coh}}\Bigl(\zeta,\frac{x}{\beta},t,\mu\Bigr)
=  \int_x^1\! \frac{d\zeta}{\zeta} H^{\kappa}_{2/L}\Big(\frac{x}{\zeta}, Q,\mu\Big)  f_\kappa\left(\zeta,\mu\right)\,.
\end{align}
Changing variables from $\beta$ to $\zeta' = x \zeta/\beta$, and then from $\zeta$ to $\xi=x/\beta=\zeta'/\zeta$ 
gives
\begin{align}
\frac{\pi}{4}\int_x^1 \!\!d\beta \int\! dt \frac{x}{\beta^2}\,&  \int_\beta^1 \frac{d\zeta}{\zeta}\, 
  H^{\kappa}_{2/L}\Big(\frac{\beta}{\zeta}, Q,\mu\Big) \,
  f_\kappa^{D\,{\rm coh}}\left(\zeta,\frac{x}{\beta},t,\mu\right)\\
&=  \int_x^1\! \frac{d\zeta'}{\zeta'}\, H^{\kappa}_{2/L}\Big(\frac{x}{\zeta'}, Q,\mu\Big) \left[\frac{\pi}{4}\zeta' \int_{\zeta'}^1 \frac{d\zeta}{\zeta^2}\int dt f_\kappa^{D\,{\rm coh}}\left(\zeta,\frac{x}{\beta},t,\mu\right)\right]\nn\\
&=   \int_x^1 \frac{d\zeta'}{\zeta'}\, H^{\kappa}_{2/L}\Big(\frac{x}{\zeta'}, Q,\mu\Big) \left[\frac{\pi}{4} \int_{\zeta'}^1 d\xi\int dt \,f_\kappa^{D\,{\rm coh}}\left(\frac{\zeta'}{\xi},\xi,t,\mu\right)\right]\,.\nn
\end{align}
Using \eq{hardrel} this implies that 
\begin{align}
\frac{\pi}{4} \int_{\zeta'}^1 d\xi\int dt \,f_\kappa^{D\,{\rm coh}}
 \left(\frac{\zeta'}{\xi},\xi,t,\mu\right)
&=
 \frac{1}{2}\int\frac{d^3{p'}}{2E'_p} \,f_\kappa^{D\,{\rm coh}}
 \biggl( \frac{\zeta'}{\xi}=\frac{{p}_\kappa^-/p^-}{\tau^-/p^-} ,
 \xi=\frac{\tau^-}{p^-} ,t= -\frac{\vec{p}_\perp^{\,'2}}{1-\xi},\mu\biggr)
\nn\\
&= f_\kappa\left(\zeta',\mu\right)\,.
\end{align}
This is consistent with the operator definitions 
\begin{align}
f_q\left(\zeta',\mu\right) 
&= \int \frac{d y^+}{4 \pi}\, e^{-i  \zeta' p^- y^+} \Bigl\langle p \Bigl| \bar{\psi}_q \Bigl(y^+\frac{\nbar}{2}\Bigr) \slashed{\bar{n}} \psi_q (0) \Bigr|p\Bigr\rangle
 \,,\\
f_q^{D\,{\rm coh}}\left(\frac{\zeta'}{\xi},\xi,t,\mu\right)
 &= \frac{2}{(2\pi)^3} \sum_X \hspace{-0.56cm} \large\int\normalsize \,\,
  \int \frac{d y^+}{4 \pi} \, e^{-i  \zeta' p^- y^+}
 \Bigl\langle p\Bigl| 
\bar{\psi}_q \Bigl(y^+\frac{\nbar}{2}\Bigr)  \slashed{\bar{n}}
 \Bigl|p' X\Bigl\rangle \Bigr\langle p' X \Bigr| \psi_q(0)
 \Bigr|p\Bigr\rangle\,, 
\nn
\end{align}
for the quark PDF and dPDF, and similarly for the gluon case.

\section{Large Logarithms in (Quasi-)Diffraction and Sudakov Suppression}
\label{sec:logarithms}

\begin{figure}[t!]
 \hspace{3cm} {\bf Diffractive scales} \hspace{2cm} {\bf Quasi-diffractive scales}\\
	\begin{center}
		\includegraphics[width = 0.33\textwidth]{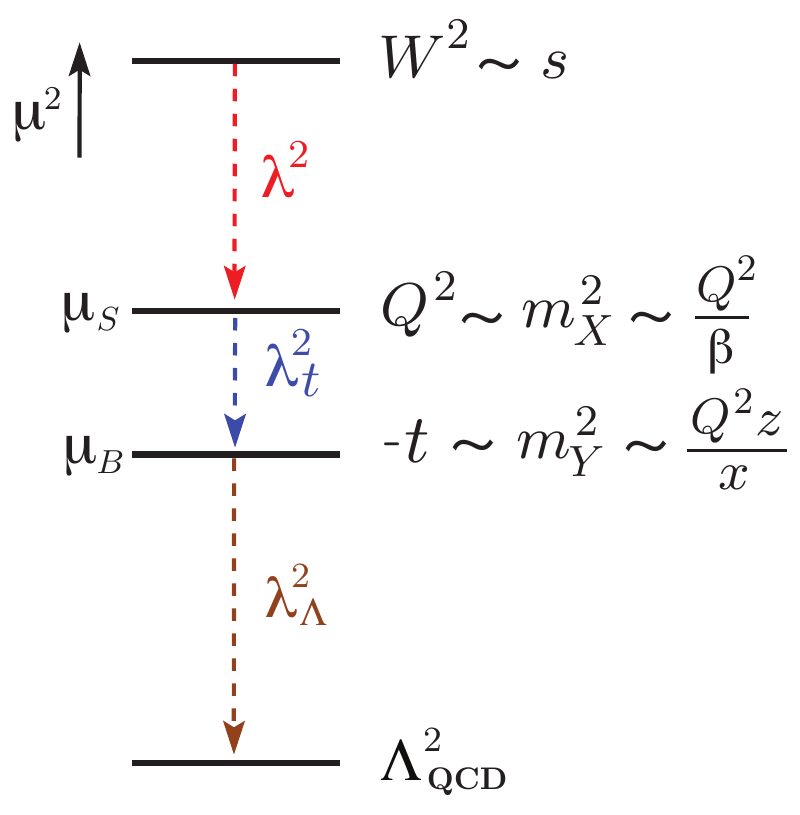} 
		\hspace{0.7cm} 
		\raisebox{1.5cm}{\includegraphics[width = 0.15\textwidth]{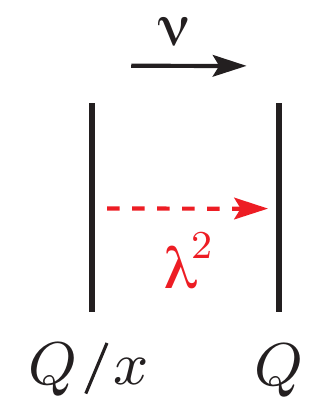}}
         \hspace{0.7cm} 
		\includegraphics[width = 0.33\textwidth]{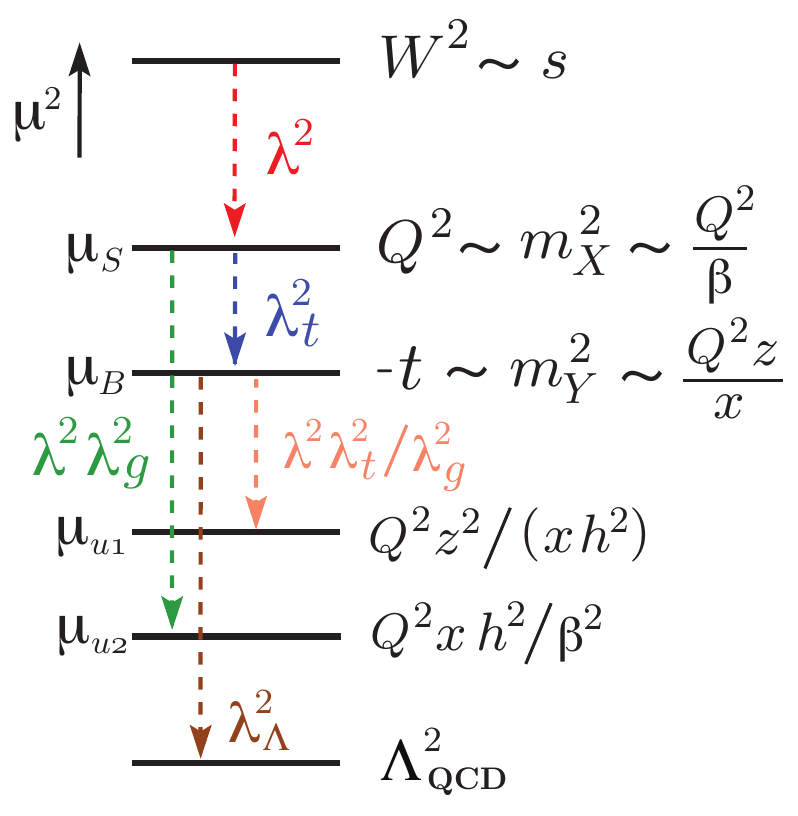} 
\caption{
Scale hierarchies in diffraction (left and middle graphics) and quasi-diffraction (middle and right graphics).  The power counting parameters $\lambda_i$ characterize ratios of scales indicated by the arrows (see also \eq{power-counting-params}), and are shown for the case $\beta\sim 1$, $\rho \sim 1$, $\lambda_t \lesssim 1$ and $\lambda\lambda_t\lesssim \lambda_g$. Here $\mu^2$ indicates a separation in invariant mass, while $\nu$ indicates a separation in longitudinal momentum, which tracks a separation in rapidity. Some of these hierarchies produce large logarithms, as discussed in the text.}
		\label{fig:scales} 
	\end{center}
\end{figure}

As discussed in \secs{diffractive-constraints}{power-counting}, (quasi-)diffractive processes depend on a significant number of hierarchical momentum scales.  
\Fig{scales} depicts these  scales and their associated power counting expansion parameters.
Quantum fluctuations lead to large logarithms of ratios of these physical scales, which can spoil a strict perturbative expansion in $\alpha_s(\mu)\ll 1$ even when $\mu\gg \LQCD$.  To obtain convergence, we can sum up the large logarithms to all orders in perturbation theory using RG techniques in the EFT. 

We begin our discussion of the structure of  large logarithms in the situation where   $\lambda\ll 1$ with $\lambda_t\sim 1$ (whose factorization we discussed in \sec{lambda-fact}). We also analyze the additional large logarithms present when $\lambda_t\ll 1$ (whose factorization we discussed in \sec{smalllambdat}). 
Additionally, we address the special role played by the rapidity-cut-induced parameter $\lambda_g$ in \eq{power-counting-params}.
Throughout this section, we treat $\lambda_\Lambda\sim 1$, and hence do not address additional logarithms that would appear from a hierarchy $\LQCD^2 \ll -t$.  We also do not consider the large logarithms that arise in the case of $\lambda_t\gg 1$ with $\lambda\lambda_t\ll 1$.

\subsection{Diffraction}
\label{sec:difflogs}

For diffraction with color-singlet Glauber exchange, the $\lambda\ll 1$ limit yields the factorization formula in \eq{FDfactdiff}. In this factorization only rapidity logarithms appear, which correspond to $\ln(1/x)$ terms, as shown in the center panel of \fig{scales}. We can resum these logs using RG equations (RGEs) in the rapidity scale $\nu$, as rapidity evolution sums logarithms between the beam function scale $\nu_n = Q/x$ and the soft function scale $\nu_s=Q/\beta$. At lowest order, this sums a single logarithmic series  $\sim \sum_k a_k (\alpha_s\ln x)^k$. In the language of Pomeron exchange used in the diffraction literature \cite{Frankfurt:2022jns}, this resummation corresponds to solving the BFKL equation for the Pomeron at the cross section level~\cite{Kovchegov:2012mbw}. 
For coherent diffraction, our beam function $B$ in \eq{Bdefn} 
is simply a squared amplitude (see \sec{coherent-and-beams}). 
There is a large community studying the Regge limit of amplitudes, using modern field theory methods based on Reggeons%
\footnote{The original Reggeon field theory approach was formulated by Gribov and colleagues in Refs.~\cite{Gribov:1967vfb,Gribov:2003nw,Abarbanel:1975me,Baker:1976cv}.}%
~\cite{Caron-Huot:2013fea,Caron-Huot:2017fxr,Falcioni:2020lvv,Falcioni:2021buo,Falcioni:2021dgr,Caola:2021izf}, the bootstrap approach~\cite{Correia:2025uvc}, and Glauber  SCET~\cite{Rothstein:2016bsq,Moult:2022lfy,Rothstein:2023dgb,Gao:2024qsg,Rothstein:2024fpx,Gao:2024fyz}.
Interestingly, at the level of color-singlet two-Glauber exchange, our beam function involves the same operators as the amplitude-level collinear functions in Ref.~\cite{Gao:2024qsg}, for which the BFKL Pomeron RGE was derived.
By RG consistency, our diffractive soft function $S$ (which involves a cross section-level sum over final states $X$) must be governed by the same Pomeron BFKL evolution. Furthermore, in multi-particle (incoherent) diffraction, this same $S$ appears, and its rapidity logarithms must also be governed by the same amplitude-level Pomeron BFKL evolution (as must $B$ for multi-particle diffraction). 
We leave further numerical exploration of this resummation to future work.

Further power expansions in other power-counting parameters $\lambda_i \ll 1$ or $\lambda_i \gg 1$ induce new logarithms as well. In \sec{smalllambdat}, we examined how a $\lambda_t \ll 1$ expansion factorizes the soft function into a hard function that depends on $Q^2$ and $\mu$, and a soft-collinear function that depends on $t$ and $\mu$. This diffractive hard function is related to the inclusive DIS hard function, and hence its RGE in $\mu$ is governed by the DGLAP equations. The resummation associated to the $Q^2\gg -t$ scale separation thus produces a single-logarithmic series for diffraction, which at leading order takes the form $\sim \sum_k b_k [\alpha_s \ln(-t/Q^2)]^k$. 
These invariant mass scales are pictured in the left graphic in \fig{scales}. 

The factorization formula for diffraction in \eq{FDfactdiff} does not depend on an ultrasoft collinear function $U$, and thus diffraction exhibits no logarithms coming from evolution between $U$ and other scales. This applies for the small-$m_X$ and coherent forward detector  
methods used for diffractive measurements, as discussed in \sec{gap-radiation}. 
In contrast, the ``rapidity gap-method'' induces sensitivity to an additional scale $E_{\rm gap}^{\rm lab}$ in the soft function, as mentioned in \sec{lambda-fact}, but we  do not treat the factorization for this case here. Depending on the relative hierarchies between $\sqrt{-t}$, $Q$, and $E_{\rm gap}^{\rm lab}$, this can induce further factorization of the soft function and require resummation of additional logarithms.

\subsection{Quasi-diffraction}
\label{sec:qdifflogs}

Like diffraction, the $\lambda\ll 1$ factorization formula for quasi-diffraction in \eq{FDfactquasi} exhibits logarithms associated to the beam and soft rapidity scales $\nu_c$ and $\nu_s$. The corresponding rapidity RGEs sum single-logarithmic series of $\alpha_s\ln x$ terms, which correspond to BFKL-type equations for each contributing color channel at the cross section level. Quasi-diffraction also has $\alpha_s \ln(-t/Q^2)$ contributions in the $\lambda_t \ll 1$ regime from DGLAP evolution, just like we discussed for diffraction in \sec{difflogs}. 

However, unlike diffraction, the quasi-diffractive factorization in \eq{FDfactRen} includes a non-trivial ultrasoft-collinear function $U$ and thus $\mu$-evolution between $U$ and $B$, as well as between $U$ and $S_i$.  
The presence of $U$ induces sensitivity to invariant mass scales $\mu_{u1}$ and $\mu_{u2}$  that we can read off from the factorization formula in \eq{FDfactquasi}:
\begin{align} \label{eq:Uscales}
  \Bigl( \frac{p_g^+}{h\sqrt{x}}\Bigr)^2 
   &\sim  \mu_{u1}^2 \simeq \frac{Q^2z^2}{x h^2}
   \sim  \frac{\lambda^4\lambda_t^4}{\lambda_g^2} \,, 
 &\big( h\sqrt{x} p_g^- \big)^2 
   &\sim \mu_{u2}^2 \simeq \frac{x h^2 Q^2}{\beta^2} 
   \sim  \lambda^4 \lambda_g^2 
  \,.
\end{align}
These two additional scales are shown in the right graphic in \fig{scales}.
Note that due to the constraint $z\ll h^2$ from \eq{hconstraint} that we always have $\mu_{u1}^2 \ll \mu_B^2\sim -t \sim m_Y^2\sim Q^2 z/x$.
Similarly, $xh^2\ll 1$ from~\eq{softhconstraint} implies that  $\mu_{u2}^2 \ll \mu_S^2 \sim Q^2\sim m_X^2 \sim Q^2/\beta$. This condition corresponds to
\begin{align}
  x^2 \ll \frac{\beta^2 \mu_S^2}{\gamma^2 s}\:  e^{2\etalabcut} 
  \,,
\end{align}
which is always satisfied for large $\etalabcut$, but starts to constrain the phase space for smaller $\etalabcut$. For example, with $\etalabcut = 3$ and HERA kinematics, it corresponds to $x\ll 0.2$. The orange and green arrows in the right graphic of \fig{scales} correspond to large double (Sudakov) logarithms that arise due to the $\mu_{u1}^2 \ll \mu_B^2$ and $\mu_{u2}^2 \ll \mu_S^2$ scale hierarchies. 

The form of the quasi-diffractive factorization implies that the RG evolution of $U$ in $\mu$ involves contributions 
related to both its $p_g^+/(h\sqrt{x})$ and $h\sqrt{x}p_g^-$ variables, 
which factorize to all orders in perturbation theory. This directly generalizes results for the hemisphere soft function for the dijet invariant mass spectrum in  \refscite{Hoang:2007vb,Fleming:2007xt} and implies that
\begin{align}
  U_{(N,N')}^{R_A R_B}(p_H^+,p_H^-,\mu) &=\!\! \int\!\! d\ell^+ d\ell^-\,
    {\cal U}(p_H^+-\ell^+,\mu,\mu_0) \, {\cal U}(p_H^- -\ell^-,\mu,\mu_0) \,
     U_{(N,N')}^{R_A R_B}(\ell^+,\ell^-,\mu_0), 
 \nn\\
  \tilde U_{(N,N')}^{R_A R_B}(y^-,y^+,\mu) &=
    \tilde {\cal U}(y^-,\mu,\mu_0) \, \tilde {\cal U}(y^+,\mu,\mu_0) \,
     \tilde U_{(N,N')}^{R_A R_B}(y^-,y^+,\mu_0) 
 \,.
\end{align}
Here ${\cal U}$ and its Fourier transform $\tilde{\cal U}$ are evolution kernels that sum large logarithms (whose dependence on $(N,N')$ and $R_A R_B$ we suppress for brevity). We see from the second line that in Fourier space, evolution is multiplicative.  

It is convenient to factorize the boundary condition by utilizing different $\mu_0$ values in the two evolution kernels ${\cal U}$~\cite{Hoang:2007vb}, specifically, $\mu_0=\mu_{u1}$ and $\mu_0=\mu_{u2}$. This helps us avoid unphysical $\ln(\mu_{u1}/\mu_{u2})$ terms that can be induced by choosing a single $\mu_0$; see \refscite{Kelley:2011ng,Hornig:2011iu}.  
When combining the RG evolution of $U$ with that of $B$ and $S$, we also must run the two ${\cal U}$'s up to two different scales $\mu=\mu_B$ and $\mu=\mu_S$, as appropriate for $B$ and $S$.  For the $\lambda\ll 1$ expansion without an expansion in $\lambda_t$, these correspond to
\begin{align}
\label{eq:muBmuS}
  \mu_B^2 \simeq \{ -t, m_Y^2 \} \,, \qquad \qquad
  \mu_S^2 \simeq \{ Q^2, Q^2/\beta, -t \} \,.
\end{align}
Taken together, the RG evolution in $\mu$ sums large logarithms through the evolution factors ${\cal U}(\ell^+,\mu_B,\mu_{u1})$ and ${\cal U}(\ell^-,\mu_S,\mu_{u2})$. The right panel of \fig{scales} illustrates the ratio of scales associated to the summation of these logs using green and orange dashed arrows. The leading logarithms (LL) are an infinite series of Sudakov double logarithms, of the form $\sim \sum_k c_k [\alpha_s \ln^2(\mu_{B}/\mu_{u1})]^k$ and $\sim \sum_k c_k [\alpha_s \ln^2(\mu_{S}/\mu_{u2})]^k$. The resummed result is a Sudakov form factor, and has the physical interpretation of suppressing quasi-diffraction due to a restriction forbidding radiation into a (gapped) region of phase space.

When we consider additional scale hierarchies beyond those induced by $\lambda\ll 1$, new large logarithms arise. Let us consider the $\lambda_t\ll 1$ case discussed in \sec{smalllambdat}. Large logarithms appear in the soft function, which can be resummed by considering its refactorization into hard and soft-collinear functions. Just as we saw in diffraction in \sec{difflogs}, this involves a series of single logarithms at leading order $\sim \sum_k b_k' [\alpha_s \ln(-t/Q^2)]^k$, determined by DGLAP.  
There are a number of allowed hierarchies involving the parameters $\lambda_g$ and $\lambda_t$. 
These can induce additional logarithms for $U$ called non-global logarithms (NGLs)~\cite{Dasgupta:2001sh}, which start with a term $\sim \alpha_s^2 \ln^2(\mu_{u1}/\mu_{u2})$. These are large logarithms unless $\lambda_t\sim \lambda_g$ where $\mu_{u1}\sim \mu_{u2}$.
NGLs arise due to correlated emissions, where radiation into the phase space region $X$ influences radiation that enters the region $Y$. The nonglobal structure of an analogue of our $U$ (in a triplet rather than octet color channel) is known at ${\cal O}(\alpha_s^2)$ from studies of the dihemisphere soft function~\cite{Kelley:2011ng,Hornig:2011iu}, with resummation at next-to-leading-logarithmic (NLL) order~\cite{Dasgupta:2001sh,Banfi:2002hw,Schwartz:2014wha,Khelifa-Kerfa:2015mma,Becher:2015hka,Neill:2015nya}.
Note that in contrast, diffraction does not exhibit NGLs due to the absence of $U$ (and thus $\lambda_g$) as well as the control of $\lambda_t\ll 1$ logs by DGLAP.

Below, we study in greater detail the Sudakov suppression of quasi-diffraction, which is one of the key features necessary to quantify the magnitude of color-nonsinglet contamination relative to the color-singlet signal in the diffractive cross section in \sec{predictions}. At leading logarithmic (LL) order, we only need to calculate the Sudakov double logarithms from the ultrasoft collinear function and can neglect NGLs and BFKL logarithms, which only enter beyond LL order. To our knowledge, the precise size of this quasi-diffractive background has not been previously quantified in the literature.  

\subsubsection{One Glauber amplitudes in $q\qbar$ scattering}

We can extract a substantial amount of information about diffractive anomalous dimensions from existing results in the SCET literature.
First, let us build off a set of one-loop \SCETa $2\to 2$ amplitude calculations from \refcite{Rothstein:2016bsq}; in particular, of forward scattering an $n$-collinear quark and an $\nbar$-collinear antiquark  in the presence of virtual soft and ultrasoft gluons and quarks. 
For a single Glauber exchange, the $n$-collinear amplitude in \refcite{Rothstein:2016bsq} is identical to the one appearing in our beam function $B$, $A^n=A^n_{\text{\tiny\cite{Rothstein:2016bsq}}}$, the ultrasoft amplitude is identical to our $U$, $A^{uc}=A^{us}_{\text{\tiny\cite{Rothstein:2016bsq}}}$, and remaining $\nbar$-collinear and soft amplitudes can be combined to provide a proxy for our soft amplitude, $A^s\simeq A^{\nbar}_{\text{\tiny\cite{Rothstein:2016bsq}}}
+A^{s}_{\text{\tiny\cite{Rothstein:2016bsq}}}$. Although the combination of these forward scattering objects governs a different process than our desired soft amplitude (which produces a soft quark pair at tree level), the equivalence of the other amplitudes and RG consistency requires the structure of their logarithms to be related, and hence it provides a useful proxy for us to see how various scales can appear.

Using dimensional regularization $d=4-2\epsilon$, the $\eta$ rapidity regulator, and offshellness IR regulators $p^2=m_Y^2$ and $\bar p^2 =m_X^2$~\cite{Rothstein:2016bsq} (see \S 7.3), the amplitudes have logarithmic terms
\begin{align} \label{eq:Ampl}
  A^{uc} &=\! A_G^{\rm tree} \frac{C_A\alpha_s(\mu)}{2\pi} 
  \biggl( -\frac{1}{\epsilon^2} -\frac{1}{\epsilon} \ln \frac{\mu^2 s}{m_X^2 m_Y^2} 
     -\frac12 \ln^2 \frac{\mu^2 s}{m_X^2 m_Y^2}  \biggr)+ \ldots 
   \,,  \\
  A^{n} &=\! A_G^{\rm tree} \frac{C_A\alpha_s(\mu)}{2\pi} 
   \biggl( \frac{g_\epsilon}{\eta} 
   -\ln\frac{\nu}{p^-} \Bigl[ \frac{1}{\epsilon} \!+\! \ln\frac{\mu^2}{-t} \Bigr] 
   \!+\! \frac{1}{\epsilon^2} \!+\! \frac{1}{\epsilon} \ln\frac{\mu^2}{m_Y^2}
   \!+\! \frac12 \ln^2\frac{\mu^2}{m_Y^2}
   +\frac12 \ln^2\frac{m_Y^2}{-t}   
  \biggr)   +\ldots
   ,  \nn\\
  A^{s} &=\! A_G^{\rm tree} \frac{C_A\alpha_s(\mu)}{2\pi}
   \biggl( -\frac{g_\epsilon}{\eta} 
    -\ln \frac{m_X^2}{\nu p_s^+} \Bigl[\frac{1}{\epsilon} +\ln\frac{\mu^2}{-t}  \Bigr]
   +   \ln^2 \frac{-t}{m_X^2} 
   +\frac{\beta_0}{2C_A}\ln\frac{\mu^2}{-t} \biggr)  +\ldots
  \,, \nn
\end{align}
where $A_G^{\rm tree}$ stands for the tree-level amplitude, the coefficient of the $1/\eta$ rapidity divergence is $g_\epsilon =e^{\epsilon\gamma_E} \cos(\pi\epsilon)\Gamma(1+\epsilon)\Gamma(-\epsilon)\big(\frac{\mu^2}{-t}\big)^\epsilon$, and $p_s^+\simeq s/p^-$. 
The ellipses in \eq{Ampl} denote non-logarithmic terms or terms with other color factors that are not needed for our examination of scales. 
Each of these functions has its logarithms minimized at a common $(\mu,\nu)$ scale choice, which we label as $(\mu_i,\nu_i)$ with an index $i=\{S,U,B\}$ corresponding to the choice of function.
In $A_U^{uc}$ we would pick $\mu_{uc}^2\simeq m_X^2 m_Y^2/s$, in $A_B^n$ we would pick $\mu_B^2\simeq -t\sim m_Y^2$ and $\nu_B\simeq p^-$, and in $A_S^s$ we would pick $\mu_s^2\simeq m_X^2$ and $\nu_s\simeq m_X^2/p_X^+\sim -t/p_X^+$.
The RG evolution in $\mu$ and $\nu$ allows us to choose these canonical scales $(\mu_i,\nu_i)$ as the boundary conditions for the various functions, and then evolve them to a common $(\mu,\nu)$ choice. 

In our $\lambda$ power counting, we have that $\mu_{uc}^2\sim \lambda^4$, $\mu_B^2\sim \lambda^2$, $\nu_B\sim \lambda^{-1}$, $\mu_s^2\sim \lambda^2$, and $\nu_s\sim \lambda$, which match with the assignments in \fig{scales}. Evolving between these scales sums large logarithms.
When summing these amplitudes at one-loop, all explicit dependences on $\mu$ and $\nu$ cancel, except for the $\ln\mu$ associated with the running coupling. We have 
\begin{align}
	A^{us}+A^{n}+A^{s} 
	= A_G^{\rm tree} \frac{C_A\alpha_s}{2\pi} \Big[ -\frac12 \ln^2\frac{m_X^2m_Y^2}{-st} + 
 	\ln^2\frac{m_X^2}{-t} +\ln^2\frac{m_Y^2}{-t}
 	+\frac{\beta_0}{2C_A}\ln\frac{\mu^2}{-t}\: \Big]
	\,,
\end{align}
giving Sudakov double logarithms. 
Although we only present the pattern of logarithms for the virtual one-loop amplitudes from a single Glauber exchange here, the canonical scales apply equally well once we include additional logarithms from emission diagrams in the full beam, soft, and uc functions. 
For the rapidity divergences $1/\eta$ and their associated $\nu$-dependent logarithms, the results exactly match with the expected gluon Regge exponent $\propto [1/\epsilon +\ln\frac{\mu^2}{-t}]$ multiplying the $\ln\nu$ terms, and the real emission graphs simply generalize this to the full BFKL equation for the $8_A$ color channel. (Note that the $1/\epsilon$ here is UV due to ultrasoft 0-bin subtractions~\cite{Rothstein:2016bsq}.)  

\subsubsection{Sudakov suppression from $U$ in quasi-diffraction}

Due to RG consistency, we can carry out resummation at leading double-logarithmic order, using the $\mu$-RGE for the ultrasoft collinear function $U$.  This enables us to predict the precise form of the LL Sudakov form factor, including which scales enter its double logarithms. In the region of perturbative invariant mass scales $Q^2, -t, m_X^2, m_Y^2 \gg \LQCD^2$, color-octet single Glauber exchange ($N=N'=1$) dominates $U$; hence, this is our focus in this section. Note that we expect other color channels to have similar LL Sudakov factors, differing simply in the leading color Casmir factor present in the Sudakov exponentiation. 

For $N=N'=1$, the function $U^{\mathbf{8}}_{(1,1)}(p_H^+,p_H^-)$ is identical to the well-studied di-hemisphere invariant mass soft function, which appears in the di-hemisphere invariant mass distribution for dijet events in $e^+e^-$ collisions~\cite{Fleming:2007qr,Fleming:2007xt}, except that it involves adjoint rather than fundamental Wilson lines, and hence will have
a color factor of $C_A=3$ rather than $C_F=4/3$. 
As explained in \sec{qdifflogs}, the evolution for the beam and soft function parts of $U$ can be carried out independently, as shown by the orange and green dashed lines in \fig{scales}. 
Therefore, the LL solution for quasi-diffraction is:
\begin{align} \label{eq:ULL}
	  U^{\mathbf{8}}_{(1,1)}(p_H^+,p_H^-) \Big|_{LL} 
	  &=  N_{\textbf{8}}\,  \delta(p_H^+) \: \delta(p_H^-)\: U^{\rm LL}
	 \,, \qquad
	  U^{\rm LL}= e^{K_1} \, e^{K_2} \,,
\end{align} 
where $N_{\textbf{8}}$ is a normalization factor from the color projection. These $\delta$-functions collapse the $dp_g^+ dp_g^-$ integrals in \eq{FDfactquasi}. 
The Sudakov kernels are
\begin{align}
	 K_1 &= K(\mu_B,\mu_{u1}) \,, \qquad\qquad 
	 K_2 = K(\mu_S,\mu_{u2}) \,,
\end{align}
where at LL order,
\begin{align} \label{eq:KLL}
   K(\mu,\mu_0) &= \frac{8\pi C_A}{\beta_0^2} 
   \bigg[ \frac{1}{\alpha_s(\mu_0)} -\frac{1}{\alpha_s(\mu)}  
    - \frac{1}{\alpha_s(\mu_0)} \ln \frac{\alpha_s(\mu)}{\alpha_s(\mu_0)}  \bigg] \,,
\end{align}
with $\beta_0=11 C_A/3 - 2 n_f/3$ for $n_f$ light quarks.
\Eq{KLL} includes the leading double logarithm, $K(\mu,\mu_0) = -C_A \frac{\alpha_s(\mu)}{\pi} \ln^2(\mu/\mu_0) + {\cal O}(\alpha_s^2)$, plus the correct infinite tower of logarithms produced by the running coupling constant. 
We have $K(\mu,\mu_0)< 0$, so summing these Sudakov logarithms always causes exponential suppression of quasi-diffraction. 

\begin{figure}[t!]
	\begin{center}  
		\includegraphics[width = 0.49\textwidth]{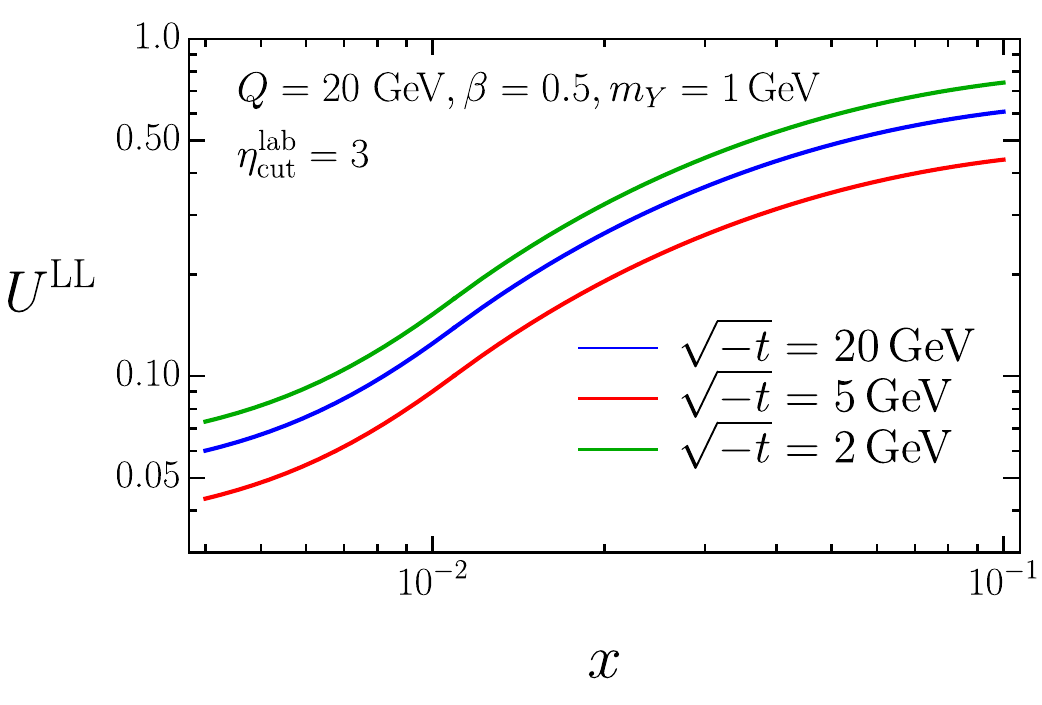}  \ 
        \includegraphics[width = 0.49\textwidth]{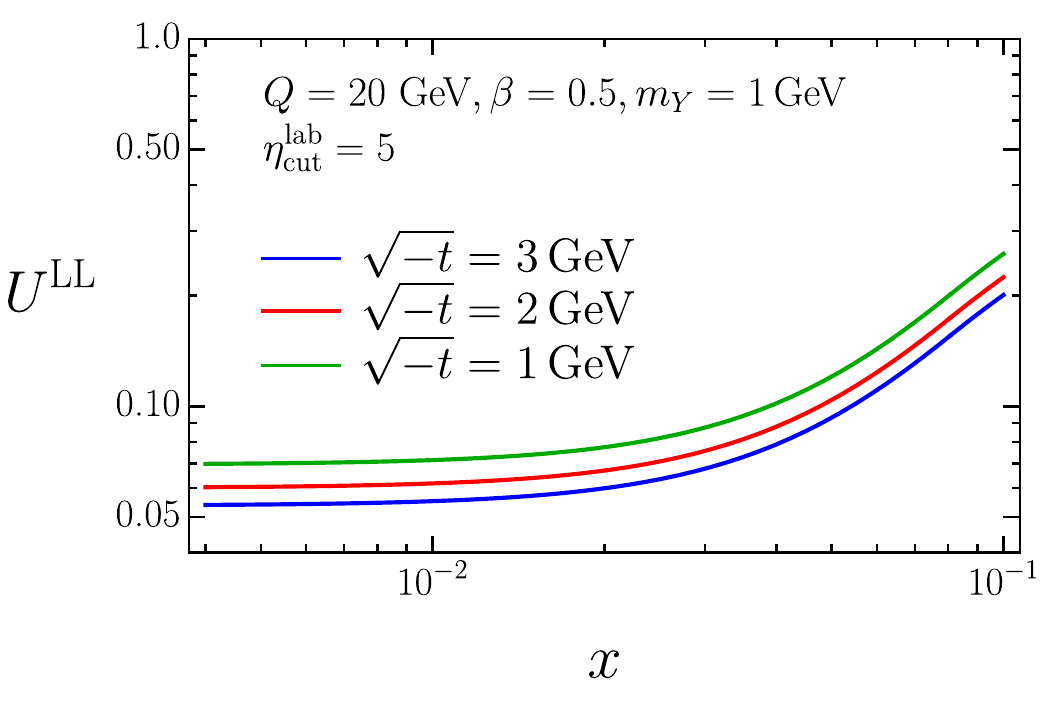}   
		\caption{
	Plot of the Sudakov factor that suppresses the quasi-diffractive ultrasoft-collinear function for representative phase space points and HERA kinematics for $s$ and $\gamma$.
	The left panel shows a smaller $\etalabcut=3$ with a less stringent cut on the radiation in $Y$, 
	allowing for larger values of $-t$.  The right plot shows $\etalabcut=5$, where only smaller values of $-t$ are allowed to ensure that the forward scattered proton or its remnants remain in $Y$.
}
		\label{fig:sudakov} 
	\end{center} 
\end{figure} 

Choosing our canonical scales to be 
\begin{align} 
  &\mu_B^2 = m_Y^2 - t ,
  && \mu_S^2 = Q^2 + \frac{Q^2}{\beta} ,  
 \nn\\
  &\mu_{u1}^2 =  \frac{Q^2 z^2}{x h^2} 
     = \frac{e^{2\etalabcut} (m_Y^2-t)^2 }{\gamma^2 s} \,,
  && \mu_{u2}^2 = \frac{x h^2 \,Q^2}{\beta^2}
     = \frac{x^2 \gamma^2 e^{-2\etalabcut}\,s}{\beta^2}\,,
\end{align}
the large logarithms we resum are $\ln^2(\mu_B^2/\mu_{u1}^2)\sim \ln^2[e^{2\etalabcut}(m_Y^2-t)/(\gamma^2 s)]$ and $\ln^2(\mu_S^2/\mu_{u2}^2)\sim \ln^2[x^2\gamma^2e^{-2\etalabcut}s)/(\beta^2 Q^2)]$. The second one has  double logarithms in $x$,
so the damping from the Sudakov exponents increases as $x$ decreases. 
Since we can only trust our Sudakov calculation for perturbative couplings, $\alpha_s(\mu_i) \ll 1$, we include a profile which freezes out all scales $\mu_i$ at $1\,{\rm GeV}$. 

In \fig{sudakov} we plot the leading Sudakov suppression from $U^{\rm LL}$ in \eq{ULL}, which corresponds to the leading damping of the color octet exchange channel. 
The smallest $x$ shown here corresponds to the kinematic limit $x_{\rm min} = Q^2/s$ for HERA. 
While the choice of kinematics and $\etalabcut$ impact the value of the damping, 
it is clear that this quasi-diffractive background is \textit{not} always strongly damped by $U$. 
For example, the left panel of \fig{sudakov} shows that for $x=0.01$, $U$ damps quasi-diffraction by only $\sim$ 10\%; indeed, it only drops to the $\sim$ 5\% level near $x=x_{\rm min}\sim 0.004$.
The right panel of \fig{sudakov} shows the Sudakov damping for a larger value of $\etalabcut$, which is more in line with those used at HERA.  Here we find more damping, but again, it plateaus at the 5-7\% level for the smaller values of $x$. This occurs because the scale $\mu_2$ hits the $1\,{\rm GeV}$ threshold, where $\alpha_s(\mu_2)$ is frozen, and there is no increase in the perturbative suppression from the $(\mu_S,\mu_{u2})$ hierarchy. Sudakov exponentials are predictions based on perturbative QCD emission probabilities.  
In this frozen coupling region there may be further suppression from nonperturbative effects (like the shape functions, whose form is known for the hemisphere $U$).  

Typically, perturbative uncertainties enter from higher-order resummation, which at LL order may amount to as much as 30\%, depending on the  parameter choices. This provides strong motivation for improving the perturbative accuracy of our results in the future, along with considering shape function models for nonperturbative effects.

Of course, to determine the overall contamination of the color-singlet diffractive signal in a cross section by the color-nonsinglet quasi-diffractive background, we must consider information about the soft and beam factors as well, since schematically
\begin{align}
	\frac{F_i^{\rm D\,quasi}}{F_i^{\rm D\, diff}} = \frac{B^{\rm quasi}\otimes U \otimes S_i^{\rm quasi}}{B^{\rm diff} \otimes S_i^{\rm diff}}  \,,
\end{align}
where exact expressions are given in~\eqs{FDfactdiff}{FDfactquasi}.
In \sec{predictions}, we combine the Sudakov suppression results from this section with perturbative soft function calculations in \sec{soft-results} to obtain a more accurate estimate contamination of the diffractive cross section by quasi-diffraction. 

\section{Soft Function in the Perturbative Regime}
\label{sec:soft-results}

In this section, we perturbatively expand the soft function $S_i$ in \eq{Sdefn} in $\alpha_s(\mu)$ to understand (quasi-)diffraction at perturbative invariant mass scales $\mu^2 \sim Q^2,m_X^2, -t\gg \LQCD^2$.
This large region of kinematic phase space provides powerful tests of the universality of hadronic dynamics in (quasi-)diffraction and probes the dependence of the factorization formulas in \eqs{FDfactRen}{FDfactdiff} on the choice of structure function $i=\{L,2,3,4\}$.
Interestingly, HERA analyses largely focused on nonperturbative kinematics with $-t\lesssim 1\,{\rm GeV}^2 \sim \LQCD^2 \ll Q^2$; in this case, we can still match the soft function onto the soft-collinear function at $\mu\sim Q$ with perturbative hard coefficients ${\cal H}_i^\kappa$; see \sec{smalllambdat}.

In \sec{soft-setup}, we begin by defining notation. We write $S_i$ in terms of a squared amplitude and phase space integrals, a form that is valid to all orders in perturbation theory. Our main focus is to calculate $S_i$ at tree level in the soft sector ($q\qbar$) with any number of Glauber exchanges, and we determine its amplitude in \sec{NglaubAmp}.
Because a $q\bar{q}$ state consists of color-singlet and octet components ($3\otimes \bar{3} = 1\oplus8$), our calculations probe both the singlet diffractive signal and the octet quasi-diffractive background.   In \sec{soft-1g}, we calculate the soft function with one perturbative Glauber exchange; this proceeds through the octet channel.
In \sec{one-pert-n-non}, we calculate the soft function with one perturbative Glauber exchange and $N$ nonperturbative Glauber exchanges, proving that only octet exchanges contribute. Finally, in \sec{two-perturbative}, we discuss the leading color-singlet contributions to the soft function.

\subsection{General setup and notation}
\label{sec:soft-setup}

When we perturbatively expand the soft function in \eq{Sdefn}, we must consider diagrams with various numbers of Glauber exchanges and soft particles, which we categorize as:
\begin{align} \label{eq:Sexpnk}
	S_{i(N,N')}^{R} & =  \sum_{k=0}^\infty  \Big(  \frac{\alpha_s(\mu)}{2\pi}\Big)^k \: S_{i(N,N')}^{R[k]}
	\,.
\end{align}
Here, $[k]$ denotes the perturbative order in the soft sector and $(N,N')$ indicates the number of Glauber operators inserted on each side of the final-state cut (see \fig{factthm}).  In \eq{Sexpnk}, $k=0$ is tree level in the soft sector, with an outgoing soft $q\qbar$ state; $k=1$ has either an extra real emission with a $q\qbar g$ state or a one-loop virtual correction to the $q\qbar$ state; and so forth. The color channel of the exchange is $R=R_B^{N\!N'}$. The requirement that $-t\gg \LQCD^2$ implies that at least one Glauber exchange has a perturbative coupling. The soft function scales $\mu^2\sim Q^2,m_X^2, -t\gg \LQCD^2$ set the expansion scale $\alpha_s(\mu)$.  

\begin{figure}[t!]
	\begin{center}
a) \hspace{2.4cm} 
b) \hspace{2.4cm} 
c) \hspace{2.4cm} 
d) \hspace{2.4cm} 
e) \hspace{1cm} \phantom{x} \\
		\includegraphics[width = 0.19\textwidth]{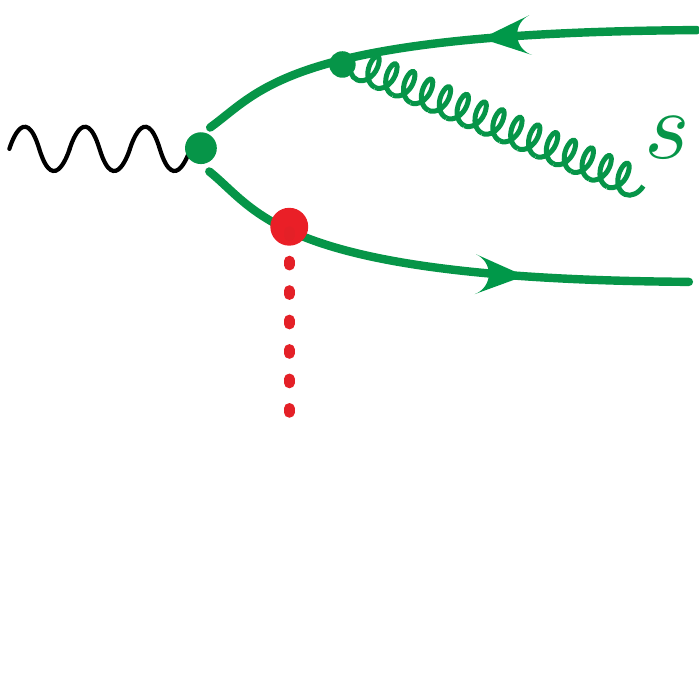}  
        \includegraphics[width = 0.19\textwidth]{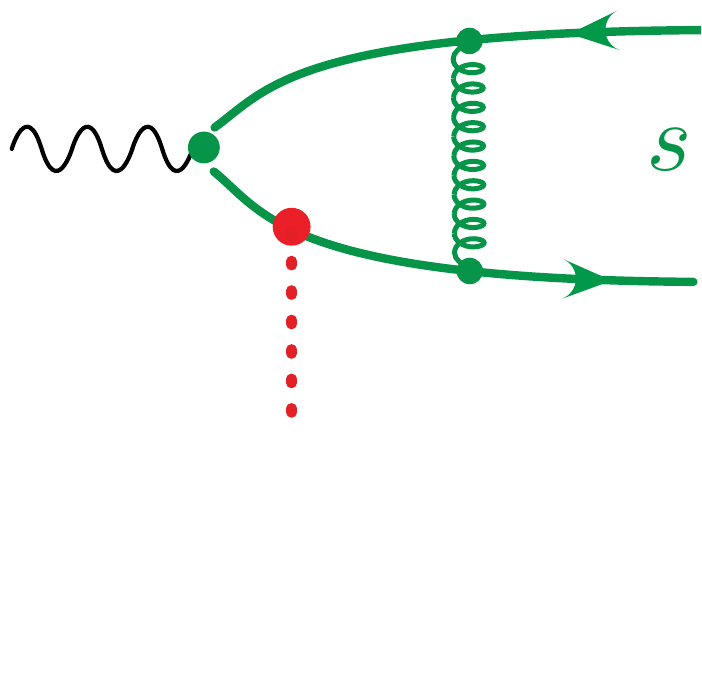}   
		\includegraphics[width = 0.19\textwidth]{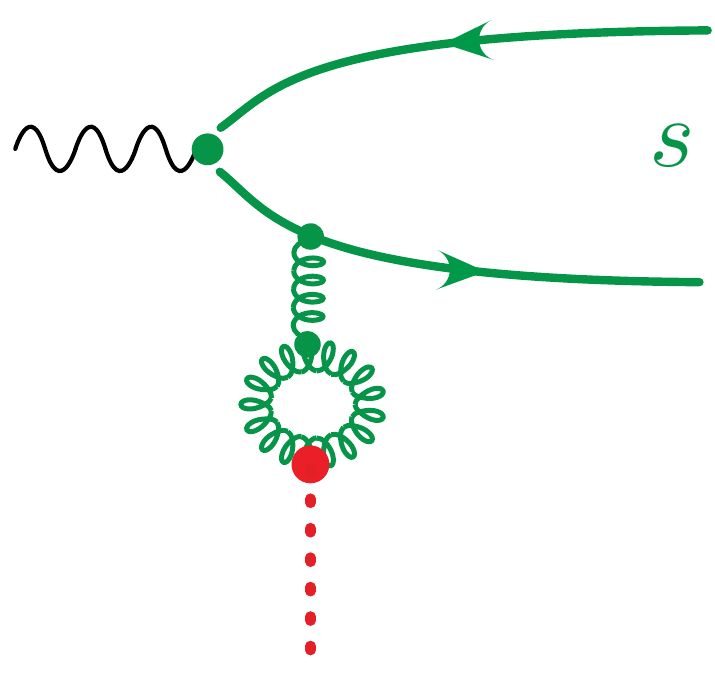}  
        \includegraphics[width = 0.19\textwidth]{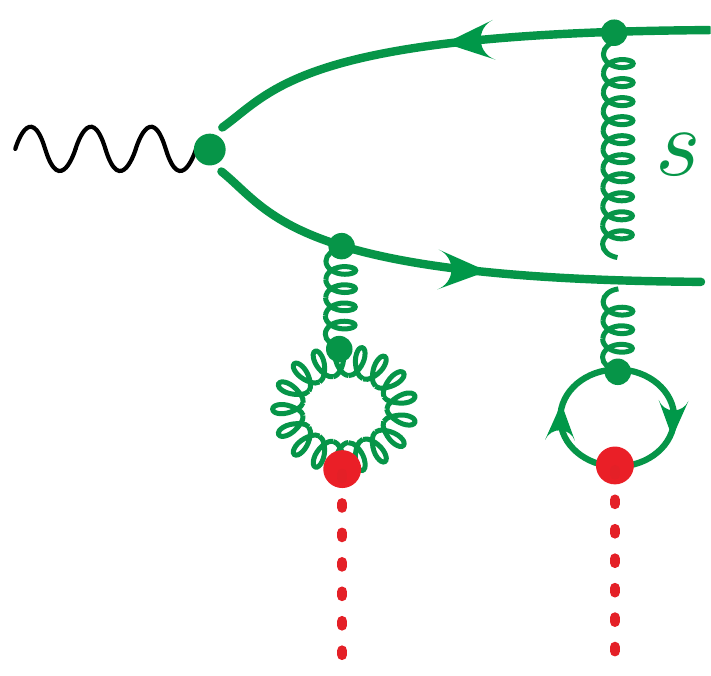}   
        \includegraphics[width = 0.19\textwidth]{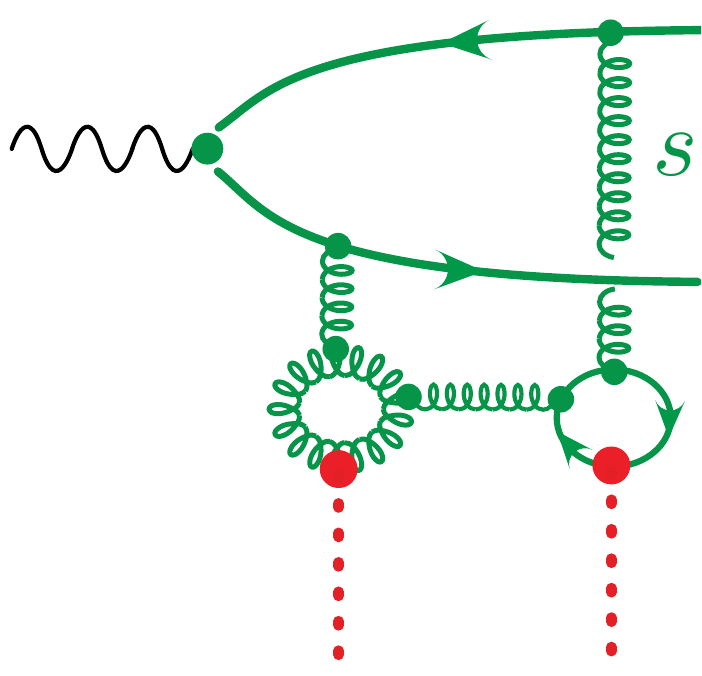}    
		\caption{Examples of radiative corrections to the soft function. The real emission and virtual correction in graphs a) and b) are sensitive to the scales $Q$ and $m_X$, while the Glauber potential radiative corrections in c) and d) are only sensitive to scales $\tau_{i\perp}^2$. Graph e) vanishes due to the Glauber collapse rule. }
		\label{fig:Gptnlrad} 
	\end{center} 
\end{figure}

\Eq{Sexpnk} only explicitly writes $\alpha_s^k$ factors emanating from the soft sector. 
Each $S_{i(N,N')}^{R[k]}$ also contains additional factors of $\alpha_s(\mu)$ associated to the Glauber operators ${\cal O}_s^{q_nB}$ and ${\cal O}_s^{g_nB}$ from \eq{2secop} and their associated virtual loop diagrams; the scale $\mu$ reflects the transverse momentum $\mu^{2} \sim \vec\tau_{j\perp}^{\,2}$ of the corresponding Glauber exchange.
Each exchange is either perturbative ($\vec \tau_{j\perp}^{\,2} \gg \LQCD^2$) or nonperturbative ($\vec\tau_{j\perp}^{\,2} \sim \LQCD^2$). 
That is, each $S_{i(N,N')}^{R[k]}$ carries an implicit factor of $\alpha_s^{N+N'}(\mu)$ along with multiplicative radiative corrections associated to the Glauber potentials; see \fig{Gptnlrad}a,b for examples. 
To factor out these radiative corrections, we write
\begin{align} \label{eq:hatS}
 S_{i(N,N')}^{R[k]} &= \hat S_{i(N,N')}^{R[k]}\:  \frac{1}{(2\pi)^d} \prod_{j=1}^{N+N'}
   \Big[8\pi \alpha_s(\mu)\mu^{2\epsilon} \iota^\epsilon\, G(\tau_{j\perp},\mu) \Big] 
 \,,
\end{align}
where $\hat S_{i(N,N')}^{R[k]}$ are perturbatively calculable functions and $G$ encodes the radiative corrections, which may be perturbative or nonperturbative.
Note that $\iota = \exp(\gamma_E)/(4\pi)$ is a factor arising from using the $\overline{\rm MS}$ scheme, and we define each $G$ to depend on only one scale $\tau_{j\perp}$. The factors in square brackets are $\mu$-independent. We also note that  Glauber collapse forces graphs like \fig{Gptnlrad}e to be zero.  

\paragraph{Phase space integrals.}
To calculate $S_{i(N,N')}^{R} $ from \eq{Sexpnk},
we first massage \eq{Sdefn} into a form involving a phase space integral and a squared amplitude,
\begin{align} \label{eq:SasPS}
 & S_{i(N,N')}^{R} (k_s^-, \{\tau_{i\perp},\tau_{j\perp}'\}, Q,t )
  = \Bigl(\frac{Q}{n\cdot q}\Bigr)^{2} \cP^\prime_{i\,\mu\nu}\: \sumint_{X_s}
  \int\! [d \tilde y] \, [d\tilde y^{\prime}]  \, \ddslash^{d}\!z \:   
  \  e^{\frac{i}{2} (y^+ - y^{\prime +}) k_s^-}\,  e^{+ i z\cdot q}\,
\nn  \\*
 & \ \times
   \big\langle 0 \big| \bar T\,  J_s^\mu(z) 
  \collapsel\! \text{\footnotesize $\prod_{i=1}^{N}$} 
   \cO_s^{B_i}\!(\tilde y,-\tau_{i\perp}) \!\!\collapser 
  P_{N R_B} \big| X_s \big\rangle  
   \big\langle X_s \big| P_{N'R_{B'}} T\,
   J_s^\nu(0) 
  \collapsel\! \text{\footnotesize $\prod_{j=1}^{N'}$} 
   \cO_s^{B_j'}\!(\tilde y^{\prime},-\tau_{j\perp}^\prime) \!\!\collapser
   \big| 0 \big\rangle
  \nn \\
 &= \Bigl(\frac{Q}{n\cdot q}\Bigr)^{2} \cP^\prime_{i\,\mu\nu}\: \sumint_{X_s}
  \int\! [d \tilde y] \, [d\tilde y^{\prime}]  \, \ddslash^{d}\!z
  \  e^{\frac{i}{2} (y^+ - y^{\prime +}) k_s^-}\,  e^{+ i z\cdot q}\,
  e^{-iz\cdot p_{Xs}} \, e^{-i z_\perp\cdot \sum_i \tau_{i\perp}}
  \nn \\*
 & \ \times
   \big\langle 0 \big| \bar T\,  J_s^\mu(0) 
  \collapsel\! \text{\footnotesize $\prod_{i=1}^{N}$} 
   \cO_s^{B_i}\!(\tilde y-\tilde z,-\tau_{i\perp}) \!\!\collapser 
  P_{N R_B} \big| X_s \big\rangle  
   \big\langle X_s \big| P_{N'R_{B'}} T\,
   J_s^\nu(0) 
  \collapsel\! \text{\footnotesize $\prod_{j=1}^{N'}$} 
   \cO_s^{B_j'}\!(\tilde y^{\prime},-\tau_{j\perp}^\prime) \!\!\collapser
   \big| 0 \big\rangle
  \nn \\
 &=
  \Bigl(\frac{Q}{n\cdot q}\Bigr)^{2} 
 \cP^\prime_{i\,\mu\nu} \sumint_{X_s}   \delta^d(q+\check\tau - p_{X_s})  
  \Big\langle 0 \Big| \!\int\! [d \tilde y]  \:  
   e^{i\tilde y\cdot \check \tau}\,
   \bar T\,  J_s^\mu(0) 
  \collapsel\! \text{\footnotesize $\prod_{i=1}^{N}$} 
   \cO_s^{B_i}\!(\tilde y,-\tau_{i\perp}) \!\!\collapser 
  P_{N R_B} \Big| X_s \Big\rangle  
  \nn \\*
 & \qquad\qquad \times
   \Big\langle X_s \Big| P_{N'R_{B'}}\! \int\! [d\tilde y^{\prime}] \: 
   e^{-i\tilde y'\cdot \check \tau}\,
   T\, J_s^\nu(0) 
  \collapsel\! \text{\footnotesize $\prod_{j=1}^{N'}$} 
   \cO_s^{B_j'}\!(\tilde y^{\prime},-\tau_{j\perp}^\prime) \!\!\collapser
   \Big| 0 \Big\rangle 
   \,, 
\end{align}
where as before $[d\tilde y]=dy^+dy^-/2$. To obtain the second equality, we shift the coordinates in the first matrix element by $-z$, and to obtain the last line we shift $\tilde y\to \tilde y + \tilde z$ and perform the $\ddslash^{d}\!z$ integral.  
The final expression in \eq{SasPS} makes clear that the soft function corresponds to a scattering cross section with incoming momenta $q$ and $\check\tau$, where 
\begin{align}
  \check \tau^\mu = \frac{1}{2} k_s^- n^\mu - \sum_{i=1}^N \tau_{i\perp}^\mu
  =  \frac{1}{2} k_s^- n^\mu  + \tau_\perp^\mu
   \,.
\end{align}
See \fig{CoordMom} for our sign conventions.
If uc radiation is not present then $\check\tau^- = k_s^- = \tau^-$ and $\check\tau_\perp^\mu = \tau_\perp^\mu$; note that $\check\tau^+=0$.

For $N=N'=1$, it is instructive to write the Glauber operators in position space, 
\begin{align}
 & S_{i(1,1)}^{8} 
 = \Bigl(\frac{Q}{n\cdot q}\Bigr)^{2} \cP^\prime_{i\,\mu\nu}\: 
  \sumint_{X_s} \ \delta^d(q+\check\tau - p_{X_s})  
  \ \Big\langle 0 \Big| \!\int\! d^dy  \:  
   e^{i y\cdot \check \tau}\,
   \bar T\,  J_s^\mu(0) 
   \cO_s^{B_1}\!(y) 
  \Big| X_s \Big\rangle  
  \nn \\*
 & \qquad\qquad \times
   \Big\langle X_s \Big| \int\! d^d y^{\prime} \: 
   e^{-i y'\cdot \check \tau}\,
   T\, J_s^\nu(0) 
   \cO_s^{B_1'}\!(y^{\prime}) 
   \Big| 0 \Big\rangle
   \,,
\end{align}
where $R=8$ because a single gluon is always a color octet. Only the four-vector $\check \tau$ appears because there is only one relevant transverse momentum $\tau_{1\perp}=-\tau_\perp = -\tau_{1\perp}'$. 

Following \eq{hatS}, we introduce a notation for the reduced amplitudes after stripping off Glauber potential factors from \eq{SasPS}, 
\begin{align} \label{eq:colorAmp}
 & \hat S_{i(N,N')}^{R} 
  = \Bigl(\frac{Q}{n\cdot q}\Bigr)^{2} \cP^\prime_{i\,\mu\nu}\: 
  \sumint_{X_s} \ (2\pi)^{d}\, \delta^d(q+\check\tau - p_{X_s})  
  \: \Tr \Big[{\cal M}_N^{\mu\{B_i\}\dagger}\: {\cal M}_{N'}^{\nu\{B'_j\}} \Big]
  \,,
\end{align}
traced over final-state spin and color indices. Like in \eq{Sexpnk}, we will sometimes add a superscript $[k]$ to both $\hat S_{i(N,N')}^{R}$ and ${\cal M}_{N}^{\nu\{B_j\}}$ to denote order in the soft sector.  

\subsection{Leading order soft amplitude with $N$ Glaubers}
\label{sec:NglaubAmp}

Let us examine the soft function amplitude ${\cal M}_{N}^{\mu\{B_i\}[0]}$ at lowest order ($q\qbar$) in the soft sector and with $N$ Glaubers. As this section only considers the LO case, we drop superscripts $[0]$ for concision.
Because $q\bar q$ can only be in a $1$ or $8$ color representation, we project the amplitudes onto these two color channels using quark color indices $\alpha\beta\gamma\delta$ as
\begin{align}
\label{eq:Mcolorproj}
{\cal M}_{N\alpha \beta}^{\mu\{B_i\}\dagger}\: \delta_{\alpha\gamma}\delta_{\beta\delta} \: {\cal M}_{N'\gamma\delta}^{\nu\{B'_j\}} 
 &=
{\cal M}_{N\alpha \beta}^{\mu\{B_i\}\dagger}\: 
 \Big(\frac{1}{N_c} \delta_{\alpha\beta}\delta_{\gamma\delta}
    + 2 (T^C)_{\alpha\beta} (T^C)_{\gamma\delta} \Big) 
 \: {\cal M}_{N'\gamma\delta}^{\nu\{B'_j\}} 
\nn\\
 &\equiv 
 \frac{1}{N_c} {\cal M}_{N(1)}^{\mu\{B_i\}\dagger}\: 
    {\cal M}_{N'(1)}^{\nu\{B'_j\}} 
    + 2 {\cal M}_{N(8)}^{\mu\{B_i C\}\dagger}\: 
    {\cal M}_{N'(8)}^{\nu\{B'_j C\}} 
 \,.
\end{align}
This color projection onto the final $q\bar{q}$ state differs from the soft color projection from the Glaubers in \eq{proj_compl}. 
While these two different color projections must have irreducible representations of the same dimension to obtain a color singlet, their overlap may take the form of a linear combination when one of the projections has more than one irrep of the same dimension. 
For example, three Glaubers have eight octet irreps, and the single octet $q\bar{q}$ state irrep projects onto a linear combination of these octet Glauber irreps.

\begin{figure}[t!]
	\begin{center}
		\includegraphics[width =4.5 in]{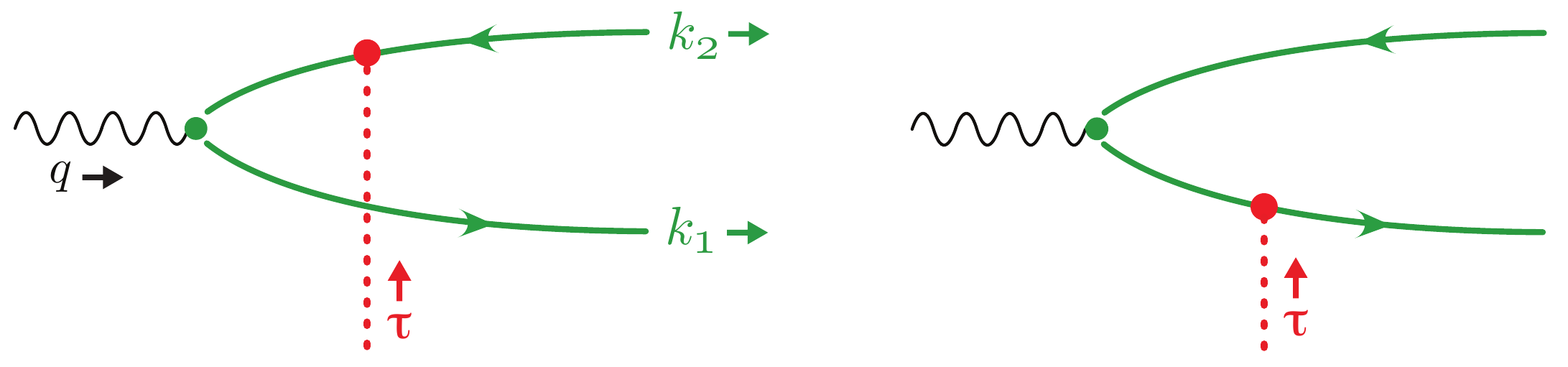}
	\end{center}
	\caption{The soft function amplitude at tree level in the soft sector ($q\qbar$, green solid lines) with one Glauber exchange (red dotted lines).}\label{fig:oneglauberamp}
\end{figure}

We start with the case $N=1$ in \fig{oneglauberamp}, which has only an octet channel. We have
\begin{align}
\label{eq:AN1soft}
{\cal M}^{\nu}_{1(8)}
=& e_q\, {\rm tr}[T^C T^{B_1'}]\,
  \bar{u}(k_1)\bigg[ \gamma^\nu
 \frac{(\slashed{\tau}_{\!\perp} -\slashed{k}_2)\frac{\slashed{n}}{2}}{(\tau-k_2)^2}
 + 
 \frac{\frac{\slashed{n}}{2}(\slashed{k}_1 
  -\slashed{\tau}_{\!\perp})}{(k_1-\tau)^2}
 \gamma^\nu \bigg]  v(-k_2)
 \\
=& e_q\, T_F\delta^{CB_1'} \bar{u}(k_1)\bigg[ - \gamma^\nu
\frac{(n\cdot k_2-\slashed{\tau}_{\!\perp}\frac{\slashed{n}}{2})}{(-2\tau\cdot k_2+\tau^2)}
+
\frac{(n\cdot k_1-\frac{\slashed{n}}{2}\slashed{\tau}_{\!\perp})}{(-2\tau\cdot k_1+\tau^2)}
 \gamma^\nu \bigg]  v(-k_2)
\nn\\
=& e_q\, T_F\delta^{CB_1'} \bar{u}(k_1)\bigg[ \gamma^\nu
\frac{(1-\slashed{\tau}_{\!\perp}\frac{\slashed{n}}{2n\cdot k_2})}
 {\tau^- +\frac{\vec\tau_\perp^{\,2}}{n\cdot k_2}-\frac{2\vec\tau_\perp\cdot\vec k_{2\perp}}{n\cdot k_2} }
-
\frac{(1-\frac{\slashed{n}}{2n\cdot k_1}\slashed{\tau}_{\!\perp})}
 {\tau^- +\frac{\vec\tau_\perp^{\,2}}{n\cdot k_1}-\frac{2\vec\tau_\perp\cdot\vec k_{1\perp}}{n\cdot k_1} }
 \gamma^\nu \bigg]  v(-k_2)
 \,, \nn
\end{align}
where we have stripped the prefactor $8\pi \alpha_s$ as discussed above and color-stripped the spinors. Here, $e_q$ is the quark electric charge. To obtain the second line, we use the on-shell equations of motion $\bar u(k_1)\slashed{k}_1=\slashed{k}_2 v(-k_2)=0$ and $k_1^2=k_2^2 =0$.

\begin{figure}[t!]
\begin{center}
	\includegraphics[width =6 in]{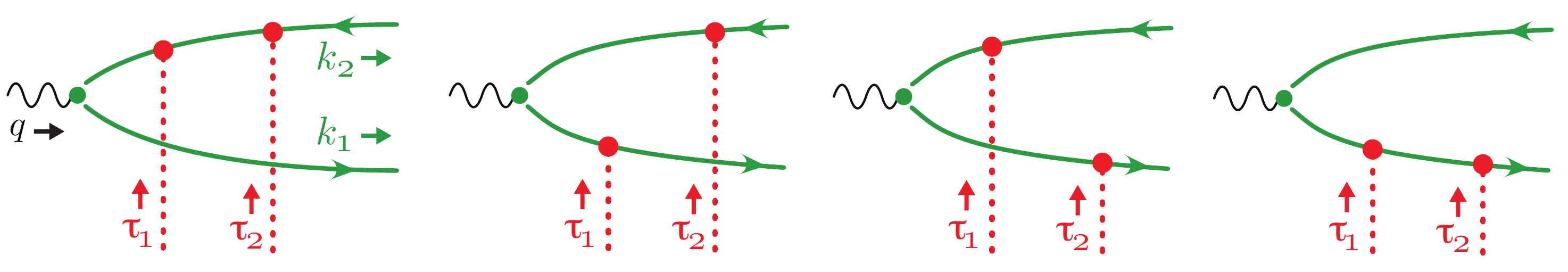}
\end{center}
\caption{The soft function amplitude at tree level in the soft sector (green solid lines) with two Glauber exchanges (red dotted lines).}\label{fig:twoglauberamp}
\end{figure}

Next, we consider $N=2$, which has both color-singlet and octet contributions. In \fig{twoglauberamp}, we see that four diagrams contribute. We find that 
\begin{align}\label{eq:AN2soft}
{\cal M}^{\nu}_{2(r)}
&= e_q\, {\cal T}_{(r)a} \,
  \bar{u}(k_1) \bcollapsel\!
 \frac{\gamma^\nu(\slashed{k}_2  \!-\!\slashed{\tau}_{\!\perp}\!)\frac{\slashed{n}}{2}
  (\slashed{k}_2\!-\! \slashed{\tau}_{\!2\perp}\!) \frac{\slashed{n}}{2} }
  {(k_2-\tau)^2\, (k_2-\tau_2)^2}
 -
\frac{\frac{\slashed{n}}{2} (\slashed{k}_1 \!-\!\slashed{\tau}_{\!1\perp}\!) \gamma^\nu
  (\slashed{k}_2\!-\! \slashed{\tau}_{\!2\perp}\!) \frac{\slashed{n}}{2} }
  {(k_1-\tau_1)^2\, (k_2-\tau_2)^2}
\! \bcollapser  v(-k_2)
 \\
&-
e_q\, {\cal T}_{(r)b}\,
  \bar{u}(k_1) \bcollapsel\!
\frac{\frac{\slashed{n}}{2} (\slashed{k}_1 \!-\!\slashed{\tau}_{\!2\perp}\!) \gamma^\nu
  (\slashed{k}_2\!-\! \slashed{\tau}_{\!1\perp}\!) \frac{\slashed{n}}{2} }
  {(k_1-\tau_2)^2\, (k_2-\tau_1)^2}
 -
 \frac{\frac{\slashed{n}}{2} (\slashed{k}_1  \!-\!\slashed{\tau}_{\!2\perp}\!)
  \frac{\slashed{n}}{2} (\slashed{k}_1\!-\! \slashed{\tau}_{\!\perp}\!)  \gamma^\nu }
  {(k_1-\tau_2)^2\, (k_1-\tau)^2}
\! \bcollapser v(-k_2)\nn
\end{align}
The first and last pairs of diagrams in \fig{twoglauberamp} each carry the same color factors
\begin{align} \label{eq:T2glauber}
	{\cal T}_{(1)a} &= T_F \delta^{B_1'B_2'} \,,
	& {\cal T}_{(8)a} &= {\rm tr}[T^C T^{B_1'} T^{B_2'} ] \,,
	\nn\\ 
	{\cal T}_{(1)b} &= T_F \delta^{B_1'B_2'} \,,
	& {\cal T}_{(8)b} &= {\rm tr}[T^C T^{B_2'} T^{B_1'} ] 
	\,,
\end{align}
where the subscripts $a$ and $b$ indicate color factors appearing for $N=2$. 
To simplify \eq{AN2soft}, we carry out the Glauber collapse operation $\scollapsel \cdots \scollapser$, which sets $\tau_1^\pm =\tau^\pm -\tau_2^\pm$ and requires integrating over $\tau_2^\pm$. 
To perform the integrations properly, we must include collinear propagators that arise from Glauber interactions in the beam function. However, the value of $\cM$ is independent of which collinear propagators the Glaubers attach to in the beam function. Thus,
\begin{align}\label{eq:AN2softA}
	{\cal M}^{\nu}_{2(r)}
	&= e_q\, {\cal T}_{(r)a}\,
	  \bar{u}(k_1)\Bigg[\!
	 -\frac{\gamma^\nu \Bigl(1  \!-\!\frac{\slashed{\tau}_{\!\perp}\slashed{n}}{2k_2^+}\Bigr) }
	  {\tau^- \!+\!\frac{\vec\tau_\perp^{\,2}}{k_2^+}
	  \!-\! \frac{2\vec\tau_\perp\cdot\vec k_{2\perp}}{k_2^+} }
	 +
	\frac{ \Bigl(1\!-\!\frac{\slashed{n}\slashed{\tau}_{\!1\perp}}{2k_1^+} \! \Bigr) 
	 \gamma^\nu \Bigl(1\!-\! \frac{\slashed{\tau}_{\!2\perp}\slashed{n}}{2k_2^+} \!\Bigr) }
	  { \tau^- \!+\! \frac{\vec\tau_{1\perp}^{\,2}}{k_1^+}
	  \!-\! \frac{2\vec\tau_{1\perp}\cdot\vec k_{1\perp}}{k_1^+}
	  \!+\! \frac{\vec\tau_{2\perp}^{\,2}}{k_2^+}
	  \!-\! \frac{2\vec\tau_{2\perp}\cdot\vec k_{2\perp}}{k_2^+} }
	\Bigg] \! v(-k_2)
	 \nn\\
	&\hspace{-0.55 cm}+
	  e_q\, {\cal T}_{(r)b}\,
	  \bar{u}(k_1) \Bigg[ \!
	\frac{ \Bigl(1\!-\!\frac{\slashed{n} \slashed{\tau}_{\!2\perp}}{2k_1^+}\!\Bigr) 
	 \gamma^\nu \Bigl(1\!-\!  \frac{\slashed{\tau}_{\!1\perp}\slashed{n}}{2k_2^+} \!\Bigr)  }
	  {\tau^- \!+\! \frac{\vec\tau_{2\perp}^{\,2}}{k_1^+}
	  \!-\! \frac{2\vec\tau_{2\perp}\cdot\vec k_{1\perp}}{k_1^+}
	  \!+\! \frac{\vec\tau_{1\perp}^{\,2}}{k_2^+}
	  \!-\! \frac{2\vec\tau_{1\perp}\cdot\vec k_{2\perp}}{k_2^+} }
	 -
	\frac{ (1\!-\! \frac{\slashed{n}}{2n\cdot k_1} \slashed{\tau}_{\!\perp}\!)  \gamma^\nu}
	  {\tau^- \!+\! \frac{\vec\tau_\perp^{\,2}}{k_1^+} \!-\!
	    \frac{2\vec\tau_\perp\cdot\vec k_{1\perp}}{k_1^+} }
	\Bigg]\! v(-k_2)
	 \,.
\end{align}
When we collapse the two Glaubers, a factor of $1/(2!)$ arises.  We choose to group this factor into the $\perp$-convolution between the soft and beam functions in~\eq{iint_perp}, and hence it does not appear in \eq{AN2softA}.

\begin{figure}
	\begin{center}
		\includegraphics[width =4.5 in]{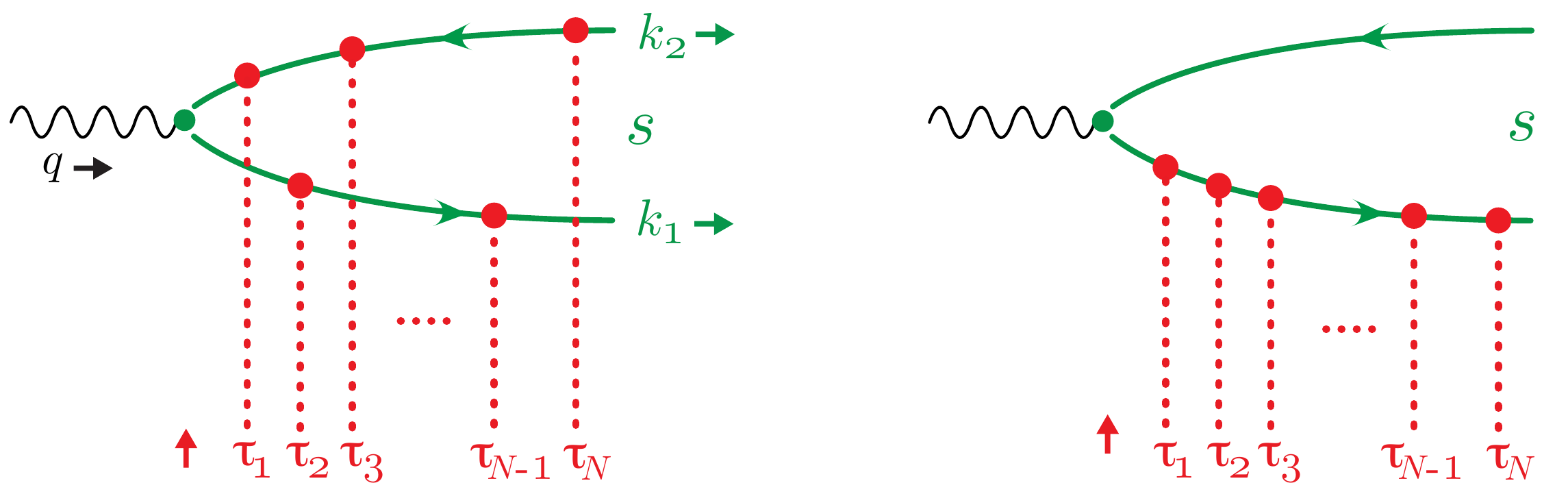}
	\end{center}
	\caption{Soft function at tree level in the soft sector ($q\qbar$, green lines) and at arbitrary Glauber order $N$ (red dotted). Each Glauber gluon may attach to either the quark or the anti-quark. We say that a diagram is planar when all Glaubers attach to the same particle in the soft sector (right), and nonplanar when they attach to different soft particles (left).}\label{fig:Nglauberamp}
\end{figure}

The soft amplitude with a $q\qbar$ soft sector and $N$ Glaubers has contributions from a multitude of diagrams, as shown in \fig{Nglauberamp}. We carry out full calculations in  \app{Amultiglauber} and present the result here:
\begin{align}\label{eq:NglauberAmp}
		& \mathcal{M}^{\nu}_{N(r)} =e_q \sum_\sigma (-1)^{N_q}
  	\frac{\bar{u}(k_1)\Big(1-\frac{\slashed{n}}{2n\cdot k_1}\slashed{\tau}_{\sigma_q \perp}\Big)\gamma^\nu
	 \Big(1-\slashed{\tau}_{\sigma_{\qbar}\perp}\frac{\slashed{n}}{2n\cdot k_2}\Big)
	 \mathcal{T}^\sigma_{(r)} \, v(-k_2)}
	 {\tau^- + \frac{1}{n\cdot k_1}(\vec{\tau}_{\sigma_q\perp}-2\vec{k}_{1\perp})\cdot \vec{\tau}_{\sigma_q\perp} +\frac{1}{n\cdot k_2}(\vec{\tau}_{\sigma_{\qbar}\perp}-2\vec{k}_{2\perp})\cdot \vec{\tau}_{\sigma_{\qbar}\perp}}
	 \,.
\end{align}
Here, the notation $\sigma_X$ indicates permutations of Glauber attachments to each particle $X=\{q,\bar{q}\}$ in the soft sector. For example, for $N=5$ Glaubers we have $2^5$ permutations for attachments to $q\qbar$, which include cases like $\sigma = \{\sigma_q,\sigma_{\qbar}\} = \big\{\{1,3,4\},\{2,5\}\big\}$ and $\sigma = \big\{\emptyset,\{1,2,3,4,5\} \big\}$. Note that for consistency, we always number Glauber momenta $\tau_i$ from left to right, as in \fig{Nglauberamp}. We let $N_X$ denote the number of elements in $\sigma_X$. The total momentum of the particles in $\sigma_X$ is then $\tau_{\sigma_X} = \sum_{i\in \sigma_X} \tau_{i}$.  The singlet and octet color matrices for each $\sigma$ are
\begin{align}  \label{eq:TNglauber}
\mathcal{T}^\sigma_{(1)} &=
 {\rm tr}\Big[ T^{A({\sigma_q^{\mbox{\tiny $N_q$}} })}
  \cdots 
  T^{A({\sigma_q^2})}\, T^{A(\sigma_q^1)} \
  T^{A({\sigma_{\bar q}^1})} \,  T^{A(\sigma_{\bar q}^2)}
  \cdots 
  T^{A(\sigma_{\bar q}^{\mbox{\tiny $N_{\bar q}$}} )} \Big]
  \,, \nn\\
\mathcal{T}^\sigma_{(8)} &=
 {\rm tr}\Big[ T^C\, T^{A(\sigma_q^{\mbox{\tiny $N_q$}})}
  \cdots 
  T^{A(\sigma_q^2)}\, T^{A(\sigma_q^1)} \
  T^{A(\sigma_{\bar q}^1)} \,  T^{A(\sigma_{\bar q}^2)}
  \cdots 
  T^{A(\sigma_{\bar q}^{\mbox{\tiny $N_{\bar q}$}})} \Big]
  \,.
\end{align}
Here $\sigma_X^i$ indicates element $i$ in $\sigma_X$. $T^{A(\sigma_X^i)}$ is the color matrix for the corresponding Glauber attachment. Note that sometimes different permutations $\sigma$ have identical color factors, which is why \eq{TNglauber} has a more complicated notation than \eq{T2glauber}.

\subsection{One perturbative Glauber (quasi-diffraction)}
\label{sec:soft-1g}

At leading soft order ($q\qbar$) and one perturbative Glauber exchange, the soft function is
\begin{align}
\label{eq:S11def}
	\hat{S}_{i(1,1)}^{8[0]} 
  &= \cP'_{i\mu\nu}  \Bigl(\frac{Q}{n\cdot q}\Bigr)^{2} \int d\Phi_2^d\:
     2\, \mathcal{M}_{1(8)}^{\mu \dagger} \mathcal{M}_{1(8)}^\nu  \\
  &=  \cP'_{i\mu\nu}  \Bigl(\frac{Q}{n\cdot q}\Bigr)^{2} \int [d^dk_1]_+ [d^dk_2]_+
  (2\pi)^d
 \delta^d(q+\tau-k_1-k_2) \,2\,  \mathcal{M}_{1(8)}^{\mu\dagger} \mathcal{M}_{1(8)}^\nu  
  \nn\\
  &\equiv \cP'_{i\mu\nu}\ \hat{S}_{(1,1)}^{8[0]\mu\nu}  
  \nn \,,
\end{align}
where we have used $ \sumint_{X_s} =  \int d\Phi_2^d$ to obtain an integral over two-body phase space, $[d^d k_i]_+ = d^dk_i \delta(k_i^2) \theta(k_i^0)/(2\pi)^{d-1}$. 
Also recall that $n\cdot\tau=0$ for the soft function; see \sec{diffract-factorization}.
Note that we use $[0]$ in the soft function, but we continue dropping $[0]$ from $\cM$. To simplify the calculation, we use the tensor structure decomposition from \eq{primeprojectors}:
\begin{align}
  \hat{S}_{(1,1)}^{8[0]\mu\nu} &= \sum_i w_i^{\prime\mu\nu} \hat{S}_{i(1,1)}^{8[0]}  
  \,.
\end{align}
Here, we use $w_i^{\prime\mu\nu}$ rather than $w_i^{\mu\nu}$ because
the soft function is $x$-independent. 
(This decomposition also applies for higher-order soft functions and any $N$ and $N'$.)

Some structures $w_i^{\prime\mu\nu}$ are defined in terms of $p^\mu$ (via $U^\mu$). While $p^\mu$ cannot directly enter $S_i$, as its only incoming momenta are $q^\mu$ and $\tau^\mu$, $S_i$ also depends on $n^\mu$ through its soft Glauber operators. Using $p^\mu = \nbar \cdot p\, n^\mu/2= Q^2 n^\mu /(2x n\cdot q)$, we write
\begin{align}
	U^\mu = \frac{Q}{n\cdot q}\left(n^\mu - \frac{n\cdot q}{q^2}q^\mu\right)\,,
\end{align}
which is an allowed basis element for the soft function.
Note that in the Breit frame, $x \bar{n}\cdot p/Q=Q/n\cdot q=1$. By RPI invariance, $S$ is the same in any frame related to the Breit frame by a boost along the $n$-$\bar{n}$ axis.

Our calculations will use dimensional regularization $d=4-2\epsilon$ 
for both matrix elements and phase space integrals, so to work with the structures $w_i^{\prime\mu\nu}$ we also need $d$-dimensional versions of the projection operators in \eq{primeprojectors} and~\eq{P4A}: 
\begin{align}\label{eq:projectorsd}
	&\cP_2^{\prime\mu\nu} = -\frac{1}{d-3}\left[g^{\mu\nu} +(5-2d) U^\mu U^\nu - (d-4) X^\mu X^\nu\right] \,,
	&&\cP_L^{\prime\mu\nu} = 2\, U^\mu U^\nu \,,
	\\
	&\cP_3^{\prime\mu\nu} = \frac{1}{d-3}\,\left[g^{\mu\nu} - U^\mu U^\nu+(d-2) X^\mu X^\nu \right] \,,
	&&\cP_4^{\prime\mu\nu} = -\,( U^\mu X^\nu + X^\mu U^\nu)\,,\nn\\
	&\cP_{4A}^{\prime\mu\nu} =i\,( U^\mu X^\nu - X^\mu U^\nu)\,.
	\nonumber
\end{align}
These projectors obey $\cP^\prime_{i\mu\nu} w_j^{\prime\mu\nu} =\delta_{ij}$. Note that the five additional structure functions associated with a polarized hadron beam affect only the beam functions and therefore share the same soft functions.

The soft matrix elements in \eq{S11def} are simply those for $\gamma^*(q) g(\tau)\to q(k_1) \bar q(k_2)$. We use back-to-back kinematics for the photon and gluon, and parametrize their momenta as
\begin{align}\label{eq:soft-qq1-vectors}
	&q = (q_0,0,0,-\tau_z)\,,
	&&\tau = (\tau_0,0,0,\tau_z)\,,
	\nonumber\\*
	&k_1= (k_0,k_0\sin\theta,0,k_0\cos\theta)\,,
	&&k_2 = (k_0,-k_0\sin\theta,0,-k_0\cos\theta)\,,\nonumber\\*
	&n = (n_0, |n_\perp| \cos\phi , |n_\perp| \sin\phi, n_z)\,.
\end{align}
Here $n^\mu$ from the Breit frame 
transforms as a four-vector to leave dot products like $n\cdot \tau$ and $n\cdot k_1$ fixed.  We take the outgoing quarks to be back-to-back in the $x$-$z$ plane, so that only $n^\mu$ depends on the azimuthal angle $\phi$. 
The momentum components in \eq{soft-qq1-vectors} are
\begin{align}
  &q_0 = \frac{Q}{2}\frac{1-2\beta}{\sqrt{\beta(1-\beta+\beta\frac{t}{Q^2})}}\,,
 &&\tau_0 =\frac{Q}{2}\frac{1+2\beta\frac{t}{Q^2}}
     {\sqrt{\beta(1-\beta+\beta\frac{t}{Q^2})}} \,,
 &&\tau_z = \frac{Q}{2}\sqrt{\frac{1+4\beta^2\frac{t}{Q^2}}
     {\beta(1-\beta+\beta\frac{t}{Q^2})}}\,,
	\nonumber\\
 &n_0 = \frac{n\cdot q}{Q}
    \sqrt{\frac{\beta}{1-\beta+\beta\frac{t}{Q^2}}}\,,
 &&k_0 = \frac{Q}{2}\sqrt{\frac{1-\beta+\beta\frac{t}{Q^2}}{\beta}}\,,
 &&n_z = n_0 \frac{1+2\beta\frac{t}{Q^2}}{\sqrt{1+4\beta^2\frac{t}{Q^2}}}\,, 
	\nn\\
 &|n_\perp| = \frac{n\cdot q}{Q} 
    \frac{\sqrt{-t}}{Q} \frac{ 2\beta}{\sqrt{1+4\beta^2\frac{t}{Q^2}}}\,.
\end{align}
Next, the phase space factor in \eq{S11def} simplifies to
\begin{align}
\label{eq:Phi2}
	\int\!\! d\Phi_2^d
	& =
	\int [d^dk_1]_+ [d^dk_2]_+ (2\pi)^{d}\delta^d(q+\tau-k_1-k_2)=\frac{(4\pi)^\epsilon}{32\pi^2} \frac{ k_0^{-2\epsilon} }{\Gamma(1-\epsilon)} \int\!\!\frac{d\cos\theta\, d\phi}{(1-\cos^2\theta)^\epsilon}\,.
\end{align}
Carrying out the $\theta$ and $\phi$ integrals, we can evaluate \eq{S11def}
to obtain
\begin{align}\label{eq:lo-soft-final}
	\hat{S}_{i(1,1)}^{8[0]} =\delta^{AB} T_F \frac{1 }{ \pi} \frac{\beta^2\lambda_t^2  }{(1-4\beta^2\lambda_t^2)^4} \sum_j e_j^2 \bigg(\!
   \hat{ S}_{i,\rm{r}}
	\!+\! 
	\hat{ S}_{i,\rm{t}} \frac{\text{tanh}^{-1}\sqrt{1-4\beta^2\lambda_t^2}}{\sqrt{1-4\beta^2\lambda_t^2}}\bigg),
\end{align}
where $j$ sums over quark and anti-quark flavors in the fermion loop, and the terms $\hat S_{i,r/t}$ capture the rational ($r$) and transcendental ($t$) terms in the soft function:
\begin{align}
\hat S_{L,\rm{r}} =& 4\beta(1-\beta)+4 \beta(-47 \beta^3+72 \beta^2-30 \beta+1) \lambda_t^2 
  \\*
	& -4\beta^2(6 \beta^4-116 \beta^3+180 \beta^2-72 \beta+1) \lambda_t^4-4\beta^4(60 \beta^2-116 \beta+47)\lambda_t^6-24 \beta^6 \lambda_t^8\,,
	\nonumber\\
\hat S_{L,\rm{t}} = & 8 \beta^2(18 \beta^2-24 \beta+7) \lambda_t^2
    +48 \beta^3(6 \beta^3-18 \beta^2+17 \beta-4) \lambda_t^4 \nn\\
	& -48\beta^4(2 \beta^4+4 \beta^3-18 \beta^2+18 \beta-3) \lambda_t^6+32 \beta^6(2 \beta^2-6 \beta+9) \lambda_t^8-96 \beta^8 \lambda_t^{10}\,,
	\nonumber\\
\hat S_{2,\rm{r}} =& -2(6 \beta^2-6 \beta+1)-6\beta(39 \beta^3-64 \beta^2
    +28 \beta-2)\lambda_t^2 
     \nn\\
	& - 12\beta^2(11 \beta^4-66 \beta^3+90 \beta^2-32 \beta+1) \lambda_t^4-2\beta^4(212 \beta^2-396 \beta+117) \lambda_t^6-132 \beta^6 \lambda_t^8\,,
	\nonumber\\
\hat S_{2,\rm{t}} =&
	2(2 \beta^2-2 \beta+1)+4 \beta(48 \beta^3-66 \beta^2+17 \beta-1) \lambda_t^2
    \nn\\
	& +4 \beta^2(108 \beta^4-324 \beta^3+318 \beta^2-66 \beta+1) \lambda_t^4
    -16\beta^4(\beta^4+26 \beta^3-73 \beta^2+81 \beta-12) \lambda_t^6
    \nn\\
    &+16 \beta^6(22 \beta^2-26 \beta+27) \lambda_t^8-16 \beta^8 \lambda_t^{10}
     \,,\nonumber\\
\hat S_{3,\rm{r}} =& -\beta^2-4\beta^2(18 \beta^2-25 \beta+6) \lambda_t^2
     +\beta^2(-116 \beta^4+440 \beta^3-468 \beta^2+100 \beta-1) \lambda_t^4
    \nn \\
	&- 8\beta^4(33 \beta^2-55 \beta+9) \lambda_t^6-116 \beta^6 \lambda_t^8
    \,,\nn\\
\hat S_{3,\rm{t}} =& 8 \beta^2(6 \beta^2-6 \beta+1) \lambda_t^2
    +24 \beta^3(9 \beta^3-24 \beta^2+16 \beta-2) \lambda_t^4 \nn\\
	& +48 \beta^4(\beta^4-6 \beta^3+18 \beta^2-12 \beta+1) \lambda_t^6-8\beta^6(4 \beta^2+36 \beta-27) \lambda_t^8+48 \beta^8 \lambda_t^{10}\,,\nn\\
\hat S_{4,\rm{r}} =& 2 \beta(25 \beta^2-28 \beta+6) \lambda_t+4 \beta^2(55 \beta^3
     -138 \beta^2+96 \beta-14) \lambda_t^3 \nn\\
	&+ 2 \beta^3(-128 \beta^3+396 \beta^2-276 \beta+25) \lambda_t^5+4 \beta^5(55-64 \beta)  \lambda_t^7
    \,,\nn\\
 \hat S_{4,\rm{t}} =& -4\beta(6 \beta^2-6 \beta+1) \lambda_t+24 \beta^2(-12 \beta^3
    +22 \beta^2-11 \beta+1) \lambda_t^3 \nn\\
	& -24\beta^3(6 \beta^4-36 \beta^3+54 \beta^2-22 \beta+1) \lambda_t^5-32\beta^5(13 \beta^2-27 \beta+9) \lambda_t^7-144 \beta^7 \lambda_t^9\,,\nn\\
\hat S_{4A,\rm{r}} =& \,0    \,,\nn\\
\hat S_{4A,\rm{t}} =& \,0\,,\nn
\end{align}
where as usual, $\lambda_t = \sqrt{-t}/Q$. 
For $N=N'=1$ one finds that $\hat{S}_{(1,1)}^{8[0]\mu\nu}$ is symmetric in $\mu\nu$, and hence 
$S^{8[0]}_{4A(1,1)}$ vanishes.

\paragraph{Case of $\lambda_t\ll 1$.}\label{sec:soft-lo-1g-lt}
So far, we have considered the one-Glauber soft function for $\lambda \ll 1$ and $\lambda_t \sim 1$. 
To study the soft function for $\lambda, \lambda_t \ll 1$, we restore the appropriate factors of $\alpha_s$ and $\pi$ from \eq{hatS} into \eq{lo-soft-final} and expand in $\lambda_t = \sqrt{-t}/Q \ll1$, giving:
\begin{align}
\label{eq:Slambdat}
	S_{L(1,1)}^{8[0]} &=\delta^{AB} T_F \frac{-t}{Q^2}\frac{\beta^2}{ \pi}  \frac{[8\pi \alpha_s(\mu)]^2}{(2\pi)^4} \sum_j e_j^2  \: 4\beta(1-\beta)\times[1+\mathcal{O}(\lambda_t)] \,,
	\\
	S_{2(1,1)}^{8[0]}   &=\delta^{AB} T_F \frac{-t}{Q^2} \frac{\beta^2}{\pi} \frac{[8\pi \alpha_s(\mu)]^2}{(2\pi)^4} \sum_j e_j^2	\, \nn\\
	&\qquad\times\left[-(1-2\beta+2\beta^2)\ln\left(\frac{-t}{Q^2}\beta^2\right)-2(1-6\beta+6\beta^2)\right]\times[1+\mathcal{O}(\lambda_t)] \,,
	\nonumber\\
	S_{3(1,1)}^{8[0]} &=\delta^{AB} T_F\frac{-t}{Q^2} \frac{  \beta^2}{ \pi}  \frac{[8\pi \alpha_s(\mu)]^2}{(2\pi)^4} \sum_j e_j^2  \: (-\beta^2) \times [1+\mathcal{O}(\lambda_t)]\,,
	\nonumber\\
	S_{4(1,1)}^{8[0]} &=\delta^{AB} T_F \left(\frac{-t}{Q}\right)^{3/2} \frac{ \beta^2}{ \pi} \frac{[8\pi \alpha_s(\mu)]^2}{(2\pi)^4} \sum_j e_j^2 
	\, \nn\\
	&\qquad \times2\beta\left[(1-6\beta+6\beta^2)\ln\left(\frac{-t}{Q^2}\beta^2\right) + 6 - 28 \beta + 25 \beta^2\right]\times [1+\mathcal{O}(\lambda_t)]\,.\nonumber
\end{align}
Interestingly, $S_{L,2,3}$ have $\cO(\lambda_t^2)$ contributions, whereas $S_4$ starts at $\cO(\lambda_t^{3})$. We do not consider $S_{4A(1,1)}^{8[0]}$, which is zero at this order even before the expansion in $\lambda_t$.
Because the soft function is dimensionless, it can only depend on $\lambda_t =\sqrt{-t}/Q$ through a single logarithm (or power), while its $\beta$-dependence is less constrained. For $\lambda_t \ll 1$, our fixed-order expression in \eq{Slambdat} develops large logarithms $\ln(-t/Q^2)$ that require the  further hard-collinear factorization from \sec{smalllambdat}, and then DGLAP resummation. 

\paragraph{Hard-collinear factorization for $\lambda_t \ll 1$.} 
The $\lambda_t$-refactorization in \eq{refactS} refactorizes $S_i$ as the convolution of a hard and a soft-collinear (sc) function.\footnote{Recall that this $\lambda_t$-refactorization is valid even for $-t\sim \LQCD^2$, as long as $Q^2\gg -t$. In this case, the soft-collinear function becomes a nonperturbative function and the perturbative partonic calculations carried out in this section still suffice to determine the ${\cal H}_i$, but not $S_{\rm c(1,1)}$.}
In the $q\qbar$ case, either both particles can carry $\lambda_t$-hard scaling, or one is $\lambda_t$-hard and the other is $\lambda_t$-collinear; see \eq{four-vector-scales3}. Thus, for $i=L,2,3$ we can write \eq{refactS} as\footnote{In \eq{Slambdat}, we observe that $S_4$ is subleading in $\lambda_t$; therefore, the leading $\lambda_t$ hard-collinear factorization described by the soft-collinear function here does not necessarily describe it.}
\begin{align}
\label{eq:expansionS}
 S_{i(1,1)}^{8[0]}\Bigl(\frac{Q^2}{\beta} &, Q,t,\mu,\frac{\nu}{Q/\beta} \Bigr)
  = \frac{\beta^2}{Q^2}\Big[ \mathcal{H}_i^{q(0)}\otimes_{-\:}  S^{q; 8(2)}_{\rm c(1,1)}
   +\mathcal{H}_i^{g(1)} \otimes_{-\:} S^{g; 8(1)}_{\rm c(1,1)}  \Big] \times [1+\mathcal{O}(\lambda_t)]
 \nonumber \\*
   & =\frac{\beta^2}{Q^2}
  \int_{\beta}^1 \frac{d\zeta}{\zeta}\: \bigg[
    \mathcal{H}^{q(0)}_{i}\Bigl(\frac{\beta}{\zeta}, Q,\mu\Bigr)\: 
  S^{q; 8(2)}_{\rm c(1,1)}\left( \zeta,t,  \frac{\nu}{Q/\beta},\mu \right)\nonumber\\
  &\hspace{2cm}+\mathcal{H}^{g(1)}_{i}\Bigl(\frac{\beta}{\zeta}, Q,\mu\Bigr)\: 
  S^{g; 8(1)}_{\rm c(1,1)}\left( \zeta,t,  \frac{\nu}{Q/\beta},\mu \right)\bigg] \times [1+\mathcal{O}(\lambda_t)]\,.
\end{align}
$S_{i(1,1)}^{8[0]}$ starts at $\alpha_s^2$,
and taking $\lambda_t\ll 1$ factorizes it into quark and gluon terms whose organization of $\alpha_s$ factors differs.
For the quark, $\mathcal{H}_i^{q} \sim \alpha_s^0$ and $S_{c}^q\sim \alpha_s^2$, whereas for the gluon, $\mathcal{H}_i^{g}\sim \alpha_s$ and $S_{c}^g\sim \alpha_s$. This mirrors the usual convention for hard functions in inclusive DIS.  

We can determine the first term on the RHS of \eq{expansionS} by expanding the matrix elements and two-body phase space in~\eq{Phi2} when either $q$ or $\bar{q}$ has collinear-$\lambda_t$ scaling, 
and including the $\overline{\rm MS}$ counterterm for the $P_{qg}$ splitting function:
\begin{align}
\hspace{-1cm}
\label{eq:HS2}
	& \mathcal{H}_i^{q(0)} \otimes_- S^{q; 8(2)}_{\rm c(1,1)} \\
	&\ \ =\begin{cases}
		0\,, &i\!=\!L,3\\
		\delta^{AB} T_F \frac{(-t)}{\pi} \frac{(8\pi \alpha_s(\mu))^2}{(2\pi)^4} \sum e_j^2 \left[(1-2\beta+2\beta^2)\ln \frac{\mu^2}{(1-\beta)\beta (-t)}-(1-2\beta)^2\right]\,,
   &i=2\\
	\end{cases}
	\,. 
\nonumber
\end{align}
We must be careful with the $\epsilon$-dependence in the $\overline{\rm MS}$ factors, $[\alpha_s(\mu) \mu^{2\epsilon} \iota^\epsilon]^2$.  To mirror the usual
choice for hard functions in inclusive DIS, we expand the $\epsilon$ dependence of one of these two factors, while keeping the other unexpanded together with
the $(2\pi)^{-2\epsilon}$ from the Glauber potential normalization $(2\pi)^d$ in~\eq{hatS}.\footnote{We can trace this extra $(2\pi)^d$ to the difference between the hadronic tensor definitions in DIS and diffraction in~\eq{tensors}, where the diffractive case lacks a $(2\pi)^d$.}
Different choices affect each term on the RHS of~\eq{expansionS}, but not their overall sum on the LHS.  

The leading-order soft-collinear function $S^{j; 8[1]}_{\rm c(1,1)}$ occurs for $j=g$  and is expected to be $\delta_{jg} \delta(1-\zeta)$ up to a normalization.  This overall factor is fixed by the normalization choice in the operator definition of the soft-collinear function. Here, we take
\begin{align}
\label{eq:ScsLO}
    S^{j; 8(1)}_{\rm c(1,1)} 
  = \delta^{AB}\,\frac{(8\pi )^2 \alpha_s(\mu)}{(2\pi)^4}(-t)
   2\delta_{jg}\delta(1-\zeta)\,.
\end{align}
From these results, we can use \eqs{Slambdat}{expansionS} to derive that 
\begin{align}\label{eq:hardcoeffdg}
	\mathcal{H}^{g(1)}_L(\zeta,Q,\mu) &= \sum_j \frac{\alpha_s T_F e_j^2}{2\pi} 4 \zeta(1-\zeta) \,,
	\\
	\mathcal{H}^{g(1)}_2(\zeta,Q,\mu) &=\sum_j \frac{\alpha_s T_F e_j^2}{2\pi}   \left[ (1-2\zeta+2\zeta^2)\ln\frac{Q^2(1-\zeta)}{\mu^2 \zeta}-1+8\zeta-8\zeta^2\right]\,,
	\nn  \\
	\mathcal{H}^{g(1)}_3(\zeta,Q,\mu) &=\sum_j \frac{\alpha_s T_F e_j^2}{2\pi}  (-\zeta^2)\,.\nn
\end{align}
As expected from \sec{smalllambdat}, the hard coefficients ${\cal H}_{2,L}^g$ are identical to those of DIS. We also present a new hard coefficient ${\cal H}_3^g$. We expect that the number of Glaubers $(N,N')$ and color channel do not affect these results, as the $\lambda_t$-refactorization is insensitive to them.

We leave to future work an independent perturbative computation of the soft-collinear function from its operator definition. Here, we instead take as input the known DIS coefficients ${\cal H}_{2,L}^{q[0]}=H_{2,L}^{q[0]}$ from \eq{hardcoeff}, and derive the form of the soft-collinear function from \eq{HS2}:
\begin{align}
\label{eq:cs1loop}
S^{q; 8(2)}_{\rm c(1,1)} (\zeta,t,\mu)= \delta^{AB} T_F 
\frac{(-t)}{\pi} \frac{[8\pi \alpha_s(\mu)]^2}{(2\pi)^4} \left[(1-2\zeta+2\zeta^2)\ln \frac{\mu^2}{(1-\zeta)\zeta (-t)}-\zeta(1-2\zeta)^2\right]\,.
\end{align}
This NLO soft-collinear function corresponds to the operator definition choice that yields the normalization in~\eq{ScsLO} for its LO expression.
We see that the soft function logarithm $\ln \big(\frac{-t}{Q^2}\beta^2\big)$ in \eq{Slambdat}
gets decomposed into  $\ln \big(\frac{\mu^2}{Q^2(1-\beta)/\beta}\big )$ in the hard function in \eq{hardcoeffdg}  and 
$\ln \big(\frac{\mu^2}{-t \beta (1-\beta)}\big)$ in the soft-collinear function in \eq{cs1loop}.
Since the standard DIS hard functions obey the DGLAP renormalization group equations, we see that the $\lambda_t$-refactorization resums large DGLAP logarithms  $\ln (-t/Q^2) = \ln \lambda_t^2$ as anticipated.

\subsubsection{One perturbative and $N-1$ nonperturbative Glaubers}\label{sec:one-pert-n-non}

We now consider diagrams with a single perturbative Glauber and an arbitrary number $(N-1)$ of nonperturbative Glauber exchanges, which have $\tau_{i\perp}\sim \Lambda_{\rm QCD}$ and $\alpha_s\sim 1$.  
These are important contributions as nothing suppresses them in principle, unlike contributions with additional perturbative Glaubers that are suppressed by factors $\alpha_s \ll 1$. 
We will see that these contributions do not spoil the predictive power of our leading-order soft function calculation.  
This analysis corresponds to carrying out a $\lambda_\Lambda\ll 1$ expansion for the soft function.

To assist in our calculations, let us define some notation. As introduced in \sec{NglaubAmp}, we label Glauber momenta $\tau_i$ in order from left to right, and we classify diagrams by the permutation $\sigma = \{\sigma_q,\sigma_{\bar{q}}\}$ indicating which Glaubers $i$ attach to the quark and anti-quark lines, respectively.  In this section, we tack an extra subscript $p$ onto each $\sigma_p$ that indicates which of the $N$ Glauber gluons is perturbative. 
The key expansion that we can exploit to simplify the perturbative soft amplitudes in the presence of nonperturbative Glaubers is $\tau_{j,\perp}\ll \tau_{p,\perp}$ for $j\neq p$, while $\tau_\perp \simeq \tau_{p,\perp}$.
There are two qualitatively different types of diagrams: those with $p=1$ and those with $p\neq 1$.

\begin{figure}
	\begin{center}
		\includegraphics[width = 5in]{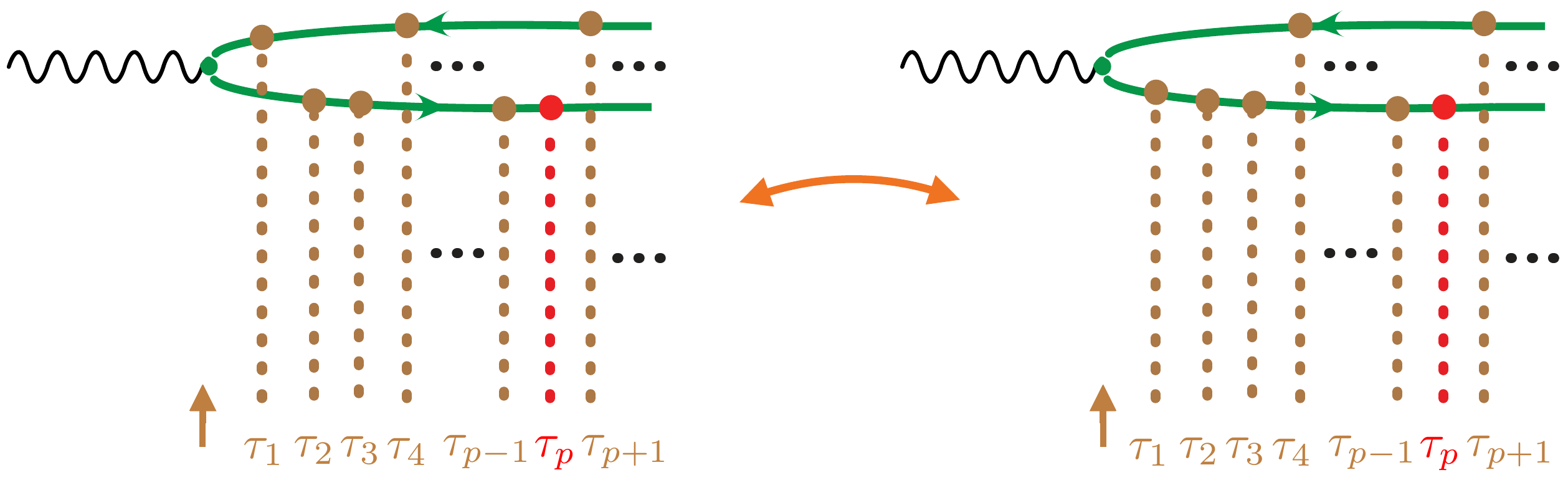}
	\end{center}
	\caption{Two representative diagrams with 1 perturbative Glauber and $N-1$ nonperturbative Glauber gluons. In both diagrams, the perturbative Glauber $\tau_{p\neq 1}$ attaches to the quark. The left panel and the right panel differ only by whether $\tau_1$ attaches to the $q$ or $\qbar$. These two diagrams sum to zero, as explained in \eq{cancel-glaubers}.}
     \label{fig:permutations}
\end{figure}

\paragraph{Vanishing of $p\neq 1$ diagrams.} Let us discuss $p \neq 1$ first, considering an arbitrary permutation $\sigma_{p\neq 1}=\{\sigma_{q},
\sigma_{\bar q}\}$ that attaches the perturbative $p$ to the quark line. 
Expanding in $\tau_{j,\perp}\ll \tau_{p,\perp}$,
we can write the soft amplitude as 
\begin{align}
\label{eq:ineq1}
\mathcal{M}_N^{\sigma_{p\neq1}}=&(-1)^{N_q-1}\bar{u}(k_1)\frac{\slashed{n}}{2}\frac{(\slashed{k}_1-\slashed{\tau}_{\perp})}{(k_1 - \tau)^2}\gamma^\mu
  \\*
&\times T^{A(\sigma_{q}^{N_q})}T^{A(\sigma_{q}^{N_q-1})}
 \cdots T^{A_p}\cdots 
  T^{A(\sigma_{q}^{2})} T^{A(\sigma_{q}^{1})} T^{A({\sigma_{\bar{q}}^1})} \,  T^{A(\sigma_{\bar{q}}^2)}
  \cdots 
  T^{A(\sigma_{\bar{q}}^{\mbox{\tiny $N_{\bar q}$}} )}
v(-k_2)\,,\nn
\end{align}
where $A_p$ is the color index for the perturbative Glauber. 
Because the first Glauber is nonperturbative, we can consider how \eq{ineq1} differs
when that Glauber attaches to the quark or antiquark lines, as shown in \fig{permutations}. 
These two amplitudes are identical except for the change of $(-1)^{N_q-1} \to(-1)^{N_q}$.
Thus
\begin{align}\label{eq:cancel-glaubers}
  \sum_{\sigma_{p\neq 1}} 
	\mathcal{M}_N^{\sigma_{p\neq 1}} =0 
  \,.
\end{align}
It is straightforward to see that this statement holds for all $N$, arbitrary color projections, and regardless of whether $p$ attaches to the quark or anti-quark.
Therefore, the lowest-order soft function vanishes for one perturbative Glauber inserted \emph{after} one or more nonperturbative Glauber exchanges.

\paragraph{Vanishing of color-singlet exchange with $p=1$.} We now consider the $p=1$ case, with the perturbative Glauber attached to the quark line,
which has the amplitude
\begin{align}
\label{eq:ieq1}
\mathcal{M}_N^{\sigma_1}=&(-1)^{N_q-1}\bar{u}(k_1)\frac{\slashed{n}}{2}\frac{(\slashed{k}_1-\slashed{\tau}_{\perp})}{(k_1 - \tau)^2}\gamma^\mu\nn\\
&\times T^{A(\sigma_{1,q}^{N_q})}T^{A(\sigma_{1,q}^{N_q-1})}\cdots T^{A(\sigma_{1,q}^{2})}T^{A_1}T^{A({\sigma_{1,\bar{q}}^1})} \,  T^{A(\sigma_{1,\bar{q}}^2)}
  \cdots 
  T^{A(\sigma_{1,\bar{q}}^{\mbox{\tiny $N_{\bar q}$}} )}
v(-k_2)\,.
\end{align}
Swapping the position of the first Glauber no longer induces a cancellation as in \eq{cancel-glaubers},
as the two diagrams have different Dirac structures, as in \eq{AN1soft}.
We could instead consider swapping whether the final nonperturbative Glauber attachment attaches to the $q$ or $\qbar$, which corresponds to a color structure in the second line of \eq{ieq1} of either $T^{A(\sigma_{1,q}^{N_q})} = T^{A_N}$ or $T^{A(\sigma_{1,\bar{q}}^{\mbox{\tiny $N_{\bar q}$}} )}= T^{A_N}$.  
\begin{align}
	&T^{A_N}T^{A(\sigma_{1,q}^{N_q-1})}\cdots T^{A(\sigma_{1,q}^{2})}T^{A_1}T^{A({\sigma_{1,\bar{q}}^1})} \,  T^{A(\sigma_{1,\bar{q}}^2)}
	  \cdots 
	  T^{A(\sigma_{1,\bar{q}}^{\mbox{\tiny $N_{\bar q}-1$}} )}T^{A(\sigma_{1,\bar{q}}^{\mbox{\tiny $N_{\bar q}$}} )} \nn
	\\ \to 
	&T^{A(\sigma_{1,q}^{N_q-1})}\cdots T^{A(\sigma_{1,q}^{2})}T^{A_1}T^{A({\sigma_{1,\bar{q}}^1})} \,  T^{A(\sigma_{1,\bar{q}}^2)}
	  \cdots 
	  T^{A(\sigma_{1,\bar{q}}^{\mbox{\tiny $N_{\bar q}-1$}} )}T^{A(\sigma_{1,\bar{q}}^{\mbox{\tiny $N_{\bar q}$}} )}T^{A_N}\,,
\end{align}
where we assumed the last Glauber attaches to the quark before the swap. A similar relation follows for the case where the last Glauber attaches to the anti-quark. The rest of the matrix element only changes by an overall factor $(-1)^{N_q-1} \to (-1)^{N_q}$; that is, a relative sign. 
Now, if the Glaubers form a color-singlet exchange, then the original and swapped diagrams cancel one another out by cyclicity of the color trace. Thus, when $-t \gg \LQCD^2$, a single perturbative Glauber is \emph{insufficient} to produce a color-singlet exchange for the lowest-order soft function, no matter how many nonperturbative Glauber exchanges occur.
For the color-nonsinglet channel the sum does not vanish, but there are still simplifications.

\paragraph{Color octet exchange with $p=1$.} The only non-vanishing contribution to the soft function with one perturbative and any number nonperturbative Glaubers arises from the color-octet channel, with $p=1$.  
From \eq{AN1soft}, the $N=1$ octet amplitude is
\begin{align}
\label{eq:AN1softagain}
	{\cal M}^{\nu}_{1(8)}
	=& e_q\, {\rm tr}[T^C T^{B_1'}]\,
	  \bar{u}(k_1)\bigg[ -\gamma^\nu
	 \frac{(\slashed{\tau}_{\!\perp} -\slashed{k}_2)\frac{\slashed{n}}{2}}{(\tau-k_2)^2}
	 + 
	 \frac{\frac{\slashed{n}}{2}(\slashed{k}_1 
	  -\slashed{\tau}_{\!\perp})}{(k_1-\tau)^2}
	 \gamma^\nu \bigg]  v(-k_2)\nn\\
	 \equiv& {\rm tr}[T^C T^{B_1'}] \: C_{N=1}^\nu
\end{align}
where the terms in the first line correspond to a single Glauber exchange with an anti-quark and quark, respectively. If we now add $N-1$ nonperturbative Glauber exchanges after the initial perturbative Glauber exchange, the Dirac structure remains identical. We can see this, for example, when comparing the second term of~\eq{AN1soft} with~\eq{ieq1}. This motivates us to factor out the color factors in the second line of \eq{AN1softagain}. 

It is straightforward to show that summing all Glauber diagrams with 1 perturbative and $N-1$ nonperturbative Glauber exchanges leads to a nested commutator structure
\begin{align} \label{eq:ANsoftagain}
\mathcal{M}_{N=\mathrm{1P+(}N\mathrm{-1)NP} (8)}^\nu =&\text{Tr}[[\cdots [[[T^{A_1},T^{A_2}],T^{A_3}],T^{A_3}],\cdots ,T^{A_N}] T^C] \: C_{N=1}^{\nu}\,.
\end{align}
In turn, we can write the soft function with 1 perturbative and $N-1$ nonperturbative on one side, as well as 1 perturbative and $N'-1$ nonperturbative Glauber exchanges on the complex-conjugate amplitude side, as
\begin{align}\label{eq:lo-soft-NN'}
\hat{S}_{i(\mathrm{1P+(}N\mathrm{-1)NP} ,\mathrm{1P+(}N'\mathrm{-1)NP} )}^{8[0]}  =&\text{Tr}[[\cdots [[[T^{A_1},T^{A_2}],T^{A_3}],T^{A_4}],\cdots ,T^{A_N}] T^C]\nn\\
& \times\text{Tr}[[\cdots [[[T^{B_1},T^{B_2}],T^{B_3}],T^{B_4}],\cdots ,T^{B_{N'}}] T^C] \nonumber\\
&  \times \frac{1}{\text{Tr}[T^A T^C]\text{Tr}[T^B T^C]}\,
  \hat{S}_{i(1,1)}^{8[0]}\,,
\end{align}
where the inverse color factor in the last line cancels the color factor in $\hat{S}_{i(1,1)}^{8[0]}$.
That is, adding nonperturbative Glaubers modifies the overall color factor, but not the underlying dependence on kinematic variables. 
We can group these nonperturbative Glauber-induced color factor changes into the beam function; thus, they do not modify the predictive power of the perturbative soft function at this order.

\subsection{Two perturbative Glauber exchanges and color-singlet diffraction}\label{sec:two-perturbative}

Here, we offer a brief discussion of the two Glauber exchange case in color-singlet diffraction, leaving the presentation of explicit results for future work. Let us first consider the $-t \gg \LQCD^2$ regime. In \sec{one-pert-n-non}, we demonstrated that at LO ($q\qbar$) in the soft sector, there must be at least two perturbative Glauber exchanges for the soft function to be nonzero and for color-singlet diffraction to occur. 
This case corresponds to the so-called \textit{hard Pomeron} (or BFKL Pomeron) in the literature \cite{Lipatov:1985uk,Fadin:1993wh}. 
At NLO in the soft sector (with $q\qbar g$ and radiative corrections to $q\qbar$), 
it is not known if a single perturbative Glauber gluon with additional nonperturbative Glaubers will give a nonzero color-singlet soft function; this can be determined by future diagrammatic calculations. 
We remark that this discussion holds for both coherent and incoherent diffraction,
which have the same universal soft function, as both utilize the same color-singlet projection. (The soft functions for diffraction and quasi-diffraction differ, however, as they have different color projections.)
For further discussion we refer to \sec{one-pert-n-non} on the leading quasi-diffractive soft function, \sec{coherent-and-beams} on coherent diffraction, and \sec{quasibkgnd} on incoherent diffraction. 

In the $-t \simeq \LQCD^2$ regime, all Glauber exchanges become nonperturbative and are on an equal footing.  In this case, 
color-singlet diffraction corresponds to the so-called \textit{soft Pomeron} in the literature~\cite{Donnachie:1992ny,Donnachie:2002en}.
For $Q^2 \gg -t \simeq \LQCD^2$, our soft function is no longer entirely perturbative, but we can still carry out the $\lambda_t \ll 1$ hard-collinear factorization in \sec{smalllambdat} to decompose this soft function into a perturbative hard coefficient and a nonperturbative soft-collinear function.
Unfortunately, there is no reason \textit{a priori} to suspect that this soft-collinear function will appear universally in other processes, like $pp$ diffractive scattering. 
We will take up a dedicated analysis of the universality properties of diffractive cross sections in $ep$ and $pp$ scattering in a separate publication.

\section{Theoretical Predictions and Experimental Targets}\label{sec:predictions}

In this section, we provide some first predictions for diffractive structure functions following from our formalism, and we discuss targets for experimental measurements. 
In particular, our Regge factorization derived in \sec{diffract-factorization} demonstrates that for incoherent diffraction, there is both a color-singlet diffractive signal and an irreducible nonsinglet quasi-diffractive background. 
We provide results for these components using our calculations of Sudakov suppression and the soft function from \secs{logarithms}{soft-results}.

In \sec{predict-f34} we discuss the expected size of various diffractive structure functions, including the common $F_{2,L}^D$ and oft-neglected $F_{3,4}^D$, based on power counting considerations. 
In \sec{predict-quasiratio} we use our perturbative calculation of the soft function to make predictions for the $\beta$- and $t$-dependence of ratios of quasi-diffractive structure functions. 
In \sec{predict-phase-space} we discuss prospects for studying non-traditional regions of diffractive phase space, and in \sec{coherent-and-beams} we discuss the steps required to  extract the coherent beam functions from experimental data.  
Finally in \sec{quasibkgnd}, we discuss the size of the quasi-diffractive background for incoherent diffraction, arguing that it may be much more significant in size than has previously been assumed.

\subsection{Size of structure functions: $F_{3,4}^D$ are comparable to $F_{2,L}^D$}\label{sec:predict-f34}

In \eq{power-counting-params}, we provide a list of dimensionless parameters $\{\lambda, \lambda_t, \rho, \lambda_g, \lambda_\Lambda\}$ that enable us to distinguish different regions of phase space, and to distinguish diffraction from non-diffractive processes. As discussed in \sec{power-counting},
a (quasi-)diffractive process requires $\lambda^2 = Q^2/s \ll 1$; however, there is greater freedom for the remaining parameters. In particular, they need only satisfy $\lambda^2\lambda_t^2\ll 1$, $\lambda^2\lambda_t^2\rho^2\ll 1$, $x\ll \beta$, and $\lambda^2 \lambda_t^2 \ll \lambda_g^2 \ll \lambda^{-2}$.  For example, this allows for various choices for the size of $\lambda_t^2=-t/Q^2$; e.g., $\lambda_t \sim 1$ or $\lambda_t \ll 1$. 

By expanding in $\lambda$ and considering all leading power contributions, we derived the Regge factorizations in \eqs{FDfactdiff}{FDfactquasi}.  These factorizations predict that the five diffractive structure functions for an unpolarized proton arise at the same order in the power expansion for both incoherent diffraction and quasi-diffraction,
\begin{align} \label{eq:Flambdascaling}
\text{incoherent:}\quad  F_2^D \sim F_L^D \sim F_3^D \sim F_4^D \sim F_{4A}^D\sim 
 \lambda^{-6} 
 \,.
\end{align}
All five structure functions are thus of interest for probing the dynamics of 
diffraction in $e p$ colliders, including the interference terms $F_{3,4,4A}^D$, which are often neglected by analyses in the literature. See \sec{experiments-xbar} for experimental prospects for $F_{3,4,4A}^D$.

To derive the $\lambda$-scaling in \eq{Flambdascaling}, we analyze each component of the factorization formula. 
The prefactor $Q^2 \BoostInv^2/x\sim \lambda^{0}$ and $\IInt^\perp_{(N,N')}\sim \lambda^{-4}$.
For the soft function, we have $[d\tilde y][d\tilde y'] d^dz \sim \lambda^{-8}$; two $J_s\sim \lambda^3$ factors; and two collapsed products of Glauber operators that each scale the same, as a single operator $O_s^B(\tilde y,\tau_\perp)\sim \lambda$; and $\sumint_{X_s} |X_s\rangle\langle X_s| \sim \lambda^0$. All together, this means the soft function scales as $S\sim \lambda^0$.  
For the beam function, we have that $dv^-/(p^-)^3 \sim \lambda^{0}$, 
each of the two 
$\bar O_n^A(v^-\frac{n}{2}) \sim \lambda $,
and the collapsed product of Glauber operators does not change this scaling.  Additionally, each proton state $|p\rangle \sim \lambda^{-1}$, and the $p'$ momentum constrains $\sumint_{Y_n} |Y_n\rangle\langle Y_n| \sim |p'\rangle\langle p'| \sim \lambda^{-2}$. All together, this implies the beam function scales as 
$B\sim \lambda^{-2}$; 
this is due to the fact that all terms appearing in $B$ scale
as $1/(p^- p^{\prime +})\sim \lambda^{-2}$, arising from  $\delta(m_Y^2-\ldots)\sim \frac{1}{p^-} \delta(p^{\prime +}- \ldots)$.  The scaling $U\sim \lambda^{-4}$ cancels against that of $dp_g^+ dp_g^-\sim \lambda^4$. All together this yields \eq{Flambdascaling}.  Note that two powers $\lambda^{-2}$ arise through $B$ because the $F_i^D$ in \eqs{FDfactdiff}{FDfactquasi} are differential in $m_Y^2$. 

For the case of $\lambda_t\ll 1$, we can also determine how the ${\cal O}(\lambda^0)$ structure functions scale in $\lambda_t$, utilizing the additional hard-collinear factorization in \eq{refactS}.
This yields
\begin{align} \label{eq:Fscaling}
	F_L^{D\,{\rm quasi}} \sim 
	F_2^{D\,{\rm quasi}} \sim 
	F_3^{D\,{\rm quasi}} \sim \lambda^{-6} \lambda_t^{-4} \,, \qquad
	F_4^{D\,{\rm quasi}} \sim \lambda^{-6} \lambda_t^{-3}   \,.
\end{align}
Interestingly, $F_4^{D\,{\rm quasi}}$ has a $\lambda_t$-suppression relative to the other structure functions in the $\lambda_t \ll 1$ region of phase space. 
To derive this scaling in $\lambda_t$, we  again analyze each component of the factorization formula. For the soft function, the explicit calculations in \sec{soft-results} give $S_{i(1,1)}^8\sim \lambda_t^2$ for $i=2,3,L$ and $S_{4(1,1)}^8\sim \lambda_t^3$. 
The extra factor of $\lambda_t$ for $i=4$ arises because its projection operator in \eq{projectorsd} involves $U^\mu$ and $X^\mu$ together. In \eq{basis} we show that $U^\mu$ and $X^\mu$ can be thought of as time- and transverse-component unit vectors, respectively; thus, the projection for $F_4^D$ forces the fermion loop in the soft function to pick up a balancing transverse component proportional to $\lambda_t$. We explicitly observe this $\lambda_t$-suppression in our LO soft function in~\eq{Slambdat}.
The presence of nonperturbative Glaubers does not change these results. For $N,N'>1$ we can infer that the $\lambda_t$ scaling is of this same size, unless there are further cancellations in particular color channels that lead to higher powers of $\lambda_t$. 
This would also imply that $F_{4A}^{D\,{\rm quasi}}\sim \lambda^{-6} \lambda_t^{-3}$ for its first nonvanishing contribution. 
Since $F_{4A}^{D\,{\rm quasi}}$ vanishes for $N=N'=1$, our above results do not suffice to confirm this explicitly.
The beam function only depends on the scales  $p^- k_n^+\sim t\sim \tau_{i\perp}^{2}\sim \tau_{j\perp}^{\prime\,2} \sim \lambda^2\lambda_t^2$, so 
$\lambda_t$ and $\lambda$ enter with the same power, which gives $B\sim \lambda^{-2}\lambda_t^{-2}$. The scaling of $U$ cancels against $dp_g^+ dp_g^-$, and $x\IInt^\perp_{(N,N')}\sim \lambda_t^{-4}$.  All together this yields \eq{Fscaling}.  Again, two of the powers $\lambda_t^{-2}$ arise through $B$ because the structure functions are differential in $m_Y^2$. 

It is also useful to consider how each structure function contributes to the differential cross section $d\sigma = \sum_i \sigma_i^D \propto \sum_i c_i F_i^D$, which is impacted by the size of the coefficient $c_i$ in \eq{unpolarized-coefficients} and the cross section prefactor in \eq{sigmaLW}.
We have that:
\begin{align} \label{eq:cFscaling}
	&   d\sigma_L^D \sim d\sigma_0\: \frac{y^2}{2}  \lambda^{-6} \lambda_t^{-4}  ,
	&& d\sigma_2^D \sim 
	   d\sigma_3^D \sim   d\sigma_0\: \lambda^{-6} \lambda_t^{-4} , 
	&& d\sigma_4^D \sim   d\sigma_0\: \lambda^{-6} \lambda_t^{-3} ,
\end{align} 
where there is a common kinematic prefactor 
$d\sigma_0 = 1/(Q^4\beta^2 N_\sigma)$ with $N_\sigma \sim x\lambda_t$; 
see \eq{diffcross}. 
Because $F_L^D$ appears with a prefactor of $y^2/2$, separating it from $F_2^D$ requires measurements differential in $y$ values to disentangle the two structure functions. In practice, this is experimentally challenging, so most HERA analyses~\cite{H1:2006zyl, ZEUS:2008xhs, Aaron:2010aa} report the diffractive ``reduced cross section'', which is linear combination of $F_2^D$ and $F_L^D$ integrated over a bin in $y$. This is often taken to be $F_2^D$ to good approximation, due to the suppression of $F_L^D$ by $y^2/2$. A notable exception that extracts $F_L^D$ is \refcite{H1:2011jpo}. 
As discussed in \sec{experiments-xbar}, the structure functions $F_{3,4}^D$ require precision reconstruction of $\bar x$, but do not have additional suppression from $y$, and hence are important experimental targets.
With the scaling for $F_{4A}^D$ discussed above, we also have $d\sigma_{4A}^D \sim   d\sigma_0\: \lambda^{-6} \lambda_t^{-3}$.

\paragraph{Hadron-polarized structure functions.} 
Recall from \eq{diffcrosspol} that up to a factor of $\tilde S_T'$, the leading-power hadron-polarized structure functions $F_{iP}^D$ have the same tensor structure as their five hadron-unpolarized counterparts $F_i^D$ with $i=L,2,3,4,4A$. Thus,
\begin{align} \label{eq:polarized-Fscaling}
	&F_{iP}^D \sim F_i^D \,,
	&&c_{iP} \sim c_i \,.
\end{align}
These structure functions $F_{iP}^D$ provide additional targets for diffractive measurements at the EIC beyond those considered in the  literature \cite{AbdulKhalek:2021gbh}.

\subsection{Perturbative predictions for $F_i^{\rm D\,quasi}$ ratios}\label{sec:predict-quasiratio}

\begin{figure}
	\begin{center}
		\includegraphics[width = 2.9 in]{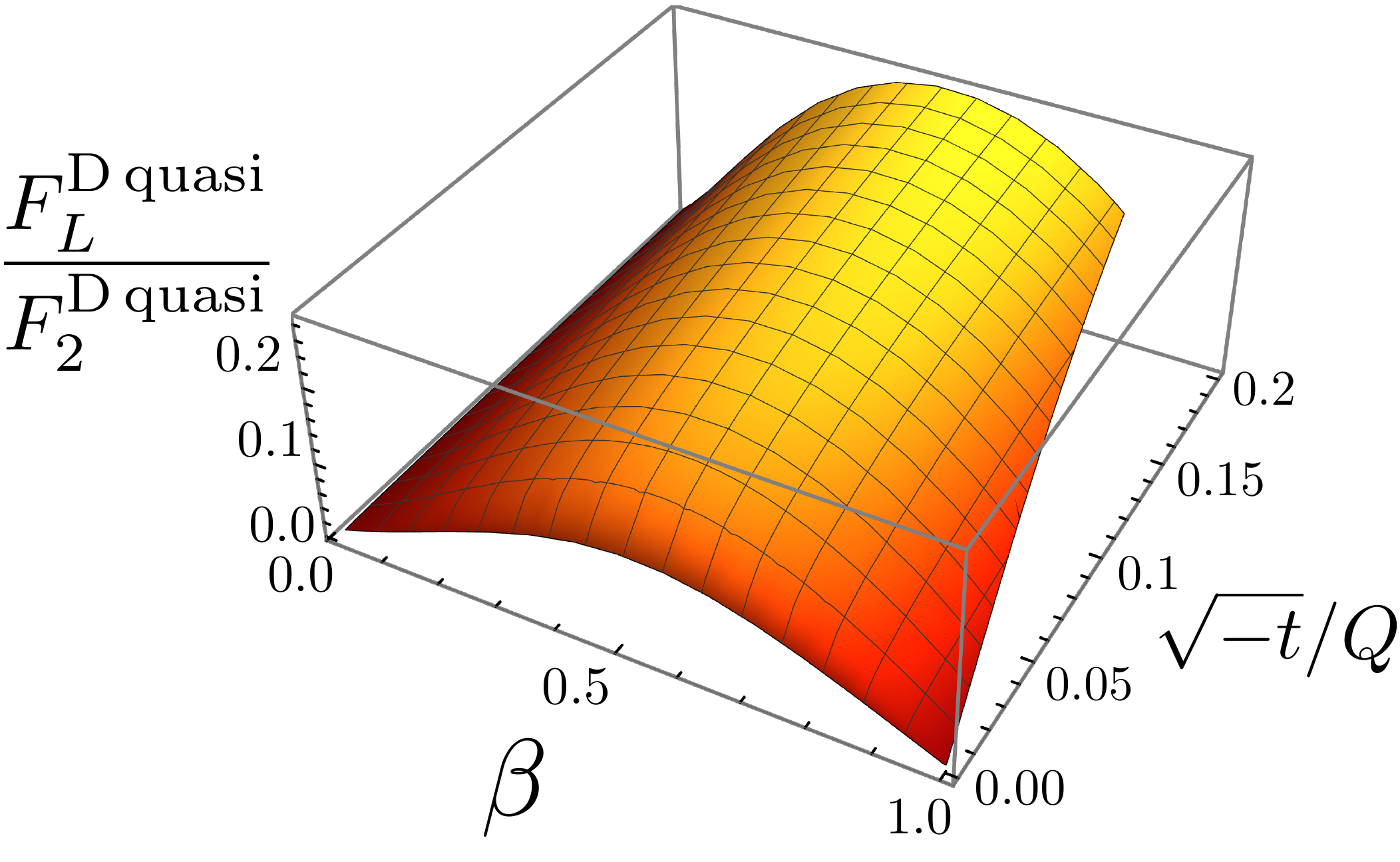}\quad
		\includegraphics[width = 2.9in]{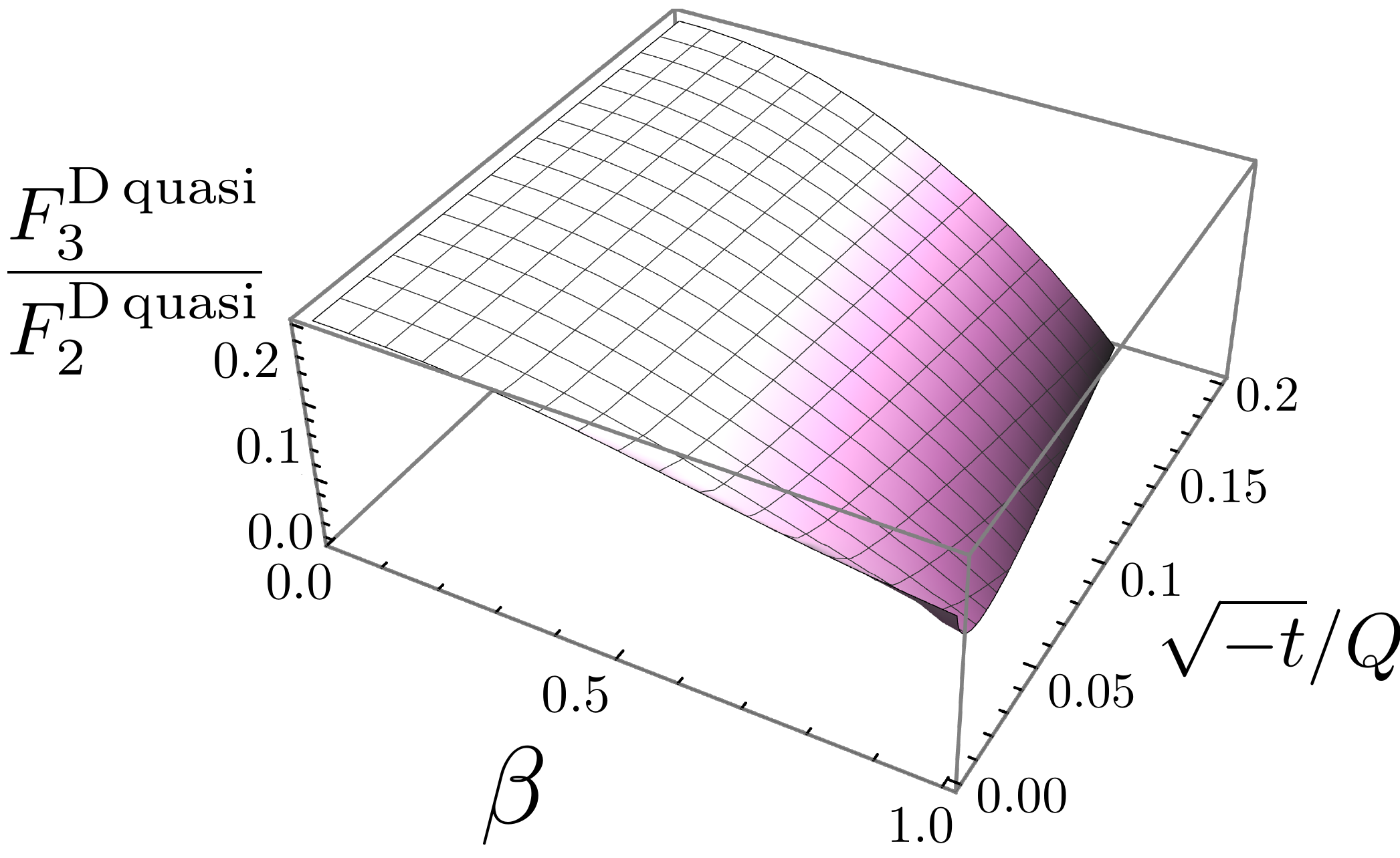}\quad
		\includegraphics[width = 2.9 in]{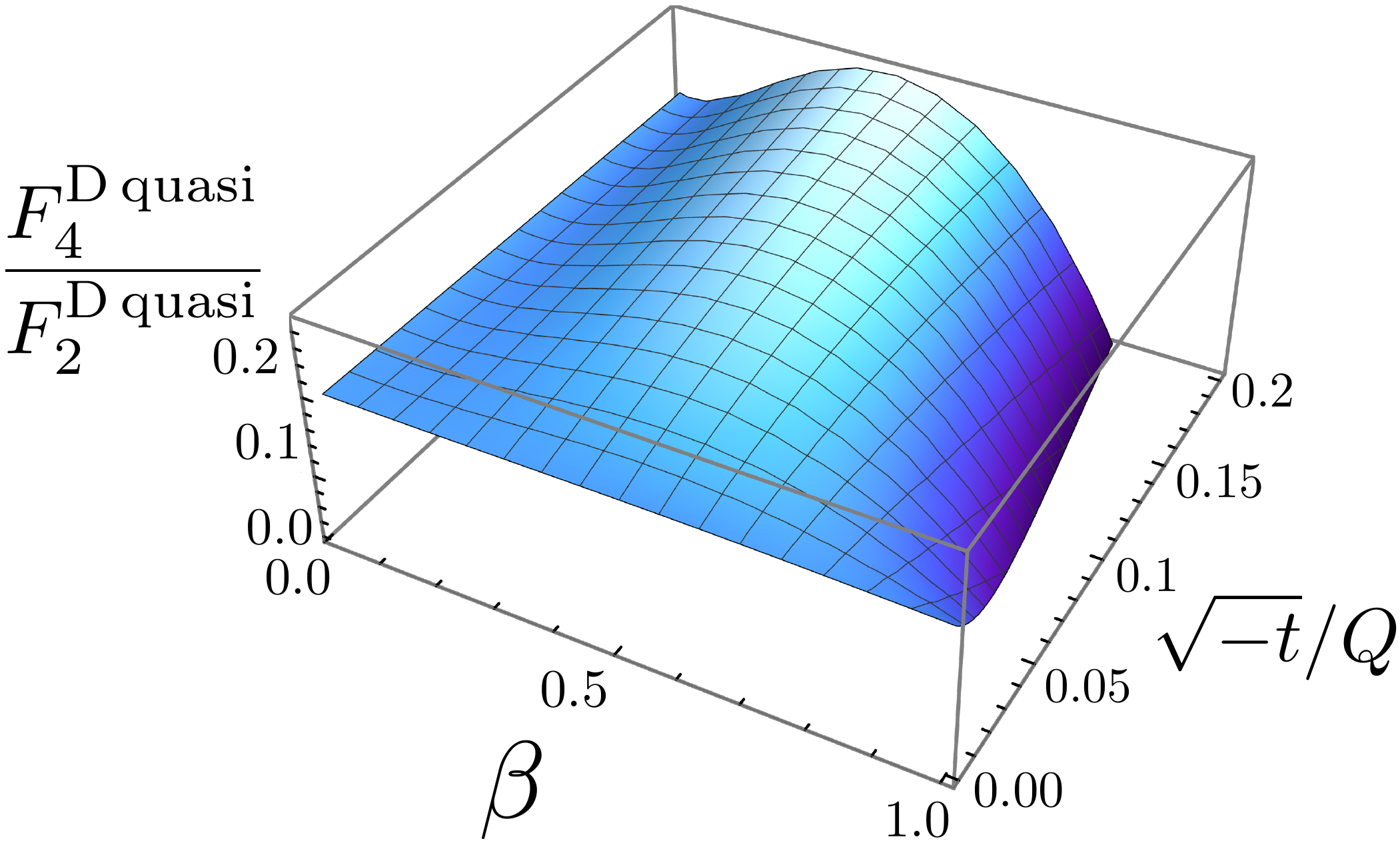}
	\end{center}
	\caption{ Leading order perturbative predictions for the ratios of quasi-diffraction structure functions in \eq{s-ratios}, which are a function of $\beta$ and $\lambda_t = \sqrt{-t}/Q$.  The kinematic bounds on $\beta$ from~\eq{beta-bounds} carry an implicit $x$-dependence, which constrains the allowed region but does not alter the value of the ratios. In all plots shown here, we fix $x = 0.02$.
}
   \label{fig:quasi-ratio}
\end{figure}

Recall from \sec{soft-results} that in the perturbative regime $-t\gg \Lambda_{\rm QCD}^2$, the dominant contribution to quasi-diffraction is given by a single soft $q\qbar$ pair, with one perturbative Glauber exchange and any number of non-perturbative Glaubers. Furthermore, in \sec{logarithms}, we saw that the leading Sudakov double logarithms in $U$ are captured by a multiplicative factor with a trivial $p_g^\pm$ convolution.  At this order, the factorization  in \eq{FDfactquasi} becomes:
\begin{align}\label{eq:simplify-f-ratios}
	F_i^{\rm D\,quasi} 
    & = \sum_{N,N'} \sum_{R_X \ne 1} 
    (B_{(N,N')}^R \otimes_\perp S_{i(N,N')}^{R'}) \otimes_\pm U_{(N,N')}^{RR'}  
     \nn\\
    & 
=  \frac{\hat{S}_{i(1,1)}^{8[0]}}{T_F^2\, \delta^{AB}}    \times  B_{\Lambda}^8
      \times U^{LL}  + \ldots \,,
\end{align}
In the first line, the convolutions $\otimes_\perp$ and $\otimes_\pm$ are shorthands for the full set of integrals and other factors in \eq{FDfactquasi}; see \eq{FDfact-shorthand}.
In the second line, we see that for a single perturbative Glauber exchange, these convolutions reduce to simple multiplications.  The $T_F^2\delta^{AB}$ in the final line cancels the color factor in $\hat{S}_{i(1,1)}^{8[0]}$. Note that here, we absorb color and momentum factors associated with non-perturbative Glaubers from \eqs{hatS}{lo-soft-NN'} into a modified beam function:
\begin{align} \label{eq:BL8}
	  B_{\Lambda}^8\Big(m^2,t,\mu,\frac{\nu}{Q/x}\Big) &= \sum_{N,N'} \IInt^\perp_{(N,N')}  
		B_{(N,N')}^{R_A^{N\!N'}}\Big(m^2,\{\tau_{i\perp},\tau_{j\perp}'\},t,\mu,\frac{\nu}{Q/x}\Big)
	    \: G(\tau_{1\perp})\cdots G(\tau_{N'\perp}) 
	  \nn\\
	&  \times \frac{1}{(2\pi)^4} [8\pi\alpha_s(\mu)]^{N+N'} \, 
	\text{Tr}[[\cdots [[[T^{A_1},T^{A_2}],T^{A_3}],T^{A_4}],\cdots ,T^{A_N}] T^C]
	  \nn\\
	& \times\text{Tr}[[\cdots [[[T^{B_1},T^{B_2}],T^{B_3}],T^{B_4}],\cdots ,T^{B_{N'}}] T^C]
	  \,.
\end{align}
Because $B_\Lambda^8$ and $U^{LL}$ are independent of the structure function choice $i$, ratios of structure functions are equivalent to ratios of soft functions at this order:
\begin{align}\label{eq:s-ratios}
	\frac{F_i^{\rm D\,quasi} }{F_2^{\rm D\,quasi} } 
   &= \frac{\hat S_{i(1,1)}^{8[0]}}{\hat S_{2(1,1)}^{8[0]}} 
    +  \ldots 
   \equiv {r}^s_i(\beta,-t/Q^2) + \ldots \,,
\end{align}
where we define the ratio of soft functions as $r^s_i$, and the ellipses denote higher-order terms. for convenience.
These ratios are identical for the quasi-diffractive hadron-polarized structure functions in \eq{tensor-structures-polarized}, due to their identical tensor structures:
\begin{align}\label{eq:polarized-ratios}
	\frac{F_{iP}^{\rm D\,quasi} }{F_{2P}^{\rm D\,quasi} } &=
	\frac{\hat S_{i(1,1)}^{8[0]}}{\hat S_{2(1,1)}^{8[0]}} 
    + \ldots \,.
    =r^s_i(\beta,-t/Q^2) + \ldots \,,
\end{align}
for $i = L,2,3,4,4A.$

\Eqs{s-ratios}{polarized-ratios} imply that the kinematic dependence of the quasi-diffractive structure function ratios is analytically calculable at leading order. Using our perturbative results in \sec{soft-1g},  we show predictions in \fig{quasi-ratio} for $r^s_i(\beta,-t/Q^2)$ for $i = L,3,4$.  These non-trivial predictions are relevant for incoherent diffraction, which has color-nonsinglet quasi-diffractive contributions. 
Note that these ratios depend on $\beta$ and $\lambda_t^2=-t/Q^2$, but not on $Q^2$ alone. To test these predictions at a given $Q$, one must ensure that we remain in the perturbative regime $-t\gg \LQCD^2$. For example, for a smaller $Q^2=20\,{\rm GeV}^2$ at the EIC, $-t \gtrsim 1\,{\rm GeV}$ implies that one should only trust these predictions for $-t/Q^2 \gtrsim 0.05$.
We caution that testing these ratios also requires identifying a region where quasi-diffraction dominates over diffraction, which may be difficult.  A further discussion of the relative size of quasi-diffractive and diffractive contributions can be found in \sec{quasibkgnd}. 

Another application of \eqs{s-ratios}{polarized-ratios} is to construct linear combinations of diffractive structure functions where the quasi-diffraction contributions will partially cancel. For example, for the hadron-unpolarized case we can consider
\begin{align}
  & F_L^D  - r^s_L F_2^D \,,
  && F_3^D  - r^s_3 F_2^D \,,
  &&  F_4^D  - r^s_4 F_2^D \,,
  &&  F_{4A}^D  - r^s_{4A} F_2^D \,.
\end{align}
By construction the leading quasi-diffractive contributions cancel in these combinations, which is likely to make their diffractive contributions dominate (though strictly speaking, we have not proven  that the leading diffractive contributions are not suppressed in these combinations as well). Note from \eq{polarized-ratios} that parallel conclusions hold for linear combinations of leading-power hadron-polarized structure functions. 

\begin{figure}
	\begin{center}
	\includegraphics[width= 6in]{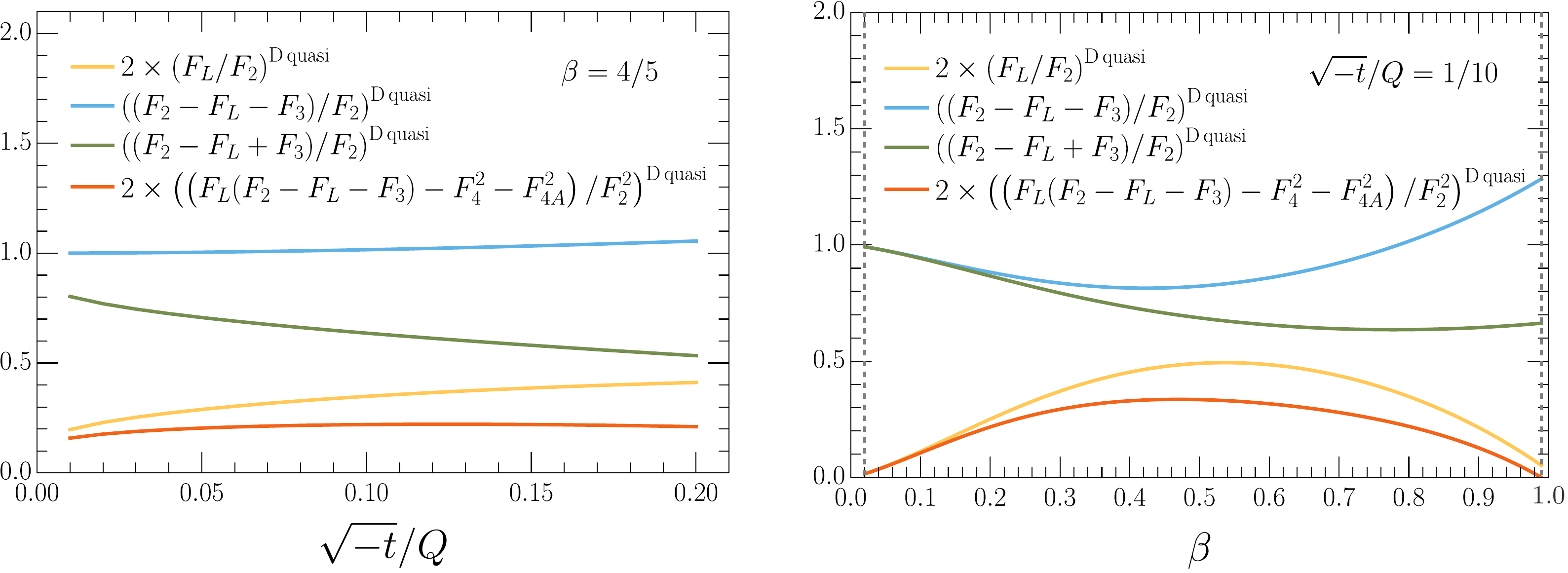}
	\end{center}
	\caption{
		Linear combinations of (quasi-)diffractive structure functions satisfy positivity bounds; see~\eq{positive-fi}.  Here, we verify that our LO soft-function predictions in~\eq{s-ratios} satisfy these bounds by plotting them as a function of $\sqrt{-t}/Q$ at fixed $\beta=0.8$ and as a function of $\beta$ at fixed $\sqrt{-t}/Q=0.1$.  
        At LO, the structure function ratios only depend on $\beta$ and $\lambda_t = \sqrt{-t}/Q$, while $x$-dependence only affects the lower limit of the allowed range of $\beta$ from \eq{beta-bounds}, plotted here as gray vertical dashed lines. 
	}\label{fig:positivity-constraints}
\end{figure}

\paragraph{Positivity constraints.}  Certain linear combinations of the four unpolarized diffractive structure functions are directly proportional to cross sections for physical scattering processes of specific helicities, and hence must be positive. These constraints have been discussed in \Refcite{Arens:1996xw}, and yield relations in terms of our structure functions
\begin{align}\label{eq:positive-fi}
	&F_L^D \geq 0\,,
	&&F_2^D - F_L^D \pm F_3^D \geq 0\,,
	&&F_L^D(F_2^D-F_L^D+F_3^D) - F_4^{D\,2}-F_{4A}^{D\,2} \geq 0\,,
\end{align} 
which also implies $F_2^D\ge 0$. From our calculation of soft functions involving single-Glauber exchange, we see that $\hat S_{2(1,1)}^{8[0]}>0$, 
so from our ratio predictions in \eq{s-ratios}, these positivity conditions translate to equivalent statements for the soft functions
\begin{align}\label{eq:positive-soft}
	&S_{L(1,1)}^{8[0]} \geq 0\,,
	\qquad\qquad 
  S_{2(1,1)}^{8[0]} - S_{L(1,1)}^{8[0]} \pm S_{3(1,1)}^{8[0]} \geq 0\,,
   \nn\\*
	&S_{L(1,1)}^{8[0]}(S_{2(1,1)}^{8[0]}-S_{L(1,1)}^{8[0]}+S_{3(1,1)}^{8[0]}) - S_{4(1,1)}^{8[0]\,2}-S_{4A(1,1)}^{8[0]\,2} \geq 0
   \,.
\end{align} 
In \fig{positivity-constraints}, we demonstrate that at one-Glauber order, the positivity constraints in \eq{positive-soft} are satisfied, providing a nontrivial cross-check on our results.

\subsection{Diffraction beyond the traditional regions of phase space}\label{sec:predict-phase-space}

Diffraction experiments at HERA focused on the kinematic region with $-t\lesssim 1\,{\rm GeV}^2$ and $\etalabcut\simeq 6$, 
in which the hadronic system $Y$ enters the far-forward detector.
This region has $\lambda_t = \sqrt{-t}/Q \ll 1$ and predominantly nonperturbative $-t$ values.  As seen in \sec{gap-radiation}, 
we can still maintain $\lambda_t\ll 1$ and allow for larger (perturbative) values of $-t$ by decreasing $\etalabcut$. 
This region is interesting to explore because the soft function in our factorization formula becomes perturbative, allowing for more predictive power.

Further potential targets include diffractive kinematic regimes characterized by $\lambda_t \sim 1$ or $\lambda_t \gg 1$ (where we still have $\lambda\lambda_t\ll 1$), which are relevant for smaller values of $Q$. So far only  $\lambda_t\gg 1$ photoproduction processes with $Q^2 \approx 0$ have been studied at HERA \cite{ZEUS:1999ptu, Cox:1999hw, ZEUS:2002vvv, H1:2003ksk, H1:2006ogl, H1:2008jdn, ZEUS:2009ixm}, which lies outside the scope of our analysis.  
It would be interesting to consider diffractive processes with larger  $-t\sim Q^2 \gg \LQCD^2$ and $-t\gg Q^2 \gg \LQCD^2$, but kept small enough to retain a significant statistical sample 
and $\lambda^2\lambda_t^2\ll 1$.
The choice of $\etalabcut$ may also need to be adjusted to enable larger $-t$.

One may also consider other general regions of (quasi-)diffractive phase space, although we do not explore them in detail here.
For the reader's convenience, we provide a pointer to the more general discussion specifying these kinematic regimes:
\begin{itemize}\setlength\itemsep{-5 pt}
	\item General kinematic bounds from positivity constraints in eqs.~\eqref{eq:xy-bounds} through \eqref{eq:my-bounds}
	\item Experimental requirements for measuring $\xbar$ from the lab frame polar angle of the hadronic system $Y$ in \eq{xbar-theta} 
	\item Constraints that must be satisfied for a diffractive process in eqs.~\eqref{eq:collimated-jets}, \eqref{eq:rapidity-gap}, \eqref{eq:forward-scatter}
    \item Restriction on the rapidity cut used to define the hadronic systems $X$ and $Y$ 
    which are given in \eqref{eq:beamconstraint} and \eqref{eq:softhconstraint}.
\end{itemize}

\subsection{Coherent scattering and beam function extraction}
\label{sec:coherent-and-beams}

A clean way of probing the dynamics of a pure color-singlet exchange is by measuring an intact forward proton. Here, $m_Y^2=m_p^2$, so we can integrate \eq{diffcross} over $m_Y^2$ and specialize to coherent structure functions
\begin{align}
\label{eq:diffcrosscoh}
\frac{d^6\sigma^{\rm coh}}{dx\, dQ^2\, d\beta\, dt\, d\bar x}
&= \frac{\alpha^2 \pi }{2Q^4\beta^2\, N_\sigma} \Biggl\{ -\frac{y^2}{2} F_L^{D\,{\rm coh}} +  \left(1-y+\frac{y^2}{2}\right)F_2^{D\,{\rm coh}}
\\
& \quad
+ \bigg[ \frac{2(k\cdot X)^2 y^2}{Q^2 } - 1+y  \bigg] F_3^{D\,{\rm coh}}
+ \frac{2y^2(k\cdot U)(k\cdot X)}{Q^2}F_4^{D\,{\rm coh}} \Biggr\}
 ,\nn
\end{align}
where
\begin{align}
  F_i^{D\,{\rm coh}}(x,Q^2,\beta,t) 
    = \lim_{\delta\to 0} \int_{m_p^2-\delta}^{m_p^2+\delta} dm_Y^2  
   \ F_i^{D}(x,Q^2,\beta,t,m_Y^2)  \,,
\end{align}
which picks out the part of $F_i^{D}(x,Q^2,\beta,t,m_Y^2)$ that behaves as $\delta(m_Y^2-m_p^2)$. 
From \sec{diffract-factorization}, we recall that the factorization for coherent diffraction has no quasi-diffractive contribution, so here the state $Y_n$ is a single proton with momentum $p'$.
In the diffractive factorization formula in \eq{FDfactdiff}, the integral over $m_Y^2$ only influences the beam function.  Using translation invariance on $B$ from \eq{Bdefn} 
gives
\begin{align}
	  & B_{(N,N')}^{R_A^{N\!N'}}
	  \bigl(p^-k_n^+, \{\tau_{i\perp},\tau_{j\perp}'\}, t\bigr)
	   =  
\frac{1}{(\nbar\cdot p)^2}
 \sumint_{Y_{n}}  
	  \int\! \frac{dv^-}{2p^-}\:
	  e^{\frac{i}{2} v^- k_n^+ } \: e^{-\frac{i}{2} v^- p_Y^+}
	  \\*
	 & \ \times
	    \big\langle p \big| \!
	   \collapsel\text{\footnotesize $\prod_{i=1}^{N-1}$}
	   \cO_n^{A_i}(0,\tau_{i\perp}) \bar \cO_n^{A_N}(0)
	   \!\collapser P_{N R_A}
	   \big| Y_{n} \big\rangle
	   \big\langle Y_{n} \big|  P_{N' R_{A}}
	  \collapsel \text{\footnotesize $\prod_{j=1}^{N'-1}$}
	  \cO_n^{A'_j}(0,\tau_{j\perp}^\prime) \bar \cO_n^{A'_{N'}}(0)
	  \!\collapser
	   \big| p \big\rangle
	   \nn\\
	 &\ = 
\frac{1}{(\nbar\cdot p)^2} 
\sumint_{Y_{n}}
	  \: \delta\big( p^- k_n^+ - p^- p_Y^+)\big) 
	 \nn\\
	 & \ \times
	    \big\langle p \big| \!
	   \collapsel\text{\footnotesize $\prod_{i=1}^{N-1}$}
	   \cO_n^{A_i}(0,\tau_{i\perp}) \bar \cO_n^{A_N}(0)
	   \!\collapser P_{N R_A}
	   \big| Y_{n} \big\rangle
	   \big\langle Y_{n} \big|  P_{N' R_{A}}
	  \collapsel \text{\footnotesize $\prod_{j=1}^{N'-1}$}
	  \cO_n^{A'_j}(0,\tau_{j\perp}^\prime) \bar \cO_n^{A'_{N'}}(0)
	  \!\collapser
	   \big| p \big\rangle
	 \nn\,,
\end{align}
where only singlet representations $R_A^{NN'}=1$ are needed for diffraction.
From this we obtain the relevant beam function for coherent diffraction
\begin{align}  \label{eq:Bdefncoh}
	  B_{(N,N')}^{R_A^{N\!N'}\,{\rm coh}}\big(\{\tau_{i\perp},\tau_{j\perp}'\},t \big) 
	 &= \lim_{\delta\to 0} \int_{m_p^2-\delta}^{m_p^2+\delta} dm^2  \ 
	   B_{(N,N')}^{R_A^{N\!N'}}\big(m^2-t,\{\tau_{i\perp},\tau_{j\perp}'\},t \big) 
	  \nn \\
	 &= 
	 \frac{1}{(\nbar\cdot p)^2}
	 \big\langle p \big| \!
	   \collapsel\text{\footnotesize $\prod_{i=1}^{N-1}$}
	   \cO_n^{A_i}(0,\tau_{i\perp}) \bar \cO_n^{A_N}(0)
	   \!\collapser P_{N R_A}
	   \big| p'\big\rangle
	 \nn\\
	 &\quad\qquad \times
	   \big\langle p' \big|  P_{N' R_{A}}
	  \collapsel \text{\footnotesize $\prod_{j=1}^{N'-1}$}
	  \cO_n^{A'_j}(0,\tau_{j\perp}^\prime) \bar \cO_n^{A'_{N'}}(0)
	  \!\collapser
	   \big| p \big\rangle
	 \,. 
\end{align}
The factorization formula for coherent diffraction then takes the same form as in \eq{FDfactdiff}, but with this coherent version of the beam function,
\begin{align} \label{eq:FDfactdiffcoh}
	F_i^{D\,{\rm coh}}\!
	&= \FiDpre
	\sum_{{\footnotesize N,N'=1}}^\infty
	  \sum_{R^{N\!N'}=1} \IInt^\perp_{(N,N')}  
	  \!\! B_{(N,N')}^{R^{N\!N'}\,{\rm coh}}
	\Bigl(\{\tau_{i\perp},\tau_{j\perp}'\}, t,\frac{\nu}{Q/x}\Bigr) 
	 \nn\\*
	 & \qquad\qquad\qquad\qquad \quad  \times
	 S_{i(N,N')}^{R^{N\!N'}} 
	  \Bigl(\frac{Q^2}{\beta}, \{\tau_{i\perp},\tau_{j\perp}'\}, Q,t, \frac{\nu}{Q/\beta} \Bigr) 
	 \nn\\
	 &= \sum_{R=1} \, B^{R\,{\rm coh}} \otimes_\perp S_i^{R}
	 \,,
\end{align}
where the second line establishes a short-hand notation.

Looking at the mass dimensions in \eq{FDfactdiffcoh}, $F_i^D$ has dimension $-2$, which equals the sum of (-4) for $\IInt^\perp_{(N,\,N')}$ setting $d'=2$, (0) for $B^{\rm coh}$, (0) for $S$, and (2) for the prefactor. Due to the extra $m_Y^2$ integration, the scaling from \sec{predict-f34} becomes $B^{\rm coh}\sim \lambda^0\lambda_t^0$, and hence 
\begin{align}
\label{eq:cohscaling}
	& F_{2,L,3}^{D\,{\rm coh}}\sim \lambda^{-4}\lambda_t^{-2}\,,
	&&  F_4^{D\,{\rm coh}}\sim \lambda^{-4}\lambda_t^{-1}\,,
\end{align}
and we would anticipate that $F_{4A}^{D\,{\rm coh}}\sim \lambda^{-4}\lambda_t^{-1}$.

In general, the object in a factorization formula that is the most universal across different processes is the beam function, which contains the initial proton states. For $t\sim \Lambda_{\rm QCD}^2$ the function $B$ is nonperturbative, and it would be interesting to develop a method to compute even the simplest coherent Glauber matrix elements $\langle p | O \bar O| p' \rangle$ in lattice QCD. It is also interesting to ask how we might separate the soft and beam functions and make measurements to extract $B^{\rm coh}$. The separation of $S_i$ can potentially be achieved in the perturbative regime discussed in \sec{soft-results}. In this situation, we may hope to factorize the nonperturbative Glauber exchanges to define a modified beam function $B_\Lambda^1$, analogous to the $B_\Lambda^8$ for quasi-diffraction in \eq{BL8}, and then target the measurement of this $B_\Lambda^1$ from this kinematic scenario.  From such calculations we anticipate obtaining a simple multiplicative soft function, but it remains to be demonstrated
that the nonperturbative Glaubers can be factorized into the beam function in the same manner that we achieved for quasi-diffraction at lowest order. 

\subsection{Incoherent diffraction: background versus signal }
\label{sec:quasibkgnd}
Currently, the diffraction literature largely assumes that gapped small-$x$ processes are always mediated by color-singlet exchanges \cite{AbdulKhalek:2021gbh}. 
For coherent scattering with an intact forward proton, the process must indeed be mediated by a color-singlet state. However, for incoherent diffraction, our factorization from \sec{diffract-factorization} has both color-singlet (diffractive) and color-nonsinglet (quasi-diffractive) contributions, which are impossible to distinguish experimentally; see \fig{diffraction-chart}.  
Recall from \sec{intro} that the definition of incoherent diffraction excludes the coherent signal. In contrast, the factorization prediction for $W^2\gg m_Y^2\gg \Lambda_{\rm QCD}^2$, has a jet-like state $Y$ that includes both the incoherent and coherent contributions.
From our factorization results in \sec{diffract-factorization}, we can 
write the sum of the coherent and incoherent structure functions as:
\begin{align}\label{eq:inc-contributions}
   F_i^{D} &=
   \sum_{R, R'} \, B^R \otimes_\perp S_i^{R'} \otimes_\pm U^{RR'} 
	\nonumber\\*
	&= 
\sum_{R,R' =1} \, B^R \otimes_\perp S_i^{R'}
     + \sum_{R,R'\ne1} \, B^R \otimes_\perp S_i^{R'}\otimes_\pm  U^{RR'} 
   \nn\\
   &= \Big( F_i^{D\,\rm coh} \delta(m_Y^2-m_p^2) +  F_i^{D\,\rm inc\: diff}\Big)  + F_i^{D\,\rm quasi}
\end{align}
where the two terms in the second line are
the diffractive (singlet) and quasi-diffractive (non-singlet) color exchange components. 
The last two terms in \eq{inc-contributions} are the contributions to incoherent diffraction.
Subtracting the result for coherent diffraction from \eq{FDfactdiffcoh}, we thus have
\begin{align}\label{eq:inc-only-contribution}
   F_i^{D\,\rm inc}
	&= 
\sum_{R,R' =1} \, \Big(B^R - B^{R\,{\rm coh}}\delta(m_Y^2-m_p^2) \Big) \otimes_\perp S_i^{R'}
     + \sum_{R,R'\ne1} \, B^R \otimes_\perp S_i^{R'}\otimes_\pm  U^{RR'} 
 \,. 
\end{align}
Although we do not yet know the size of the nonperturbative ingredients in our factorization formula, we can estimate the sizes of the diffractive and quasi-diffractive contributions based on power counting, large logarithms, and (where applicable) expansions in $\alpha_s$. First, consider the regime $-t\gg \LQCD^2$, where we can 
perturbatively expand $S_i$.  In this case, quasi-diffraction is suppressed relative to diffraction by the Sudakov exponential $U^{\rm LL}$ (see \sec{logarithms}), but enhanced by the fact that it starts with one less perturbative $\alpha_s$ than diffraction (see \sec{soft-results}). This yields 
\begin{align}   \label{eq:contamination-ratio}
	\frac{F_i^{D\,\rm quasi}}{F_i^{D\,\rm inc\, diff}}
   \bigg|_{\rm -t\gg\LQCD^2}
  & \sim\ \frac{\alpha_s(-t)\: U^{\rm LL}}{\alpha_s^2(-t)} = \frac{U^{\rm LL}}{\alpha_s(-t)} 
	  \,,
\end{align}
which relies on an estimate that the beam functions for diffraction and quasi-diffraction are of similar numerical size.  
This estimate could be significantly improved by including exact numerical factors; i.e.,~from a future calculation of the singlet perturbative soft function. 
In \eq{contamination-ratio}, we see that the Sudakov suppression through $U^{\rm LL}$ (plotted in \fig{sudakov}) competes against an $\alpha_s$ suppression. 
The exact size of each contribution depends on the kinematics, but in general quasi-diffractive contamination is sizable. For example, the left panel of \fig{sudakov} takes $\etalabcut=3$, and the ratio in \eq{contamination-ratio} varies from
$\sim$25-100\%. The right panel has $\etalabcut=5$, and contamination of $\sim$25\% in most of phase space.

HERA performed many diffractive measurements in a more nonperturbative region with $-t\sim \LQCD^2$.  In this case, we lose the $\alpha_s$ suppression factors and our estimate for the ratio is entirely given by the Sudakov factor,
\begin{align}\label{eq:contamination-ratio2}
\frac{F_i^{D\,\rm quasi}}{F_i^{D\,\rm inc\, diff}}
 \bigg|_{\rm -t\sim\LQCD^2}
  & \sim\ U^{\rm LL}
	  \,.
\end{align}
Here, we can more directly apply the results shown in \fig{sudakov}, which yields a ratio of $\sim$5--10\% in the favorable region of phase space at smaller $x$. 
We emphasize that this estimate is surprisingly large, given that the literature currently assumes that nonsinglet exchanges do not contribute \textit{any} background to incoherent diffractive processes.

The beam functions differ in diffraction and quasi-diffraction, so their extraction from incoherent experimental data is more challenging than for the coherent case discussed in \sec{coherent-and-beams}.  Nevertheless it would be interesting to build specific models for the nonperturbative parts of these beam functions, and utilize these for fits to experimental data. 

Furthermore, incoherent diffraction shares the same color-singlet soft function as coherent diffraction. 
Consequently, in a kinematic region where the quasi-diffractive background can be 
neglected, one can look at ratios $F_i^{D\,{\rm coh}}/F_i^{D\,{\rm inc}}$, which measure
the ratio of coherent and incoherent beam functions integrated over the same transverse momentum 
projection.%
\footnote{This depends on whether one can factorize nonperturbative Glaubers into the beam functions for the color-singlet case; see the discussion at the end of \sec{coherent-and-beams}.}
This will immediately provide information on how large these two beam functions are. 

We conclude that color-nonsinglet quasi-diffraction is likely to be a prominent irreducible background for the diffractive signal for large parts of the incoherent diffractive kinematics.

\section{Comparison with Existing Methods}
\label{sec:comparisons}

The primary result of this paper was to derive factorization formulas for inclusive (quasi-) diffractive DIS, which occurs in the kinematic regime $\lambda=Q/\sqrt{s}\ll 1$. 
We can summarize the results schematically as follows. 
For coherent diffraction we have from \eq{FDfactdiffcoh}
\begin{align}\label{eq:fact-schematic-coh}
	F_i^{D\,{\rm coh}} &= \sum_{R=1} \, B^{R\,{\rm coh}} \otimes_\perp S_i^{R}\,,
\end{align}
while for incoherent diffraction we have from \eqs{FDfactdiff}{FDfactquasi}
\begin{align}\label{eq:fact-schematic-incoh}
	F_i^{D\,{\rm inc}} &= \sum_{R=1} \, \big(B^R - B^{R\,{\rm coh}} \big) \otimes_\perp S_i^{R}
     + \sum_{R,R'\ne 1} \, B^R \otimes_\perp S_i^{R'}\otimes_\pm  U^{RR'} \,,
\end{align}
where the first term corresponds to color-singlet diffraction and the second to nonsinglet quasi-diffraction. 
Additionally, for the $\lambda_t = \sqrt{-t/s}\ll 1$ regime, we saw in \eq{refactS} that we can refactorize the soft function $S_i^R$, which can be written schematically as
\begin{align}\label{eq:fact-schematic2}
	S_i^{R'}&= \sum_\kappa H_i^\kappa \otimes_{-}S_{\rm c}^{\kappa\,R'}
\end{align}
In this section, we compare our results to the variety of theoretical approaches used to study diffractive DIS over the years, and discuss how EFT may help resolve some outstanding questions from these approaches.  We can group the main existing methods into three broad categories: 
hard-collinear factorization methods (\sec{HCfact}), the Ingelman-Schlein model (\sec{ISmodel}), and dipole methods (\sec{dipole}).
  
\subsection{Hard-collinear factorization}\label{sec:HCfact}

As discussed in \sec{smalllambdat},  \refscite{Berera:1995fj, Collins:1997sr} 
successfully derived a hard-collinear factorization formula for diffraction applicable in the $\lambda_t \ll 1$ region of diffractive phase space. 
The result is similar to the factorization of inclusive DIS, except with diffractive PDFs (dPDFs) $f_\kappa^D(\zeta,x/\beta,t,\mu)$ replacing the standard inclusive PDFs $f_\kappa(\zeta,\mu)$. 
Many modern diffractive DIS analyses are based on this framework.
For coherent diffraction this factorization can be written as
\begin{align} \label{eq:collins-factorization2}
 \frac{1}{x}
 F^{D\,{\rm coh}}_{2/L} =  \sum_\kappa \int_\beta^1 \frac{d\zeta}{\zeta} H^{\kappa}_{2/L}\Big(\frac{\beta}{\zeta}, Q,\mu\Big)  f_\kappa^{D\,{\rm coh}}\Big(\zeta,\frac{x}{\beta},t,\mu\Big)+ \cO\bigg( \frac{-t}{Q^2},\frac{ \LQCD^2}{Q^2} \bigg) 
 ,
\end{align}
where $H^{\kappa}_{2/L}$ are the same hard coefficients as in inclusive DIS.
Due to RG consistency of the factorization, the dPDFs evolve with the same DGLAP equations as PDFs.
The primary predictions of this factorization are that the $Q$-dependence of diffractive DIS is calculable, and that the dPDFs are universal for both $F_2^D$ and $F_L^D$. 
The operator definition of dPDFs is identical to that of PDFs, except the final state has an observed hadron state $Y(p')$. 

In the literature, dPDFs are regarded as universal distributions describing hadronic structure, and hence have been extracted in numerous global fits~\cite{H1:2006zyl, ZEUS:2009uxs,Salajegheh:2022vyv, Salajegheh:2023jgi}. Projections for precision dPDF extraction are often cited as a motivation for future colliders~\cite{Armesto:2019gxy,LHeC:2020van,AbdulKhalek:2021gbh}.
The hard-collinear diffractive and inclusive DIS factorizations take similar forms; thus, many less-inclusive processes are also expected to exhibit similar factorizations in the diffractive and inclusive cases, 
with the same hard coefficient functions and PDFs replaced by dPDFs.
Examples of such processes include diffractive central dijet production~\cite{ZEUS:2007yji, H1:2007oqt,H1:2011kou,H1:2014pjf,H1:2015okx} and diffractive heavy-quark production~\cite{H1:2006zxb,H1:2017bnb}.%
\footnote{
Recall that to ensure a process involves diffractive scattering requires $\lambda\ll 1$, and this remains true for these less-inclusive diffractive processes. 
We should also guarantee that the measurement imposes a rapidity gap between the forward and central sectors, in the style of the conditions discussed in \secs{power-counting}{gap-radiation}.
For hard-collinear factorization one must also take $\lambda_t\ll 1$.
}
These cases will be dominated by the gluon dPDF, while quark dPDFs only contribute as part of higher-order perturbative corrections.
In contrast, diffractive photoproduction has been analyzed in terms of a dPDF and photon PDF to show that factorization is broken~\cite{Guzey:2016awf,H1:2007jtx,H1:2010xdi}. 
It would be interesting to study this process directly with Glauber exchange in SCET to provide further insight into these observations.

The literature widely recognizes that~\eq{collins-factorization} accounts for the hard-collinear sector that appears in diffraction when $\lambda_t \ll1$, but does not factorize the forward scattering (Regge) dynamics that are 
associated with $\lambda \ll 1$~\cite{Collins:1997sr,Berera:1995fj,Collins:2001ga,Berger:1986iu,Arneodo:2005kd}.
Our results in \eqs{FDfactdiff}{FDfactquasi}
resolve this longstanding challenge by deriving the forward (Regge) factorization of diffraction using SCET techniques, as described in \sec{smalllambdat}. We also remark that this Regge factorization applies for diffraction even  when $\lambda_t\sim 1$, and for incoherent diffraction for non-singlet exchanges (quasi-diffraction); see \sec{lambda-fact}. 
Without this Regge factorization, the dPDFs can only be treated as non-perturbative objects to be fitted at an input scale $\mu_0^2 \sim 1~\rm{GeV}^2$ from data. 

\subsection{Ingelman-Schlein model}\label{sec:ISmodel}

The Ingelman-Schlein model is motivated by Regge trajectories and parton model dynamics\footnote{The Ingelman–Schlein model appears most often in the context of diffractive DIS. See~\cite{ZEUS:1995sar} for early analyses comparing different Regge-theory–based phenomenological models. 
Another model commonly used in the literature is the Good-Walker model \cite{Good:1960ba, Frankfurt:2022jns,ParticleDataGroup:2022pth,Mantysaari:2020axf}, which conceptualizes diffraction as the result of fluctuations in the internal structure of a nucleon. 
},
and conceptualizes diffraction as being mediated by the exchange of an object called a Pomeron ($\bP$)~\cite{Ingelman:1984ns, Frankfurt:2022jns}. 
Although originally the Ingelman-Schlein model was designed for $pp$ scattering, in modern times
it has been prominently used as input for the dPDFs $f_\kappa^D$ at a low scale
$\mu_0^2 \sim 1~\mathrm{GeV}^2$.%
\footnote{Historically, the first applications of the Ingelman-Schlein model to diffractive DIS used these models directly for the diffractive structure functions $F_{2,L}^D$, rather than for the dPDFs, see eg.~\cite{H1:1997bdi,ZEUS:1995sar,Golec-Biernat:1995ryr}. 
}
For values $\xi = x/\beta > 0.1$, global fits with Pomeron exchange alone failed to provide an accurate description of the data, and hence were extended to include a term representing color-singlet Reggeon ($\bR$) exchanges. This gives~\cite{H1:2006zyl,ZEUS:2009uxs,Goharipour:2018yov,Monfared:2011xf,Salajegheh:2022vyv}
\begin{align}
\label{eq:ISdPDF}
  f_\kappa^D(\zeta,\xi,t,\mu_0) = f_{\bP/p}(\xi,t)\,f_{\kappa/\bP}(\zeta,\mu_0) +f_{\bR/p}(\xi,t)\,f_{\kappa/\bR}(\zeta,\mu_0)\,,
\end{align}
where each contribution factorizes into a \textit{flux factor} $f_{\cQ/p}(\xi,t)$ describing the emission of a Pomeron/Reggeon from the proton, and a parton density $f_{i/\cQ}(\zeta,\mu_0)$ describing the distribution of a parton $\kappa$ inside the Pomeron/Reggeon.
Here $\xi = x/\beta$ and $\cal Q = \{\bP, \bR\}$. 
These flux factors are parameterized in terms of Pomeron/Reggeon trajectories, $\alpha_{\cal Q}(t) = \alpha_{\cal Q}(0) + \alpha'_{\cal Q} t$, a normalization factor $A_{\cal Q}$, and a slope parameter $B_{\cal Q}$:
\begin{align}\label{eq:PomeronFlux}
	&f_{\bP/p}(\xi,t)  = A_{\bP} \frac{e^{B_\bP\, t }}{\xi^{2\alpha_\bP(t) - 1}}
	&&	f_{\bR/p}(\xi,t)  = A_{\bR} \frac{e^{B_\bR\, t }}{\xi^{2\alpha_\bR(t) - 1}}
  \,,
\end{align}
where the eight parameters $\{  A_\cQ,\, B_\cQ,\, \alpha_\cQ(0),\, \alpha'_\cQ\}$ are fit to experimental data.

Let us compare and contrast the Ingelman-Schlein model in \eq{ISdPDF} to the Regge factorization 
in \eq{FDfactdifflambdat}.
First, \eq{ISdPDF} is based on a parton model picture with longitudinal momentum fractions and a simple product structure, whereas the full Regge factorization
in \eq{FDfactdifflambdat} involves soft and beam functions tied together by transverse momentum integrals and includes an infinite number of terms, encapsulating all possible Glauber exchanges that can mediate color-singlet exchange. 
In the full Regge factorization, both objects depend on $t$, unlike the $f_{\kappa/Q}(\zeta,\mu_0)$ objects in the Ingelman-Schlein model. The dependence on $\xi=x/\beta$ arises in the full factorization from the resummation of large logarithms, whose solutions induce terms that go beyond those of a simple Regge trajectory. Finally, for incoherent scattering the full factorization also involves color-nonsinglet contributions (quasi-diffraction) which are predicted by a related factorization formula.
These differences do not change predictions for the $Q$-dependence of diffractive DIS in the $\lambda_t \ll 1$ regime, but suggest modifications for input dPDFs that will impact predictions for other parameter dependences. 
Note that our conclusions about the Ingelman-Schlein model also apply if the dPDF is replaced by structure functions for the case $\lambda_t\sim 1$, since our Regge factorization also holds for this case; see \eqs{FDfactdiff}{FDfactquasi}. 

\subsection{Dipole methods} \label{sec:dipole} 

The dipole framework ~\cite{Mueller:1989st,Mueller:1993rr,Mueller:1994gb} is a common approach for studying small-$x$ dynamics, including DIS and saturation phenomena. 
It has also been applied to inclusive~\cite{Nikolaev:1990ja,Nikolaev:1991et,Mueller:1994jq, Golec-Biernat:1998zce,Kowalski:2008sa} and exclusive \cite{Hentschinski:2005er, Kowalski:2006hc, Armesto:2014sma,Altinoluk:2015dpi,Mantysaari:2016ykx, Hatta:2019ixj, Mantysaari:2019csc,Mantysaari:2020lhf, Mantysaari:2021ryb, Iancu:2022lcw} diffractive processes; for reviews, see~\refscite{Morreale:2021pnn,Frankfurt:2022jns, Penttala:2023udn,Gelis:2010nm}.
In this framework, the virtual photon fluctuates into a $q\bar q$ dipole that probes the gluon  structure of a fast-moving hadron in the high-energy limit $W^2\gg Q^2$. The scattering amplitude is decomposed into three components:
a light-cone wave function $\Psi_{\gamma^*\to n}$ describing the $\gamma^*$-to-$n$ transition (with $n=q\bar{q}$ a dipole at LO), 
a ``dipole-amplitude'' $\langle\hat{\mathcal{O}}_n\rangle$ describing the interaction between the Fock state $n$ and the target, 
as well as a light-cone wave function $\Psi_{n\to X}$ describing how the $n$-parton configuration projects onto the final state $X$. At LO, the amplitude takes the form
\begin{align}
\label{eq:dipolefact}
	-i \mathcal{M}_{\gamma^*\to q\bar{q}\to X}\!\sim\! \int \!d^2b_\perp \,d^2 r_\perp\, dz \, e^{-i \mathbf{b_\perp} \cdot \tau_\perp}\: \Psi_{\gamma^*\to q \bar{q}}(r_\perp, z)\: \langle \hat{O}_{q\bar{q}} \rangle(r_\perp, b_\perp) \: \Psi_{q \bar{q} \to X}(r_\perp, z).
\end{align} 
Often, one works in a frame where $\gamma^*$ has a large $+$-momentum $q^+$. 
The momentum fraction  $z$ is the ratio of the quark $+$-momentum over $q^+$. 
The dipole amplitude $\langle\hat{O}_{q\bar{q}}\rangle$  is 
a matrix element of 
Wilson lines of the classical gauge field evaluated at the transverse positions  $r_{i\perp}$ of the  particles $i=q, \qbar$. Equivalently, one may use as basis coordinates the \textit{dipole vector} $r_\perp = r_{q\perp} - r_{\bar{q}\perp}$ and the \textit{impact parameter} $b_\perp = ({r_{q\perp} + r_{\bar{q}\perp}})/{2}$, which is conjugate to $\tau_\perp$.
Physically, \eq{dipolefact}  
encodes that the fast-moving hadron is Lorentz-contracted so that its interaction with the dipole 
takes the form of a shockwave, occurring over a shorter timescale than the photon lifetime.  

Since its development in 2016, Glauber SCET has successfully reproduced and built upon several results in the dipole model. 	
For example,  \S 9.1 of ~\refcite{Rothstein:2016bsq} showed that Glauber collapse gives rise to a shockwave in position space.
\Refcite{Stewart:2023lwz} developed a picture of saturation within Glauber SCET. Drawing further connections between the two communities is of great interest.

With this motivation, let us compare the dipole picture in~\eq{dipolefact} with the Glauber SCET‐based factorization result of \sec{lambda-fact} for inclusive diffractive DIS.
Importantly, the dipole literature focuses on color-singlet diffraction, so we can only compare to this case.
The first distinction we notice between the dipole and SCET literature is that 
for the part of the process sensitive to the virtual photon,
the former carries out factorization at the amplitude level (via wavefunctions)
whereas the latter factorizes at the cross section level (via a soft function).
Nonetheless, we see many overlapping features.
In particular, one can Fourier transform our $\tau_{i\perp}$ and $\tau_{j\perp }'$ integrals in \eq{FDfactdiffcoh} to transverse position space, giving variables $r_{i\perp }$ and $r_{j\perp}^\prime$. 
One would then expect the rough correspondence to be 
\begin{align}
\label{eq:BSreltoDipole}
   B_{(N,N')}^{R_A^{NN'}\: \rm coh} \Big(\{r_{i\perp }, r_{j\perp}^\prime\},t,\mu,\frac{\nu}{Q/x}\Big)
  & \stackrel{?}{\sim}  \big|\langle \hat O_{n}\rangle (r_\perp,b_\perp,\ldots) \big|^2
    \,, \\
   S_{(N,N')}^{R_A^{NN'}}\big(Q/\beta,\{r_{i\perp }, r_{j\perp}^\prime\},Q,t \big)
  & \stackrel{?}{\sim} \sumint_X
\bigg| \int dz\: \Psi_{\gamma^*\to n}(r_\perp, \ldots, z)\: 
        \Psi_{n \to X}(r_\perp, \ldots, z) \bigg|^2
   \,.\nn
\end{align}
The ellipses on the RHS indicate further impact parameters that become relevant for higher Fock states $n$. To make this correspondence more precise, we would have to connect the dipole formula involving a sum over states $n$ to the Glauber factorization involving a sum over $N,N'$ and all possible color-singlet channels $R_A^{NN'}$. 

For our LO soft function calculation in \sec{soft-results}, we fix $n=q\bar q$ and see that the dipole variables $\{r_{q\perp},\,r_{\bar q\perp}\}$ in \eq{dipolefact} are Fourier-conjugate to the SCET momenta $\{\tau_{\sigma_q\perp},\,\tau_{\sigma_{\bar q}\perp}\}$ in \eq{NglauberAmp}.
In \eq{dipolefact}, only $\langle\hat{O}_{q\bar{q}}\rangle$ depends on $b_\perp$, so the part involving light-cone wavefunctions has no $t$ dependence. This differs from our result, where both the beam and soft functions depend on $t$.  The reason that our soft function depends on $t$ is because we cannot always use an eikonal approximation for the interaction Glauber exchanges.  (This $t$-dependence was also crucial for reproducing the hard-collinear factorization in \secs{smalllambdat}{soft-1g}.)
At NLO, the light-cone wavefunctions involve the tripole $q\bar{q}g$,
which was calculated in the large-$Q^2$ limit~\cite{Wusthoff:1997fz,Golec-Biernat:1999qor,Golec-Biernat:2001gyl,Kowalski:2008sa}, in the large-$M_X^2$ limit~\cite{Bartels:1999tn,Kopeliovich:1999am,Kovchegov:2001ni,Munier:2003zb,Golec-Biernat:2005prq}, and recently in exact kinematics~\cite{Beuf:2022kyp,Beuf:2024msh}. 
Our comparison here is somewhat schematic, and 
it would be interesting to work out in more detail how these LO and NLO results are related to perturbative computations of our soft function for $-t\gg\Lambda_{\rm QCD}$. 

It is also interesting to consider where nonperturbative physics appears in our results and the dipole formalism.  In our factorization, $B$ always involves some nonperturbative physics, but will factorize into perturbative and nonperturbative pieces for $-t\gg \LQCD^2$; 
$S$ is purely perturbative for 
$-t\gg \LQCD^2$ but involves a nonperturbative component when $-t\sim \LQCD^2$ 
(i.e., the soft-collinear function).  
However, in the dipole formalism 
the light-cone wavefunctions $\gamma^*\to n$ and $n\to X$ are always treated as perturbative objects\footnote{Depending on the observables, the light-cone wave function for $n\to X$ may be non-perturbative. For example, quarkonium production~\cite{Cheung:2024qvw} involves long-distance matrix elements of non-relativistic QCD.}, which is another potential difference from our soft functions.
Additionally, the dipole amplitude is considered to be purely non-perturbative, unlike $B$.
The dipole amplitude is universally used for different processes, and we anticipate that our $B$ will also appear universally in different processes.

Calculations of the dipole amplitude have been carried out
using various models~\cite{Golec-Biernat:1998zce,Golec-Biernat:1999qor,Kowalski:2003hm,Kowalski:2006hc,McLerran:1993ka,McLerran:1993ni,
Iancu:2000hn,
JalilianMarian:1996xn,
Jalilian-Marian:1997qno,
Watt:2007nr}.
For example, in the CGC  framework~\cite{McLerran:1993ka,McLerran:1993ni,Iancu:2000hn},
the dipole amplitude is given by the functional integral $\langle\hat{\mathcal{O}}_Y\rangle=\int \mathcal{D} \rho\, \mathcal{W}_Y[\rho]\, \mathcal{O}_Y[\rho]$, where $\mathcal{W}_Y$ is the probability density of the color density $\rho$, 
$\mathcal{O}_Y[\rho]$ contains Wilson lines formed by classical fields encoding interactions between $n$ and the target,
and $Y$ is the rapidity evolution scale. 
Here $\mathcal{O}_Y$ depends on the saturation scale~$Q_s$, which 
grows with decreasing $b_\perp$ and as $Q_s^2 \sim A^{1/3}$, where $A$ is the nucleon number. 
The $Y$-dependence of $\mathcal{W}_Y[\rho]$ is either governed by the linear BFKL equations~\cite{Balitsky:1978ic, Kuraev:1977fs} (for $r_\perp Q_s \ll 1$) or the non-linear BK/KL/BJIMWLK equations~\cite{Balitsky:1995ub,Kovchegov:1999yj,Kovchegov:1999ji,Jalilian-Marian:1997qno,Iancu:2001ad} (for $r_\perp Q_s \gg 1$).  
Just as in the CGC, our SCET soft and beam functions evolve by BFKL-style equations; see \sec{difflogs}. 
It will be interesting to compare CGC computations to understand the relationship between $\langle \mathcal{O}_n\rangle$  and our beam functions. 
Additionally, \refcite{Stewart:2023lwz} used SCET to demonstrate how 
$Q_s$ arises as an emergent scale in DIS, 
showing how the transitions to saturation and non-linear evolution are distinct from one another. 
An important future direction will be to combine our factorization framework with a full treatment of saturation and small-$x$ evolution.

\section{Conclusions}\label{sec:diffract-outlook}

In this paper, we use effective field theory (EFT) techniques to derive the first factorization formula for the forward (Regge) physics in electron-proton diffraction,
enabling a connection between diffractive cross sections and the fundamental underlying hadronic dynamics. Specifically, we set up a systematic power counting for diffraction involving multiple dimensionless parameters, and we derive an all-orders Regge factorization of diffraction using Glauber SCET. In the case of incoherent diffraction, our analysis reveals a new irreducible background from color-nonsinglet exchange, which we dub quasi-diffraction (see \fig{diffraction-chart}). While quasi-diffractive contributions to the cross section are Sudakov suppressed by radiation below the detection threshold penetrating the gap, they are nonetheless still numerically relevant.  Along the way, we identify the regimes where (quasi-)diffractive processes dominate over gapped hard scattering. 
One must have $xy\sim Q^2/s = \lambda^2 \ll 1$ for a process to be diffractive, but one may consider wide ranges for the other kinematic variables, like $t$. 
We also advocate for measuring diffractive structure functions $F_3^D$ and $F_4^D$ associated with asymmetry measurements, which we find contribute at leading power in the small-$x$ limit, though they are often neglected in the literature. Finally, we give some first phenomenological predictions and make connections with previous literature. 

We review some key highlights from the paper. In 
\sec{diffract-kinematics}, we present the full six-dimensional differential diffractive cross section with Lorentz-invariant variables. 
We call particular attention to the independent variable $\bar x$, which is essential for studying  $F_{3,4}^D$, and we provide guidance for extracting these structure functions from experimental data. 
We also provide the full set of kinematic bounds, which,  to our knowledge, have not appeared fully in the literature.  \Sec{diffract-eft} sets up our EFT tools in Glauber SCET, showing that our analysis provides a first phenomenological application of the $\SCETa$ version of this theory, which has both soft and ultrasoft modes. 

In \sec{lambda-fact}, we derive the long-sought Regge factorization of diffraction as well as of our newly-identified quasi-diffraction, working at leading power and to all orders in $\alpha_s$.  
Notably, the factorization includes contributions from both perturbative and nonperturbative regimes of momentum space, and provides a setup where they can be separated. 
For diffraction, the factorization formula involves a
convolution between beam functions $B$ that encode the forward hadronic state and soft functions $S$ that describe well-separated central hadronic radiation. The formula involves a sum over any number of $t$-channel Glauber exchanges on each side of the cut.  
For quasi-diffraction, the factorization involves not only $B$ and $S$, but also an ultrasoft‐collinear function $U$ capturing low‐energy emissions that may enter the rapidity gap. 
We give explicit field theory matrix element definitions of $B$, $S$, and $U$. 
Our proof that $U$ vanishes for diffractive color-singlet channels validates the physical intuition that singlet channels are less likely to radiate into the rapidity gap. 
We also predict that coherent and incoherent diffraction have the same $S$ function, but different $B$. 
In contrast, the choice of structure function $F_i^D$ only impacts $S$, but not $B$ and $U$. 
In~\sec{logarithms}, we calculate the Sudakov suppression from renormalization group equations for $U$, summing the associated leading double logarithms. We also identify other types of  (subleading) logarithms in both diffraction and quasi-diffraction, and outline the EFT program needed to extend their calculation to higher orders.

In \sec{smalllambdat}, we consider an additional expansion in $\sqrt{-t}/Q = \lambda_t \ll 1$.  This leads to a hard-collinear factorization of our soft function. 
The widely-known factorization of diffractive structure functions into a hard function and a diffractive PDF (dPDF)~\cite{Berera:1995fj, Collins:1997sr} arises from considering this $\lambda_t\ll 1$ limit \textit{without} carrying out the $\lambda\ll 1$ expansion. Importantly, the Regge factorization of these dPDFs is widely noted as missing in the literature~\cite{Collins:1997sr,Berera:1995fj,Collins:2001ga,Berger:1986iu,Arneodo:2005kd}. Our $\lambda,\lambda_t\ll 1$ results enable us to predict this Regge factorization of the dPDFs: it involves a convolution between a soft-collinear function $S_c$ and $B$, while separating out hard dynamics. The hard functions associated with $F_2^D$ and $F_L^D$ are known to be the same as those in inclusive DIS; the hard function of $F_3^D$ is new, and that of $F_4^D$ is higher order in the $\lambda_t\ll 1$ expansion. 
We also provide an analogous hard-collinear factorization formula for dPDFs in quasi-diffraction (where $U$ also appears).

In \sec{soft-results}, we compute the leading-order soft function $S$ 
in the perturbative regime $Q^2,\, m_X^2,\, -t \gg \LQCD^2$, obtaining the amplitude for any number of Glauber exchanges.  For $\lambda_t\ll 1$, $\lambda_\Lambda = \LQCD/\sqrt{-t}\ll 1$, and a single perturbative Glauber exchange dressed by arbitrary nonperturbative Glaubers, we explicitly confirm our predicted $\lambda_t \ll 1$ refactorization involving the known NLO DIS hard coefficients. 

We use our factorization results to make several important observations, as detailed in \sec{predictions}.  Our power counting and factorization prove that, in the small-$x$ limit, all four unpolarized structure functions $F_i^D$, the four corresponding polarized structure functions $F_{iP}^D$ ($i=2,L,3,4$), and two additional antisymmetric structure functions, $F_{4A}^D$ and $F_{4AP}^D$, are leading order.
For quasi-diffraction, we use our $\lambda_{t,\Lambda}\ll 1$ soft function calculations to fully predict the ratios $F_i^{D\,{\rm quasi}}/F_2^{D\,{\rm quasi}}$ at leading order, including their $\beta$ and $-t/Q^2$ dependence. 
This provides an avenue for suppressing quasi-diffractive backgrounds by using linear combinations of structure functions to reveal purely diffractive dynamics. 
We discuss prospects for extracting hadronic information about the proton transition occurring in coherent diffraction, and we point out interesting new experimental targets using kinematic regimes that are not usually considered in diffractive analyses.
Finally, we use our results to estimate the relative size of diffractive and quasi-diffractive contributions to incoherent cross sections.  In particular, for $\lambda_{\Lambda}\ll 1$, the former is $\alpha_s$ suppressed, while the latter is Sudakov suppressed, indicating that for realistic kinematics they are often both numerically relevant contributions. 

In \sec{comparisons}, we discuss implications of our results for commonly used diffractive frameworks, including those favored by global fits. For $\lambda_t\ll 1$, the hard-collinear factorization into dPDFs correctly describes the $Q$-dependence of diffractive structure functions $F_{2,L}^D$, and we provide new hard-collinear factorization predictions for $F_3^D$. Our results rule out the Ingelman-Schlein model as a rigorous QCD description of the Regge factorization in all kinematic regimes. We also provide a first look at the correspondence with the color-singlet exchange dipole formalism, and emphasize some interesting puzzles in the correspondence.  

\paragraph{Outlook.}
Our establishment of a rigorous framework for factorizing diffraction suggests several avenues for future work, including: improving perturbative precision for $ep$ diffraction, factorizing diffractive processes with different initial states ($eA$, $pp$, and $pA$), and factorizing processes with more exclusive final states or additional rapidity gaps. For $eA$ diffraction, it is important to incorporate contributions associated to saturation, which is possible using the same SCET-based framework~\cite{Stewart:2023lwz}. Being differential in the kinematics associated to $p'$ provides greater access to small-$x$ phenomena like saturation than one can achieve from small-$x$ inclusive DIS.  

Achieving a meaningful comparison between our theoretical results and experimental data will require improved perturbative accuracy for various functions. This includes computing the  color‐singlet soft function in the perturbative regime, incorporating BFKL‐type resummation%
\footnote{As shown in \sec{logarithms}, the rapidity evolution equations for our color-singlet $S$ and $B$ functions are exactly those of the amplitude-level Pomeron BFKL equation, providing a connection to the study of Regge amplitudes.
As our factorization holds to all orders in $\alpha_s$, it could provide input for extending BFKL-type equations to NNLL order.
Note that Regge behavior is a general target in QFT even beyond QCD~\cite{Caron-Huot:2017vep,Caron-Huot:2020ouj,Maldacena:2015waa,Costa:2012cb,Brower:2006ea}; for example, in conformal field theory, there is growing interest in formulating Regge trajectories using light-ray operators and asymptotic detectors~\cite{Homrich:2024nwc,Caron-Huot:2022eqs,Homrich:2022cfq,Henriksson:2023cnh,Kravchuk:2018htv,Caron-Huot:2017vep}.},
implementing higher‐order Sudakov resummation, performing DGLAP evolution in the $\lambda_t\ll1$ region, 
and considering the $\lambda_t\sim 1$ and $\lambda_t\gg 1$ regimes in more detail. 
We also plan to investigate the factorization for the (quasi-)diffractive beam functions from the $-t\gg \LQCD^2$ limit, which may enable connections to other hadronic distributions.
Together, these steps will produce more realistic predictions across a range of kinematic regimes.  It would also be very interesting to incorporate these improvements into global fits of diffractive DIS processes to study their impact.

Factorizing other diffractive processes is important not only in its own right, but also will enable us to determine the universality of the nonperturbative functions that appear.
The dPDFs defined for $ep$ diffraction appear to be universal in certain DIS processes (see~\sec{HCfact}).
There is also a common belief that they are not universal in hadron colliders,
which is primarily based on their inability to describe experimental data and an  extrapolation of examples of factorization violation for TMD dihadron production in hadron colliders~\cite{Collins:2007nk,Collins:2007jp,Mulders:2011zt}, where the final-state hadrons are replaced by jets.
Extending our EFT framework to hadron collisions and investigating the nature of the beam functions will clarify whether there is any universality
for diﬀractive processes across diﬀerent collider environments.
It is also possible to use our framework to study exclusive diffractive processes in the small-$x$ regime. 
For example, we can apply the framework to the well-studied process of deeply virtual Compton scattering (DVCS) at small-$x$~\cite{Donnachie:2000px, Favart:2003cu, Kowalski:2006hc, Watt:2007nr, Kumericki:2009uq, Rezaeian:2012ji, Hatta:2017cte, Boussarie:2023xun, Guo:2024wxy}
to provide a proof of
Regge factorization of GPDs at small $x$~\cite{Diehl:2003ny,Radyushkin:1997ki}. 
Likewise, one could also provide a full Regge analysis of diffractive SIDIS and the associated diffractive TMDs~\cite{Hatta:2022lzj,Hatta:2024vzv}.
Another topic for future study is the Regge factorization of event shapes in the forward regime, like small-angle energy correlators and their generalizations~\cite{Basham:1978zq,Basham:1978bw, Liu:2023aqb,Kang:2023oqj,Chen:2024bpj, Mantysaari:2025mht,Moult:2025nhu, Chen:2025rjc}. 
Beyond these examples, our Glauber SCET framework will enable the systematic study of a wide breadth of compelling physical questions for the first time.

\acknowledgments
We thank Philipp Aretz, Markus Diehl, Anjie Gao, Henry Klest, Chris Lee, Jani Penttala, Farid Salazar, Dave Soper, Raju Venugopalan, and Ivan Vitev for helpful discussions. 
The authors appreciate the hospitality of DESY, the Erwin Schrödinger Institute, the Mainz Institute for Theoretical Physics, Nikhef, the University of Amsterdam, and/or Washington University in St. Louis during long-term visits or sabbaticals while completing this manuscript.
This work was supported by the U.S. Department of Energy, Office of Science, Office of Nuclear Physics from DE-SC0011090, 
and the Quark-Gluon Tomography Topical Collaboration with award number DE-SC0023646.   
K.L. and I.S. were supported in part by the Delta ITP consortium, a program of the Netherlands Organisation for
Scientific Research (NWO) that is funded by the Dutch Ministry of Education, Culture and Science (OCW).
I.S. was also supported in part by the Simons Foundation through the Investigator grant 327942. S.T.S. was also supported by the U.S. National Science Foundation through a Graduate Research Fellowship under Grant No. 1745302; fellowships from the MIT Physics Department and School of Science; and the Hoffman Distinguished Postdoctoral Fellowship through the LDRD Program of Los Alamos National Laboratory under Project 20240786PRD1. Los Alamos National Laboratory is operated by Triad National Security, LLC, for the National Nuclear Security Administration of the U.S. Department of Energy (Contract Nr. 892332188CNA000001). 

\clearpage
\appendix
\section{Appendices}
\subsection{Kinematics and structure functions}
\subsubsection{Polarized structure functions}\label{app:structures}

Here, we define tensor structures for the full set of 18 structure functions addressed in \eqs{tensor-structures}{tensor-structures-polarized} of the main text. We use the definitions of $U^\mu$, $X^\mu$ in \eq{ux-vectors} and $\tilde{S}_T$ in \eq{tensor-structures-polarized}. We also define:
\begin{align}
	&\Ubar^\mu = \frac{1}{Q}\epsilon^{\mu S q X} \,,
	&& \Xbar^\mu = \frac{1}{Q}\epsilon^{\mu S q U} \,,
\end{align}
where we note that $\Ubar\cdot \Xbar = 0$, and $U\cdot \Ubar = X \cdot \Xbar \neq 0$. Note that $\bar X^\mu$ has only transverse spin.  In the rest frame, $\bar U^\mu$ has both longitudinal and transverse spin.
\\ 

\begin{center}
	\begin{tabular}{|c|c|c|}
		\hline
		\rowcolor{WhiteSmoke} & \makecell{\null\\\null \hspace{1 in}{\bf Symmetric}\hspace{1 in}\null \\\null}&  \null \hspace{0 in}{\bf Antisymmetric}\hspace{0 in}\null
		\\ \hline 
		\cellcolor{WhiteSmoke} {\bf Unpolarized}
		& \makecell[l]{
			\\
			$w_L^{\mu\nu} =  \frac{1}{2x} \Big( g^{\mu\nu} - \frac{q^\mu q^\nu}{q^2}\Big)$
			\vspace{0.05 in}\\
			$w_2^{\mu\nu} = \frac{1}{2x} \Big( U^\mu U^\nu - g^{\mu\nu} +\frac{q^\mu q^\nu}{q^2}\Big)$
			\vspace{0.05 in}\\
			$w_3^{\mu\nu} = \frac{1}{2x} \Big( 2 X^\mu X^\nu - U^\mu U^\nu + g^{\mu\nu} 
			- \frac{q^\mu q^\nu}{q^2}\Big)$
			\vspace{0.05 in}\\
			$w_4^{\mu\nu} = \frac{1}{2x}(U^\mu X^\nu + X^\mu U^\nu ) $
			\\ \null
		} & \makecell[l]{$w_{4A}^{\mu\nu} = \frac{i}{2x}(U^\mu X^\nu - X^\mu U^\nu)$}
		\\ \hline
		\cellcolor{WhiteSmoke} {\bf Polarized} 
		& \makecell[l]{	\\
			$w_{LP}^{\mu\nu} = \tilde{S}_T \, w_L^{\mu\nu}$
			\vspace{0.05 in}\\
			$w_{2P}^{\mu\nu} = \tilde{S}_T \, w_2^{\mu\nu}$
			\vspace{0.05 in}\\
			$w_{3P}^{\mu\nu} = \tilde{S}_T \, w_3^{\mu\nu}$
			\vspace{0.05 in}\\
			$w_{4P}^{\mu\nu} = \tilde{S}_T \, w_4^{\mu\nu}$
			\vspace{0.05 in}\\
			\null\\
			$w_{5}^{\mu\nu} = \frac{1}{2x}(U^\mu \Ubar^\nu +  \Ubar^\mu U^\nu)$
			\vspace{0.05 in}\\
			$w_{6}^{\mu\nu} = \frac{1}{2x}(U^\mu \Xbar^\nu + \Xbar^\mu U^\nu )$
			\vspace{0.05  in}\\
			$w_{7}^{\mu\nu} = \frac{1}{2x}(X^\mu \Ubar^\nu + \Ubar^\mu X^\nu )$
			\vspace{0.05  in}\\
			$w_{8}^{\mu\nu} = \frac{1}{2x}(X^\mu \Xbar^\nu + \Xbar^\mu X^\nu )$
			\vspace{0.05  in}\\
			\null}
		&
		\makecell[l]{	\\
			$w_{4AP} =  \tilde{S}_T \, w_{4A}^{\mu\nu}$\vspace{0.1 in}
			\\
			\null\\
			$w_{5A}^{\mu\nu} = \frac{i}{2x}(U^\mu \Ubar^\nu - \Ubar^\mu U^\nu )$\vspace{0.05  in}
			\\
			$w_{6A}^{\mu\nu} = \frac{i}{2x}(U^\mu \Xbar^\nu -  \Xbar^\mu U^\nu)$ 
			\vspace{0.05  in}\\
			$w_{7A}^{\mu\nu} = \frac{i}{2x}(X^\mu \Ubar^\nu - \Ubar^\mu X^\nu )$
			\vspace{0.05 in}\\
			$w_{8A}^{\mu\nu} = \frac{i}{2x}(X^\mu \Xbar^\nu - \Xbar^\mu X^\nu )$ 
				\\
			\null}
		\\ \hline
	\end{tabular}
\end{center}

\clearpage

\subsubsection{Helicity decomposition of structure functions}\label{app:photon}

Here, we explore a frame presented in \refcite{Arens:1996xw} which lends a particularly neat interpretation to the diffractive structure functions $F_i^D$. We build off the more complete construction of this basis widely used in SIDIS; see e.g. \refcite{Ebert:2021jhy} and references therein. 
(We start by using a $W_i$ notation for diffractive structure functions, following~\cite{Ebert:2021jhy}, and then relate this to the $F_i^D$'s used in the main text.)
In general, we can parametrize independent hadronic structure functions in terms of the hadron beam and exchanged photon polarizations as 
\begin{align}
	\label{eq:Wpolparam}
	W^{\mu \nu}\!&=\! \sum_{m, n} \rho_{n m} W_{\mu \nu}^{m n}=\Big[W_{\mu\nu}^U + S_L W_{\mu\nu}^L + S_T \cos(\phi_{J} - \phi_{S}) W_{\mu\nu}^{T_x} + S_T \sin (\phi - \phi_{S}) W_{\mu\nu}^{T_y}\Big]\nonumber\\ 
	&=\sum_{i=-1}^7w_{i,\gamma}^{\mu \nu} \Big[W_i^U + S_L W_i^L + S_T \cos(\phi_{J} - \phi_{S}) W_i^{T_x} + S_T \sin (\phi - \phi_{S}) W_i^{T_y}\Big]\,,
\end{align}
where a hadronic structure function $W_i^{X}$ carries hadron beam polarization $X$ and photon polarization structure $i$, which we define below. 
Note that for simplicity of notation, we drop the diffraction subscript $D$, $W_D^{\mu\nu} \to W^{\mu\nu}$, throughout this appendix. The first line of \eq{Wpolparam} decomposes the structure functions in terms of the hadron beam polarization $S$, with the spin-density matrix $\rho$ containing the polarization information.
Then $W_{\mu \nu}^{m n}$ is the hadronic tensor with helicity information
\begin{align} \label{eq:Wtensor}
	W_{\mu\nu}^{mn} &
	= \sum_X\!\!\!\!\!\!\!\! \int \delta^4(q + p - p' - p_X)
	\braket{P(p,m) | J^{\dagger}_{\mu} | J(p'),X} \braket{J(p'),X | J_\nu | P(p,n)}
	\,.\end{align}
Here, we define the various $W_i^X$ as 
\begin{align}
	&W_i^U=\frac{1}{2}(W_i^{++}+W_i^{--}), 
	&& W_i^L=\frac{1}{2}(W_i^{++}-W_i^{--}), 
	\nonumber\\
	&W_i^{T x}=\frac{1}{2}(W_i^{+-}+W_i^{-+}),
	&& W_i^{T y}=\frac{1}{2}(W_i^{+-}-W_i^{-+}),
\end{align}
where we decompose the hadronic tensor using the polarization of the exchanged photon. 
We parametrize the beam polarizations in convenient frames below. 
As there is a covariant way to write each frame, we can use different frames for 
each beam. However, the physical interpretation of the polarization holds only within 
each specific frame.

To parameterize the photon polarization, we use the Breit frame, where $q^\mu$ and $\vec{p}\,^{\prime}_\perp$ point along the $z$- and $x$-axes, respectively. We write the longitudinal and transverse polarization vectors of the exchanged photon in terms of the basis vectors $\tilde n_i^\mu$ of this frame as
\begin{align}
	\label{eq:polvec}
   \epsilon_0^\mu=\tilde{n}_t^\mu \,, 
   \qquad\qquad 
   \epsilon_{ \pm}^\mu=\frac{1}{\sqrt{2}}\left(\mp \tilde{n}_x^\mu+\mathrm{i} \tilde{n}_y^\mu\right)\,,
\end{align} 
where the basis vectors take the covariant form 
\begin{align}\label{eq:HBbases}
	&\tilde{n}_t^\mu=\frac{2 x p^\mu+q^\mu}{Q \sqrt{1+\gamma^2}}\,,
	&& \tilde{n}_x^\mu=\frac{1}{|\vec{p}\,'_\perp|}\Big({p'}^\mu-z \hm q^\mu+2 x z \frac{\hm-1}{\gamma^2} p^\mu\Big)\,,
	\nonumber \\
	&\tilde{n}_z^\mu=-\frac{q^\mu}{Q}\,,\
	&& \tilde{n}_y^\mu=\epsilon^{\mu \nu \rho \sigma} \tilde{n}_{t \nu} \tilde{n}_{x \rho}\tilde{n}_{z \sigma}\,,
\end{align}
with transverse coordinates 
\begin{align}
	&\vec{p}\,^{\prime 2}_\perp = -p'_\mu p'_\nu g_\perp^{\mu\nu} \,,
	&& g_\perp^{\mu\nu} = g^{\mu\nu} - \frac{q^\mu p^\nu + p^\mu q^\nu}{p\cdot q(1+\gamma^2)} + \frac{\gamma^2}{1+\gamma^2}\left(\frac{q^\mu q^\nu}{Q^2}-\frac{p^\mu p^\nu}{p^2}\right)
   \,.
\end{align}
The parameters encoding hadronic mass corrections are
\begin{align}
	&\hm= \frac{1}{\sqrt{1+\gamma^2}} \left[1-\frac{m_Y^2 + p_\perp^{\prime 2}}{z^2Q^2}\gamma^2\right]^{1/2}\,,
	&&\gamma =\frac{2xm_p}{Q}\,.
\end{align}
Note that in the limit of zero target mass $m_p^2 \to 0$ presented in the main body of the text, we have that $\gamma \to 0$ and $\hm \to 1$.
Now, we can write the nine independent photon polarization structures in \eq{Wpolparam} in terms of the polarization vectors in \eq{polvec}: 
\begin{align}\label{eq:trento-frame}
	w_{-1,\gamma}^{\mu \nu}&=\left(\epsilon_{-}^{* \mu} \epsilon_{-}^\nu+\epsilon_{+}^{* \mu} \epsilon_{+}^\nu\right)\,,
	& w_{0,\gamma}^{\mu \nu}&=\epsilon_0^{* \mu} \epsilon_0^\nu\,,
	\nonumber\\
	w_{1,\gamma}^{\mu \nu}&=-\frac{1}{\sqrt{2}}\left(\epsilon_0^{* \mu} \epsilon_{-}^\nu-\epsilon_0^{* \mu} \epsilon_{+}^\nu+\epsilon_{-}^{* \mu} \epsilon_0^\nu-\epsilon_{+}^{* \mu} \epsilon_0^\nu\right) \,,
	&w_{2,\gamma}^{\mu \nu}&=\frac{i}{\sqrt{2}}\left(\epsilon_0^{* \mu} \epsilon_{-}^\nu-\epsilon_0^{* \mu} \epsilon_{+}^\nu-\epsilon_{-}^{* \mu} \epsilon_0^\nu+\epsilon_{+}^{* \mu} \epsilon_0^\nu\right)\,,
	\nonumber\\
	w_{3,\gamma}^{\mu \nu}&=-\left(\epsilon_{+}^{* \mu} \epsilon_{-}^\nu+\epsilon_{-}^{* \mu} \epsilon_{+}^\nu\right)\,,
	& w_4^{\mu \nu}&=-\left(\epsilon_{+}^{* \mu} \epsilon_{+}^\nu-\epsilon_{-}^{* \mu} \epsilon_{-}^\nu\right)\,,
	\nonumber\\
	w_{5,\gamma}^{\mu \nu}&=\frac{-i}{\sqrt{2}}\left(\epsilon_{-}^{* \mu} \epsilon_0^\nu+\epsilon_{+}^{* \mu} \epsilon_0^\nu-\epsilon_0^{* \mu} \epsilon_{-}^\nu-\epsilon_0^{* \mu} \epsilon_{+}^\nu\right)\,,
	&w_{6,\gamma}^{\mu \nu}&=\frac{-1}{\sqrt{2}}\left(\epsilon_0^{* \mu} \epsilon_{-}^\nu+\epsilon_0^{* \mu} \epsilon_{+}^\nu+\epsilon_{-}^{* \mu} \epsilon_0^\nu+\epsilon_{+}^{* \mu} \epsilon_0^\nu\right)\,,
	\nonumber\\
	w_{7,\gamma}^{\mu \nu}&=i\left(\epsilon_{-}^\nu \epsilon_{+}^{* \mu}-\epsilon_{+}^\nu \epsilon_{-}^{* \mu}\right)\,.
\end{align}
We can project out the components of $W_i^X$ as
\begin{align}
	\label{eq:WiX}
	W_i^X = P_i^{\mu\nu} W_{\mu\nu}^X \equiv \left(\frac{w_{i,\gamma}^{\mu\nu}}{w_{i,\gamma}\cdot w^{i,\gamma}}\right) W_{\mu\nu}^{X}\,.
\end{align}
We remark that the covariant definition of the basis vectors given in \eq{HBbases} is identical to the orthogonal vectors in \eq{ux-vectors}, 
\begin{align}
\label{eq:basis}
	&U^\mu =\tilde{n}_t^\mu\,,
	&& X^\mu =\tilde{n}_x^\mu\,,
	&& |\vec{p}\,^{\prime}_\perp| = N_X\,.
\end{align}
This relates the orthogonal vector basis and the polarization of the exchanged photon in the Breit frame. Specifically, the unpolarized structure functions discussed in \sec{structure-functions} are related to the structure functions in \eq{WiX} as:
\begin{align}
	\label{eq:unprel}
	&F_L^D=2x\,W_0^U\,,
	&& F_2^D=2x\,(W_0^U + W_{-1}^U)\,,
	&&F_3^D=2x\,W_3^U\,,
	&&F_4^D=-2x\,W_1^U\,.
\end{align}
\Eq{unprel} agrees with \eq{FinW}, which writes the photon helicity state $i$ of $W_i^U$ explicitly.

The polarization of the hadron beam is most convenient to parameterize in a rest frame, where the density matrix is given simply in terms of a Bloch spin vector. Boosting the Breit frame along the $z$-axis to  the hadron rest frame, we parameterize the hadron spin vector as
\begin{align}\label{eq:s-components}
	S^{\mu} = (0, S_T \cos (\phi - \phi_S), S_T \sin (\phi - \phi_S), S_L)_{\rm BR}\,,
\end{align}
where the subscript BR indicates the rest frame reached by boosting the Breit frame, with $\vec{p}\,^{\prime}_\perp$ along the $x$-axis. Here $\phi$ and $\phi_S$ are the azimuthal angles formed by $\vec{p}\,'_\perp$ and  $\vec{S}_\perp$ with respect to the lepton plane. The density matrix $\rho$ in \eq{Wpolparam} is then parameterized in terms of a Bloch spin vector $\vec{S}$ in the BR rest frame as
\begin{align}
	\rho_{m n}=\frac{1}{2}(1+\vec{S} \cdot \vec{\sigma})_{m n}=\frac{1}{2}\left(\begin{array}{cc}
		1+S_L & S_T e^{-\mathrm{i}\left(\phi-\phi_S\right)} \\
		S_T e^{\mathrm{i}\left(\phi-\phi_S\right)} & 1-S_L
	\end{array}\right)_{\rm BR}\,.
\end{align}
Writing the BR basis vectors as
\begin{align}\label{eq:HBRbases}
	&n_{t,\rm{BR}}^\mu=\frac{2x\, p^\mu}{Q\gamma}\,,
	&& n_{x,\rm{BR}}^\mu=\tilde{n}_x^\mu\,,	
	&&n_{z,\rm{BR}}^\mu=\frac{2x\,p^\mu - \gamma^2 q^\mu}{Q\gamma\sqrt{1+\gamma^2}}\,,
	&& n_{y,\rm{BR}}^\mu=\tilde{n}_y^\mu\,,
\end{align}
we can express the spin components in the BR frame as
\begin{align}
	&S_T \cos(\phi - \phi_S) =  - S_\mu \cdot n_{x,\rm{BR}}^{\mu}\,,
	&&S_T=\sqrt{ -g_T^{\mu\nu} S_\mu S_\nu } = \sqrt{(n_{x,\rm{BR}}^{\mu}n_{x,\rm{BR}}^{\nu}+n_{y,\rm{BR}}^{\mu}n_{y,\rm{BR}}^{\nu})S_\mu S_\nu}\,,
	\nonumber\\
	&S_T \sin(\phi - \phi_S) =  - S_\mu \cdot n_{y,\rm{BR}}^{\mu} \,,
	&&S_L = - S_{\mu} \cdot n_{z,\rm{BR}}^{\mu}\,.
	\label{eq:hadspin}
\end{align}
Note that as the Breit frame and BR are related by a boost along the $z$-axis, the transverse spin components remain identical in both frames.

If we now define
\begin{align}
\label{eq:Li}
	L_i\left(k,k', p, p',s_l\right)&\equiv w_{i,\gamma}^{\mu \nu}\left(q, p,p'\right) L_{\mu \nu}\left(k,k',s_l\right)=\frac{2 Q^2}{1-\epsilon} g_i\left(\epsilon, s_l, \phi\right)\,,
\end{align}
then we can express the contraction of the leptonic and hadronic tensor as
\begin{align}\label{eq:fulldecomp}
	 &L_{\mu \nu}(k, {k'}) W^{\mu \nu}(q, p, {p'})
	\\
	& =\frac{2Q^2}{1-\epsilon}\sum_{i=-1}^7 g_i\left(\epsilon, \lambda_{\ell}, \phi\right) \Big[W_i^U + S_L W_i^L  + S_T \cos(\phi_{J} - \phi_{S}) W_i^{T_x}+S_T \sin (\phi - \phi_{S}) W_i^{T_y}\Big]
	\nonumber\\
	&=  \frac{2Q^2}{1-\epsilon} \bigg[ \wdl_{\!U U}+S_L\wdl_{\!U L}+\lambda_{\ell}\wdl_{\!L U} 
 	+\lambda_{\ell} S_L\wdl_{\!L L}
+S_T\wdl_{\!U T}+\lambda_{\ell} S_T\wdl_{\!L T} \bigg]\,,
	 \nonumber
\end{align}
where $g_i(\epsilon, \lambda_{\ell}, \phi)$ takes the form:
\begin{align}
	\label{eq:Lfactors}
	&g_{-1}=1\,, 
	& & g_0=\epsilon\,,
	&&g_1=\sqrt{2 \epsilon(1+\epsilon)} \cos \phi\,,
	\nonumber\\
	& g_2=\lambda_{\ell} \sqrt{2 \epsilon(1-\epsilon)} \sin \phi\,,
	&& g_3=\epsilon \cos \left(2 \phi\right)\,,
	& & g_4=\lambda_{\ell} \sqrt{1-\epsilon^2}\,,
	\nonumber \\ 
	&g_5=\sqrt{2 \epsilon(1+\epsilon)} \sin \phi\,,
	& & g_6=\lambda_{\ell} \sqrt{2 \epsilon(1-\epsilon)} \cos \phi\,,
	&&g_7=\epsilon \sin (2 \phi)\,.
\end{align}
and $\epsilon$ is the longitudinal-to-transverse photon flux factor
\begin{align} \label{eq:epsilon}
	\epsilon=\frac{1-y-\frac{1}{4} y^2 \gamma^2}{1-y+\frac{1}{4} y^2 \gamma^2+\frac{1}{2} y^2}\,.
\end{align}
The last line of \eq{fulldecomp} decomposes the cross section in terms of $(W\cdot L)_{XY}$, where $X$ and $Y$ are the leptonic and hadronic beam polarizations, respectively; i.e., one can read off the coefficients of $S_{L,T}$ and $\lambda_l$ of \eq{tensors}. We can then further decompose each piece into components with different exchanged photon helicities.  This leads to 18 independent structure functions with different characteristic azimuthal asymmetries:
\begin{align}
	\
	(W \cdot L)_{U U}= & W_{U U, T}+\epsilon W_{U U, L}+\sqrt{2 \epsilon(1+\epsilon)} \cos \left(\phi\right) W_{U U}^{\cos \left(\phi\right)} \nonumber\\
	& +\epsilon \cos \left(2 \phi\right) W_{U U}^{\cos \left(2 \phi\right)}\,, \nonumber\\
	(W \cdot L)_{U L}= & \sqrt{2 \epsilon(1+\epsilon)} \sin \left(\phi\right) W_{U L}^{\sin \left(\phi\right)}+\epsilon \sin \left(2 \phi\right) W_{U L}^{\sin \left(2 \phi\right)}\,, \nonumber\\
	(W \cdot L)_{L U}= & \sqrt{2 \epsilon(1-\epsilon)} \sin \left(\phi\right) W_{L U}^{\sin \left(\phi\right)}\,, \nonumber\\
	(W \cdot L)_{L L}= & \sqrt{1-\epsilon^2} W_{L L}+\sqrt{2 \epsilon(1-\epsilon)} \cos \left(\phi\right) W_{L L}^{\cos \left(\phi\right)}\,, \nonumber\\
	(W \cdot L)_{U T}= & \sin \left(\phi-\phi_S\right)\left[W_{U T, T}^{\sin \left(\phi-\phi_S\right)}+\epsilon W_{U T, L}^{\sin \left(\phi-\phi_S\right)}\right] \nonumber\\
	& +\epsilon\left[\sin \left(\phi+\phi_S\right) W_{U T}^{\sin \left(\phi+\phi_S\right)}+\sin \left(3 \phi-\phi_S\right) W_{U T}^{\sin \left(3 \phi-\phi_S\right)}\right] \nonumber\\
	& +\sqrt{2 \epsilon(1+\epsilon)}\left[\sin \left(\phi_S\right) W_{U T}^{\sin \left(\phi_S\right)}+\sin \left(2 \phi-\phi_S\right) W_{U T}^{\sin \left(2 \phi-\phi_S\right)}\right]\,,\nn\\
	(W \cdot L)_{L T}= & \sqrt{1-\epsilon^2} \cos \left(\phi-\phi_S\right) W_{L T}^{\cos \left(\phi-\phi_S\right)} \nn\\
& +\sqrt{2 \epsilon(1-\epsilon)}\left[\cos \left(\phi_S\right) W_{L T}^{\cos \left(\phi_S\right)}+\cos \left(2 \phi-\phi_S\right) W_{L T}^{\cos \left(2 \phi-\phi_S\right)}\right]\,.
	\label{eq:WLgeneral}
\end{align}
Eq.~(2.37) of \refcite{Ebert:2021jhy} gives the explicit relationship between linear combinations of 
$W^{mn}_{\lambda \lambda'}\equiv W^{mn}_{\mu\nu} \epsilon_{\lambda}^{*\mu}\epsilon_{\lambda'}^{*\nu}$ 
and the structure functions $W_i^X$ in \eq{WiX}. 
To reduce down to the 18 independent structure functions in \eq{WLgeneral}, we impose Hermiticity and parity\footnote{We parametrize the hadron beam polarization in the BR rest frame, but the photon polarization uses the Breit frame. These frames are related by a boost along the $z$-axis, so the transverse components (off-diagonal in the helicity basis) remain identical in both frames. Diagonal components do not play a role in the Hermiticity and parity relations, so there is no inconsistency in using separate frames in these relations.}
\begin{align}
	\text{Hermiticity:}& \quad W_{\lambda \lambda^{\prime}}^{m n}=\left(W_{\lambda^{\prime} \lambda}^{n m}\right)^*,\nonumber\\
	\text{Parity:}& \quad W_{\lambda \lambda^{\prime}}^{m n}=(-1)^{\lambda+\lambda^{\prime}+(n-m) / 2} W_{-\lambda,-\lambda^{\prime}}^{-m,-n}.
\end{align}
As an example of a relationship between the $W_i$'s and the diffractive structure functions $F_i^D$, we can write the unpolarized case in terms of $W_i^U$ as
\begin{align}
	\label{eq:Wuu}
	(W \cdot L)_{U U}= & W_{-1}^U+\epsilon W_0^U+\sqrt{2 \epsilon(1+\epsilon)} \cos \left(\phi\right) W_1^U+\epsilon \cos \left(2 \phi\right) W_3^U\,.
\end{align}
Using the relations in \eq{unprel}, in terms of the basis used for the structure functions in the body of the paper, we have
\begin{align}  \label{eq:LWcosphiUNPOL}
(W\cdot L)_{UU} 
&= \frac{\epsilon-1}{2x} F_L^D + \frac{1}{2x} F_2^D + \frac{ \epsilon \cos(2\phi)}{2x} F_3^D - \frac{\sqrt{2\epsilon(1+\epsilon)}}{2x} \cos\phi F_4^D\,,
\end{align}
which agrees with \eq{LWcosphi} in the massless limit up to the overall $2Q^2/(1-\epsilon)$ that was factored out in \eq{Li}. Note that using \eqs{unpolarized-coefficients}{LWcosphiUNPOL}{}, we express $\cos\phi$ in terms of the Lorentz invariants from \sec{lorentz} in \eq{cosine}. By computing the coefficients of polarized scattering involving $\sin\phi$ in the Lorentz-invariant form, we can derive the Lorentz-invariant form of $\sin\phi$ as
\begin{align}\label{eq:sine}
	\sin \phi =  \frac{1}{Q^2}\frac{y}{\sqrt{1-y}} \epsilon^{\mu\nu\rho\sigma} k_\mu U_\nu X_\rho q_\sigma = -\frac{y}{\sqrt{1-y}} 
  \frac{k\cdot \tilde{n}_y}{Q}
\,.
\end{align}
As a cross-check, \eqs{cosine}{sine} satisfy $\cos^2\phi + \sin^2\phi = 1$.

In \app{structures}, we decompose the hadronic tensor into 18 independent structure functions using orthogonalized four-vectors $\{U,X,\bar{U},\bar{X}\}$. As explained in \sec{structure-functions}, ten of these are leading power (four unpolarized and six polarized), which arise in combinations
\begin{align}
	\label{eq:comb}
	F_i^D + \tilde{S}_T' F_{iP}^D\,,
\end{align}
for $i = \{L,2,3,4,4A\}$ and with $\tilde{S}_T'$ from \eq{tensor-structures-polarized}. 
Just as above, we can derive the relationship between $\{F_{4A}^D, F_{iP}^D\}$ and
$W_i^{X}$. Four of these leading polarized pieces come from $(W \cdot L)_{U T}$ of \eq{WLgeneral} as
\begin{align}\label{eq:structure-relationship-pol}
	S_T\,(W \cdot L)_{U T}= 
  & S_T\biggl\{ \sin \left(\phi\!-\!\phi_S\right)\left[W_{-1}^{T_y}+\epsilon W_{0}^{T_y}\right]  
  + \frac{\epsilon}{2} 
 \left[\sin \left(3 \phi\!-\!\phi_S\right) W_3^{T_y}-\sin \left(\phi\!+\!\phi_S\right) W_3^{T_y}\right]
 \nonumber\\
	& +\frac12 \sqrt{2 \epsilon(1+\epsilon)}
  \left[\sin \left(2 \phi-\phi_S\right) W_{1}^{T_y} - \sin \left(\phi_S\right) W_1^{T_y}\right]\biggr\}+\cdots\nn\\
	= & S_T\sin \left(\phi\!-\!\phi_S\right)\Big\{W_{-1}^{T_y}+\epsilon W_{0}^{T_y} +\sqrt{2 \epsilon(1+\epsilon)}\cos(\phi) W_{1}^{T_y}+\epsilon\cos(2\phi) W_3^{T_y}\Big\}
  \nn\\
  & +\cdots\,,
\end{align}
where ellipses indicate subleading terms. Comparing the final line of \eq{structure-relationship-pol} with the unpolarized case in \eq{Wuu}, we indeed find a parallel structure, grouping each term into
\begin{align}
  (W \cdot L)_{U U}+S_T\,(W \cdot L)_{U T} 
  =&\left(W_{-1}^U + S_T\sin \left(\phi\!-\!\phi_S\right) W_{-1}^{T_y} \right) +  \epsilon \left(W_{0}^U +S_T\sin \left(\phi\!-\!\phi_S\right) W_{0}^{T_y}\right)\nn\\
  &+\sqrt{2 \epsilon(1+\epsilon)}\cos(\phi)  \left(W_{1}^U +S_T\sin \left(\phi\!-\!\phi_S\right) W_{1}^{T_y}\right)\nn\\ 
  &+\epsilon\cos(2\phi)  \left(W_{3}^U +S_T\sin \left(\phi\!-\!\phi_S\right) W_{3}^{T_y}\right)+ \ldots \,,
\end{align} 
reminiscent of \eq{comb}. Indeed, the overall $S_T\sin \left(\phi-\phi_S\right)$ 
in \eq{hadspin} is identical to $\tilde{S}'_T$ in \eq{tensor-structures-polarized},
\begin{align}
	S_T \sin(\phi - \phi_S) &=  - S_\mu \cdot n_{y,\rm{BR}}^{\mu} =  \frac{2x}{|\vec{p}\,'_\perp|Q^2 \sqrt{1+\gamma^2}} S_\mu p_\nu p\,'_\rho q_\sigma \epsilon^{\mu\nu\rho\sigma} = \tilde{S}'_T\,,
\end{align}
and leads to an extension of \eq{unprel} to the leading polarized structure functions
\begin{align}
	&F_{LP}^D=2x\,W_0^{T_y}\,,
	&& F_{2P}^D=2x\,(W_0^{T_y} + W_{-1}^{T_y})\,,
	&&F_{3P}^D=2x\,W_3^{T_y}\,,
	&& F_{4P}^D=-2x\,W_1^{T_y}\,.
\end{align}
Two additional leading structures arise from $(W \cdot L)_{L U}$ and $(W \cdot L)_{L T}$, which combine naturally as
\begin{align}
\lambda_\ell (W \cdot L)_{L U} +\lambda_\ell S_T\,(W \cdot L)_{L T} 
 = & \lambda_\ell\, \sqrt{2 \epsilon(1-\epsilon)} \sin \left(\phi\right)\left( W_{2}^{U}+S_T\sin \left(\phi\!-\!\phi_S\right) W_{2}^{T_y}\right) + \cdots\,, 
\end{align}
where ellipses again indicate subleading terms. Comparing the decomposition~\eq{WiX} with~\eq{hadronic-tensor-2}, we find that
\begin{align}
F_{4A}^D = 2x\,W_2^U\,,\qquad F_{4AP}^D = 2x\,W_2^{T_y}\,.
\end{align}
The orthogonal vector construction in the main text is useful for constructing independent structure functions in a frame-independent way and to identify the leading power structures. The helicity decomposition here offers a physical interpretation of these structure functions.

\subsubsection{Lab and $e$-$p$ CM frames}  \label{app:epCMframe}

Let us define the electron-proton center-of-mass and lab frames, which are related by a simple boost along the beam direction.  
We let $k$ and $p$ carry momenta in the $\nbar$- and $n$-directions defined in \sec{Breit} respectively, where neglecting the proton mass, we have
\begin{align}\label{eq:kp-vectors2}
	k^\mu &= \frac{1}{\gamma}\frac{\sqrt s}{2} \, \nbar^\mu 
     = \frac{1}{\gamma} {\sqrt s}\, (1,0,0_\perp)  \,,
	&p^\mu&= \gamma\frac{\sqrt s}{2} \, n^\mu 
     = \gamma {\sqrt s}\, (0,1,0_\perp)  \,.
\end{align} 
Here, $s = 4 E_p E_e$ and the boost factor $\gamma=\sqrt{E_p/E_e}$ allows us to simultaneously write formulas for both the $e^-$-$p$ CM frame ($\gamma=1$) and the lab frame ($\gamma=\sqrt{E_p/E_e} \neq 1$). 
Using \eq{kp-vectors}, it is straightforward to express the components of the remaining momenta in \fig{diffraction} in terms of the Lorentz invariants in \eqs{lorentz-invariants}{supplemental-lorentz-invariants}:
\begin{align}\label{eq:four-vectors2}
	q & =\sqrt{s}\Big(\frac{y}{\gamma},\,-\gamma x y,\, \sqrt{x y(1-y)} \,\Big)
	\nonumber\\
	p_X & =\sqrt{s}\Bigg(\frac{y(1-z)}{\gamma},\, \gamma(\xbar-x y),\, \sqrt{y (\bar{x}-xy)(1-z)-xy\frac{1-\beta}{\beta }-\frac{t}{s}}\,\Bigg)
	\nonumber\\
	\tau & =\sqrt{s}\Big(-\!\frac{yz}{\gamma},\, \gamma\bar{x}, \,\sqrt{-\bar{x} y z-t / s} \, \Big)
	\nonumber\\
	p' & =\sqrt{s}\Big(\frac{yz}{\gamma},\, \gamma(1-\bar{x}),\, \sqrt{-\bar{x} y z-t / s}\,\Big)\,.
\end{align}
Note that the final component of a lightcone coordinate is a magnitude $|\vec{p}_\perp|$, and does not represent the full perpendicular vector component of a momentum. Hence, the final components of $q$ and $\tau$ do not directly sum to the final component of $p_X$ because in general, $|q_\perp| + |\tau_\perp| \neq |p_{X\perp}|$.  Note that the bounds in \sec{lorentz} imply that the factors in square roots here are all positive, and that $0 < xy < \bar x < 1$. 

\paragraph{Measurement of $\xbar$ in the lab frame.}

In the lab frame, experimental measurements of $\xbar$ are achieved through measurements of $-t$ and the polar angle $\theta$ of the recoiling forward system $Y$ relative to the incoming proton direction. Writing $p'=(E', \vec p_\perp^{\,\prime}, p_z')$ and $p_z'=\kappa E'\cos\theta$, where $\kappa = \sqrt{1-m_Y^2/E^{\prime 2}}$, we find
\begin{align}
  1-\bar x = \frac{(m_Y^2-t)}{\gamma^2 s} \frac{(1+\kappa \cos\theta)}{(1-\kappa\cos\theta)}
   \,,
\end{align}
and $E'=\big[(1-\bar x) +(m_Y^2-t)/(\gamma^2 s)\big] E_p$. In the limits $\lambda\ll 1$ and $\lambda\lambda_t\ll 1$ relevant for diffraction, we can expand in $\bar x, m_Y^2/s, -t/s
\ll 1$, giving the formula quoted in \sec{experiments-xbar}:
\begin{align}
  1-\bar x = \frac{m_Y^2-t}{\gamma^2 s \tan^2(\theta/2)+m_Y^2} 
   \,.
\end{align}
Interestingly, in the lab frame, $\xbar = \tau^-/p^-$ carries the physical interpretation of a longitudinal momentum fraction for the struck parton(s) inside the proton.

\paragraph{Lorentz transformation between lab and Breit frames.}

Next, we verify the kinematic relationship between momenta in the Breit and lab frames, as discussed in \sec{gap-radiation}.
We begin by recalling the parametrization of the lab frame momentum of a massless particle $r^\mu=(r_t,r_x,r_y,r_z)$ in terms of its transverse momentum magnitude $r_T$, a transverse momentum angle $\phi$, and (pseudo)rapidity $\mathcal{Y}_{\rm Lab} = \frac{1}{2}\ln(r^-/r^+) = \eta_{\rm Lab}$:
\begin{align}\label{eq:rmulab}
r^\mu = r_T\bigl( \cosh \eta_{\rm Lab} \,, \cos \phi \,, \sin\phi\,, \sinh \eta_{\rm Lab} 
   \: \bigr)_{\rm Lab}\,.
\end{align}
In lightcone coordinates $(r^+,r^-, \vec r_\perp )$, we have
\begin{align}
\label{eq:rmulabLC}
r^\mu = r_T\bigl(e^{-\eta_{\rm Lab}} \,,\, e^{\eta_{\rm Lab}} \,,\, 
 \hat e_x \cos\phi + \hat e_y\sin\phi\: \bigr)_{\rm Lab}\,,
\end{align}
where $\hat e_x$ and $\hat e_y$ are unit vectors in the $x$ and $y$ directions. 

The incoming lepton has no $\perp$-momentum in the lab frame but does in the Breit frame, so transforming between these frames requires both a rotation and a boost.
Here we orient our Breit frame such that the lepton $\perp$-momenta in \eq{lepton-4-vectors} are along the negative $x$-axis. Transforming the lab frame momentum in \eq{rmulab} to the Breit frame gives
\begin{align}
\label{eq:rmuBreit}
	r^\mu = r_T\bigg(&\frac{e^{\eta_{\rm Lab}}}{2} \sqrt{\frac{y}{x\gamma^2}}
	   +  e^{-\eta_{\rm Lab}} \sqrt{\frac{x\gamma^2}{y}} \Big(1-\frac{y}{2}\Big) 
	   -\sqrt{1-y} \cos (\phi )
	  \,,\nonumber\\
	& \cos (\phi ) - e^{-\eta_{\rm Lab} } \sqrt{\frac{x\gamma^2(1-y)}{y}}    \,,\:
	  \sin (\phi ) \,, \nonumber\\
	&\frac{e^{\eta_{\rm Lab}}}{2} \sqrt{\frac{y}{x\gamma^2}}
	  -\frac{y}{2}  e^{-\eta_{\rm Lab} } \sqrt{ \frac{x\gamma^2}{y} }
	  -\sqrt{1-y} \cos (\phi )\bigg)_{\rm Breit}\,.
\end{align}
In lightcone coordinates $(r^+,r^-,\vec r_\perp )$, we have
\begin{align}
	r^\mu = r_T\Bigg(&  e^{-\eta_{\rm Lab}} \sqrt{\frac{x\gamma^2}{y}}\,,\:
	   e^{-\eta_{\rm Lab}} \sqrt{\frac{x\gamma^2}{y}} (1-y) 
	   + e^{\eta_{\rm Lab}} \sqrt{\frac{y}{x\gamma^2}}
	   -2\sqrt{1-y} \cos (\phi )
	  \,,\nonumber\\
	& \hat e_x \bigg[ \cos\phi - e^{-\eta_{\rm Lab}} \sqrt{\frac{x\gamma^2(1-y)}{y}} \: \bigg]
	  + \hat e_y \sin\phi
	  \Bigg)_{\rm Breit}\,.
\end{align}
Thus the relation between the Breit and lab frame rapidities is
\begin{align}
  e^{2\eta_{\rm Breit}} 
&= 
\frac{y}{x\gamma^2}\, \bigg(  e^{2\eta_{\rm Lab}}
  -2 \sqrt{\frac{x\gamma^2 (1-y)}{y}} \cos (\phi )\:  e^{\eta_{\rm Lab}}   
   + \frac{x\gamma^2(1 -y)}{y} \bigg)
  \,,
\end{align}
and the energy in the Breit frame is
\begin{align}\label{eq:breit-lab-relation}
E_{\rm Breit} = &r_T\bigg(\frac{e^{\eta_{\rm Lab}}}{2} \sqrt{\frac{y}{x\gamma^2}}
  + e^{-\eta_{\rm Lab}} \sqrt{\frac{x\gamma^2}{y}} \Big(1-\frac{y}{2}\Big) 
   -\sqrt{1-y} \cos (\phi )\bigg)\,.
\end{align}
These expressions are functions of $\sqrt{x\gamma^2} = \sqrt{x E_p/E_e}$. Note that at HERA and the EIC, $\gamma \gg 1$ due to $E_p \gg E_e$. In contrast, we always have $x\ll 1$ in (quasi-)diffraction. 

\subsection{Soft function amplitude for $N$-Glauber exchange}\label{app:Amultiglauber}

Here, we carry out detailed calculations of the soft function  in \sec{soft-setup}, at tree level in the soft sector ($q\qbar$) and with an arbitrary number $N$ of Glauber exchanges. For convenience we work in the Breit frame.
As we see in \fig{Nglauberamp}, Glauber gluons can attach to both the soft quark and soft anti-quark. We write $A_{(N_q,N_{\bar{q}})}$ to denote an amplitude involving $(N_q,N_{\qbar})$ Glauber exchanges attached to ($q$, $\qbar$), respectively. We begin by discussing the simple planar cases $A_{(N,0)}$ and $A_{(0,N)}$. 
Explicitly writing down a beam function can assist us in our calculations, and we can factor out the beam contributions from our final result. The beam function can be proven to not affect the value of the soft function due to the collapse rules (though it is a bit subtle how the correct combinatoric factors arise in different cases), so we choose it to be a simple fermion line. 
For the planar case with all Glaubers attaching to the quark line, we have
\begin{align}
	A_{(N,0)}=&(8\pi \alpha_s \iota^\epsilon \mu^{2 \epsilon})^N \int \prod_i \left(\frac{\ddslash\! \tau_i^+ \ddslash\!\tau_i^-}{2} |2\tau_i^z|^{-\eta}\right)\delta(\tau^+-\sum_{i=1}^N \tau_i^+)\delta(\tau^--\sum_{i=1}^N \tau_i^-)\nu^{(N-1)\eta}\nn\\
	&\hspace{-1.5cm}\times\bar{u}(k_1)
	T^{A_N}\frac{k^+_1
	}{(k_1\!-\!\tau_N)^2}\cdots 
	T^{A_{3}}\frac{k^+_1
	}{(k_1\!-\!\sum_{i=3}^N\tau_i)^2}
	T^{A_{2}} \frac{k^+_1
	}{(k_1\!-\!\sum_{i=2}^N\tau_i)^2}
	\frac{\slashed{n}}{2}T^{A_1}
	  \frac{\slashed{k}_1 \!-\! \slashed{\tau}_{\perp}}
	   {(k_1\!-\! \tau)^2}\gamma^\mu v(-k_2)
	  \nn\\
	&\hspace{-1.5cm}\textcolor{gray}{\times \bar{u}_n(p-\tau)
	T^{A_{N}}\frac{p^- 
	}{(p-\tau+\tau_N)^2}\cdots
	T^{A_{3}}\frac{p^- 
	}{(p-\tau_1-\tau_2)^2}
	T^{A_{2}}\frac{p^- 
	}{(p-\tau_1)^2}
	\frac{\slashed{\bar{n}}}{2}
	T^{A_{1}}u_n(p)}
	\nn\\
		&\hspace{-1.5cm}\textcolor{gray}{\times\int \prod_i \left(\ddslash^{d-2}\!\tau_{i,\perp} \frac{1}{\vec{\tau}_{i\perp}^2}\right) \delta^{d-2}(\tau_\perp-\sum_{i=1}^N \tau_{i\perp})}\,,
\end{align}
where as before $\iota = \exp(\gamma_E)/(4\pi)$.
Note that we drop terms that are subleading in $\lambda$ from the numerators and use the projection relations 
$(\slashed{n}/2) \slashed{k_i} (\slashed{n}/{2}) = k_i^+ (\slashed{n}/{2})$, 
$(\slashed{n}\slashed{\bar{n}}/4)(\slashed{\bar{n}}/2) = (\slashed{\bar{n}}/2)$,
and 
$(\slashed{\bar{n}}\slashed{n}/4)(\slashed{n}/2) = (\slashed{n}/2)$. The final two lines in gray are associated with the beam function and $\perp$-convolution.  Only the total Glauber momentum $\tau = \sum \tau_i$ is constrained, so we must take $N-1$ Glauber loop momentum integrals. For the $\pm$ integrals, we route the Glauber loop momenta as 
\begin{align}
	&\tau_1 = \tau + \tau_1'\,,
	&&\tau_2 = \tau_2' - \tau_1'\,,
	&&\tau_3 = \tau_3' - \tau_2'\,,
	&&\hdots\,,
	&&\tau_{N} = - \tau_{N-1}'\,,
\end{align}
which ensures that loop momenta do not flow through multiple propagators. This parametrization makes clear that two different types of propagators involve loop momenta; namely,
\begin{align}
	d_{ji} &\equiv(k_j+\tau_i')^2 = k_j^+\Big(\tau_i'^{ -} - \frac{\vec{\tau}_{i\perp}^{\prime 2} + 2\vec{k}_{j\perp}\cdot \vec{\tau}_{i\perp}^\prime}{k_j^+} + i0\Big) \equiv k_j^+\left(\tau_i'^{ -} -\Delta_{ji} + i0\right)\,,
	\\
	b_i &\equiv(p-\tau - \tau_i')^2 =p^-\Big(- \tau_i'^{+} -\tau^+ - \frac{(\vec{\tau}_\perp+\vec{\tau}'_{i\perp})^2}{p^-} + i0\Big) \equiv p^-\Big(- \tau_i'^{+} -\bar{\Delta}_i + i0\Big)\,.
	\nn
\end{align}
We can then carry out the momentum integral as (see Sec.9 of~\refcite{Rothstein:2016bsq}):
\begin{align}
&\prod_{i=1}^{N-1}\Big(\int \frac{\ddslash\!\tau_i'^+ \ddslash\!\tau_i'^-}{2} \frac{1}{d_{ji} b_i}\Big)|2\tau_1'^z|^{-\eta}|2\tau_1'^z-2\tau_2'^z|^{-\eta}\cdots |2\tau_{N-2}'^z-2\tau_{N-1}'^z|^{-\eta}|2\tau_{N-1}'^z|^{-\eta} \nonumber\\
&= \Big(\frac{-1}{2\,k_1^+ p^-}\Big)^{N-1}\frac{1}{N!}\,.
\end{align}
This simplifies our amplitude to
\begin{align}
\label{eq:AN0soft}
A_{(N,0)}=&\frac{(8\pi \alpha_s\iota^\epsilon \mu^{2 \epsilon})^N}{N!}(-1)^{N-1}\bar{u}(k_1)\frac{\slashed{n}}{2}\frac{\slashed{k}_1-\slashed{\tau}_\perp}{(k_1-\tau)^2}\gamma^\mu T^{A_N}\cdots T^{A_3}T^{A_2}T^{A_1}v(-k_2)\nn\\
&\textcolor{gray}{\times I_\perp^{(N-1)}(\tau_\perp)\times \bar{u}_n(p-\tau)\frac{\slashed{\bar{n}}}{2}T^{A_{N}}\cdots T^{A_3}T^{A_2}T^{A_{1}}u_n(p)\frac{1}{2^{N-1}}}\,,
\end{align}
where
\begin{align}
\label{eq:Iperp}
I_\perp^{(N-1)}(\tau_\perp)\equiv\int \frac{\ddslash^{d-2}\!  \tau'_{1 \perp} \cdots \ddslash^{d-2}\! \tau'_{N-1 \perp}}{\left(\vec{\tau}'_{1 \perp}+\vec{\tau}_{\perp}\right)^2\left(\vec{\tau}'_{2 \perp}-\vec{\tau}'_{1 \perp}\right)^2 \cdots\left(\vec{\tau}'_{N-1 \perp}-\vec{\tau}'_{N-2 \perp}\right)^2 \vec{\tau}_{{N-1} \perp}^2}\,.
\end{align}
The first line of \eq{AN0soft} without the factor $(8\pi \alpha_s\iota^\epsilon \mu^{2 \epsilon})^N/N!$ is the soft function amplitude associated with $(N,0)$ Glauber exchange.\footnote{The $(8\pi \alpha_s\iota^\epsilon \mu^{2 \epsilon})^N$ is part of the Glauber potential as discussed in~\eq{hatS}. On the other hand, the $1/N!$ is part of the transverse momentum convolution between the soft and beam functions in the factorization, as shown in~\eq{iint_perp}, and thus is not part of the soft function amplitude.} The factor $\bar{u}_n \cdots u_n$ encodes the simple fermion line beam function. Finally, $I_\perp^{(N-1)}(\tau_\perp)$ captures the transverse momentum exchange between the soft and beam sectors.

Likewise, $A_{(0,N)}$ takes the form
\begin{align}
	A_{(0,N)}=&(8\pi \alpha_s\iota^\epsilon \mu^{2 \epsilon})^N \int \prod_i \left(\frac{\ddslash\! \tau_i^+ \ddslash\!\tau_i^-}{2} |2\tau_i^z|^{-\eta}\right)\delta(\tau^+-\sum_{i=1}^N \tau_i^+)
    \delta(\tau^--\sum_{i=1}^N \tau_i^-)\nu^{(N-1)\eta}\nn\\
	&\hspace{-1.1cm}
	\times
	   \bar{u}(k_1)\gamma^\mu
	  \frac{\slashed{\tau}{}_{\!\perp}\!-\!\slashed{k}_2}{(\tau\!-\!k_2)^2}
	  \frac{\slashed{n}}{2}T^{A_1}
	  \frac{-k^+_2 T^{A_2} }{(\sum_{i=2}^N\tau_i\!-\!k_2)^2}
	 \cdots \frac{-k^+_2  T^{A_{N-1} }}{(\tau_{N-1}\!+\!\tau_N\!-\!k_2)^2}
	\frac{-k^+_2}{(\tau_N\!-\!k_2)^2}
	T^{A_N}v(-k_2)
	\nn\\
	&\hspace{-1.1cm}
	\textcolor{gray}{\times \bar{u}_n(p-\tau)
	T^{A_{N}}\frac{p^- 
	}{(p-\tau+\tau_N)^2}\cdots\frac{\slashed{\bar{n}}}{2}T^{A_{3}}\frac{p^- \frac{\slashed{n}}{2}}{(p-\tau_1-\tau_2)^2}\frac{\slashed{\bar{n}}}{2}T^{A_{2}}\frac{p^- \frac{\slashed{n}}{2}}{(p-\tau_1)^2}\frac{\slashed{\bar{n}}}{2}T^{A_{1}}u_n(p)}\nn\\
	&\hspace{-1.1cm}
	\textcolor{gray}{\times\int \prod_i \left(\ddslash^{d-2}\!\tau_{i,\perp} \frac{1}{\vec{\tau}_{i\perp}^2}\right) \delta^{d-2}(\tau_\perp-\sum_{i=1}^N \tau_{i\perp})}\,.
\end{align}
Carrying out Glauber collapse, we see that
\begin{align}
\label{eq:A0Nsoft}
	A_{(0,N)}=&\frac{(8\pi \alpha_s\iota^\epsilon \mu^{2 \epsilon})^N}{N!}\bar{u}(k_1)\gamma^\mu\frac{\slashed{\tau}_\perp-\slashed{k}_2}{(\tau-k_2)^2}\frac{\slashed{n}}{2}T^{A_1}T^{A_2}T^{A_3}\cdots T^{A_N}v(-k_2)\nn\\
	&\textcolor{gray}{\times I_\perp^{(N-1)}(\tau_\perp)\times \bar{u}_n(p-\tau)\frac{\slashed{\bar{n}}}{2}T^{A_{N}}\cdots T^{A_3}T^{A_2}T^{A_{1}}u_n(p)\frac{1}{2^{N-1}}}\,.
\end{align}

Finally, we consider the mixed case $N_q, N_{\bar{q}} \neq 0$. Before Glauber collapse, we separate Glaubers that connect to quark and anti-quark lines, giving
\begin{align}
\label{eq:mixed-amplitude-1}
	 & A_{(N_q,N_{\bar{q}})}
	 =(8\pi \alpha_s\iota^\epsilon \mu^{2 \epsilon})^N \!\!\!\!
	 \sum_{\sigma=\{\sigma_q,\sigma_{\bar{q}}\}}
	 \int \prod_{i=1}^{N_q} \biggl(\frac{\ddslash\! \tau_{i,\sigma_q}^+ \ddslash\!\tau_{i,\sigma_q}^-}{2} 
	 |2\tau_{i,\sigma_q}^z|^{-\eta}\biggr)
	 \prod_{j=1}^{N_{\bar{q}}} \biggl(\frac{\ddslash\! \tau_{j,\sigma_{\bar{q}}}^+ \ddslash\!\tau_{j,\sigma_{\bar{q}}}^-}{2} 
	 |2\tau_{j,\sigma_{\bar{q}}}^z|^{-\eta}\biggr)
	  \nonumber\\
		&\times
	   \nu^{(N-1)\eta} \: \frac{\ddslash\! \tau_{\sigma_q}^+ \ddslash\!\tau_{\sigma_q}^-}{2}  |2\tau_{\sigma_q}^z|^{-\eta}
	   \frac{\ddslash\! \tau_{\sigma_{\bar{q}}}^+\ddslash\!\tau_{\sigma_{\bar{q}}}^-}{2} 
	   |2\tau_{\sigma_{\bar{q}}}^z|^{-\eta}
	   \delta(\tau_{\sigma_{q}}^+-\sum_{k=1}^{N_q} \tau_{k,\sigma_{q}}^+)
	   \delta(\tau_{\sigma_q}^--\sum_{k=1}^{N_q} \tau_{k,\sigma_q}^-)
	   \nonumber\\
		&\times 
	  \delta(\tau_{\sigma_{\bar{q}}}^+-\sum_{k=1}^{N_{\bar{q}}} \tau_{k,\sigma_{\bar{q}}}^+)
	   \delta(\tau_{\sigma_{\bar{q}}}^--\sum_{k=1}^{N_{\bar{q}}} \tau_{k,\sigma_{\bar{q}}}^-)
	   \delta(\tau^+-\tau_{\sigma_q}^+-\tau_{\sigma_{\bar{q}}}^+)\delta(\tau^--\tau_{\sigma_q}^--\tau_{\sigma_{\bar{q}}}^-)
	   \nn\\
		&\times
	   \bar{u}(k_1)
	\frac{T^{A(\sigma_q^{N_q})} k^+_1}{(k_1-\tau_{N,\sigma_q})^2}\cdots 
	\frac{T^{A(\sigma_q^{3})} k^+_1}{(k_1-\sum_{i=3}^N\tau_{i,\sigma_q})^2}
	\frac{T^{A(\sigma_q^{2})}  k^+_1}{(k_1-\sum_{i=2}^N\tau_{i,\sigma_q})^2}
	\frac{\slashed{n}}{2}T^{A(\sigma_q^{1})}
	 \frac{\slashed{k}_1-\slashed{\tau}_{q,\perp}}{(k_1-\tau_{\sigma_q})^2}
	  \nonumber\\
		&\times
	 \gamma^\mu
	 \frac{\slashed{\tau}_{\bar{q},\perp}-\slashed{k}_2}{(\tau_{\sigma_{\bar{q}}}-k_2)^2}
	 \frac{\slashed{n}}{2}T^{A(\sigma_{\bar{q}}^{1})}
	 \frac{-k^+_2 T^{A(\sigma_{\bar{q}}^{2})} }{(\sum_{i=2}^N\tau_{i,\sigma_{\bar{q}}}-k_2)^2}
	\cdots \frac{-k^+_2 T^{A(\sigma_{\bar{q}}^{N-1})}}{(\tau_{N-1,\sigma_{\bar{q}}}+\tau_{N,\sigma_{\bar{q}}}-k_2)^2}
	   \frac{-k^+_2 T^{A(\sigma_{\bar{q}}^{N})}}{(\tau_{N,\sigma_{\bar{q}}}-k_2)^2}
	v(-k_2)\nn\\
		&\textcolor{gray}{\times \bar{u}_n(p-\tau)
	T^{A_{N}}\frac{p^-}{(p-\tau+\tau_N)^2}\cdots
	T^{A_{3}}\frac{p^-}{(p-\tau_1-\tau_2)^2}
	T^{A_{2}}\frac{p^- }{(p-\tau_1)^2}
	  \frac{\slashed{\bar{n}}}{2}T^{A_{1}}u_n(p)}
	\nn\\
	 &\textcolor{gray}{\times
	   \int \prod_i \left(\ddslash^{d-2}\!\tau_{i,\perp} \frac{1}{\vec{\tau}_{i\perp}^2}\right) \delta^{d-2}(\tau_\perp-\sum_{i=1}^N \tau_{i\perp})}\,.
\end{align}
As discussed in \sec{NglaubAmp}, we order Glauber momenta $\tau_i$ from left to right for convenience. Additionally, we describe which Glaubers attach to the $q$ and $\qbar$ in a given permutation $\sigma$ by defining sets $\sigma = \{\sigma_q,\sigma_{\bar q}\}$. We can relabel momenta $\tau_i \to \tau_{j,\sigma_X}$ by where they lie in the set $\sigma_X$. For example, if in a permutation $\sigma$ we take all odd-numbered momenta to attach to the quark $X=q$, then we have $\sigma_q = \{1,3, ..., 2m-1\}$ and thus 
\begin{align}
	\{\tau_1,\tau_3,\cdots, \tau_{2m-1}\} \equiv& \{\tau_{1,\sigma_q},\tau_{2,\sigma_q},\cdots, \tau_{m,\sigma_q}\}\,,
\end{align}
and similarly for $\sigma_{\bar{q}}$. We also relabel $T^{A}$ in this manner.

Performing the collapse in \eq{mixed-amplitude-1}, we have
\begin{align}\label{eq:mixed-amplitude-2}
	A_{(N_q,N_{\bar{q}})}=&\textcolor{gray}{I_\perp^{(N-1)}(\tau_\perp)\times}\frac{(8\pi \alpha_s\iota^\epsilon \mu^{2 \epsilon})^N}{N!}\frac{1}{k_1^+k_2^+}(-1)^{N_q}\sum_\sigma\bar{u}(k_1)\frac{\slashed{n}}{2}(\slashed{k}_1-\slashed{\tau}_{\sigma_q,\perp})\gamma^\mu(\slashed{\tau}_{\sigma_{\bar{q}}\perp}-\slashed{k}_2)\frac{\slashed{n}}{2}\nn\\
	&\times T^{A(\sigma_{q}^{N_q})}T^{A(\sigma_{ q}^{N_q-1})}\cdots T^{A(\sigma_{ q}^{2})}T^{A(\sigma_{ q}^{1})}T^{A({\sigma_{\bar{q}}^1})} \,  T^{A(\sigma_{\bar{q}}^2)}
  \cdots 
  T^{A(\sigma_{\bar{q}}^{\mbox{\tiny $N_{\bar q}-1$}} )}T^{A(\sigma_{\bar{q}}^{\mbox{\tiny $N_{\bar q}$}} )}v(-k_2)\nn\\
	&\times\frac{1}{\tau^- + \frac{1}{k_1^+}(\vec{\tau}_{\sigma_q,\perp}-2\vec{k}_{1,\perp})\cdot \vec{\tau}_{\sigma_q,\perp} +\frac{1}{k_2^+}(\vec{\tau}_{\sigma_{\bar{q}},\perp}-2\vec{k}_{2,\perp})\cdot \vec{\tau}_{\sigma_{\bar{q}},\perp}}\nn\\
	&\textcolor{gray}{\times \bar{u}_n(p-\tau)\frac{\slashed{\bar{n}}}{2}T^{A_{N}}\cdots T^{A_3}T^{A_2}T^{A_{1}}u_n(p)\frac{1}{2^{N-1}}}\,,
\end{align}
where $\tau_{\sigma_X} \equiv \sum_{i\in \sigma_X}\tau_{i}$. Note that $I_\perp^{(N-1)}(\tau_\perp)$ is a short-hand notation for the $\perp$-integration in \eq{Iperp}, which cannot be factored out completely due to the nontrivial $\perp$-dependence in the $(N_q,N_{\bar{q}})$-soft amplitude. For this reason, we write $I_{\perp}^{(N-1)}(\tau_\perp)$ before the soft amplitude in \eq{mixed-amplitude-2}.
Removing the overall beam function component and the $\perp$-integrals, the $N$-Glauber soft part of the amplitude is
\begin{align}\label{eq:mixed-amplitude-3}
	\mathcal{M}^{\mu, \{A\}}_{N \alpha\beta} = &\sum_{N_q=0}^N \frac{1}{k_1^+k_2^+}(-1)^{N_q}\sum_\sigma\bar{u}(k_1)\frac{\slashed{n}}{2}(\slashed{k}_1-\slashed{\tau}_{\sigma_q,\perp})\gamma^\mu(\slashed{\tau}_{\sigma_{\bar{q}}\perp}-\slashed{k}_2)\frac{\slashed{n}}{2}\nn\\
	&\times T^{A(\sigma_{q}^{N_q})}T^{A(\sigma_{ q}^{N_q-1})}\cdots T^{A(\sigma_{ q}^{2})}T^{A(\sigma_{ q}^{1})}T^{A({\sigma_{\bar{q}}^1})} \,  T^{A(\sigma_{\bar{q}}^2)}
  \cdots 
  T^{A(\sigma_{\bar{q}}^{\mbox{\tiny $N_{\bar q}-1$}} )}T^{A(\sigma_{\bar{q}}^{\mbox{\tiny $N_{\bar q}$}} )}v(-k_2)\nn\\
	&\times\frac{1}{\tau^- + \frac{1}{n\cdot k_1}(\vec{\tau}_{\sigma_q,\perp}-2\vec{k}_{1,\perp})\cdot \vec{\tau}_{\sigma_q,\perp} +\frac{1}{n\cdot k_2}(\vec{\tau}_{\sigma_{\bar{q}},\perp}-2\vec{k}_{2,\perp})\cdot \vec{\tau}_{\sigma_{\bar{q}},\perp}}\,.
\end{align}
Performing a color projection like \eq{Mcolorproj} and removing the factor $(8\pi \alpha_s\iota^\epsilon \mu^{2 \epsilon})^N/N!$  gives us the final $N$-Glauber 
soft amplitude expression in \eq{NglauberAmp}.

\bibliographystyle{JHEP}
\bibliography{main_diff}

\end{document}